\documentclass[onecolumn,showpacs,preprintnumbers,amsmath,amssymb,nofootinbib]{revtex4}
\usepackage{graphicx}
\usepackage{epsfig}
\usepackage{dcolumn}
\usepackage{bm}
\usepackage[english]{babel}
\usepackage{amsmath}
\usepackage{amsfonts}
\usepackage{amssymb}
\usepackage{delarray}
\numberwithin{equation}{section}

\begin{document}

\title{GRAVITON, GHOST AND INSTANTON CONDENSATION\\ ON HORIZON SCALE OF UNIVERSE.
 DARK ENERGY AS\\ À MACROSCOPIC EFFECT OF QUANTUM GRAVITY}

\author{Leonid Marochnik}
\email{lmarochn@umd.edu}
 \affiliation{Physics Department, University of Maryland, College Park, MD 20742, USA}
\author{Daniel Usikov}%
 \email{dusikov@gmail.com}
\affiliation{36477 Buckeye St., Newark, CA 94560, USA}
\author{Grigory Vereshkov}
\email{gveresh@gmail.com} \affiliation{Research Institute of
Physics, Southern Federal University, Rostov-on-Don 344090,
Russia}

\begin{abstract}

We show that cosmological acceleration (Dark Energy effect) falls
into to a special class of quantum phenomena that occur on a
macroscopic scale. Dark Energy is the third macroscopic quantum
effect, the first after the discovery of superfluidity and
superconductivity.   Dark Energy is a phenomenon of quantum
gravity on the scale of the Universe as a whole at any stage of
its evolution, including the contemporary Universe.  The effect is
a direct consequence of the zero rest mass of gravitons, conformal
non-invariance of the graviton field, and one-loop finiteness of
quantum gravity, i.e. it is a direct consequence of first
principles only. Therefore, no hypothetical fields or "new
physics"\ are needed to explain the Dark Energy effect.  This
macroscopic effect of one--loop quantum gravity takes place in the
empty isotropic non--stationary Universe as well as in such a
Universe filled by a non--relativistic matter or/and radiation.
The effect is due to graviton--ghost condensates arising from the
interference of quantum coherent states. Each of coherent states
is a state of gravitons and ghosts of a wavelength of the order of
the horizon scale and of different occupation numbers. The state
vector of the Universe is a coherent superposition of vectors of
different occupation numbers. One--loop approximation of quantum
gravity is believed to be applicable to the contemporary Universe
because of its remoteness from the Planck epoch.  To substantiate
the reliability of macroscopic quantum effects, the formalism of
one--loop quantum gravity is discussed in detail. The theory is
constructed as follows:  Faddeev -- Popov -- De Witt  gauged path
integral $\longrightarrow$ factorization of classical and quantum
variables, allowing the existence of a self--consistent system of
equations for gravitons, ghosts and macroscopic geometry
$\longrightarrow$ transition to the one--loop approximation,
taking into account that contributions of ghost fields to
observables cannot be eliminated in any way $\longrightarrow$
choice of ghost sector, satisfying the condition of one--loop
finiteness of the theory off the mass shell. The
Bogolyubov--Born--Green--Kirckwood--Yvon (BBGKY) chain for the
spectral function of gravitons renormalized by ghosts is used to
build a self--consistent theory of gravitons in the isotropic
Universe. It is the first use of this technique in quantum gravity
calculations. We found three exact solutions of the equations,
consisting of BBGKY chain and macroscopic Einstein's equations. It
was found that these solutions describe virtual graviton and ghost
condensates as well as condensates of instanton fluctuations. All
exact solutions, originally found by the BBGKY formalism, are
reproduced at the level of exact solutions for field operators and
state vectors. It was found that exact solutions correspond to
various condensates with different graviton--ghost compositions.
Each exact solution corresponds to a certain phase state of
graviton--ghost substratum. Quantum--gravity phase transitions are
introduced. In the formalism of the BBGKY chain, the generalized
self--consistent theory of gravitons is presented, taking into
account the contribution of non--relativistic matter in the
formation of a common self--consistent gravitational field. In the
framework of this theory, it is shown that the era of
non--relativistic matter dominance must be replaced by an era of
dominance of graviton--ghost condensate. Pre--asymptotic state of
Dark Energy is a condensate of virtual gravitons and ghosts with a
constant conformal wavelength. The asymptotic state predicted by
the theory is a self--polarized graviton--ghost condensate of
constant physical wavelength in the De Sitter space. The Dark
Energy phenomenon of such a nature is presented in the form of
$\Lambda$GCDM model that interpolates the exact solutions of
equations of one--loop quantum gravity. The proposed theory is
compared with existing observational data on Dark Energy extracted
from the Hubble diagram for supernovae SNIa. We show that
$\Lambda$GCDM model has advantages over $\Lambda$CDM model if
criteria for the statistical probability are in use. Result of
processing of observational data suggests that the graviton--ghost
condensate is an adequate variable component of Dark Energy. We
show that its role was significant during the era of large--scale
structure formation in the Universe.

\end{abstract}

\pacs{Dark energy 95.36.+x; Quantum gravity 04.60.-m}

\maketitle

\newpage

\tableofcontents

\newpage

\section{\ Introduction}\label{Intr}

Macroscopic quantum effects are quantum phenomena that occur on a
macroscopic scale. To date, there are two known macroscopic
quantum effects: superfluidity at the scale of liquid helium
vessel and superconductivity at the scale of superconducting
circuits of electrical current. These effects have been thoroughly
studied experimentally and theoretically understood. A key role in
these effects is played by coherent quantum condensates of
micro-objects with the De Broglie wavelength of the order of
macroscopic size of the system. The third macroscopic quantum
effect under discussion in this paper is condensation of gravitons
and ghosts in the self--consistent field of the expanding
Universe. Hypotheses on the possibility of graviton condensate
formation in the Universe proposed by Hu \cite{01} and
Antoniadis--Mazur--Mottola \cite{02} in a general form. A
description of these effects by an adequate mathematical formalism
is the problem at the present time.

We show that condensation of gravitons and ghosts is a consequence
of quantum interference of states forming the coherent
superposition. In this superposition, quantum fields have a
certain wavelength, and with different amplitudes of probability
they are in states corresponding to different occupation numbers
of gravitons and ghosts. Intrinsic properties of the theory
automatically lead to a characteristic wavelength of gravitons and
ghosts in the condensate. This wavelength  is always of the order
of a distance to the horizon of events\footnote{Everywhere in this
paper we discuss quantum states of gravitons and ghosts that are
self--consistent with the evolution of macroscopic geometry of the
Universe. In the mathematical formalism of the theory, the ghosts
play a role of a second physical subsystem, the average
contributions of which to the macroscopic Einstein equations
appear on an equal basis with the average contribution of
gravitons. At first glance, it may seem that the status of the
ghosts as the second subsystem is in a contradiction with the
well--known fact that the Faddeev--Popov ghosts are not physical
particles. However the paradox, is in the fact that we have no
contradiction with the standard concepts of quantum theory of
gauge fields but rather full agreement with these. The
Faddeev--Popov ghosts are indeed not physical particles in a
quantum--field sense, that is, they are not particles that are in
the asymptotic states whose energy and momentum are connected by a
definite relation. Such ghosts are nowhere to be found on the
pages of our work. We discuss only virtual gravitons and virtual
ghosts that exist in the area of interaction. As to virtual
ghosts, they cannot be eliminated in principle due to lack of
ghost--free gauges in quantum gravity. In the strict mathematical
sense, the non--stationary Universe as a whole is a region of
interaction, and, formally speaking, there are no real gravitons
and ghosts in it. Approximate representations of  real particles,
of course, can be introduced for  shortwave quantum modes.  In our
work,  quantum states of shortwave ghosts are not introduced and
consequently are not discussed. Furthermore,  macroscopic quantum
effects,  which are discussed in our work, are formed by the most
virtual modes of all virtual modes. These modes are selected by
the equality $\lambda H = 1$, where $\lambda$ is the wavelength,
$H$ is the Hubble function. The same equality also characterizes
the intensity of interaction of the virtual modes with the
classical gravitational field, i.e. it reflects the essentially
non--perturbative nature of the effects. An approximate transition
to real, weakly interacting particles, situated on the mass shell
is impossible for these modes, in principle (see also the footnote
$^2$ on p. 6).}.

In this fact, a common feature of macroscopic quantum effects is
manifested: such effects are always formed by quantum
micro--objects, whose wavelengths are of the order of macroscopic
values. With this in mind, we can say that macroscopic quantum
gravity effects exist across the Universe as a whole. The results
of this work suggest that the existence of the graviton--ghost
condensate is directly responsible for the Dark Energy effect,
i.e. for the observational data of the acceleration of the
expansion of the Universe \cite{03, 04}. Most significantly, a
graviton--ghost condensate formation is direct consequence of the
first principles of the theory of gravity and quantum field
theory, so that no hypothetical fields are needed to explain the
Dark Energy effect.

Quantum theory of gravity is a non--renormalized theory and for
this reason it is impossible to calculate effects with an
arbitrary accuracy in any order of the theory of perturbations.
The program combining gravity with other physical interactions
within the framework of supergravity or superstrings theory
assumes the ultimate formulation of the theory containing no
divergences. Today we do not have such a theory;  nevertheless, we
can hope to obtain physically meaningful results. Here are the
reasons for this assumption.

First, in all discussed options for the future theory, Einstein's
theory of gravity is contained as a low energy limit. Second, from
all physical fields, which will appear in a future theory
(according to present understanding), only the quantum component
of gravitational field (graviton field) has a unique combination
of zero rest mass and conformal non--invariance properties. Third,
physically meaningful effects of quantum gravity can be identified
and quantified in one--loop approximation. Fourth, as was been
shown by t'Hooft and Veltman \cite{3}, the one--loop quantum
gravity {\it with ghost sector} and without fields of matter is
finite. For the property of one--loop finiteness, proven in
\cite{3} on the graviton mass shell, we add the following key
assertion. {\it All one--loop calculations in quantum gravity must
be done in such a way that the feature of one--loop finiteness
(lack of divergences in terms of observables) must automatically
be implemented not only on the graviton mass shell but also
outside it.}

Let us emphasize the following important fact. Because of
conformal non--invariance and zero rest mass of gravitons, no
conditions exist in the Universe to place gravitons on the mass
shell precisely. Therefore, in the absence of one--loop
finiteness, divergences arise in observables. To eliminate them,
the Lagrangian of Einstein's theory must be modified, by amending
the definition of gravitons. In other words, in the absence of
one--loop finiteness, gravitons generate divergences, contrary to
their own definition. Such a situation does not make any sense, so
{\it the one--loop finiteness off the mass shell is a prerequisite
for internal consistency of the theory.}

These four conditions provide for the reliability of theory
predictions. Indeed, the existence of quantum component of the
gravitational field leaves no doubt. Zero rest mass of this
component means no threshold for quantum processes of graviton
vacuum polarization and graviton creation by external or
self--consistent macroscopic gravitational field. The combination
of zero rest mass and conformal non--invariance of graviton field
leads to the fact that these processes are occurring even in the
isotropic Universe at any stage of its evolution, including the
contemporary Universe. Vacuum polarization and particle creation
belong to effects  predicted by the theory already in one--loop
approximation. In this approximation, calculations of quantum
gravitational processes involving gravitons are not accompanied by
the emergence of divergences. Thus, the one--loop finiteness of
quantum gravity allows uniquely describe mathematically graviton
contributions to the macroscopic observables. Other one--loop
effects in the isotropic Universe are suppressed either because of
conformal invariance of non--gravitational quantum fields, or (in
the modern Universe) by non--zero rest mass particles, forming
effective thresholds for quantum gravitational processes in the
macroscopic self--consistent field.

Effects of vacuum polarization and particle creation in the sector
of matter fields of $J=0,\ 1/2,\ 1$ spin were well studied in the
1970's by many authors (see \cite{52} and references therein). The
theory of classic gravitational waves in the isotropic Universe
was formulated by Lifshitz in 1946 \cite{13}. Grishchuk \cite{40}
considered a number of cosmological applications of this theory
that are result of conformal non--invariance of gravitational
waves. Isaacson \cite{41} has formulated the task of
self--consistent description of gravitational waves and background
geometry. The model of Universe consisting of short gravitational
waves was described for the first time in \cite{20}. The
energy--momentum tensor of classic gravitational waves of super
long wavelengths was constructed in \cite{42, 43}. The canonic
quantization of gravitational field was done in \cite{11,
44,HuParker1977}. The local speed of creation of shortwave
gravitons was calculated in \cite{45}.  In all papers listed
above, the ghost sector of graviton theory was not taken into
account. One--loop quantum gravity in the form of the theory of
gravitons defined on the background spacetime was described by De
Witt \cite{7}. Calculating methods of this theory were discussed
by Hawking \cite{37}. For the first time, the approach based on
the fact that the ghost sector of graviton theory is determined by
the condition of one--loop finiteness off the mass shell is
presented in the present paper.

One--loop finiteness provides the simplicity and elegance of a
mathematical theory that allows, in turn, discovering a number of
new approximate and exact solutions of its equations. This paper
is focused on three exact solutions corresponding to three
different quantum states of graviton--ghost subsystem in the space
of the non--stationary isotropic Universe with self--consistent
geometry. The first of these solutions describes a coherent
condensate of virtual gravitons and ghosts; the second solution
describes a coherent condensate of instanton fluctuations. The
third solution describes the self--polarized condensate in the De
Sitter space. This solution allows interpretation in terms of
virtual particles as well as in terms of instanton fluctuations.
All three solutions are directly related to the physical nature of
Dark Energy.

The principal nature of macroscopic quantum gravity effects, the
need for strict proof of their inevitability and reliability
impose stringent requirements for constructing a mathematical
algorithm of the theory. In our view, existing versions of the
theory of gravitons in macroscopic spacetime with self--consistent
geometry do not meet these requirements. In this connection, note
the following fact. Because of conformal non--invariance, the
trace of the graviton energy--momentum tensor is not zero simply
by definition of graviton field. Naturally, the information on
macroscopic quantum effects (that are the subject of study in this
paper) is contained in this trace. After the first publication of
preliminary results of our research \cite{MUV} we feel that a
number of problems under discussion needs a much more detail
description. Due to some superficial similarities, effects of
graviton and ghost condensation in the De Sitter space are
perceived sometime as another method of description of conformal
anomalies, which are calculated by a traditional method of
regularization and renormalization. (Calculations of such
anomalies see in, e.g., \cite{48,49,50,51, TW, MV}.) We would like
to emphasize that such analogies have neither physical nor
mathematical basis. Conformal anomalies describe the effect of
reconstruction of the spectrum of zero oscillations of quantum
fields. Their contributions to the energy--momentum tensor are
independent of state vector of the quantum field. They are
parameterized by numerical coefficients of the order of unity that
are factors of quadratic form of the curvature. Conformal
anomalies that are microscopic quantum effects are able to
contribute to the macroscopic evolution of the Universe only if
its parameters are close to Planck ones. In contrast, the graviton
and ghost condensation is a macroscopic quantum effect in the
contemporary Universe. It is quantitatively described by
macroscopic parameters that govern the structure of a state
vector. These parameters are averaged numbers of quanta in
coherent superpositions. Note also that in all papers known to us,
calculations of anomalies were conducted in the framework of
models with no one--loop finiteness off the graviton and ghost
mass shell. It is shown in Appendices \ref{ren}, \ref{nonren} that
such models are internally inconsistent. The rigorous theory of
one--loop quantum gravity presented here shows that the effect of
conformal anomalies on gravitons must be zero (see Appendix
\ref{an-d}). In the empty Universe where the only gravitons exist,
the De Sitter space can be formed by only a graviton--ghost
condensate.

Sections \ref{qg} and \ref{scgt} are devoted to the derivation of
the equations of the theory with a discussion of all the
mathematical details. In Section \ref{qg}, we start with exact
quantum theory of gravity, presented in terms of path integral of
Faddeev--Popov \cite{5} and De Witt \cite{4}. Key ideas of this
Section are the following. (i) The necessity  to gauge the full
metric (before its separation into the background and
fluctuations) and the inevitability of appearance of a ghost
sector in the exact path integral and operator Einstein's
equations (Sections \ref{pi} and \ref{OEE}); (ii) The principal
necessity to use normal coordinates (exponential parameterization)
in a mathematically rigorous procedure for the separation of
classical and quantum variables is discussed in Sections \ref{fac}
and \ref{var}; (iii) The derivation of differential identities,
providing the consistency of classical and quantum equations
performed jointly in any order of the theory of perturbations is
given Section \ref{equiv}. Rigorously derived equations of gauged
one--loop quantum gravity are presented in Section \ref{1loop}.

The status of properties of ghost sector generated by gauge is
crucial to properly assess the structure of the theory and its
physical content. Let us immediately emphasize that the standard
presentation on the ghost status in the theory of  $S$--matrix can
not be exported to the theory of gravitons in the macroscopic
spacetime with self--consistent geometry. Two internal
mathematical properties of the quantum theory of gravity make such
export fundamentally impossible. First, there are no gauges that
completely eliminate the diffeomorphism group degeneracy in the
theory of gravity. This means that among the objects of the
quantum theory of fields inevitably arise ghosts interacting with
macroscopic gravity. Secondly, gravitons and ghosts cannot be in
principle situated precisely on the mass shell because of their
conformal non--invariance and zero rest mass. This is because
there are no asymptotic states, in which interaction of quantum
fields with macroscopic gravity could be neglected. Restructuring
of vacuum graviton and ghost modes with a wavelength of the order
of the distance to the horizon of events takes place at all stages
of cosmological evolution, including the contemporary Universe.
Ghost trivial vacuum, understood as the quantum state with zero
occupation numbers for all modes, simply is absent from physically
realizable states. Therefore, direct participation of ghosts in
the formation of macroscopic observables is
inevitable\footnote{Once again, we emphasize that the equal
participation of virtual gravitons and ghosts in the formation of
macroscopic observables in the non--stationary Universe does not
contradict the generally accepted concepts of the quantum theory
of gauge fields.  On the contrary it follows directly from the
mathematical structure of this theory. In order to clear up this
issue once and for all, recall some details of the theory of $S-$
matrix. In constructing this theory, all space--time is divided
into regions of asymptotic states and the region of effective
interaction. Note that this decomposition is carried out by means
of, generally speaking, an artificial procedure of  turning on and
off the interaction adiabatically. (For obvious reasons, the
problem of self--consistent description of gravitons and ghosts in
the non-stationary Universe with $\lambda H = 1$ by means of an
analogue of such procedure cannot be considered a priori.) Then,
after splitting the space--time into two regions, it is assumed
that the asymptotic states are ghost--free. In the most elegant
way, this selection rule is implemented in the BRST formalism,
which shows that the BRST invariant states turn out to be
gauge--invariant automatically. The virtual ghosts, however,
remain in the area of interaction, and this points to the fact
that virtual gravitons and ghosts are  parts of the Feynman
diagrams on an equal footing. In the self--consistent theory of
gravitons in the non--stationary Universe,  virtual ghosts of
equal weight as the gravitons, appear at the same place where they
appear in the theory of $ S-$matrix, i.e. at the same place as
they were introduced by Feynman, i.e. in the region of
interaction. Of course, the fact that in the real non--stationary
Universe, both the observer and virtual particles with $\lambda H
= 1$ are in the area of interaction, is  highly nontrivial. It is
quite possible that this property of the real world is manifested
in the effect of dark energy. An active and irremovable
participation of virtual ghosts in the formation of macroscopic
properties of the real Universe poses the question of their
physical nature. Today, we can only say with certainty that the
mathematical inevitability of ghosts provides the one--loop
finiteness off the mass shell, i.e. the mathematical consistency
of one-loop quantum gravity without fields of matter. Some
hypothetical ideas about the nature of the ghosts are briefly
discussed in the final Section \ref{con}.}.

Section \ref{scgt} is devoted to general discussion of equations
of the theory of gravitons in the isotropic Universe. It focuses
on three issues: (i) The gauge invariant procedure for eliminating
gauge non--invariant modes by conditions imposed on the state
vector (Section \ref{3vs}); (ii) Construction of the state vector
of a general form as a product of normalized superpositions
(Section \ref{stvec}); (iii) The allocation of class of legitimate
gauges that are invariant with respect to transformations of the
symmetry group of the background spacetime  while providing the
one--loop finiteness of macroscopic observables (Sections
\ref{fin} and \ref{fininv}). The main conclusion is that {\it the
quantum ghost fields are inevitable and unavoidable components of
the quantum gravitational field.}  As noted above, one--loop
finiteness is seen by us as a universal property of quantum
gravity, which extends off the mass shell. The requirement of
compensation of divergences in terms of macroscopic observables,
resulting from one--loop finiteness, uniquely captures the dynamic
properties of quantum ghost fields in the isotropic Universe.

The treatment set out in Sections  \ref{qg} and \ref{scgt}, in
essence, is a sequence of transformations of equations defined by
original gauged path integral. Basically, these transformations
are of mathematical identity nature. There are only three elements
of the theory, missing in the original integral:

(i) {\it The hypothesis of the existence of classical spacetime
with the deterministic but self--consistent geometry;}

(ii) {\it A transition to the one--loop approximation in the
self--consistent system of classical and quantum equations;}

(iii) {\it A class of gauges that automatically provides the
one--loop finiteness of self--consistent theory of gravitons in
the isotropic Universe.}

Conceptually discussing of the first two additional elements was
not necessary. Their introduction to the formalism is to the
factorization of measure of path integral and expansion of
equations of the theory in a series of powers of the graviton
field. The existence of appropriate correct mathematical
procedures is not in doubt. Agreement on choosing of a gauge is
also mathematically consistent. Moreover, a gauge is necessary for
the strict definition of path integral as a mathematical object.

The existence of class of gauges, automatically providing
one--loop finiteness off the mass shell is itself a nontrivial
property of the theory. The assertion that only such gauges can be
used in the self--consistent theory of gravitons in the isotropic
Universe is actually a condition for the internal consistency of
the theory. In Appendix \ref{nonren}, we present the formal proof
that an alternative formulation of the theory (with no one--loop
finiteness) does not exist.

From the requirement of the one--loop finiteness it follows that
the quantum component of the gravitational field is of a
heterogeneous graviton--ghost structure. In further Sections of
work, it appears that this new element of the theory prejudge its
physical content.

Sections \ref{swg} and \ref{lgw} contain approximate solutions to
obtain quantum ensembles of short and long gravitational waves. In
Section \ref{sce} it is shown that approximate solutions obtained
can be used to construct scenarios for the evolution of the early
Universe. In one such scenario, the Universe is filled with
ultra--relativistic gas of short--wave gravitons and with a
condensate of super--long wavelengths, which is dominated by
ghosts. The evolution of this Universe is oscillating in nature.

At the heart of cosmological applications of one--loop quantum
gravity is the Bogolyubov--Born--Green--Kirckwood--Yvon (BBGKY)
chain (or hierarchy) for the spectral function of gravitons,
renormalized by ghosts. We present the first use of this technique
in quantum gravity calculations. Each equation of the BBGKY chain
connects the expressions for neighboring moments of the spectral
function. In Section \ref{chain} the BBGKY chain is derived by
identical mathematical procedures from graviton and ghost operator
equations. Among these procedures is averaging of bilinear forms
of field operators over the state vector of the general form,
whose mathematical structure is given in Section \ref{stvec}. The
need to work with state vectors of the general form is dictated by
the instability of the trivial graviton--ghost vacuum (Section
\ref{vac}). Evaluation of mathematical correctness of procedures
for BBGKY structure is entirely a question of the existence of
moments of the spectral function as mathematical objects. A
positive answer to this question is guaranteed by one--loop
finiteness (Section \ref{fin}). The set of moments of the spectral
function contains information on the dynamics of operators as well
as on the properties of the quantum state over which the averaging
is done. The set of solutions of BBGKY chain contains all possible
self--consistent solutions of operator equation, averaged over all
possible quantum ensembles.

{\it A nontrivial fact is that in the one--loop quantum gravity
BBGKY chain can formally be introduced at an axiomatic level.}
Theory of gravitons provided by BBGKY chain, conceptually and
mathematically corresponds to the axiomatic quantum field theory
in the Wightman formulation (see Chapter 8 in the monograph
\cite{6}). Here, as in Wightman, the full information on the
quantum field is contained in an infinite sequence of averaged
correlation functions. Definitions of these functions clearly
relate to the symmetry properties of manifold on one  this field
is  defined. {\it Once the BBGKY chain is set up, the existence of
finite solutions for the observables is provided by inherent
mathematical properties of equations of the chain.} This means
that the phenomenology of BBGKY chain is more general than field
operators, state vectors and graviton--ghost compensation of
divergences that were used in its derivation.

Exact solutions of the equations, consisting of BBGKY chain and
macroscopic Einstein's equations are obtained in Sections
\ref{ggh_cond} and \ref{Sitt}. Two solutions given in
\ref{ggh_cond}, describe heterogeneous graviton--ghost
condensates, consisting of three subsystems. Two of these are
condensates of spatially homogeneous modes with the equations of
state $p=-\varepsilon/3$ and $p=\varepsilon$. The third subsystem
is a condensate of quasi--resonant modes with a constant conformal
wavelength corresponding to the variable physical wavelength of
the order of the distance to the horizon of events. The equations
of state of condensates of quasi--resonant modes differ from
$p\sim-\varepsilon/3$  by logarithmic terms, through which the
first solution is $p\gtrsim -\varepsilon/3$, while the second is
$p\lesssim-\varepsilon/3$. Furthermore, the solutions differ by
the sign of the energy density of condensates of spatially
homogenous modes. The third solution describes a homogeneous
condensate of quasi--resonant modes with a constant physical
wavelength. The equation of state of this condensate is
$p=-\varepsilon$  and its self--consistent geometry is the De
Sitter space. The three solutions are interpreted as three
different phase states of graviton--ghost system. The problem of
quantum--gravity phase transitions is discussed in
Section\ref{pt}.

Solutions obtained in Section \ref{Bog} in terms of moments of the
spectral function, are reproduced in Sections \ref{exact} and
\ref{inst} at the level of dynamics of operators and state
vectors. A microscopic theory provides details to clarify the
structure of graviton--ghost condensates and clearly demonstrates
the effects of quantum interference of coherent states. In Section
\ref{K} it is shown that the condensate of quasi--resonant modes
with the equation of state  $p\gtrsim -\varepsilon/3$ consists of
virtual gravitons and ghosts. In Section \ref{S} a similar
interpretation is proposed for the condensate in the De Sitter
space, but it became necessary to extend  the mathematical
definition of the moments of the spectral function.

New properties of the theory, whose existence was not anticipated
in advance, are studied in Section \ref{inst}. In Section
\ref{git} we find that the self--consistent theory of gravitons
and ghosts is invariant with respect to the Wick turn. In this
Section we also construct the formalism of quantum theory in the
imaginary time and discuss the physical interpretation of this
theory. The subjects of the study are correlated fluctuations
arising in the process of tunnelling between degenerate states of
graviton--ghost systems, divided by classically impenetrable
barriers. The level of these fluctuations is evaluated by
instanton solutions (as in Quantum Chromodynamics).  In Section
\ref{ins1} it is shown that the condensate of quasi--resonant
modes with the equation of state $p\lesssim-\varepsilon/3$  is of
purely instanton nature. In Section \ref{ins2} the instanton
condensate theory is formulated for the De Sitter space.

Potential use of the results obtained to construct scenarios of
cosmological evolution was briefly discussed in Sections
\ref{approx} --- \ref{inst} to obtain approximate and exact
solutions. {\it A main application of the theory of macroscopic
effects of quantum gravity is to explain the physics of Dark
Energy. As a carrier of Dark Energy, quantum gravity offers
graviton--ghost   condensate of quasi--resonant modes.} In Section
\ref{gm} the theory of gravitons and ghosts is formulated in a
self--consistent field, in the formation of which heavy particles
(baryons and particles of Dark Matter (neutralino as a example))
are involved. In Section \ref{I} it is shown that the cosmological
solution describing the evolution of the scale factor,
non--relativistic matter and graviton--ghost subsystem,
automatically satisfies the chain integral identities contained in
the equations of this theory. From these identities it follows
that the stage of the Universe evolution, which is dominated by
non--relativistic matter, must inevitably be replaced by a stage
at which substantive contribution to the energy density of the
cosmological substratum will produce graviton--ghost condensate of
quasi--resonant modes of a constant conformal wavelength. This
stage of the Universe evolution we consider as the pre--asymptotic
one. Conversion of pre--asymptotic condensate to a condensate of
constant physical wavelength occurs in the process of
quantum--gravity phase transition.

In Section \ref{LGCDM} the exact solutions of equations of the
one--loop quantum gravity are used to construct model of Dark
Energy, intended for interpretation of experimental data. The
model is based on simple interpolation formulae, describing the
energy density and pressure of graviton--ghost condensates at
various stages of cosmological evolution. For the cosmological
model based on the physical considerations presented here, we
suggest the abbreviation "$\Lambda$GCDM  model."\  Here "CDM"\
means Cold Dark Matter; "G"\ means a graviton--ghost condensate in
the role of Dark Energy; symbol $\Lambda$ indicates an asymptotic
state of the Universe in which the energy density of cosmological
substratum goes to a constant value. The following cosmological
Einstein equations correspond to the  $\Lambda$GCDM model
\begin{equation}
 \begin{array}{c}
 \displaystyle
 3H^2=\varkappa\left(\varepsilon_{g}+\frac{M}{a^3}\right)
 , \qquad
 6Q=-\varkappa\left(\varepsilon_{g}+3p_{g}+\frac{M}{a^3}\right)\ ,
    \end{array}
 \label{1.1}
 \end{equation}
where
\begin{equation}
 \begin{array}{c}
 \displaystyle
 \varepsilon_{g}=\Lambda_{\infty}+\frac{C_g}{a^2}\ln\frac{a_0}{a}\ , \qquad
 p_{g}=-\Lambda_{\infty}-\frac{C_g}{3a^2}\ln\frac{a_0}{ea}\ .
    \end{array}
 \label{1.2}
 \end{equation}
are energy density and pressure of graviton--ghost medium; $M/a^3$
is the density of non--relativistic matter; $H=\dot a/a$ is Hubble
function; $Q=\ddot a/a$ is acceleration of the Universe expansion.
The symbol $\Lambda_\infty$ indicates the asymptotic value of
total energy density of graviton--ghost condensate and equilibrium
vacuum of non--gravitational physical fields. Note that
$\Lambda_\infty\ne 0$, even if the non--gravitational vacuum
energy ("standard"\ Einstein's cosmological constant) vanishes by
some compensation mechanisms of the type of supersymmetry.

As can be seen from  (\ref{1.2}), the theory of graviton--ghost
condensates in an extremely simplified version predicted by a
three--parameter model of Dark Energy. A quantitative comparison
of theory with the observational data is conducted in Section
\ref{de}. Section \ref{prob} devoted to a brief review of the
physics of Dark Energy problems. Synopsis of observational data on
supernovae SNIa is given in Section \ref{data}. Our focus is on
the choice between $\Lambda$CDM and  $\Lambda$GCDM models based on
general principles of fundamental physics without involving
hypothetical elements. In Section \ref{fit} a comparative analysis
of the results of processing the Hubble diagram for SNIa by
formulas for both models is conducted. Here we show that the
$\Lambda$GCDM model not only explains quantitatively the effect of
Dark Energy, but also reproduces a number of specific details of
observational data.

The results and problems of the theory are briefly discussed in
the Conclusion (Section \ref{con}). Appendix \ref{Lambda} is
devoted to the cosmological constant problem within the framework
of its interpretation as the energy density of equilibrium vacuum
of non--gravitational physical fields. An internal inconsistency
of the theory which lacks one--loop finiteness off the mass shell
is proved in Appendices \ref{ren},\ref{nonren}. In \ref{an-d}, it
is shown by the method of dimensional transmutation that the
one--loop self--consistent theory of gravitons and ghosts is not
only finite but is also free of anomalies.

A system of units is used, in which the speed of light is  $c=1$,
Planck constant is $\hbar=197.327$ MeV$\cdot$fm; Einstein's
gravity constant is    $\varkappa\equiv 8\pi G=8\pi\cdot
1.324\cdot 10^{-42}$ MeV$^{-1}\cdot$fm.

\section{Basic Equations}\label{qg}

According to De Witt \cite{7}, one of formulations of one--loop
quantum gravity (with no fields of matter) is reduced to the zero
rest mass quantum field theory with spin $J=2$, defined for the
background spacetime with classic metric. The graviton dynamics is
defined by the interaction between quantum field and classic
gravity, and the background space geometry, in turn, is formed by
the energy--momentum tensor (EMT) of gravitons.

In the current Section we describe how to get the self--consistent
system of equations, consisting of quantum operator equations for
gravitons and ghosts and classic $C$--number Einstein equations
for macroscopic metrics with averaged EMT of gravitons and ghosts
on the right hand side. The theory is formulated without any
constrains on the graviton wavelength that allows the use of the
theory for the description of quantum gravity effects at the long
wavelength region of the specter. The equations of the theory
(except the gauge condition) are represented in 4D form which is
general covariant with respect to the transformation of the
macroscopic metric.

The mathematically consistent system of 4D quantum and classic
equations with no restrictions with respect to graviton
wavelengths is obtained by a regular method for the first time.
{\it The case of a gauged path integral with ghost sector} is seen
as a source object of the theory. Important elements of the method
are {\it exponential parameterization of the operator of the
density of the contravariant metric; factorization of path
integral measure; consequent integration over quantum and classic
components of the gravitational field.} Mutual compliance of
quantum and classic equations, expressed in terms of fulfilling of
the conservation of averaged EMT at the operator equations of
motion is provided by the virtue of the theory construction
method.

\subsection{Path Integral and Faddeev--Popov Ghosts}\label{pi}

Formally, the exact scheme of quantum gravity is based on the
amplitude of transition, represented by path integral \cite{4, 5}:
\begin{equation}
\begin{array}{c}
 \displaystyle
 \langle \mbox{\rm out}|\mbox{\rm in}\rangle=
\int \exp\left(\frac{i}{\hbar}\int
(L_{grav}+L_\Lambda)d^4x\right)\left(\mbox{\rm det}\, \hat M^i_{\;
 k}\right)\prod_x\left(\prod_i\delta(\hat A_k\sqrt{-\hat g}\hat
 g^{ik}-B^i)\right)d\hat\mu=
 \\[5mm] \displaystyle
 =\int \exp\left(\frac{i}{\hbar}\int (L_{grav}+L_\Lambda+L_{ghost})d^4x\right)
 \prod_x\left(\prod_i\delta(\hat A_k\sqrt{-\hat g}\hat
 g^{ik}-B^i)\right)d\hat\mu d\mu_\theta\ ,
 \end{array}
 \label{2.1}
 \end{equation}
where
\[
\displaystyle L_{grav}+L_\Lambda=-\frac{1}{2\varkappa}\sqrt{-\hat g}\hat
g^{ik}\hat R_{ik}-\sqrt{-\hat g}\Lambda
\]
is the density of gravitational Lagrangian, with cosmological
constant included;  $L_{ghost}$  is the density of ghost
Lagrangian, explicit form of which is defined by localization of
$\mbox{\rm det}\, \hat M^i_{\; k}$;  $\hat A_k$ is gauge operator,
$B^i(x)$ is the given field;   $\hat M^i_{\; k}$ is an operator of
equation for infinitesimal parameters of transformations for the
residual degeneracy $\eta^i=\delta x^i$;
\begin{equation}
 \displaystyle
  d\hat\mu=\prod_x\left\{(-\hat g)^{5/2}\prod_{i\leqslant
 k}d\hat g^{ik}\right\}
 \label{2.2}
 \end{equation}
is the gauge invariant measure of path integration over
gravitational variables; $d\mu_\theta$ is the measure of
integration over ghost variables. Operator $\hat M^i_{\; k}$ is of
standard definition:
\begin{equation}
 \displaystyle
 \hat M^i_{\;  k}\eta^k\equiv \hat A_k(\delta\sqrt{-\hat g}\hat
 g^{ik})=0,
 \label{2.3}
 \end{equation}
where
\begin{equation}
 \displaystyle
\delta\sqrt{-\hat g}\hat  g^{ik}=-\partial_l(\sqrt{-\hat g}\hat
g^{ik}\eta^l)+\sqrt{-\hat g}\hat  g^{il}\partial_l\eta^k+
\sqrt{-\hat g}\hat  g^{kl}\partial_l\eta^i
 \label{2.4}
 \end{equation}
is variation of metrics under the action of infinitesimal
transformations of the group of diffeomorphisms. According to
(\ref{2.1}), the allowed gauges are constrained by the condition
of existence of the inverse operator $(\hat M^i_{\;  k})^{-1}$.

The equation (\ref{2.1}) explicitly manifests the fact that the
source path integral is defined as a mathematical object only
after the gauge has been imposed. In the theory of gravity, there
are no gauges completely eliminating the degeneracy with respect
to the transformations (\ref{2.4}). Therefore, the sector of
nontrivial ghost fields, interacting with gravity, is necessarily
present in the path integral. {\it This aspect of the quantum
gravity is important for understanding of its mathematical
structure, which is fixed before any approximations are
introduced.} By that reason, in this Section we discuss the
equations of the theory, by explicitly defining the concrete
gauge.

In cosmological applications of the quantum gravity it is
convenient to use synchronous gauges of type:
\begin{equation}
 \displaystyle
 \sqrt{-\hat g}\hat g^{00}=\sqrt{\bar\gamma}\ , \qquad \sqrt{-\hat g}\hat
 g^{0\alpha}=0\ .
 \label{2.5}
 \end{equation}
For that gauge
\begin{equation}
 \displaystyle
 \hat A_k=(1,\, 0,\, 0,\, 0)\ , \qquad B^i=(\sqrt{\bar\gamma},\, 0,\, 0,\, 0),
 \label{2.6}
 \end{equation}
where  $\bar\gamma=\bar\gamma({\bf x})$ is the metric determinant
of the basic 3D space of constant curvature (for the plane
cosmological model $\bar\gamma=1$). More general approach to the
choice of gauge used for cosmological problems is discussed in
Section \ref{fininv}.

The construction of the ghost sector, i.e. finding of the
Lagrangian density $L_{ghost}$, is reduced to two operations.
First, $\mbox{\rm det}\, \hat M^i_{\; k}$  is represented in the
form, factorized over independent degrees of freedom for ghosts,
and then the localization of the obtained expression is conducted.
Substitution of (\ref{2.6}) and (\ref{2.4}) to (\ref{2.3}) gives
the following system of equations
\begin{equation}
 \displaystyle
 -\partial_\alpha\sqrt{\bar\gamma}\eta^\alpha+
  \sqrt{\bar\gamma}\frac{\partial\eta^0}{\partial t}=0,\qquad
 \sqrt{-\hat g}\hat  g^{\alpha\beta}\partial_\beta\eta^0+
 \frac{\partial\sqrt{\bar\gamma}\eta^\alpha}{\partial t}=0.
 \label{2.7}
 \end{equation}
According to (\ref{2.7}), with respect to variables $\eta^0$,
$\sqrt{\bar\gamma}\eta^\alpha$  the operator--matrix $\hat M^i_{\;
k}$  reads
\begin{equation}
 \displaystyle
   \hat M^i_{\; k}=
   \begin{pmatrix}
		\displaystyle\sqrt{\bar\gamma}\frac{\partial}{\partial t} & \quad -\partial_\alpha 
		\\[5mm] \displaystyle
		\sqrt{-\hat g} \hat g^{\alpha\beta}\partial_\beta & \quad \displaystyle\delta_\beta^\alpha\frac{\partial}{\partial t}
   \end{pmatrix}
\label{2.8}
\end{equation}
(Note matrix--operator is obtained in the form (\ref{2.8}) without
the substitution of transformation parameters if Leutwiller
measure $d\hat \mu_L=\hat g\hat g^{00}d\hat \mu$ is used. The
measure discussion see, e.g. \cite{8}.) Functional determinant of
matrix--operator  $\mbox{\rm det}\, \hat M^i_{\; k}$ is
represented in the form of the determinant of matrix $\hat M^i_{\;
k}$, every element of which is a functional determinant of
differential operator. As it is follows from (\ref{2.8}),
  \begin{equation}
 \displaystyle \mbox{\rm det}\, \hat M^i_{\; k}=
\left(\mbox{\rm det}\; \partial_i\sqrt{\hat g}\hat
g^{ik}\partial_k\right)\times \left(\mbox{\rm det}\;
\frac{\partial}{\partial t}\right)\times \left(\mbox{\rm det}\;
\frac{\partial}{\partial t}\right).
 \label{2.9}
\end{equation}
One can see that the first multiplier in (\ref{2.9}) is
4--invariant determinant of the operator of the zero rest mass
Klein--Gordon--Fock equation, and two other multipliers do not
depend on gravitational variables.

Localization of determinant (\ref{2.9}) by representing it in a
form of path integral over the ghost fields is a trivial
operation. As it follows from (\ref{2.9}), the class of
synchronous gauges contains three dynamically independent ghost
fields $\theta,\, \varphi,\, \chi$, two of each --- $ \varphi,\,
\chi$ do not interact with gravity. For the obvious reason, the
trivial ghosts  $ \varphi,\, \chi$ are excluded from the theory.
The Lagrangian density of nontrivial ghosts coincides exactly with
Lagrangian density of complex Klein--Gordon--Fock fields (taking
into account the Grassman character of  fields $\bar\theta,\,
\theta$):
\begin{equation}
 \displaystyle L_{ghost}=-\frac{1}{4\varkappa}\sqrt{-\hat g}\hat
g^{ik}\partial_i\bar \theta\cdot \partial_k\theta.
 \label{2.10}
\end{equation}
The normalization multiplier $-1/4\varkappa$  in (\ref{2.10}) is
chosen for the convenience. The integral measure over ghost fields
has a simple form:
\[
\displaystyle d\mu_\theta=\prod_xd\bar\theta d\theta\ .
\]

The calculations above comply with both general requirements to
the construction of ghost sector. First, path integration should
be carried out only over the dynamically independent ghost fields.
Second, in the ghost sector, it is necessary to extract and then
to take into account only the nontrivial ghost fields, i.e. those
interacting with gravity.

The extraction of dynamically independent nontrivial ghost fields
can be done not only by factorization of functional determinant
(as it made in (\ref{2.9})), but by means of researching the
equations for the ghosts as well. It is well known \cite{9}, that
from the definition of $\mbox{\rm det}\, \hat M^i_{\; k}$ it
follows that the ghost equations coincide with equations for
parameters of infinitesimal transformations of residual
degeneracy. Therefore, according to (\ref{2.3}), we can
immediately get $ \hat M^i_{\;  k}\theta^k=0$, where  $\theta^k$
are the Grassman fields. For the gauge (\ref{2.5}) with  $B=1$ we
get
\begin{equation}
 \displaystyle
 -\partial_\alpha\sqrt{\bar\gamma}\theta^\alpha+
  \sqrt{\bar\gamma}\frac{\partial\theta^0}{\partial t}=0,
 \label{2.11}
 \end{equation}
 \begin{equation}
 \displaystyle
 \sqrt{-\hat g}\hat  g^{\alpha\beta}\partial_\beta\theta^0+
 \frac{\partial\sqrt{\bar\gamma}\theta^\alpha}{\partial t}=0.
 \label{2.12}
 \end{equation}
From (\ref{2.12}) it follows that those transversal components of
vector $\theta^\alpha$  yield the equations
$\partial\theta^\alpha_\perp/\partial t=0$, i.e. these degrees of
freedom correspond to two trivial ghosts non--interacting with
gravity. Note, that in (\ref{2.11}) only the longitudinal
component  $\theta^\alpha_\parallel$ is present. Equation
(\ref{2.12}) has a status of equation connecting the longitudinal
component $\theta^\alpha_\parallel$  and function $\theta^0$. It
means that the longitudinal field $\theta^\alpha_\parallel$  is
not dynamically independent. Equation for dynamically independent
degrees  of freedom  $\theta^0$ is obtained as follows. First, the
operator $\partial_\alpha$  is applied to the equation
(\ref{2.12}), next the result is substituted into
time--differentiated equation (\ref{2.11}). After the
substitution, one gets Klein--Gordon--Fock equation,
\begin{equation}
 \displaystyle
\frac{\partial}{\partial
t}\sqrt{\bar\gamma}\frac{\partial\theta}{\partial t}+
\partial_\alpha\sqrt{-\hat g}\hat  g^{\alpha\beta}\partial_\beta\theta=
\partial_i\sqrt{-\hat g}\hat  g^{ik}\partial_k\theta=0\ ,
 \label{2.13}
 \end{equation}
where  $\theta=\theta^0$. Reconstruction of the ghost Lagrangian
(\ref{2.10}) from dynamical equation (\ref{2.13}) does not require
an additional explanation.

Thus, the prove of wave properties of ghosts in the class of
synchronous gauges is done by two methods with clear correlations
between objects and operations used in these methods.

We will return to the discussion of gauges and ghosts in the Section \ref{fininv}.

\subsection{Einstein Operator Equations}\label{OEE}

Let us take into account the fact that the calculation of gauged
path integral should be mathematically equivalent to the solution
of dynamical operator equations in the Heisenberg representation.
It is also clear that operator equations of quantum theory should
have a definite relationship with Einstein equations. In the
classic theory, it is possible to use any form of representation
of Einstein equations, e.g.
\begin{equation}
\begin{array}{c}
 \displaystyle
 (-\hat g)^n\left(\hat g^{il}\hat g^{km}\hat R_{lm}-
 \frac12\hat g^{ik}\hat g^{lm}\hat R_{lm}-\hat g^{ik}\varkappa\Lambda\right)=0\ ,\qquad (a)
 \\[3mm]
 \displaystyle (-\hat g)^n\left( \hat g^{km}\hat R_{im}-
 \frac12 \delta_i^k\hat g^{lm}\hat R_{lm}-\delta_i^k\varkappa\Lambda\right)=0\ ,\qquad\qquad (b)
 \\[3mm]
 \displaystyle (-\hat g)^n\left(\hat R_{ik}-
\frac12\hat g_{ik}\hat g^{lm}\hat R_{lm}-\hat
g^{ik}\varkappa\Lambda\right)=0\ ,\qquad\qquad\qquad (c)
 \end{array}
  \label{2.14}
 \end{equation}
where, for example,  $n=0,\; 1/2,\; 1$. Transition from one to
another is reduced to the multiplication by metric tensor and its
determinant, which are  trivial operations in case when the metric
is a $C$--number function. If the metric is an operator, then the
analogous operations will, at least, change renormalization
procedures of quantum non--polynomial theory. Thus, the question
about the form of notation for Einstein's operator equations has
first--hand relation to the calculation procedure. Now we show
that in the quantum theory one should use operator equations
(\ref{2.14}$b$) with $n=1/2$, supplemented by the energy--momentum
pseudo--tensor of ghosts.

In the path integral formalism, the renormalization procedures are
defined by the dependence of Lagrangian of interactions and the
measure of integration of the field operator in terms of which the
polynomial expansion of non--polynomial theory is defined
\cite{9}. The introduction of such an operator, i.e. {\it the
parameterization of the metric}, is, generally speaking, not
simple. Nevertheless, it is possible to find a special
parameterization for which the algorithms of renormalization
procedures are defined only by Lagrangian of interactions.
Obviously, in such a parameterization the measure of integration
should be trivial. It reads:
\begin{equation}
\displaystyle
 d\hat\mu=\prod_x\prod_{i\leqslant k}d\hat\Psi_i^k\ ,
\label{2.15}
\end{equation}
where $\hat\Psi_i^k$  is a dynamic variable. The metric is
expressed via this variable. It is shown in \cite{9} that the
trivialization of measure (\ref{2.15}) takes place for the
exponential parameterization that reads
\begin{equation}
\begin{array}{c}
 \displaystyle \sqrt{-\hat{g}}\hat{g}^{ik} = \sqrt{-\bar g}\bar g^{il}(\exp{\hat\Psi})_l^k=
  \sqrt{-\bar g}\bar g^{il}
  \left(
         \delta_l^k + \hat\Psi_l^k + \frac{1}{2}\hat\Psi_l^m\hat\Psi_m^k+\ldots
  \right)\ ,
\end{array}
\label{2.16}
\end{equation}
where $\bar g^{ik}$  is the defined metric of an auxiliary basic
space. In that class of our interest, the metric is defined by the
interval
\[
\displaystyle d\bar s^2=dt^2-\bar\gamma_{\alpha\beta}dx^\alpha dx^\beta\ ,
\]
where  $\bar\gamma_{\alpha\beta}$ is the metric of 3D space with a
constant curvature. (For the flat Universe
$\bar\gamma_{\alpha\beta}$ is the Euclid metric.)

The exponential parameterization is singled out among all other
parameterizations by the property that $\hat\Psi_i^k$  are the
normal coordinates of gravitational fields \cite{10}. In that
respect, the gauge conditions (\ref{2.5}) are identical to $\hat
\Psi_0^i=0$. The fact that the "gauged"\ coordinates are the
normal coordinates, leads to a simple and elegant ghost sector
(\ref{2.10}). The status of $\hat\Psi_i^k$, as normal coordinates,
is of principal value for the mathematical correctness while
separating the classic and quantum variables (see Section
\ref{var}). Besides, in the framework of perturbation theory the
normal coordinates allow to organize a calculation procedure,
which is based on a simple classification of nonlinearity of
quantum gravity field. It is important that this procedure is
mathematically non--contradictive at every order of perturbation
theory over amplitude of quantum fields (see Section \ref{equiv},
\ref{1loop}).

Operator Einstein equations that are mathematically equivalent to
the path integral of a trivial measure are derived by the
variation of gauged action by variables $\hat\Psi_i^k$. The
principal point is that the gauged action necessarily includes the
ghost sector because there are no gauges that are able to
completely eliminate the degeneracy. According to (\ref{2.10}), in
the class of synchronous gauges we get
\begin{equation}
\displaystyle
 S=-\int d^4x\left\{\frac{1}{2\varkappa}\sqrt{-\hat g}\hat g^{ik}\left(\hat R_{ik}+
 \frac12\partial_i\bar\theta\cdot\partial_k\theta\right)+\sqrt{-\hat g}\Lambda\right\}\ .
\label{2.17}
\end{equation}
In accordance with definition (\ref{2.16}), the variation is done
by the rule
\[
\displaystyle \delta\sqrt{-\hat g}\hat g^{ik} =\sqrt{-\hat g}\hat g^{il}\delta\hat \Psi_l^k\ .
\]
Thus, from (\ref{2.17}) it follows
\begin{equation}
\displaystyle \hat{\mathcal{G}}_i^k\equiv \sqrt{-\hat g}\hat
g^{kl}\hat R_{il}-\varkappa\left(\sqrt{-\hat g}\hat g^{kl}\hat
T^{(ghost)}_{il}-\frac12\delta_i^k\sqrt{-\hat g}\hat g^{ml}\hat
T^{(ghost)}_{ml}-\sqrt{-\hat g}\delta_i^k\Lambda\right)=0\ .
 \label{2.18}
\end{equation}
After subtraction of semi--contraction from (\ref{2.18}) we obtain
a mathematically equivalent equation
\begin{equation}
\begin{array}{c}
\displaystyle
\hat{\mathcal{E}}_i^k=\hat{\mathcal{G}}_i^k-\frac12\delta_i^k\hat{\mathcal{G}}_l^l\equiv
\\[5mm]
\displaystyle  \sqrt{-\hat g}\hat g^{kl}\hat
R_{il}-\frac12\delta_i^k\sqrt{-\hat g}\hat g^{ml}\hat
R_{ml}-\varkappa\left(\sqrt{-\hat g}\hat g^{kl}\hat
T^{(ghost)}_{il}+\sqrt{-\hat g}\delta_i^k\Lambda\right)=0\ .
\end{array}
\label{2.19}
\end{equation}
In (\ref{2.18}), (\ref{2.19}) there is an object
\begin{equation}
\displaystyle \hat
T^{(ghost)}_{ik}=-\frac{1}{4\varkappa}\left(\partial_i\bar\theta\cdot\partial_k\theta+
\partial_k\bar\theta\cdot\partial_i\theta-
\hat g_{ik}\hat g^{lm}\partial_l\bar\theta\cdot\partial_m\theta\right)\ ,
\label{2.20}
\end{equation}
which has the status of the energy--momentum pseudo--tensor of
ghosts.

In accordance with the general properties of Einstein's theory,
six spatial components of equations (\ref{2.18}) are considered as
quantum equations of motion:
\begin{equation}
\displaystyle \sqrt{-\hat g}\hat g^{\beta l}\hat R_{\alpha
l}=\varkappa\left(\sqrt{-\hat g}\hat g^{\beta l}\hat
T^{(ghost)}_{\alpha l}-\frac12\delta_\alpha^\beta\sqrt{-\hat
g}\hat g^{ml}\hat T^{(ghost)}_{ml}-\sqrt{-\hat g}\delta_\alpha^\beta\Lambda\right)\ .
\label{2.21}
\end{equation}
(Everywhere in this work the Greek metric indexes stand for
$\alpha,\ \beta=1,\ 2,\ 3$.) In the classic theory,  equations of
constraints $\hat{\mathcal{E}}_0^0=0$ and
$\hat{\mathcal{E}}_0^\alpha=0$ are the first integrals of
equations of motion (\ref{2.21}). Therefore, in the quantum theory
formulated in the Heisenberg representation four primary
constraints from (\ref{2.19}), have the status of the {\it
initial} conditions for the Heisenberg state vector. They read:
\begin{equation}
\begin{array}{c}
\displaystyle \left\{\sqrt{-\hat g}\hat g^{0l}\hat
R_{0l}-\frac12\sqrt{-\hat g}\hat g^{ml}\hat
R_{ml}-\varkappa\left(\sqrt{-\hat g}\hat g^{0l}\hat
T^{(ghost)}_{0l}+\sqrt{-\hat g}\Lambda\right)\right\}|\Psi\rangle=0\ ,
 \\[3mm]
\displaystyle \left\{\sqrt{-\hat g}\hat g^{\alpha l}\hat
R_{0l}-\varkappa\sqrt{-\hat g}\hat g^{\alpha l}\hat
T^{(ghost)}_{0l}\right\}|\Psi\rangle=0\ .
 \end{array}
 \label{2.22}
\end{equation}
If conditions (\ref{2.22}) are valid from the start, then the
internal properties of the theory must provide their validity at
any subsequent moment of time. Four secondary relations, defined
by the gauge non containing the higher order derivatives, also
have the same status:
\begin{equation}
\displaystyle \left\{\hat A_k(\sqrt{-\hat g}\hat
g^{ik})-B^i\right\}|\Psi\rangle=0\ .
 \label{2.23}
\end{equation}

The system of equations of quantum gravity is closed by the
ghosts' equations of motion, obtained by the variation of action
(\ref{2.17}) over ghost variables:
\begin{equation}
\begin{array}{c}
 \displaystyle \partial_i\sqrt{-\hat g}\hat g^{ik}\partial_k\theta=0\ ,
 \\[3mm]
\displaystyle \partial_i\sqrt{-\hat g}\hat g^{ik}\partial_k\bar\theta=0\ .
\end{array}
\label{2.24}
\end{equation}
Ghost fields  $\bar \theta$ and  $\theta$  are not defined by
Grassman scalars, therefore  $T^{(ghost)}_{ik}$ is not a tensor.
Nevertheless, all mathematical properties of equations
(\ref{2.24}) and expressions (\ref{2.20}) coincide with the
respected properties of equations and EMT of complex scalar
fields. This fact is of great importance when concrete
calculations are done (see Section \ref{scgt}).

 \subsection{Factorization of the Path Integral}\label{fac}

Transition from the formally exact scheme (\ref{2.21}) ---
(\ref{2.24})) to the semi--quantum theory of gravity can be done
after some additional hypotheses are included in the theory. The
physical content of these hypotheses consists of the assertion of
existence of classical spacetime with metric $g_{ik}$,
connectivity $\Gamma^i_{kl}$  and curvature $R_{ik}$. The first
hypothesis is formulated at the level of operators. Assume that
{\it operator of metric $\hat g^{ik}$  is a functional of
$C$--number function  $g^{ik}$  and the quantum operator $\hat
\psi_i^k$.} The second hypothesis is related to the state vector.
{\it Each state vector that is involved in the scalar product
$\langle \mbox{\rm out}|\mbox{\rm in}\rangle$, is represented in a
factorized form $|\Psi\rangle=|\Phi\rangle|\psi\rangle$, where
$|\psi\rangle$ are the vectors of quantum states of gravitons;
$|\Phi\rangle$ are the vectors of quasi--classic states of
macroscopic metric}. In the framework of these hypotheses the
transitional amplitude is reduced to the product of amplitudes:
\begin{equation}
\displaystyle
 \langle \mbox{\rm out}|\mbox{\rm in}\rangle=
 \langle \Phi_{out}|\Phi_{in}\rangle\langle
 \psi_{out}|\psi_{in}\rangle\ .
 \label{2.25}
\end{equation}
Thus, {\it the physical assumption about existence of classic
spacetime formally (mathematically) means that the path integral
must be calculated first by exact integration over quantum
variables, and then by approximate integration over the classic
metric.}

Mathematical definition of classic and quantum variables with
subsequent integrations are possible only after the trivialization
and factorization of integral measure are done. As already noted,
trivial measure (\ref{2.15}) takes place in exponential
parameterization (\ref{2.16}). The existence of $|\mbox{\rm
in}\rangle=|\Psi\rangle$ vector allows the introduction of classic
$C$--number variables as follows
\[
\displaystyle \Phi_i^k=\langle\Psi|\hat\Psi_i^k|\Psi\rangle\
,\qquad \sqrt{-g}g^{ik}=\sqrt{-\bar g}\bar g^{il}(\exp{\Phi})_l^k\
.
\]
Quantum graviton operators are defined as the difference
$\hat\psi_i^k=\hat\Psi_i^k-\Phi_i^k$. Factorized amplitude
(\ref{2.25}) is calculated via the factorized measure
 \begin{equation}
\begin{array}{c}
\displaystyle
  d\hat \mu=d\mu_g\times d\mu_{\psi}\ ,
\\[3mm]\displaystyle
 d\mu_g=\prod_x\left\{(-g)^{5/2}\prod_{i\leqslant
 k}dg^{ik}\right\}\ , \qquad
 d\mu_{\psi}=\prod_x\prod_{i\leqslant k}d\hat\psi_i^k\ .
\end{array}
\label{2.26}
\end{equation}

Factorization of the measure allows the subsequent integration,
first by $d\mu_{\psi}$, $d\mu_\theta$, then by approximate
integration over  $d\mu_g$. In the operator formalism, such
consecutive integrations correspond to the solution of
self--consistent system of classic and quantum equations.
Classical equations are obtained by averaging of operator
equations (\ref{2.19}). They read:
\begin{equation}
\begin{array}{c}
\displaystyle
  \langle \Psi|\hat{\mathcal{E}}_i^k|\Psi\rangle=0\ .
\end{array}
\label{2.27}
\end{equation}
Subtraction of (\ref{2.27}) from (\ref{2.19}) gives the quantum dynamic equations
\begin{equation}
\begin{array}{c}
\displaystyle
 \hat{\mathcal{E}}_i^k-\langle \Psi|\hat{\mathcal{E}}_i^k|\Psi\rangle=0\ .
\end{array}
\label{2.28}
\end{equation}
Synchronous gauge (\ref{2.23}) is converted to the gauge of
classical metric and to conditions imposed on the state vector:
\begin{equation}
\begin{array}{c}
\displaystyle
\sqrt{-g}g^{00}=\sqrt{\bar \gamma},\qquad  \sqrt{-g}g^{0\alpha}=0\ ,
\\[5mm]
\displaystyle
  \hat\psi_0^i|\Psi\rangle=0\ .
\end{array}
\label{2.29}
\end{equation}
Quantum equations (\ref{2.24}) of ghosts' dynamics are added to
equations (\ref{2.27}) --- (\ref{2.29}).

Theory of gravitons in the macroscopic spacetime with
self--consistent geometry is without doubt an approximate theory.
Formally, the approximation is in the fact that the single
mathematical object $\sqrt{-\hat g}\hat g^{ik}$  is replaced by
two objects --- classical metric and quantum field, having
essentially different physical interpretations. That "coercion"\
of the theory can lead to a controversy, i.e. to the system of
equations having no solutions, if an inaccurate mathematics of the
adopted hypotheses is used. The scheme described above does not
have such a controversy. The most important element of the scheme
is the exponential parameterization (\ref{2.16}), which separates
the classical and quantum variables, as can be seen from
(\ref{2.26}). After the background and quantum fluctuations are
introduced, this parameterization looks as follows:
\begin{equation}
\begin{array}{c}
\displaystyle \sqrt{-\hat{g}}\hat{g}^{ik} = \sqrt{-\bar g}\bar
g^{il}\left(\exp{(\Phi+\hat\psi})\right)_l^k=
  \sqrt{-g} g^{il}(\exp\hat\psi)_l^k\ ,
\end{array}
\label{2.30}
\end{equation}
Note that the auxiliary basic space vanishes from the theory, and
instead the macroscopic (physical) spacetime with self--consistent
geometry takes its place.

If the geometry of macroscopic spacetime satisfies symmetry
constrains, the factorization of the measure (\ref{2.26}) becomes
not a formal procedure but strictly mathematical in its nature.
These restrictions must ensure the existence of an algorithm
solving the equations of constraints in the framework of the
perturbation theory (over the amplitude of quantum fields). The
theory of gravity is non--polynomial, so after the separation of
single field into classical and quantum components, the use of the
perturbation theory in the quantum sector becomes unavoidable. The
classical sector remains non--perturbative. In the general case,
when quantum field is defined in an arbitrary Riemann space, the
equations of constraints is not explicitly solvable. The problem
can be solved in the framework of perturbation theory if
background $g_{ik}$   and the free (linear) tensor field  $\hat
\psi_i^k$ belong to different irreducible representations of the
symmetry group of the background spacetime. In that case at the
level of linear field we obtain (\ref{2.26}), because the full
measure is represented as a product of measure of integration over
independent irreducible representations. At the next order,
factorization is done over coordinates, because the classical
background and the induced quantum fluctuations have essentially
different spacetime dynamics. Note, to factorize the measure by
symmetry criterion we do not need to go to the short--wave
approximation.

Background metric of isotropic cosmological models and classical
spherically symmetric non--stationary gravitational field meet the
constrains described above. These two cases are covering all
important applications of semi--quantum theory of gravity which
are quantum effects of vacuum polarization and creation of
gravitons in the non--stationary Universe and in the neighborhood
of black holes.

\subsection{Variational Principle for Classic and Quantum Equations}\label{var}

Geometrical variables can be identically transformed to the form
of functionals of classical and quantum variables. At the first
step of transformation there is no need to fix the
parameterization. Let us introduce the notations:
\begin{equation}
\begin{array}{c}
\displaystyle \sqrt{-\hat g}\hat g^{ik}=\sqrt{-g}\hat X^{ik},\qquad
 \frac{1}{\sqrt{-\hat g}}\hat g_{ik}=\frac{1}{\sqrt{-g}}\hat Y_{ik}\ ,
 \\[5mm]
\displaystyle \hat Y_{il}\hat X^{lk}=\delta_i^k\ .
\end{array}
\label{2.31}
\end{equation}
According to (\ref{2.30}), formalism of the theory allows
definition of quantum field $\hat\psi_i^k$  as symmetric tensor in
physical space, $g_{kl}\hat\psi^l_k=\hat\psi_{ik}=\hat\psi_{ki}$.
Objects, introduced in (\ref{2.31}), have the same status. With
any parameterization the following relationships take place:
\[
\displaystyle \lim_{\hat \psi_l^m\to 0}\hat X^{ik}=g^{ik},\qquad
\lim_{\hat \psi_l^m\to 0}\hat Y_{ik}=g_{ik}\ .
\]
We should also remember that the mixed components of tensors
$\hat X_i^k,\ \hat Y_i^k$ do not contain the background metric as
functional parameters. For any parameterization, these tensors are
only functionals of quantum fields $\hat \psi_i^k$ which are also
defined in mixed indexes. For the exponential parameterization:
\begin{equation}
\begin{array}{c}
\displaystyle \hat X_i^k=\delta_i^k+\hat\psi_i^k+\frac12\hat\psi_i^l\hat\psi_l^k+...\ ,\qquad
\hat Y_i^k=\delta_i^k-\hat\psi_i^k+\frac12\hat\psi_i^l\hat\psi_l^k+...\ ,
\\[5mm]
\displaystyle \hat g=g\cdot \hat d =ge^{\hat\psi}\ ,
\end{array}
\label{2.32}
\end{equation}
where  $d=\mbox{\rm det}\,|| \hat X^i_k||$. One can seen from
(\ref{2.32}), that the determinant of the full metric contains
only the trace of the quantum field.

Regardless of parameterization, the connectivity and curvature of
the macroscopic space $\Gamma_{ik}^l$, $R^i_{\, klm}$ are
extracted from full connectivity and curvature as additive terms:
\[
\displaystyle \hat \Gamma_{ik}^l=\Gamma_{ik}^l+ \hat {\mathcal{T}}_{ik}^l,\qquad
\hat R^i_{\, klm}=R^i_{\, klm}+\hat {\mathcal{R}}^i_{\, klm}\ .
\]
Quantum contribution to the curvature tensor,
\[
\displaystyle \hat {\mathcal{R}}^i_{\, klm}=\hat
{\mathcal{T}}_{km\ ;l}^i-\hat {\mathcal{T}}_{kl\ ;m}^i+\hat
{\mathcal{T}}_{nl}^i\hat {\mathcal{T}}_{km}^n-\hat
{\mathcal{T}}_{nm}^i\hat {\mathcal{T}}_{kl}^n\ ,
\]
is expressed via the quantum contribution to the full
connectivity:
\begin{equation}
\begin{array}{c}
\displaystyle \hat{\mathcal{T}}_{ik}^l=\frac12\left(-\hat
Y_{im}\hat X^{ml}_{\;\;\; ;k} -\hat Y_{km}\hat X^{ml}_{\;\;\;
;i}+\hat Y_{ij}\hat Y_{kn}\hat X^{ml}\hat X^{jn}_{\;\;\;
;m}\right)+\frac14Y_{jn}\left(\delta_i^lX^{jn}_{\;\;\;
;k}+\delta_k^lX^{jn}_{\;\;\; ;i}-Y_{ik}X^{ml}X^{jn}_{\;\;\;
;m}\right)\ .
\end{array}
\label{2.33}
\end{equation}
The density of Ricci tensor in mixed indexes reads
\begin{equation}
\begin{array}{c}
\displaystyle \displaystyle \sqrt{-\hat g}\hat g^{kl}\hat
R_{il}=\sqrt{-g}\left\{\hat X^{kl}R_{il}+ \frac12\left[\hat
Y_{in}\left(\hat X^{ml}\hat X^{nk}_{\;\;\; ;m}-\hat X^{mk}\hat
X^{nl}_{\;\;\; ;m}\right)-\hat X^{lk}_{\;\;\;\; ;\ i} -\frac12
\delta_i^k\hat Y_{nj}\hat X^{ml}\hat X^{jn}_{\;\;\;
;m}\right]_{;l}\right.-
\\[5mm]
\displaystyle \left.-\frac14\left(\hat Y_{jn}\hat
Y_{sm}-\frac12\hat Y_{jm}\hat Y_{sn}\right)\hat X^{kl}\hat
X^{jm}_{\;\;\; ;i}\hat X^{ns}_{\;\;\; ;l}+ \frac12\hat Y_{ml}\hat
X^{km}_{\;\;\; ;n}\hat X^{nl}_{\;\;\; ;i}\right\}\ .
\end{array}
\label{2.34}
\end{equation}
Symbol ";"\ in (\ref{2.33}), (\ref{2.34}) and in  what follows
stands for the covariant derivatives in background space. The
density of gauged gravitational Lagrangian is represented in a
form which is characteristic for the theory of quantum fields in
the classical background spacetime:
\begin{equation}
\begin{array}{c}
\displaystyle
S=\int d^4x\sqrt{-g}\left({\mathcal{L}}_{grav}-\sqrt{\hat d}\Lambda-
\frac{1}{4\varkappa}\hat X^{ik}\bar\theta_{ ,i}\theta_{ , k}\right)\ ,
\\[5mm]
\displaystyle
{\mathcal{L}}_{grav}=-\frac{1}{2\varkappa}\hat X^{ik}R_{ik}+
\frac{1}{8\varkappa}\left[\hat X^{kl}\left(\hat Y_{jn}\hat Y_{sm}-
\frac12\hat Y_{jm}\hat Y_{sn}\right)
\hat X^{jm}_{\;\;\; ;k}\hat X^{sn}_{\;\;\; ;l}-
2\hat Y_{ik}\hat X^{il}_{\;\;\; ;m}\hat X^{km}_{\;\;\;\; ;l}\right]\ .
\end{array}
\label{2.35}
\end{equation}
When the expression for ${\mathcal{L}}_{grav}$  was obtained from
contraction of tensor (\ref{2.34}), the full covariant divergence
in the background space have been excluded. Formulas (\ref{2.34}),
(\ref{2.35}) apply for at any parameterization.

Let us discuss the variation method. In the exact quantum theory
of gravity with the trivial measure (\ref{2.15}), the variation of
the action over variables  $\hat \Psi_i^k$ leads to the Einstein
equations in mixed indexes (\ref{2.18}) and (\ref{2.19}). In the
exact theory, the exponential parameterization is convenient, but,
generally speaking, is not necessary. A principally different
situation takes place in the approximate self--consistent theory
of gravitons in the macroscopic spacetime. In that theory the
number of variables doubles, and with this, the classical and
quantum components of gravitational fields have to have the status
of the dynamically independent variables due to the doubling of
the number of equations. The variation should be done separately
over each type of variables. The formalism of the path integration
suggests a rigid criterion of dynamic independence: {\it the full
measure of integration, by definition, must be factorized with
respect to the dynamically independent variables.} Obviously, only
the exponential parameterization (\ref{2.30}), leading to the
factorized measure (\ref{2.26}), meets the criterion.

The variation of the action over the classic variables is done
together with the operation of averaging over the quantum
ensemble. In the result, equations for metric of the macroscopic
spacetime are obtained:
\begin{equation}
\displaystyle \langle\Psi|\frac{\delta S}{\delta g^{in}}|\Psi\rangle=
-2\varkappa \sqrt{-g}g_{nk}\langle\Psi|\hat G_i^k-\frac12\delta_i^k\hat G_l^l|\Psi\rangle=0\ ,
\label{2.36}
\end{equation}
where $\hat G_i^k=\hat {\mathcal{G}}_i^k/\sqrt{-g}$. Variation of
the action over background variables, defined as
$\Phi_i^k=\langle\Psi|\hat\Psi^k_i|\Psi\rangle$, yields the
equations:
\begin{equation}
\displaystyle \langle\Psi|\frac{\delta S}{\delta\Phi_k^i}|\Psi\rangle=
-2\varkappa \sqrt{-g}\langle\Psi|\hat G_i^k|\Psi\rangle=0\ .
\label{2.37}
\end{equation}
Equations (\ref{2.36}) and (\ref{2.37}) are mathematically
identical. We should also mention that if the variations over the
background metric are done with the fixed {\it mixed} components
of the quantum field, these equations are valid for any
parameterization.

Exponential parameterization (\ref{2.30}) has a unique property:
{\it the variations over classic  $\Phi_i^k$ (before averaging)
and quantum $\hat\psi_i^k$  (without averaging) variables lead to
the same equations.} That fact is a direct consequence of the
relations, showing that variations $\delta\Phi_l^k$  and
$\delta\hat\psi_l^k$  are multiplied by the same operator
multiplier:
\[
\begin{array}{c}
\displaystyle
\delta\sqrt{-\hat g}\hat g^{ik}=\sqrt{-\hat g}\hat g^{il}\delta\Phi_l^k\ ,\qquad
\hat\psi_i^k=const\ ,
\\[5mm]
\displaystyle
\delta\sqrt{-\hat g}\hat g^{ik}=\sqrt{-\hat g}\hat g^{il}\delta\hat\psi_l^k\ ,\qquad
\Phi_i^k=const\ .
\end{array}
\]
By a simple operation of subtraction, the identity allows the
extraction of pure background terms from the equation of quantum
field. The equations of graviton theory in the macroscopic space
with self--consistent geometry are written as follows:
\begin{equation}
\displaystyle  \langle\Psi|\hat E_i^k|\Psi\rangle\equiv
\langle\Psi|\hat G_i^k-\frac12\delta_i^k\hat G_l^l|\Psi\rangle=0\ ,
\label{2.38}
\end{equation}
\begin{equation}
\displaystyle \hat L_i^k\equiv
\hat G_i^k-\frac12\delta_i^k\hat G_l^l-\langle\Psi|\hat G_i^k-
\frac12\delta_i^k\hat G_l^l|\Psi \rangle=0\ .
\label{2.39}
\end{equation}

With the exponential parameterization, the formalism of the theory
can be expressed in an elegant form. Let us go to the rules of
differentiation of exponential matrix functions
\begin{equation}
\displaystyle \hat Y_{im}\hat X^{mk}_{\;\;\; ;l}=\hat
\psi_{i;l}^k\ ,\qquad \hat X^{ik}_{\;\;\; ;l}=\hat X^{im}\hat
\psi_{m\;\;\; ;l}^k\ . \label{2.40}
\end{equation}
Taking into account (\ref{2.40}), we get the quantum contribution
to the full connectivity  (\ref{2.33}) as follows
\begin{equation}
\displaystyle \hat{\mathcal{T}}_{ik}^l=\frac12\left(-\hat
\psi_{i;k}^l -\hat \psi_{k;i}^l+\hat Y_{kn}\hat
X^{lm}\hat\psi^n_{i;m}\right)+\frac14\left(\delta_i^l\hat\psi_{;k}+\delta_k^l\hat\psi_{;i}-
\hat Y_{ik}\hat X^{lm}\hat\psi_{;m}\right)\ . \label{2.41}
\end{equation}
Formulas (\ref{2.35}) could be rewritten as follows:
\begin{equation}
\begin{array}{c}
\displaystyle  \hat X^l_k=(\exp \hat\psi)_k^l,\qquad \sqrt{\hat d}=e^{\hat\psi/2}\ ,
\\[5mm]
\displaystyle S=\int
d^4x\sqrt{-g}\left({\mathcal{L}}_{grav}-\Lambda
e^{\hat\psi/2}-\frac{1}{4\varkappa}\hat
X_k^l\bar\theta^{;k}\theta_{; l}\right)\ ,
\\[5mm]
\displaystyle {\mathcal{L}}_{grav}=-\frac{1}{2\varkappa}\hat
X_k^lR^k_l+\frac{1}{8\varkappa}\hat X_k^l\left(\hat
\psi_n^{m;k}\hat \psi_{m;l}^n-\frac12\hat\psi^{;k}\hat\psi_{;l}
-2\hat\psi^{k;m}_n\hat\psi^n_{m;l}\right)\ .
\end{array}
\label{2.42}
\end{equation}
As is seen from (\ref{2.42}), for the exponential
parameterization, the non--polynomial structures of quantum theory
of gravity have been completely reduced to the factorized
exponents\footnote{We are using the standard definitions. Matrix
functions are defined by their expansion into power series as any
operator functions:
\[
   \hat U(\hat V)=\sum_nc_n{\hat V}^n\ .
\] .
The derivative of  $n$--th degrees of matrix by the same matrix is
defined as
 \[
\displaystyle \frac{\partial {\hat V}^n}{\partial \hat V}=n{\hat V}^{n-1}\ .
  \]
The derivative by numerical (non matrix) parameter $z$  is
\[
\displaystyle \frac{\partial {\hat V}^n}{\partial z}=n{\hat
V}^{n-1}\cdot\frac{\partial {\hat V}}{\partial z}\ .
\]
If matrix function  $ \hat U(\hat V)$ and its derivative  $\hat
W=\partial {\hat U}^n/\partial \hat V$ are elementary functions,
then
\[
\displaystyle \frac{\partial {\hat U}}{\partial z}=\hat
W\cdot\frac{\partial {\hat V}}{\partial z}\ .
\]
Formulas (\ref{2.40}) --- (\ref{2.42}) are the consequence of
these definitions. It worth to mention, that in matrix analysis in
all intermediate formulas one should be careful with the index
ordering.}.

The explicit form of the tensor, in the terms of which the
self--consistent system of equations could be written is as
follows
\begin{equation}
\begin{array}{c}
\displaystyle \hat E_i^k\equiv \hat G_i^k-\frac12\delta_i^k\hat
G_l^l=\hat X^{kl}{R_{li} -\frac12\delta_i^k\hat
X^{lm}}R_{ml}-\delta_i^k\varkappa\Lambda e^{\hat\psi/2}+
\\[5mm]
\displaystyle + \frac12\left[\hat X^{lm}\left(\hat\psi^k_{i;m}-
\hat\psi^k_{m;i}\right) - \hat X^{km} \hat
\psi^l_{i;m}+\frac12\delta_i^k\left( \hat X^{mn} \hat
\psi_{n;m}^l+ \hat X^{lm} \hat\psi_{m;n}^n\right)\right]_{;l}-
\\[5mm]
\displaystyle -\frac14\hat X^{kl}\left(\hat\psi^n_{m;i}
\hat\psi_{n;l}^m- \frac12\hat\psi_{;i}
\hat\psi_{;l}-2\hat\psi_{l;m}^n\hat
\psi_{n;i}^m\right)+\frac18\delta_i^k\hat
X^{rl}\left(\hat\psi^n_{m;r} \hat\psi_{n;l}^m-
\frac12\hat\psi_{;r} \hat\psi_{;l}-2\hat\psi_{l;m}^n\hat
\psi_{n;r}^m\right)-
\\[5mm]
\displaystyle + \frac{1}{4}\left[\hat X^{kl}\left(\bar
\theta_{;l}\theta_{;i}+\bar
\theta_{;i}\theta_{;l}\right)-\delta_i^k\hat
X^{ml}\bar\theta_{;m}\theta_{;l}\right]\ .
\end{array}
\label{2.43}
\end{equation}
Let us introduce the following notations:
\begin{equation}
\begin{array}{c}
\displaystyle \hat X_{(1)}^{ik}=\hat X^{ik}-
g^{ik}=\hat\psi^{ik}+\frac12\hat\psi^{il}\hat\psi_l^k+...\ ,
\\[3mm]
\displaystyle \hat X_{(2)}^{ik}=\hat X^{ik}-
g^{ik}-\hat\psi^{ik}=\frac12\hat\psi^{il}\hat\psi_l^k+...\ .
\end{array}
\label{2.44}
\end{equation}
With use of (\ref{2.44}), let us extract from (\ref{2.43}) the
terms not containing the quantum field, and the terms linear over
the quantum field:
\begin{equation}
\begin{array}{c}
\displaystyle \hat
E_i^k=R_i^k-\frac12\delta_i^kR-\delta_i^k\varkappa\Lambda
+\frac12\left(\hat\psi^{k\; ;l}_{i;l}-\hat\psi^{k\; ;l}_{l;i}
-\hat\psi^{l;k}_{i\;
;l}+\delta_i^k\hat\psi_{m;l}^{l;m}\right)+\hat\psi^k_lR_i^l-\frac12\delta_i^k\hat\psi^m_lR_m^l-
 \frac12\delta_i^k\varkappa\Lambda\hat\psi-\varkappa\hat T_i^k\ ,
\\[5mm]
\displaystyle \hat T_i^k=\hat T_{i(grav)}^k+\hat T_{i(ghost)}^k\ ,
\end{array}
\label{2.45}
\end{equation}
where
\begin{equation}
\begin{array}{c}
\displaystyle \varkappa\hat T_{i(grav)}^k=\frac14\hat
X^{kl}\left(\hat\psi^n_{m;i} \hat\psi_{n;l}^m-
\frac12\hat\psi_{;i} \hat\psi_{;l}-2\hat\psi_{l;m}^n\hat
\psi_{n;i}^m\right)-\frac18\delta_i^k\hat
X^{rl}\left(\hat\psi^n_{m;r} \hat\psi_{n;l}^m-
\frac12\hat\psi_{;r} \hat\psi_{;l}-2\hat\psi_{l;m}^n\hat
\psi_{n;r}^m\right)-
\\[5mm]
\displaystyle -\frac12\left[\hat
X_{(1)}^{lm}\left(\hat\psi^k_{i;m}- \hat\psi^k_{m;i}\right) - \hat
X_{(1)}^{km} \hat\psi^l_{i;m}+\frac12\delta_i^k\left( \hat
X_{(1)}^{mn} \hat\psi_{n;m}^l+ \hat X_{(1)}^{lm}
\hat\psi_{m;n}^n\right)\right]_{;l}-
\\[5mm]
\displaystyle -\hat X_{(2)}^{kl}{R_{li} +\frac12\delta_i^k\hat
X_{(2)}^{lm}}R_{ml}+\delta_i^k\varkappa\Lambda
\left(e^{\hat\psi/2}-1-\frac12\hat\psi\right)
\end{array}
\label{2.46}
\end{equation}
is the EMT of gravitons;
\begin{equation}
\begin{array}{c}
\displaystyle \varkappa\hat T_{i(ghost)}^k=-\frac{1}{4}\left[\hat
X^{kl}\left(\bar \theta_{;l}\theta_{;i}+\bar
\theta_{;i}\theta_{;l}\right)-\delta_i^k\hat
X^{ml}\bar\theta_{;m}\theta_{;l}\right]
\end{array}
\label{2.47}
\end{equation}
is the EMT of ghosts. In the averaging of (\ref{2.45}), it was
taken into account that  $\langle\Psi|\hat
\psi_i^k|\Psi\rangle\equiv 0$ by definition of the quantum field.
Averaged equations for the classic fields (\ref{2.38}) take form
of the standard Einstein equations containing averaged EMT of
gravitons, renormalized by ghosts:
\begin{equation}
\begin{array}{c}
\displaystyle \langle\Psi|\hat E_i^k|\Psi\rangle\equiv
R_i^k-\frac12\delta_i^kR-\delta_i^k\varkappa\Lambda
 -\varkappa\langle\Psi|\hat T_i^k|\Psi\rangle=0\ .
\end{array}
\label{2.48}
\end{equation}
Quantum dynamic equations for gravitons (\ref{2.39}) could be
rewritten as follows:
\begin{equation}
\begin{array}{c}
\displaystyle \hat L_i^k\equiv\frac12\left(\hat\psi^{k\;
;l}_{i;l}-\hat\psi^{k\; ;l}_{l;i} -\hat\psi^{l;k}_{i\;
;l}+\delta_i^k\hat\psi_{m;l}^{l;m}\right)+\hat\psi^k_lR_i^l-\frac12\delta_i^k\hat\psi^m_lR_m^l-
\frac12\delta_i^k\varkappa\Lambda\hat\psi-\varkappa\left(\hat
T_i^k-\langle\Psi|\hat T_i^k|\Psi\rangle\right)=0\ .
\end{array}
\label{2.49}
\end{equation}
As is seen in the equations (\ref{2.49}), in the theory of
gravitons all nonlinear effects are in the difference between the
EMT operator and its average value. System of equations
(\ref{2.48}), (\ref{2.49}) is closed by the quantum dynamic
equations for ghosts, which could be also written in 4D covariant
form:
\begin{equation}
\displaystyle (\hat X^{ik}\theta_{;k})_{;i}=0,\qquad (\hat X^{ik}\bar\theta_{;k})_{;i}=0
\label{2.50}
\end{equation}
Equations (\ref{2.50}) provide the realization of the conservative
nature of the ghosts' EMT:
\begin{equation}
\displaystyle \langle\Psi|\hat T_{i(ghost)}^k|\Psi\rangle_{;k}=0\ .
\label{2.51}
\end{equation}

\subsection{Differential Identities}\label{equiv}

In the exact theory, which is dealing with the full metric, there is an identity:
\begin{equation}
\displaystyle \hat D_k\left\{\hat g^{kl}\hat
R_{li}-\frac12\delta_i^k\hat g^{ml}\hat
R_{lm}-\delta_i^k\varkappa\Lambda+\frac14\left[\hat
g^{kl}\left(\bar \theta_{;l}\theta_{;i}+\bar
\theta_{;i}\theta_{;l}\right)-\delta_i^k\hat
g^{ml}\bar\theta_{;m}\theta_{;l}\right]\right\}=0\ , \label{2.52}
\end{equation}
where $\hat D_k$ is the covariant derivative in the space with
metric  $\hat g_{ik}$. This identity is satisfied  by Bianchi
identity and by the ghost equations of motion. In terms of
covariant derivative in the background space, identity
(\ref{2.52}) could be rewritten as follows:
\begin{equation}
\displaystyle \hat E_{i;k}^k-\frac12(\ln\hat d)_{;k}\hat
E_i^k+\hat{\mathcal{T}}_{kl}^k\hat
E_i^l-\hat{\mathcal{T}}_{ik}^l\hat E_l^k\equiv \hat
E_{i;k}^k-\hat{\mathcal{T}}_{ik}^l\hat E_l^k=0\ . \label{2.53}
\end{equation}
For the exponential parameterization, taking into account
(\ref{2.41}), the expression (\ref{2.53}) can be  transformed to
the following form
\begin{equation}
\displaystyle \hat E_{i;k}^k+\frac12\hat\psi^l_{k;i}\left(\hat
E_l^k-\frac12\delta_l^k\hat E_l^l\right)=0\ . \label{2.54}
\end{equation}
Identity transformation  $\hat E_i^k\equiv \langle\Psi|\hat
E_i^k|\Psi\rangle+\hat L_i^k$ and the subsequent averaging of
(\ref{2.54}) yields:
\begin{equation}
\displaystyle \langle\Psi|\hat
E_i^k|\Psi\rangle_{;k}+\frac12\langle\Psi|\hat\psi^l_{k;i}\left(\hat
L_l^k-\frac12\delta_l^k\hat L_m^m\right)|\Psi\rangle=0\ .
\label{2.55}
\end{equation}
Here we have used explicitly the fact that $\langle\Psi|\hat
\psi_i^k|\Psi\rangle\equiv 0$, $\langle\Psi|\hat
L_i^k|\Psi\rangle\equiv 0$,   by definition. Next, expression
(\ref{2.48}) is substituted into (\ref{2.55}). Taking into account
the Bianchi identity and the conservation of the ghost EMT, we
obtain:
\begin{equation}
\displaystyle \langle\Psi|\hat
T_{i(grav)}^k|\Psi\rangle_{;k}=\frac12\langle\Psi|\hat\psi^l_{k;i}\left(\hat
L_l^k-\frac12\delta_l^k\hat L_m^m\right)|\Psi\rangle\ .
\label{2.56}
\end{equation}
As is seen from (\ref{2.56}), quantum equations of motion
(\ref{2.49}) provide the conservation of the averaged EMT of
gravitons:
\begin{equation}
\displaystyle \langle\Psi|\hat T_{i(grav)}^k|\Psi\rangle_{;k}=0\ .
\label{2.57}
\end{equation}

Take notice, that tensors $\hat E_i^k$  and $\hat L_i^k$ in
(\ref{2.54}), (\ref{2.56}) are multiplied by the linear forms of
graviton field operators only. Such a structure of identities is
only valid for the exponential parameterization. This fact is of
key value for the computations in the framework of perturbation
theory. The order $n$   of the perturbation theory is defined by
the highest degree of the field operator in the quantum dynamic
equations for gravitons (\ref{2.49}). The EMT of gravitons which
is consistent with the quantum equation of order $n$ contains
averaged products of field operators of the order $n+1$ (e.g., the
quadratic EMT is consistent with the linear operator equation). We
see that by defining the order of the perturbation theory, we have
identity (\ref{2.56}), in which all terms are of the same maximal
order of the quantum field amplitude:
\begin{equation}
\displaystyle \langle\Psi|\hat
T_{i(grav)}^{k(n+1)}|\Psi\rangle_{;k}=\frac12\langle\Psi|\hat\psi^l_{k;i}\left(\hat
L^{k(n)}_l-\frac12\delta_l^k\hat L^{m(n)}_m\right)|\Psi\rangle\ .
\label{2.58}
\end{equation}
Such a structure of the identity automatically provides the
conservation condition (\ref{2.57}) at any order of perturbation
theory\footnote{In the framework of the perturbation theory, any
parameterization, except the exponential one, creates
mathematically contradictory models, in which the perturbative EMT
of gravitons  $\langle\Psi|\hat T_{i(grav)}^{k(n+1)}|\Psi\rangle$
is not conserved. In our opinion, a discussion of artificial
methods of solutions of this problem, appeared, for example, if
linear parameterization $g_{ik}=g_{ik}+\hat\psi_{ik}$ is used,
makes no sense. The algorithm we have suggested here is well
defined because it is based on the exact procedure of separation
between the classical and quantum variables in terms of normal
coordinates. We believe there is no other mathematically
non--contradictive scheme. }.

\subsection{One--Loop Approximation}\label{1loop}

In the framework of one--loop approximation, quantum fields
interact only with the classic gravitational field. Accordingly,
equations (\ref{2.49}) are being converted into linear operator
equations:
\begin{equation}
\begin{array}{c}
\displaystyle
        \hat L_i^k=\frac12\left( \hat\psi_{i;l}^{k;l}
         -\hat\psi_{i;\; l}^{l;k}
         -\hat\psi_{l;i}^{k;\; l}
         +\delta_i^k\hat\psi_{l;m}^{m;l}\right)
                  +\hat\psi_l^kR_i^l
         -\frac12\delta_i^k\hat\psi_m^lR_l^m-\frac12\delta_i^k\varkappa\Lambda\hat\psi
 =0\ .
\end{array}
\label{2.59}
\end{equation}
Of course, these equations are separated into the equations of
constraints (initial conditions):
\begin{equation}
\displaystyle \hat L_0^0|\Psi\rangle=0,\qquad \hat
L_0^\alpha|\Psi\rangle=0,\qquad \hat L_\alpha^0|\Psi\rangle=0\ ,
\label{2.60}
\end{equation}
and the equations of motion:
\begin{equation}
\displaystyle \hat L_\alpha^\beta-\frac12\delta_\alpha^\beta\hat L_l^l=0\ .
\label{2.61}
\end{equation}
The equations for ghosts (\ref{2.50}) are also transformed into the linear operator equations:
\begin{equation}
\begin{array}{c}
\displaystyle
         \theta_{;i}^{;i}=0\ ,\qquad \bar\theta_{;i}^{;i}=0\ .
\end{array}
\label{2.62}
\end{equation}

In the one--loop approximation, the state vector is represented as
a product of normalized state vectors of gravitons and ghosts:
\begin{equation}
\displaystyle |\Psi\rangle=|\Psi_g\rangle|\Psi_{gh}\rangle\ .
 \label{2.63}
 \end{equation}
Equations for macroscopic metric (\ref{2.48}) take the form:
\begin{equation}
\displaystyle
  R_i^k -\frac{1}{2}\delta_i^k R =
  \varkappa\left(\langle \Psi_g|\hat{T}_{i(grav)}^k|\Psi_g\rangle+
  \langle \Psi_{gh}|\hat{T}_{i(ghost)}^k|\Psi_{gh}\rangle +\delta_i^k\Lambda\right)\ .
\label{2.64}
\end{equation}
The averaged EMTs of gravitons and ghosts in equations
(\ref{2.64}) are the quadratic forms of the quantum fields.
Assuming that $\hat X^{ik}=g^{ik}$, $\hat X^{ik}_{(1)}=\hat
\psi^{ik}$, $\hat X^{ik}_{(2)}=\hat \psi^{il}\hat \psi_l^k/2$  in
(\ref{2.46}), (\ref{2.47}), we obtain:
\begin{equation}
\begin{array}{c}
\displaystyle
  \hat{T}_{i(grav)}^k = \frac{1}{4\varkappa}\left\{

        \hat\psi_{m;i}^l\hat\psi_l^{m;k}
        -\frac{1}{2}\hat\psi_{;i}\hat\psi^{;k}
        -\hat\psi_{i;m}^l\hat\psi_l^{m;k}  -\hat\psi_l^{k;m}\hat\psi_{m;i}^l
  -\frac{1}{2}\delta_i^k
  \left(
        \hat\psi_{m;n}^l\hat\psi_l^{m;n}
        -\frac{1}{2}\hat\psi_{;n}\hat\psi^{;n}
        -2\hat\psi^l_{n;m}\hat\psi_l^{m;n}
  \right)-\right.
\\[3mm] \\ \displaystyle
 \left. -
  2\left[
        \hat\psi_m^l\hat\psi_i^{k;m}
        -\hat\psi_m^k\hat\psi_i^{l;m}
        -\hat\psi^{lm}\hat\psi_{m;i}^k
         +\frac12\delta_i^k\left(\hat\psi_m^n\hat\psi_n^l\right)^{;m}
        \right]_{;l}
                -2\hat\psi_m^k\hat\psi_l^mR_i^l
        +\delta_i^k\hat\psi_l^n\hat\psi_n^mR_m^l+\frac12\delta_i^k\varkappa\Lambda\hat\psi^2
  \right\}\ ,
  \end{array}
\label{2.65}
\end{equation}

\begin{equation}
\displaystyle \hat
T_{i(ghost)}^{k}=-\frac{1}{4\varkappa}\left(\bar\theta_{;i}\theta^{;k}+\bar\theta^{;k}\theta_{;i}-
\delta_i^k\bar\theta^{;l}\theta_{;l}\right)\ . \label{2.66}
\end{equation}
Quantum equations (\ref{2.59}), (\ref{2.62}) provide the
conservation of tensors (\ref{2.65}), (\ref{2.66}) in the
background space:
\begin{equation}
\displaystyle \langle \Psi_g|\hat{T}_{i(grav)}^k|\Psi_g\rangle_{
;\ k}=0, \qquad
  \langle \Psi_{gh}|\hat{T}_{i(ghost)}^k|\Psi_{gh}\rangle_{;\ k}=0\ .
 \label{2.67}
 \end{equation}

The ghost sector of the theory (\ref{2.59}) --- (\ref{2.67})
corresponds to the gauge (\ref{2.29}). Note, however, that all
equations of the theory, except gauges, are formally general
covariant in the background space. That provides a way of
expanding the class of gauges for classic fields. Obviously, we
can move from the initial 4--coordinates, corresponding to the
classic sector of gauges (\ref{2.29}), to any other coordinates,
conserving quantum gauge condition
\begin{equation}
\displaystyle \hat\psi_0^i|\Psi\rangle=0\ .
 \label{2.68}
\end{equation}
It is not difficult to see, that in the classic sector any gauges
of synchronous type are allowed:
\begin{equation}
\displaystyle g_{00}=N^2(t), \qquad g_{0\alpha}=0\ .
 \label{2.69}
\end{equation}
where $N(t)$ is an arbitrary function of time.

An important technical detail is that in the perturbation theory
the graviton field should be consistent with an additional
identity. In one--loop approximation that identity is obtained
from the covariant differentiation of equation (\ref{2.59}):
 \begin{equation}
\displaystyle  \hat Q_i\equiv\left(R_l^k+\varkappa\Lambda\delta_l^k\right)\hat\psi_{k;i}^l=0\ .
 \label{2.70}
\end{equation}
The appearance of conditions (\ref{2.70}) reflects the fact that
we are dealing with an approximate theory. As it was already
mentioned in Section \ref{fac}, the partition of the metric into
classic and quantum components, and, respectively, the
factorization of the path integral, can be only done under the
condition that additional constrains are applied to the geometry
of background space. These constrains are manifested through the
structure of the Ricci tensor of the background space which should
provide the identity (\ref{2.70}) for the solutions of dynamic
equations for gravitons. In the Heisenberg form of quantum theory
the additional identity can be written as conditions on the state
vector:
\begin{equation}
\displaystyle  \hat
Q_i|\Psi\rangle\equiv\left(R_l^k+\varkappa\Lambda\delta_l^k\right)\hat\psi_{k;i}^l|\Psi\rangle=0\
.
 \label{2.71}
\end{equation}
Status of all constrains for the state vectors are the same and
are as follows. If (\ref{2.60}), (\ref{2.68}), (\ref{2.71}) exist
at the initial moment of time, the internal properties of the
theory should provide their existence at any following instance of
time.

While one is conducting a concrete one--loop calculation, there is
a problem of gauge invariance of the total EMT of gravitons and
ghosts. As was mentioned by De Witt \cite{7}, after the separation
of the metric into background and graviton components, the
transformations of the diffeomorphism group (\ref{2.4}) can be
represented as transformations of the internal gauge symmetry of
graviton field. In the framework of one--loop approximation, these
transformations are as follows:
\begin{equation}
 \displaystyle
\delta\hat\psi_i^k=-\delta_i^k\eta^l_{;l}+\eta^k_{;i}+\eta_i^{;k} \ .
 \label{2.72}
 \end{equation}
The problem of gauge non--invariance is twofold. First, the EMT of
gravitons (\ref{2.65}) is not invariant with respect to
transformations in (\ref{2.72}). Second, the ghost sector (the
ghost EMT), inevitably presented in the theory, depends on the
gauge. Concerning the first problem, it is known that the
operation removing gauge non--invariant terms from the EMT of
gravitons belongs to the operation of averaging over a quantum
ensemble. In the general case of arbitrary background geometry and
arbitrary graviton wavelengths we encounter a number of problems
(when conducting this operation), which should be discussed
separately.

In the particular case of the theory of gravitons in a homogeneous
and isotropic Universe, the averaging problem has a consistent
mathematical solution. It was shown in Section \ref{3vs} that {\it
removing the gauge non--invariant contributions from the EMT of
gravitons from the quantum ensemble has been set
gauge--invariantly.} To address the second aspect of the problem,
we should take into account that the theory of gravitons in the
macroscopic space with the self--consistent geometry operates with
macroscopic observables. Therefore, in this theory one--loop
finiteness, as the general property of one--loop quantum gravity,
should have a specific embodiment: {\it by their mathematical
definition, macroscopic observables must be the finite values.}
This requirement on the theory is realized in {\it the class of
allowable gauges of full metric consistent with the hypothesis of
the existence of macroscopic space and macroscopic observables}
(see Section \ref{fininv}). Gauge (\ref{2.29}) used above belongs
to this class.

\section{Self--Consistent Theory of Gravitons in the Isotropic Universe}\label{scgt}

\subsection{Elimination of 3--Vector and 3--Scalar Modes
by Conditions Imposed on the State Vector}\label{3vs}

We consider the quantum theory of gravitons in the spacetime with
the following background metric
\begin{equation}
 \displaystyle
 ds^2=g_{ik}dx^idx^k=N^2(t)dt^2-a^2(t)(dx^2+dy^2+dz^2)\ .
 \label{3.1}
 \end{equation}
In this space the graviton field is expanded over the irreducible
representations of the group of three--dimensional rotations, i.e.
over 3--tensor $\hat\psi_{\alpha(t)}^\beta$, 3--vector
$\hat\psi_{i(v)}^k=(\hat\psi_{0(v)}^\alpha,\,
\hat\psi_{\alpha(v)}^\beta)$  and 3--scalar
$\hat\psi_{i(s)}^k=(\hat\psi_{0(s)}^0,\, \hat\psi_{0(s)}^\alpha,\,
\hat\psi_{\alpha(s)}^\beta)$ modes. Equations (\ref{2.59}) are
split  into three independent systems of equations, so that each
of such systems represents each mode separately. The state vector
of gravitons is of multiplicative form that reads
 \[
 \displaystyle
|\Psi_g\rangle=|\Psi_t\rangle|\Psi_v\rangle|\Psi_s\rangle\ .
 \]
The averaged EMT (\ref{2.63}) is presented by an additive form that reads:
\begin{equation}
 \displaystyle
\langle \Psi_g|\hat T_{i(grav)}^k|\Psi_g\rangle= \langle
\Psi_t|\hat T_{i(t)}^k|\Psi_t\rangle +
 \langle \Psi_v|\hat T_{i(v)}^k|\Psi_v\rangle+\langle \Psi_s|\hat T_{i(s)}^k|\Psi_s\rangle \ .
 \label{3.2}
 \end{equation}
The averaged EMT contains no products of modes that belong to
different irreducible representations. This is because the
equality $\langle \Psi_g|\hat\psi_i^k|\Psi_g\rangle=0$ is divided
into three following three independent equalities
\begin{equation}
\displaystyle
\langle \Psi_s|\hat\psi_{i(s)}^k|\Psi_s\rangle=0,\qquad
\langle \Psi_v|\hat\psi_{i(v)}^k|\Psi_v\rangle=0, \qquad
\langle \Psi_t|\hat\psi_{i(t)}^k|\Psi_t\rangle=0\ .
\label{3.3}
\end{equation}
Equalities (\ref{3.3}) are conditions that provide the consistency
of properties of quantum ensemble of gravitons with the properties
of homogeneity and isotropy of the background. In the homogeneous
and isotropic space, the same equalities hold for Fourier images
of the graviton field. Therefore, the satisfaction of these
equalities is provided by the isotropy of graviton spectrum in the
${\bf k}$--space and by the equivalence of different
polarizations.

3--tensor modes $\hat\psi_{\alpha(t)}^\beta$ and their EMT
$\langle\Psi_t|\hat T_{i(t)}^k|\Psi_t\rangle$, respectively, are
gauge invariant objects. Gauge non--invariant modes
$\hat\psi_{i(v)}^k,\, \hat\psi_{i(s)}^k$ are eliminated by
conditions that, imposed on the state vector, read
 \begin{equation}
 \displaystyle
\hat\psi_{i(v)}^k|\Psi_v\rangle=\hat\psi_{i(s)}^k|\Psi_s\rangle=0\ .
 \label{3.4}
 \end{equation}
Note that the conditions (\ref{3.4}) automatically follow from
equations (\ref{2.59}) and conditions (\ref{2.66}). As a result of
this, a gauge non--invariant EMT of 3--scalar and 3--vector modes
is eliminated from the macroscopic Einstein equations, and we get
\begin{equation}
\displaystyle \langle \Psi_v|\hat T_{i(v)}^k|\Psi_v\rangle=0\ ,\qquad
\langle \Psi_s|\hat T_{i(s)}^k|\Psi_s\rangle=0\ .
 \label{3.5}
 \end{equation}

The important fact is that {\it in the isotropic Universe, the
separation of gauge invariant EMT of 3--tensor gravitons is
accomplished without the use of short--wave approximation}. In
connection with this, note the following fact. In the theory,
which formally operates with waves of arbitrary lengths, the
problem of existence of a quantum ensemble of waves with
wavelengths greater than the distance from horizon is open
\cite{11}. In cosmology, the existence of such an ensemble is
provided by the following experimental fact. In the real Universe
(whose properties are controlled by observational data beginning
from the instant of recombination), the characteristic scale of
casually--connected regions is much greater (many orders of
magnitude) than the formal horizon of events. The standard
explanation of this fact is based on the hypothesis of early
inflation. Taking into account these circumstances, we do not
impose any additional restrictions on the quantum ensemble.

The procedure described above is based on the existence of
independent irreducible representations of graviton modes only.
But in this procedure, gauge--non--invariant modes are eliminated
by using of a gauge, i.e. they are eliminated by using of
gauge--non--invariant procedures. The gauge--invariant procedure
of getting the same results is presented below.

\subsubsection{Elimination of Scalar Modes}

We consider equations (\ref{2.59}) with the (\ref{3.1}) background
using the conformal time: $N=a,\; Ndt\to a(\eta)d\eta$ . (Symbol
"$t$"\ belongs now to the physical time, for which $N=1$). The
metric of the 3D flat isotropic Universe is conformally similar to
Minkowski's metric, and it reads
\begin{equation}
 \displaystyle
ds^2=a^2(\eta)\left(d\eta^2-dx^2-dy^2-dz^2\right)=a^2(\eta)\bar g_{ik}dx^idx^k\ .
 \label{3.6}
 \end{equation}
Let us introduce the new variables that can be interpreted as
quantum fluctuations of covariant metric
\begin{equation}
 \displaystyle \hat h_i^k=-\hat\psi_i^k+\frac12\delta_i^k\psi\ .
 \label{3.7}
 \end{equation}
In terms of variables (\ref{3.7}), the equations (\ref{2.59}) read
(after calculations of covariant derivatives and Ricci tensor
components in the (\ref{3.6}) metric):
\begin{equation}
\begin{array}{c}
 \displaystyle
a^2\hat L_i^k=\frac12\left[\hat h_{i,l}^{l,k}+\hat
h_{l,i}^{k,l}-\hat h_{i,l}^{k,l}-\hat h_{,i}^{,k}-\delta_i^k(\hat
h_{m,l}^{l,m}-\hat h_{,l}^{,l})\right]+ \frac{a'}{a}n^l\left[\hat
h_{l,i}^k+\hat h_{li}^{\;\; ,k}-\hat h_{i,l}^k-\delta_i^k(2\hat
h_{l,m}^m-\hat h_{,l})\right]+
\\[5mm]
\displaystyle +
2\left(\frac{a''}{a}-2\frac{a'^2}{a^2}\right)\left(n_in^l\hat
h_l^k-\frac12n_in^k\hat h\right)-
\delta_i^k\left(2\frac{a''}{a}-\frac{a'^2}{a^2}\right)\left(n_mn^l\hat
h_l^m-\frac12\hat h\right)-\frac12\delta_i^k\varkappa\Lambda
a^2\hat h
 =0\ ,
\end{array}
 \label{3.8}
 \end{equation}
where
\[
\displaystyle n^i=n_i=(1,\; 0,\; 0,\; 0)\ .
\]
In equation (\ref{3.8}), operations with indexes are defined in
the Minkowski space; comas mean derivatives of metric fluctuations
over the coordinates of the Minkowski space; dashes mean
derivatives of scale factor over the conformal time  $\eta$. In
equation (\ref{3.8}) and further on in this Section, we do not
impose any gauge on metric fluctuations.

As was shown by Lifshitz in 1946 \cite{13}, fluctuations of metric
can be expanded in Fourier series over 3--scalar, transverse
3--vector and transverse 3--tensor plane waves. Projections of
general equations (\ref{3.9}) onto scalar, vector and tensor basis
functions lead to three independent systems of equations --- for
each type of mode separately. Fourier images of scalar
fluctuations of metric and parameters of gauge transformations are
defined as follows
\begin{equation}
\begin{array}{c}
\displaystyle  \hat h_0^0({\bf k})=\hat\varphi_{\bf k},\qquad \hat
h^0_\alpha({\bf k})=-\hat h_0^\alpha({\bf k})=-ik_\alpha
\hat\chi_{\bf k}\ ,\qquad \hat h_\alpha^\beta({\bf
k})=\frac{\delta_\alpha^\beta}{3}\left(\hat\mu_{\bf
k}+\hat\lambda_{\bf k}\right)-\frac{k_\alpha
k^\beta}{k^2}\hat\lambda_{\bf k}
\\[5mm]
\displaystyle  a^{-2}\eta_0({\bf k})=\eta^0({\bf k})=\omega_{\bf
k},\qquad -a^2\eta_\alpha({\bf k})=\eta^\alpha({\bf
k})=-ik^\alpha\nu_{\bf k}  \ .
\end{array}
 \label{3.9}
 \end{equation}
All operations with space indexes of vector--tensor basis in
equations (\ref{3.9}) and further on are conducted with the Euclid
metric. Gauge transformations (\ref{2.69}) for Fourier images of
scalar fluctuations read
\begin{equation}
\begin{array}{c}
\displaystyle \hat\varphi_{\bf k}\to \hat\varphi_{\bf
k}-2\left(\omega'_{\bf k}+\frac{a'}{a}\omega_{\bf k}\right),
\qquad
 \hat\chi_{\bf k}\to \hat\chi_{\bf k}-(\nu'_{\bf k}+\omega_{\bf k})\ ,
 \\[5mm]
\displaystyle
\hat\mu_{\bf k}\to \hat\mu_{\bf k} -6\frac{a'}{a}\omega_{\bf k}+2k^2\nu_{\bf k},\qquad
\hat\lambda_{\bf k}\to \hat\lambda_{\bf k} -2k^2\nu_{\bf k}\ .
\end{array}
 \label{3.10}
 \end{equation}
For brevity, the following notation is used below:
\[
\displaystyle \hat N_{\bf k}=\hat\mu_{\bf k}+\hat\lambda_{\bf
k},\qquad \hat M_{\bf k}=\hat\mu'_{\bf k}+2k^2\hat\chi_{\bf k},
\qquad \hat\psi_{\bf k}=\hat\varphi_{\bf k}+\hat\mu_{\bf k} \ .
\]
There are two liner combinations of Fourier images of metric
fluctuations that are invariant with respect to transformations
(\ref{3.10}) \cite{14}, which is an important sequence of the
theory. They read
\begin{equation}
\begin{array}{c}
\displaystyle \hat J_{\bf k}=k^2\hat\varphi_{\bf
k}+\frac{1}{a}\left[a\left(\hat N'_{\bf k}-\hat M_{\bf
k}\right)\right]', \qquad \hat J_{\bf k}\to \hat J_{\bf k}\ ,
  \\[5mm]
\displaystyle \hat I_{\bf k}=\frac{k^2}{3}\hat N_{\bf
k}+\frac{a'}{a}\left(\hat N'_{\bf k}-\hat M_{\bf k}\right), \qquad
\hat I_{\bf k}\to \hat I_{\bf k}\ ,
\end{array}
 \label{3.11}
 \end{equation}

The Fourier image of the equation of motion (\ref{2.61}) is
expanded over the tensor basis. It reads
\begin{equation}
\begin{array}{c}
\displaystyle \hat L_\alpha^\beta({\bf
k})-\frac12\delta_\alpha^\beta\hat L_l^l=\delta_\alpha^\beta\hat
L_1({\bf k})+\frac{k\alpha k^\beta}{k^2}\hat L_2({\bf k})=0\ .
\end{array}
 \label{3.12}
 \end{equation}
In accordance with (\ref{3.12}), we obtain
 \begin{equation}
\begin{array}{c}
\displaystyle \hat L_1({\bf k})=-\frac16\left(N''_{\bf
k}+k^2N_{\bf k}\right)-\frac{a'}{a}\left[\frac13N'_{\bf
k}+\frac12\left(M_{\bf k}- \varphi'_{\bf k}\right)\right]+
\\[5mm]\displaystyle
+
\left(\frac{a''}{a}+\frac{a'^2}{a^2}\right)\varphi_{\bf k}
-\frac12\left(\frac{a''}{a}+\frac{a'^2}{a^2}-\varkappa\Lambda a^2\right)\psi_{\bf k}=0\ ,
\end{array}
 \label{3.13}
 \end{equation}
\begin{equation}
\begin{array}{c}
\displaystyle \hat L_2({\bf k})=\frac12\left(N''_{\bf k}-M'_{\bf
k}-k^2\varphi_{\bf k}\right)-\frac16k^2N_{\bf
k}+\frac{a'}{a}\left(N'_{\bf k}-M_{\bf k}\right)=0\ .
\end{array}
 \label{3.14}
 \end{equation}
To eliminate gauge--non--invariant scalar fluctuations, it is
necessary to prove the existence of such initial conditions that
are independent of gauge and fix the only trivial solution of
equations (\ref{3.13}), (\ref{3.14}).

As it was mention above, the initial conditions must contain the
equations of constrains.  After Fourier transformations, primary
constrains can be written as follows:
\begin{equation}
\displaystyle \left[\frac{k^2}{3}\hat N_{\bf k}+\frac{a'}{a}\hat
M_{\bf k}-3\frac{a'^2}{a^2}\hat\varphi_{\bf
k}+\frac12\left(3\frac{a'^2}{a^2}-\varkappa\Lambda
a^2\right)\hat\psi_{\bf k}\right]|\Psi_s\rangle=0\ , \label{3.15}
\end{equation}
\begin{equation}
\begin{array}{c}
\displaystyle \left[\frac13\hat N'_{\bf k}-\frac{a'}{a}\hat
\varphi_{\bf k}\right]|\Psi_s\rangle=0\ , \qquad \left[\frac13\hat
N'_{\bf k}-\frac{a'}{a}\hat\varphi_{\bf
k}+2\left(2\frac{a''}{a}-\frac{a'^2}{a^2}\right)\hat\chi_{\bf
k}\right]|\Psi_s\rangle=0\ ,
\end{array}
\label{3.16}
\end{equation}
The condition (\ref{3.15}) is  $\hat L_0^0({\bf k})=0$, and
(\ref{3.16}) conditions are obtained from  $\hat L_\alpha^0({\bf
k})=0$ and $\hat L^\alpha_0({\bf k})=0$. Now, equations  $\hat
Q_0|\Psi\rangle=0$ and $\hat Q_\alpha|\Psi\rangle=0$ obtained from
(\ref{2.71}) are also included in the number of initial
conditions:
\begin{equation}
\displaystyle
 \left[2\left(\frac{a''}{a}-2\frac{a'^2}{a^2}\right)\varphi'_{\bf k}+
\left(\frac{a'^2}{a^2}-2\frac{a''}{a}+\varkappa\Lambda
a^2\right)\psi'_{\bf k}\right]|\Psi_s\rangle=0\ , \label{3.17}
\end{equation}
\begin{equation}
\displaystyle
\left[2\left(\frac{a''}{a}-2\frac{a'^2}{a^2}\right)\left(\varphi_{\bf
k}+2\frac{a'}{a}\chi_{\bf k}\right)+
\left(\frac{a'^2}{a^2}-2\frac{a''}{a}+\varkappa\Lambda
a^2\right)\psi_{\bf k}\right]|\Psi_s\rangle=0\ . \label{3.18}
\end{equation}
All constrains (\ref{3.15}) --- (\ref{3.18}) are contained in the
original non--gauged equations of the theory. It means that their
mathematical structure is independent of the choice of gauge.
Thus, {\it any quantum ensemble of gravitons in the isotropic
Universe must satisfy to (\ref{3.15}) --- (\ref{3.18}).} Of
course,  initial conditions cannot be fully defined by these
constrains because the quantum ensemble is still not actually
defined. What is actually defined at this point, are ensemble's
properties that follow from the isotropy of background. The full
determination of ensemble properties can be done by imposition of
gauge. Such a procedure is gauge non--invariant, and this is the
reason why it was disputable for many years. (For discussion of
the problem of gauge invariant description of scalar fluctuations
see, e.g. \cite{Ishi}.)

To solve this problem, one needs to use the (\ref{3.11}) invariant
and to impose the following {\it gauge invariant conditions on the
state vector}
\begin{equation}
\begin{array}{c}
\displaystyle \hat I_{\bf k}|\Psi_s\rangle\equiv
\left[\frac{k^2}{3}\hat N_{\bf k}+\frac{a'}{a}\left(\hat N'_{\bf
k}-\hat M_{\bf k}\right)\right]|\Psi_s\rangle=0\ ,
\end{array}
 \label{3.19}
 \end{equation}
\begin{equation}
\begin{array}{c}
\displaystyle \hat J_{\bf
k}|\Psi_s\rangle\equiv\left[k^2\hat\varphi_{\bf
k}+\frac{1}{a}\left[a\left(\hat N'_{\bf k}-\hat M_{\bf
k}\right)\right]'\right]|\Psi_s\rangle=0\ .
 \end{array}
 \label{3.20}
 \end{equation}
The relation (\ref{3.19}) can be immediately used as an initial
condition because it does not contain higher derivatives. Higher
derivatives also can be excluded from (\ref{3.20}) by non--gauged
equation (\ref{3.14}) via a simple algebraic procedure. As a
result of these operations, one gets the last initial condition
that reads
 \begin{equation}
\begin{array}{c}
\displaystyle \left[2k^2\hat\varphi_{\bf k}+\frac{k^2}{3}\hat
N_{\bf k}-\frac{a'}{a}\left(\hat N'_{\bf k}-\hat M_{\bf
k}\right)\right]|\Psi_s\rangle=0\ .
\end{array}
 \label{3.21}
 \end{equation}

Thus, we have initial conditions that are presented by the closed
system of algebraic equations (\ref{3.15}) --- (\ref{3.19}) and
equation (\ref{3.21})  with respect to Fourier images of metric
and their first derivatives over time. Because of its homogeneity,
this system of equations has the only trivial solution that reads
\begin{equation}
\displaystyle \hat N_{\bf k}|\Psi_s\rangle=\hat N'_{\bf
k}|\Psi_s\rangle=\hat M_{\bf k}|\Psi_s\rangle=\hat\chi_{\bf
k}|\Psi_s\rangle=\hat\psi_{\bf k}|\Psi_s\rangle=\hat\varphi_{\bf
k}|\Psi_s\rangle=\hat\varphi'_{\bf k}|\Psi_s\rangle=0\ .
\label{3.22}
\end{equation}
The substitution of (\ref{3.22}) into the equations of motion
(\ref{3.13}), (\ref{3.14}) shows that higher derivatives are also
zeroes at the initial instance of time, i.e.
\begin{equation}
\displaystyle \hat N''_{\bf k}|\Psi_s\rangle=\hat M'_{\bf k}|\Psi_s\rangle=0\ .
\label{3.23}
\end{equation}
It follows from (\ref{3.23}), that conditions (\ref{3.22}) defined
at the initial instance of time are valid for any future instances
of time. Thus, scalar fluctuations are excluded from the theory.
It is important to emphasize that gauge was not used in the
procedure that was described above. {\it Scalar fluctuations are
eliminated if gauge invariant conditions, which are imposed on the
state vector, are added to the equations of constrains that are
already contained in the theory itself.}

\subsubsection{Elimination of Vector Modes}

Vector fluctuations and vector parameters of gauge transformations
are presented by their Fourier images after their expansion over
3--transversal plain waves. They read:
\begin{equation}
\begin{array}{c}
\displaystyle  \hat h_0^0({\bf k})=0,\qquad \hat h^0_\alpha({\bf
k}\lambda)=-\hat h_0^\alpha({\bf k}\lambda)=-S_\alpha({\bf
k}\lambda) \hat\chi_{{\bf k}\lambda}\ ,\qquad
  \hat h_\alpha^\beta({\bf k}\lambda)=i\left[k_\alpha S^\beta({\bf k}\lambda)+
  k^\beta S_\alpha({\bf k}\lambda)\right]\hat u_{{\bf k}\lambda}\ ,
  \\[5mm]
\displaystyle -a^2\eta_\alpha({\bf k}\lambda)=\eta^\alpha({\bf
k}\lambda)=-S^\alpha({\bf k}\lambda)\nu_{{\bf k}\lambda}\ , \qquad
k_\alpha S^\alpha({\bf k}\lambda)\equiv 0\ .
\end{array}
 \label{3.24}
 \end{equation}
In (\ref{3.24}) and further, $\lambda$ is the index of
polarization of vector modes. Gauge transformations read
\begin{equation}
\displaystyle \label{3.25}  \hat\chi_{{\bf k}\lambda}\to
\hat\chi_{{\bf k}\lambda}-\nu'_{{\bf k}\lambda}\ , \qquad \hat
u_{{\bf k}\lambda}\to \hat u_{{\bf k}\lambda}-\nu_{{\bf
k}\lambda}\ .
 \end{equation}
There exists the linear superposition of Fourier images which is
invariant with respect to (\ref{3.25}) transformations. It reads
\begin{equation}
\displaystyle
 \hat I_{{\bf k}\lambda}=\hat u'_{{\bf k}\lambda}-\hat\chi_{{\bf k}\lambda}\ .
 \label{3.26}
\end{equation}
Primary constrains $\hat L_\alpha^0|\Psi_v\rangle=0$,  $\hat
L^\alpha_0|\Psi_v\rangle=0$ and additional constrains  $\hat
Q_\alpha|\Psi_v\rangle=0$ generate the following initial
conditions for the state vector
\begin{equation}
\begin{array}{c}
\displaystyle k^2\hat I_{{\bf k}\lambda}|\Psi_v\rangle=0\ ,\qquad
\left[k^2\hat I_{{\bf
k}\lambda}+4\left(\frac{a''}{a}-2\frac{a'^2}{a^2}\right)\hat\chi_{{\bf
k}\lambda}\right]|\Psi_v\rangle=0\ ,
\\[5mm]
\displaystyle
\frac{a'}{a}\left(\frac{a''}{a}-2\frac{a'^2}{a^2}\right)\hat\chi_{{\bf
k}\lambda}|\Psi_v\rangle=0\ .
\end{array}
\label{3.27}
\end{equation}
The equation of motion  $\hat L_\alpha^\beta=0$ contains only the
invariant in Eq. (\ref{3.26}). It is integrated and reads
\begin{equation}
\begin{array}{c}
\displaystyle
\hat I'_{{\bf k}\lambda}+\frac{a'}{a}\hat I_{{\bf k}\lambda}=0\ ,
\qquad \hat I_{{\bf k}\lambda}=\frac{C_{{\bf k}\lambda}}{a}\ .
\end{array}
\label{3.28}
\end{equation}
According to equations (\ref{3.27}) and (\ref{3.28}), the
following conditions are imposed on the state vector at the
initial instant of time
\[
\displaystyle \hat u_{{\bf k}\lambda}|\Psi_v\rangle= \hat u'_{{\bf
k}\lambda}|\Psi_v\rangle= \hat\chi_{{\bf
k}\lambda}|\Psi_v\rangle=0\ .
\]
They are satisfied at any further instant of time. As it can be
seen from the above consideration, to eliminate vector
fluctuations it is sufficient to take into account only the
constrains that exist in the equations of theory.

\subsection{Canonical Quantization of 3--Tensor Gravitons and Ghosts}\label{quant}

The parameters of gauge transformations do not contain terms of
expansion over transverse 3--tensor plane waves. Therefore,
Fourier images of tensor fluctuations are gauge--invariant by
definition. We have
\begin{equation}
\begin{array}{c}
\displaystyle  \hat h_0^0({\bf k})=0,\qquad \hat h^0_\alpha({\bf
k})=-\hat h_0^\alpha({\bf k})=0\ ,\qquad \hat h_\alpha^\beta({\bf
k}\sigma)=-\hat\psi_\alpha^\beta({\bf
k}\sigma)=-Q_\alpha^\beta({\bf k}\sigma)\hat\psi_{{\bf k}\sigma}\
,
  \\[5mm]
\displaystyle k_\alpha Q^\alpha_\beta({\bf k}\sigma)\equiv 0\
,\qquad Q^\alpha_\alpha({\bf k}\sigma)\equiv 0\ ,
\end{array}
 \label{3.29}
 \end{equation}
where  $\sigma$ is the index of transverse polarizations. The
operator equation for 3--tensor gravitons is
 \begin{equation}
 \displaystyle
\psi_{\alpha(t)}^\beta(t,{\bf x})=\sum_{{\bf
k}\sigma}Q_\alpha^\beta({\bf k}\sigma)\psi_{{\bf
k}\sigma}(t)e^{i{\bf kx}}\ , \qquad \ddot \psi_{{\bf
k}\sigma}+3H\dot\psi_{{\bf
k}\sigma}+\frac{k^2}{a^2}\psi_{{\bf k}\sigma}=0\ ,
 \label{3.30}
 \end{equation}
where $H=\dot a/a$  is Hubble function and dots mean derivatives
with respect to the physical time $t$.

The special property of the gauge used is the following. The
differential equation for ghosts is obtained from the equation for
gravitons by exchange of graviton operator with the ghost
operator. It reads
\begin{equation}
 \displaystyle
\theta(t,{\bf x})=\sum_{{\bf k}}\theta_{{\bf k}}(t)e^{i{\bf kx}}\
, \qquad \ddot \theta_{{\bf k}}+3H\dot\theta_{{\bf
k}}+\frac{k^2}{a^2}\theta_{{\bf k}}=0\ .
 \label{3.31}
\end{equation}
Other gauges that automatically provide finiteness of macroscopic
quantities in the one--loop quantum gravity (see Section
\ref{fininv}) have the same property.

Macroscopic Einstein equations (\ref{2.64}) read
\begin{equation}
\displaystyle
3H^2=\varkappa\left(\varepsilon_g+\Lambda\right)\ ,
\label{3.32}
\end{equation}
\begin{equation}
\displaystyle
2\dot H+3H^2=\varkappa\left(\Lambda-p_g\right)\ ,
\label{3.33}
\end{equation}
where
\begin{equation}
\begin{array}{c}
\displaystyle \varepsilon_g=\frac{1}{8\varkappa}\sum_{{{\bf
k}\sigma}}\langle\Psi_g|\dot{\hat \psi}_{{\bf k}\sigma}^+\dot{\hat
\psi}_{{\bf k}\sigma}+\frac{k^2}{a^2}\hat \psi_{{\bf
k}\sigma}^+\hat \psi_{{\bf
k}\sigma}|\Psi_g\rangle-\frac{1}{4\varkappa}\sum_{\bf
k}\langle\Psi_{gh}|\dot{\bar\theta}_{\bf k}\dot\theta_{\bf
k}+\frac{k^2}{a^2}\bar\theta_{\bf k}\theta_{\bf
k}|\Psi_{gh}\rangle\ ,
 \\[5mm]
\displaystyle p_g=\frac{1}{8\varkappa}\sum_{{{\bf
k}\sigma}}\langle\Psi_g|\dot{\hat \psi}_{{\bf k}\sigma}^+\dot{\hat
\psi}_{{\bf k}\sigma}-\frac{k^2}{3a^2}\hat \psi_{{\bf
k}\sigma}^+\hat \psi_{{\bf
k}\sigma}|\Psi_g\rangle-\frac{1}{4\varkappa}\sum_{\bf
k}\langle\Psi_{gh}|\dot{\bar\theta}_{\bf k}\dot\theta_{\bf
k}-\frac{k^2}{3a^2}\bar\theta_{\bf k}\theta_{\bf
k}|\Psi_{gh}\rangle
\end{array}
\label{3.34}
 \end{equation}
are the energy density and pressure of gravitons that are
renormalized by ghosts. Formulas (\ref{3.34})  were obtained after
elimination of 3--scalar and 3--vector modes from equations
(\ref{2.65}) and (\ref{2.66}). We also took into account the
following definitions
\[
 \displaystyle \langle\Psi|\hat T_0^0|\Psi\rangle= \varepsilon_g\ ,
\qquad \langle\Psi|\hat
T_\alpha^\beta|\Psi\rangle=\frac{\delta_\alpha^\beta}{3}\langle\Psi|\hat
T_\gamma^\gamma|\Psi\rangle=-\delta_\alpha^\beta p_g .
 \]
Also we have the following rules of averaging of bilinear forms
that are the consequence of homogeneity and isotropy of the
background
\[
\displaystyle \langle\Psi_g|\hat\psi_{{\bf
k}\sigma}^+\hat\psi_{{\bf
k'}\sigma'}|\Psi_g\rangle=\langle\Psi_g|\hat\psi_{{\bf
k}\sigma}^+\hat\psi_{{\bf k}\sigma}|\Psi_g\rangle\delta_{{\bf
kk'}}\delta_{\sigma\sigma'}\ , \qquad
\langle\Psi_{gh}|\bar\theta_{{\bf k}}\theta_{{\bf
k'}}|\Psi_{gh}\rangle=\langle\Psi_{gh}|\bar\theta_{{\bf
k}}\theta_{{\bf k}}|\Psi_{gh}\rangle\delta_{{\bf kk'}}\ .
\]

The self--consistent system of equations (\ref{3.30}) ---
(\ref{3.33}) is a particular case of general equations of
one--loop quantum gravity (\ref{2.59}), (\ref{2.62}), (\ref{2.64})
--- (\ref{2.66}). In turn, these general equations are the result
of the transition to the one--loop approximation from exact
equations (\ref{2.46}) --- (\ref{2.50}) that were obtained by
variation of gauged action over classic and quantum variables. To
canonically quantize 3--tensor gravitons and ghosts, one needs to
make sure that the variational procedure takes place for equations
(\ref{3.30}) --- (\ref{3.33}) directly. To do so, in the action
(\ref{2.42}) we keep only background terms and terms that are
quadratic over 3--tensor fluctuations and ghosts. Then, we exclude
the full derivative from the background sector and make the
transition to Fourier images in the quantum sector. As a result of
these operations, we obtain the following
\begin{equation}
\begin{array}{c}
\displaystyle S=\int dt\left(-\frac{3{\dot a}^2a}{\varkappa
N}-\Lambda a^3N+L_{grav}+L_{ghost}\right)\ ,
 \\[5mm]
 \displaystyle
L_{grav}+L_{ghost}=\frac{1}{8\varkappa}\sum_{{{\bf
k}\sigma}}\left(\frac{a^3}{N}\dot{\hat \psi}_{{\bf
k}\sigma}^+\dot{\hat \psi}_{{\bf k}\sigma}-Nak^2\hat \psi_{{\bf
k}\sigma}^+\hat \psi_{{\bf
k}\sigma}\right)-\frac{1}{4\varkappa}\sum_{\bf
k}\left(\frac{a^3}{N}\dot{\bar\theta}_{\bf k}\dot\theta_{\bf
k}-Nak^2\bar\theta_{\bf k}\theta_{\bf k}\right)\ .
\end{array}
\label{3.35}
 \end{equation}
In (\ref{3.35}), the background metric is taken to be in the form
of (\ref{3.1}), and the $N$ function is taken to be a variation
variable (the choice of this function, e.g. $N=1$, to be made
after variation of action). Here and further on, the normalized
volume is supposed to be unity, so $V=\int d^3x =1$. The terms
which are linear over the graviton field are eliminated from
(\ref{3.35}) because of zero trace of 3--tensor fluctuations.
Variations of action over $N$ and $a$ are done with the following
averaging. These procedures lead to equations (\ref{3.32}),
(\ref{3.33}) and expressions (\ref{3.34}). Variation of action
over quantum variables leads to the quantum equations of motion
(\ref{3.30}) and (\ref{3.31}).

In accordance with the standard procedure of canonical
quantization of gravitons, one introduces generalized momenta
\begin{equation}
\displaystyle \hat\pi_{{\bf k}\sigma}=\frac{\partial L}{\partial
\dot{\hat\psi}_{{\bf
k}\sigma}}=\frac{a^3}{4\varkappa}\dot{\hat\psi}_{{\bf k}\sigma}^+
\ . \label{3.36}
 \end{equation}
Then, commutation relations between operators that are defined at
the same instant of time read
 \begin{equation}
\begin{array}{c}
\displaystyle \left[\hat\pi_{{\bf k}\sigma}\, ,\ \hat\psi_{{\bf
k'}\sigma'}\right]_{-}\equiv
\frac{a^3}{4\varkappa}\left[\dot{\hat\psi}^+_{{\bf k}\sigma}\ ,\
\hat\psi_{{\bf k'}\sigma'}\right]_{-}=-i\hbar \delta_{{\bf k}{\bf
k'}}\delta_{\sigma\sigma'}\ .
\end{array}
\label{3.37}
 \end{equation}
Formulas (\ref{3.36}) and (\ref{3.37}) are presented for the $N=1$
case. Note also that the derivative in (\ref{3.36}) should be
calculated taking into account the  $\psi_{{\bf
k}\sigma}^+=\psi_{-{\bf k}-\sigma}$ condition.

The ghost quantization contains three specific issues. First,
there is the following technical detail that must be taken into
account for the definition of generalized momenta of ghost fields.
The argument in respect to which the differentiation is conducted
needs to be considered as a left co--multiplier of quadratic form.
Executing the appropriate requirement and taking into account
Grassman's character of ghost fields, we obtain
 \begin{equation}
\begin{array}{c}
 \displaystyle
{\mathcal{P}}_{\bf k}=\frac{\partial L}{\partial \dot\theta_{\bf
k}}=\frac{a^3}{4\varkappa}\dot{\bar\theta}_{\bf k}\ , \qquad
{\bar{\mathcal{P}}}_{\bf k}=\frac{\partial L}{\partial
\dot{\bar\theta}_{\bf k}}=-\frac{a^3}{4\varkappa}\dot{\theta}_{\bf
k}\ .
 \end{array}
\label{3.38}
\end{equation}
Second, the quantization of Grassman's fields is carried out by
setting the following anti--commutation relations
\begin{equation}
\begin{array}{c}
 \displaystyle
\left[{\mathcal{P}}_{\bf k}\ ,\ \theta_{\bf k}\right]_+\equiv
\frac{a^3}{4\varkappa}\left[\dot{\bar\theta}_{{\bf k}}\ ,\
\theta_{{\bf k'}}\right]_+=-i\hbar \delta_{{\bf k}{\bf k'}}\ ,
\\[5mm]
\displaystyle
\left[{\bar{\mathcal{P}}}_{\bf k}\ ,\ \bar\theta_{\bf k}\right]_+\equiv
-\frac{a^3}{4\varkappa}\left[\dot{\theta}_{{\bf k}}\ ,\
\bar\theta_{{\bf k'}}\right]_+=-i\hbar \delta_{{\bf k}{\bf k'}}\ .
\end{array}
 \label{3.39}
 \end{equation}
Third is {\it the bosonization of ghost fields}, which is carried
out {\it after quantization} of (\ref{3.39}). The possibility of
the bosonization procedure is provided by Grassman algebra, which
contains Grassman units defined by relations $\bar uu=-u\bar u=1$.
Therefore, conjunctive Grassman fields can be always presented in
the following form
 \begin{equation}
\displaystyle \theta_{\bf k}=u\vartheta_{\bf k}\ , \qquad
\bar\theta_{\bf k}=\bar u\vartheta_{\bf k}^+\ , \label{3.40}
\end{equation}
where  $\vartheta_{\bf k}$ is Fourier image of complex scalar
field which is described by the usual algebra. The substitution of
(\ref{3.40}) in (\ref{3.39})  leads to the following standard Bose
commutation relations
\begin{equation}
\begin{array}{c}
 \displaystyle
\frac{a^3}{4\varkappa}\left[\dot{\vartheta}^+_{{\bf k}}\ ,\
\vartheta_{{\bf k'}}\right]_{-}=-i\hbar \delta_{{\bf k}{\bf k'}}\ ,
\qquad
\frac{a^3}{4\varkappa}\left[\dot{\vartheta}_{{\bf k}}\ ,\
\vartheta^+_{{\bf k'}}\right]_{-}=-i\hbar \delta_{{\bf k}{\bf k'}}\ .
\end{array}
 \label{3.41}
 \end{equation}
The Hermit conjugation transforms one of them to the other.

\subsection{State Vector of the General Form}\label{stvec}

To complete the self--consistent theory of gravitons in the
isotropic Universe, one needs to present the algorithm of
introduction of the graviton--ghost ensemble into the theory.
Properties of this ensemble are defined by Heisenberg's state
vector which is expanded over the basis that has a physical
interpretation. Any possible basis is the system of eigenvectors
of an appropriate time independent Hermit operator. The existence
of such operators can be proved in a general form. Let us consider
the following operator equation which is an analog of operator
equations of gravitons and ghosts
\begin{equation}
 \displaystyle \ddot y_k+3H\dot y_k+\frac{k^2}{a^2}y_k=0\ .
  \label{3.42}
 \end{equation}
Coefficients of equation (\ref{3.42}) are continuous and
differentiated functions of time along all cosmological scales
except for the singularity. Thus, with the exception of the
singular point, the general solution of equation (\ref{3.42})
definitely exists. Below we will show that the existence of a
state vector follows only from the existence of general solution
of equation (\ref{3.42}) (see also \cite{11}).

Suppose $g_k$, $h_k$  are linear independent solutions to
(\ref{3.42}), so that their superposition with arbitrary
coefficients gives the general solution to (\ref{3.42}). With no
loss of generality, one can suppose that these solutions are
normalized in some convenient way in each concrete case. From the
theory of ordinary differential equations it is known that $g_k$,
$h_k$  functions are connected to each other by the following
relation
\begin{equation}
 \displaystyle g_k\dot h_k-h_k\dot g_k=\frac{C_k}{a^3}\ ,
 \label{3.43}
 \end{equation}
where $C_k$ is a normalization constant. The comparison of
(\ref{3.42}) with (\ref{3.30}) and (\ref{3.31}) shows that
solutions of operator equations are presented by the same
functions. For operators of graviton field we have
\begin{equation}
\displaystyle \hat\psi_{{\bf k}\sigma}=\hat A_{{\bf
k}\sigma}g_k+\hat B_{{\bf k}\sigma}h_k\ ,
 \label{3.44}
 \end{equation}
where  $\hat A_{{\bf k}\sigma},\, \hat B_{{\bf k}\sigma}$  are
operator constants of integration. Directly from these operator
constants, one needs to build the operator which gives rise to the
full set of basis vectors.

It is important to keep in mind that {\it commutation property of
operator constants $\hat A_{{\bf k}\sigma},\, \hat B_{{\bf
k}\sigma}$ and physical interpretation of basis state vectors are
determined by the choice of linear independent solutions of
equation (\ref{3.42})}. The simplest basis is that of occupation
numbers. The choice of linear independent solutions as
self--conjugated complex functions corresponds to this basis.

In accordance with (\ref{3.43}), if $g_k=f_k,\, h_k=f^*_k$ the
normalization constant is pure imaginary. Let's take $C_k=i$, so
we obtain
\begin{equation}
 \displaystyle f_k\dot f^*_k-f^*_k\dot f_k=\frac{i}{a^3}\ .
 \label{3.45}
\end{equation}
To build the graviton operator over this basis, one need to carry
out the multiplicative renormalization of operator constants
taking into account that field is real. This yield
\[
\displaystyle A_{{\bf k}\sigma}=\sqrt{4\varkappa\hbar}\hat c_{{\bf
k}\sigma}\ , \qquad B_{{\bf k}\sigma}=\sqrt{4\varkappa\hbar}\hat
c_{-{\bf k}-\sigma}^+\ .
\]
As result of these operations, we get the graviton operator and
its derivative that read
\begin{equation}
\displaystyle \hat\psi_{{\bf
k}\sigma}=\sqrt{4\varkappa\hbar}\left(\hat c_{{\bf
k}\sigma}f_k+\hat c^+_{-{\bf k}-\sigma}f^*_k\right)\ , \qquad
\dot{\hat\psi}^+_{{\bf
k}\sigma}=\sqrt{4\varkappa\hbar}\left(c^+_{{\bf k}\sigma}\dot
f^*_k+c_{-{\bf k}-\sigma}\dot f_k\right)\ . \label{3.46}
\end{equation}
Standard commutation relations for operators of graviton creation
and annihilation are obtained by the substitution of (\ref{3.46})
into (\ref{3.37}) and taking into account (\ref{3.45}). They read
\begin{equation}
\displaystyle \left[\hat c_{{\bf k}\sigma}\ ,\ \hat c^+_{{\bf
k'}\sigma'}\right]_-= \delta_{{\bf k}{\bf
k'}}\delta_{\sigma\sigma'}\ , \qquad \left[\hat c_{{\bf k}\sigma}\
,\ \hat c_{{\bf k'}\sigma'}\right]_-=0\ , \qquad \left[\hat
c^+_{{\bf k}\sigma}\ ,\ \hat c^+_{{\bf k'}\sigma'}\right]_-=0 \ .
 \label{3.47}
 \end{equation}
In accordance with (\ref{3.47}), the operator of occupation
numbers $\hat n_{{\bf k}\sigma}=\hat c^+_{{\bf k}\sigma}\hat
c_{{\bf k}\sigma}$ exists that gives rise to basis vectors
$|n_{{\bf k}\sigma}\rangle$ of Fock's space. Non--negative integer
numbers $n_{{\bf k}\sigma}=0,\ 1,\ 2, ...$ are eigenvalues of this
operator.

In accordance with (\ref{3.24}), the observables are additive over
modes with given ${\bf k}\sigma$. Therefore, the state vector is
of multiplicative structure that reads
\[
\displaystyle |\Psi_g\rangle=\prod_{{\bf k}\sigma}|\Psi_{{\bf k}\sigma}\rangle\ ,
\]
where $|\Psi_{{\bf k}\sigma}\rangle$ is state vector of ${\bf
k}\sigma$--subsystem of gravitons of momentum ${\bf p}=\hbar {\bf
k}$ and polarization $\sigma$. In turn, in a general case,
$|\Psi_{{\bf k}\sigma}\rangle$  is an arbitrary superposition of
vectors that corresponds to different occupation numbers but the
same ${\bf k}\sigma$ values. Suppose that $\mathcal{C}_{n_{{\bf
k}\sigma}}$ is the amplitude of probability of finding the  ${\bf
k}\sigma$--subsystem of gravitons in the state with the occupation
number $n_{{\bf  k}\sigma}$. If so, then {\it the state vector of
the general form is the product of normalized superpositions}
\begin{equation}
\displaystyle  |\Psi_g\rangle=\prod_{{\bf  k}\sigma}\sum_{n_{{\bf
k}\sigma}}\mathcal{C}_{n_{{\bf  k}\sigma}}|n_{{\bf
k}\sigma}\rangle\ ,\qquad \sum_{n_{{\bf
k}\sigma}}|\mathcal{C}_{n_{{\bf  k}\sigma}}|^2=1\ .
\label{3.48}
\end{equation}

After the bosonization in the ghost sector is done, one gets
equations of motion and commutation relations that are similar to
those for graviton. The same set of linear independent solutions
$f_k,\ f_k^*$ that was introduced for operators of graviton field
is used for operators of ghost fields. What is necessary to take
into account here is originally complex character of ghost fields,
which leads to $\vartheta_{\bf k}^+\ne \vartheta_{-\bf k}$. As a
result, operators of ghost and anti--ghosts creation and
annihilation appear in the theory. They read
\begin{equation}
\displaystyle \vartheta_{\bf k}=\sqrt{4\varkappa\hbar}\left(\hat
a_{\bf k}f_k+\hat b^+_{-{\bf k}}f^*_k\right)\ , \qquad
\dot\vartheta^+_{\bf k}=\sqrt{4\varkappa\hbar}\left(a^+_{\bf
k}\dot f^*_k+b_{-{\bf k}}\dot f_k\right)\ .
 \label{3.49}
 \end{equation}
The substitution of (\ref{3.49}) into (\ref{3.41}) leads to
standard commutation relations
\begin{equation}
\begin{array}{c}
\displaystyle \left[\hat a_{\bf k}\ ,\ \hat a^+_{\bf k'}\right]_-=
\delta_{{\bf k}{\bf k'}}\ ,\qquad \left[\hat a_{\bf k}\ ,\ \hat
a_{\bf k'}\right]_-= \left[\hat a^+_{\bf k}\ ,\ \hat a^+_{\bf
k'}\right]_-=0 \ ,
 \\[5mm]
\displaystyle \left[\hat b_{\bf k}\ ,\ \hat b^+_{\bf k'}\right]_-=
\delta_{{\bf k}{\bf k'}}\ , \qquad \left[\hat b_{\bf k}\ ,\ \hat
b_{\bf k'}\right]_-= \left[\hat b^+_{\bf k}\ ,\ \hat b^+_{\bf
k'}\right]_-=0 \ ,
\\[5mm]
\displaystyle \left[\hat a_{\bf k}\ ,\ \hat b_{\bf k'}\right]_-=
\left[\hat a_{\bf k}\ ,\ \hat b^+_{\bf k'}\right]_-= \left[\hat
a^+_{\bf k}\ ,\ \hat b_{\bf k'}\right]_-=\left[\hat a^+_{\bf k}\
,\ \hat b^+_{\bf k'}\right]_-=0
\end{array}
\label{3.50}
\end{equation}
Applying the reasoning which is similar to that described above,
we conclude that in the ghost sector, the state vector of the
general form is also given by product of normalized
superpositions. It reads
\begin{equation}
\begin{array}{c}
\displaystyle  |\Psi_{gh}\rangle=\prod_{\bf  k}\sum_{n_{\bf
k}}\mathcal{A}_{n_{\bf  k}}|n_{\bf k}\rangle\prod_{\bf
k}\sum_{\bar n_{\bf k}}\mathcal{B}_{\bar n_{\bf  k}}|\bar n_{\bf
k}\rangle\ ,
\\[5mm]
\displaystyle
 \sum_{n_{{\bf
k}}}|\mathcal{A}_{n_{{\bf  k}}}|^2=\sum_{\bar n_{{\bf
k}}}|\mathcal{B}_{\bar n_{{\bf  k}}}|^2=1\ .
\end{array}
\label{3.51}
\end{equation}
The set of amplitudes $\mathcal{C}_{n_{{\bf  k}\sigma}}$,
$\mathcal{A}_{n_{\bf  k}}$,  $\mathcal{B}_{\bar n_{\bf  k}}$,
which parameterizes Heisenberg's state vector actually determines
the initial condition of quantum system of gravitons and ghosts.

Formulas (\ref{3.48}) and (\ref{3.51}) can be also used in case
when real functions are chosen as linear independent solutions of
equation (\ref{3.42}). The justification for this is due to the
fact that real linear independent solutions can be obtained from
complex self--conjugated ones by the following linear
transformation
\begin{equation}
\displaystyle g_k=\frac{1}{\sqrt{2}}\left(f_k+f^*_k\right)\ ,
\qquad h_k=\frac{i}{\sqrt{2}}\left(f_k-f^*_k\right)\ .
 \label{3.52}
 \end{equation}
After transition to the basis of real functions in (\ref{3.46})
and (\ref{3.49}), we get
\begin{equation}
\begin{array}{c}
\displaystyle \hat\psi_{{\bf
k}\sigma}=\sqrt{4\varkappa\hbar}\left(\hat Q_{{\bf
k}\sigma}g_k+\hat P_{{\bf k}\sigma}h_k\right)\ ,
 \\[5mm]
\displaystyle \hat\vartheta_{\bf
k}=\sqrt{4\varkappa\hbar}\left(\hat q_{\bf k}g_k+\hat p_{\bf
k}h_k\right)\ ,
 \end{array}
 \label{3.53}
 \end{equation}
where
\begin{equation}
\begin{array}{c}
\displaystyle \hat Q_{{\bf k}\sigma}=\hat Q^+_{-{\bf
k}-\sigma}=\frac{1}{\sqrt{2}}\left(\hat c_{{\bf
k}\sigma}+c^+_{-{\bf k}-\sigma}\right)\ , \qquad \hat P_{{\bf
k}\sigma}=\hat P^+_{-{\bf k}-\sigma}=-\frac{i}{\sqrt{2}}\left(\hat
c_{{\bf k}\sigma}-c^+_{-{\bf k}-\sigma}\right)\ ,
 \\[5mm]
 \displaystyle
 \hat q_{\bf k}=\frac{1}{\sqrt{2}}\left(\hat a_{\bf k}+b^+_{-{\bf k}}\right)\ ,
 \qquad \hat p_{\bf k}=-\frac{i}{\sqrt{2}}\left(\hat a_{\bf k}-b^+_{-{\bf k}}\right)\ .
 \end{array}
 \label{3.54}
 \end{equation}
Relations  (\ref{3.54}) allow to work with real linear independent
solutions and to use simultaneously state vectors (\ref{3.48}) and
(\ref{3.51}) for the representation of occupation numbers. Note
that in the framework of the basis of real functions, operator
constants are operators of generalized coordinates and momenta:
\begin{equation}
\displaystyle \left[\hat P^+_{{\bf k}\sigma},\ \hat Q_{{\bf
k'}\sigma'}\right]_-=-i\delta_{{\bf kk'}}\delta_{\sigma\sigma'}\
,\qquad \left[\hat p^+_{{\bf k}},\ \hat q_{{\bf
k'}\sigma'}\right]_-=-i\delta_{{\bf kk'}}\ . \label{3.55}
 \end{equation}

To complete this Section, let us discuss two problems that are
relevant to intrinsic mathematical properties of the theory. First
of all, let us mention that "bosonization"\ of ghost fields is a
necessary element of the theory because only this procedure
provides the existence of state vector in the ghost sector.
Mathematically, it is because the structure of the classic
differential equation (\ref{3.42})  and properties of its solution
(\ref{3.45}) are inconsistent with the Fermi--Dirac quantization.
In terms of original ghost fields we have
\begin{equation}
\displaystyle \theta_{\bf
k}=\sqrt{4\varkappa\hbar}\left(\alpha_{\bf k}f_k+\bar\beta_{-{\bf
k}}f^*_k\right)\ , \qquad \dot{\bar\theta}_{\bf
k}=\sqrt{4\varkappa\hbar}\left(\bar\alpha_{\bf k}\dot
f^*_k+\beta_{-{\bf k}}\dot f_k\right)\ .
 \label{3.56}
 \end{equation}
Substitution (\ref{3.56}) into (\ref{3.39}) and taking into
account (\ref{3.45}) leads to anti--commutation relations for
operator constants that read
\[
\begin{array}{c}
\displaystyle \left[\bar\alpha_{\bf k}\ ,\ \alpha_{\bf
k'}\right]_+= -\delta_{{\bf k}{\bf k'}}\ ,\qquad \left[\beta_{\bf
k}\ ,\ \bar\beta_{\bf k'}\right]_+= \delta_{{\bf k}{\bf k'}}\ .
\end{array}
\]
The $\left[\beta_{\bf k}\ ,\ \bar\beta_{\bf k'}\right]_+=
\delta_{{\bf k}{\bf k'}}$  relation can formally be considered as
anti--commutation relation for operators giving rise the Fermi
space of ghost states. There is no such a possibility for
$\bar\alpha_{\bf k}\ ,\ \alpha_{\bf k}$ operators because their
anti--commutation is negative. If one considers these operators as
complete mathematical objects that are not subject to any
transformations, then it is impossible to build an operator over
them that gives rise to some space of states, and this is because
of non--standard anti--commutation relation. The problem is solved
by the fact of the existence of Grassman units which are necessary
elements of Grassman algebra. At the operator constants level, the
bosonization is reduced to the following transformation
\[
\displaystyle \alpha_{\bf k}=ua_{\bf k}\ , \qquad \bar\alpha_{\bf
k}=\bar ua^+_{\bf k}\ , \qquad \beta_{\bf k}=\bar ub_{\bf k}\ ,
\qquad \bar\beta_{\bf k}=ub^+_{\bf k}\ .
\]
This leads to operators with (\ref{3.50}) commutation properties.

The choice of basis is the most significant problem in the
interpretation of theory. In the theory of quantum fields of
non--stationary Universe, the choice of linear independent basis
$f_k,\ f^*_k$ is ambiguous, in principle. This differentiates it
from the theory of quantum fields in the Minkowski space. In the
latter, the separation of field into negative and positive
frequency components is Lorentz--invariant procedure \cite{15}. A
natural physical postulate in accordance to which the definition
of particle (quantum of field) in the Minkowski space must be
relativistically invariant leads mathematically to
$f_k=(2\omega_k)^{-1/2}e^{-i\omega_kt}$.  In the non--stationary
Universe with the (\ref{3.6}) metric, the similar postulate can be
introduced only for conformally  invariant fields and at the level
of auxiliary Minkowski space. At the same time, the graviton field
is conformally non--invariant. This can be seen from the
following. Using the conformal transformation  $y_k=\tilde y_k/a$
and transition to the conformal time $d\eta=dt/a$, one can see
that equation (\ref{3.42}) is transformed to the equation for the
oscillator with variable frequency that reads
\begin{equation}
 \displaystyle \tilde y''_k+\left(k^2-\frac{a''}{a}\right)\tilde y_k=0\ .
  \label{3.57}
 \end{equation}
Effects of vacuum polarization and graviton creation in the
self--consistent classic gravitational field correspond to
parametric excitation of the oscillator (\ref{3.57}).

The approximate separation of field on negative and positive
frequency components is possible only in the short wavelength
limit. Regardless of the background dynamic, linearly independent
solutions of equation (\ref{3.57}) exist, and they have the
following asymptotes
\begin{equation}
 \displaystyle \tilde f_k\to \frac{1}{\sqrt{2k}}e^{-ik\eta}\ , \qquad
\tilde f^*_k\to \frac{1}{\sqrt{2k}}e^{ik\eta}\ ,\qquad k^2\gg \left|\frac{a''}{a}\right|\ .
  \label{3.58}
 \end{equation}
Effects of vacuum polarization and particle creation are
negligible for the subsystem of shortwave gravitons. In this
sector, quanta of gravitational field can be considered, with a
good accuracy, as real gravitons that are situated at their mass
shell. The conservation of the number of such real gravitons takes
also place with a good accuracy. In the shortwave limit, choosing
linear independent solutions of the (\ref{3.58}) form, occupation
numbers $n_{{\bf k}\sigma}$  are interpreted as numbers of real
gravitons with energy $\varepsilon_k=\hbar k/a$, momentum   ${\bf
p}=\hbar{\bf k}/a$ and polarization  $\sigma$. The possibility of
such an interpretation is the principle and the only argument in
favor of choice of this basis. For the subsystem of shortwave
gravitons, initial conditions are permissible not in the form of
products of superpositions but in the form of products of state
vectors with determined occupation numbers. In accordance with the
usual understanding of the status of shortwave ghosts, their state
can be chosen in the vacuum form. The gas of shortwave gravitons
is described in more detail in Section \ref{swg}.

In the $k^2\sim |a''/a|$ vicinity, there is no criterion allowing
a choice of preferable basis. It is impossible to introduce the
definition of real gravitons in this region because there is no
mass shell here. This is the reason why we will use the term
"virtual graviton of determined momentum"\ in discussions of
excitations of long wavelengths. Under the term "virtual
graviton"\ we mean a graviton whose momentum is defined but whose
energy is undefined. Each set of linear independent solutions
corresponds to the distribution of energy for the determined
momentum.  This distribution can be set up, for example by the
expansion of basis function in the Fourier integral. Thus, the
choice of basis is, at the same time, the definition of virtual
graviton. One needs to mention that different sets of probability
amplitudes $\mathcal{C}_{n_{{\bf k}\sigma}}$ correspond to
different definitions of the virtual graviton for the same initial
physical state. Note also that limitations that are defined by
asymptotes (\ref{3.58}) do not fix basis functions completely.

\subsection{One--Loop Finiteness}\label{fin}

The full system of equations of the theory consists of operator
equations for gravitons and ghosts (\ref{3.30}), (\ref{3.31}),
macroscopic Einstein equations (\ref{3.32}), (\ref{3.33}) and
formulas  (\ref{3.34}) for the energy density and pressure of
gravitons. The averaging of (\ref{3.34}) is carried out over state
vectors of general form (\ref{3.48}) and (\ref{3.51}). The
one--loop finiteness is satisfied automatically in this theory.
The finiteness is provided by the structure of ghost sector, and
it is a result of the following two facts. First, in the space
with metric (\ref{3.1}) the ghost equation (\ref{3.31}) coincides
with graviton equation (\ref{3.30}). Second, the number of
internal degrees of freedom of the complex ghost field coincides
with that of 3--tensor gravitons. We will show this by direct
calculations.

Let us introduce the graviton spectral function which is
renormalized by ghosts. It reads
\begin{equation}
\displaystyle W_{{\bf k}}=\sum_\sigma\langle \Psi_g|\hat\psi^+_{{\bf
k}\sigma}\hat\psi_{{\bf
 k}\sigma}|\Psi_g\rangle-2\langle\Psi_{gh}|\bar \theta_{{\bf k}}\theta_{{\bf
 k}}|\Psi_{gh}\rangle\ .
 \label{3.59}
 \end{equation}
Zero and first moments of this function are the most important
objects of the theory. They are
\begin{equation}
\begin{array}{c}
\displaystyle W_0=\sum_{{\bf k}}\left(\sum_\sigma\langle
\Psi_g|\hat\psi^+_{{\bf k}\sigma}\hat\psi_{{\bf
 k}\sigma}|\Psi_g\rangle-2\langle\Psi_{gh}|\bar \theta_{{\bf k}}\theta_{{\bf
 k}}|\Psi_{gh}\rangle\right)\ ,
 \\[3mm]\displaystyle
 W_1=\sum_{{\bf
k}}\frac{k^{2}}{a^{2}}\left(\sum_\sigma\langle \Psi_g|\hat\psi^+_{{\bf
k}\sigma}\hat\psi_{{\bf
 k}\sigma}|\Psi_g\rangle-2\langle\Psi_{gh}|\bar \theta_{{\bf k}}\theta_{{\bf
 k}}|\Psi_{gh}\rangle\right)\ .
\end{array}
 \label{3.60}
 \end{equation}
The energy density and pressure of gravitons that are expressed
via moments (\ref{3.60}) can be obtained by transformations
identical to (\ref{3.34}) with use of equations of motion
(\ref{3.30}) and (\ref{3.31}). They read
\begin{equation}
 \begin{array}{c}
\displaystyle \varkappa\varepsilon_g=\frac{1}{16}D+\frac14W_1\ ,
\qquad \varkappa p_g=\frac{1}{16}D+\frac{1}{12}W_1\ ,
\\[3mm]
\displaystyle D=\ddot W_0+3H\dot W_0\ .
\end{array}
 \label{3.61}
\end{equation}
In addition, the following relation between moments is derived
from equations of motion
\begin{equation}
\displaystyle  \dot D+6HD+4\dot W_1+16HW_1=0\ .
\label{3.62}
\end{equation}
This relation ensures that the graviton energy--momentum tensor is
conservative:
\[
\displaystyle \dot\varepsilon_g+3H(\varepsilon_g+p_g)=0
\]

As it was shown above, field operators can always be chosen from
the basis of complex self--conjugated functions that are the same
both for gravitons and ghosts. One needs to also mention that the
interpretation of short wave gravitons as real gravitons
determines the asymptotic of basis functions (see (\ref{3.58})).
After the commutation of operators of creation and annihilation
are done, graviton contributions to the moments of the spectral
function $W_n,\, n=0,1$ can be presented in the following form
\begin{equation}
\begin{array}{c}
\displaystyle W_{n(grav)}=
\sum_{{\bf k}}\frac{k^{2n}}{a^{2n}}
\sum_\sigma\langle \Psi_g|\hat\psi^+_{{\bf k}\sigma}\hat\psi_{{\bf k}\sigma}|\Psi_g\rangle=
8\varkappa\hbar\sum_{{\bf k}}\frac{k^{2n}}{a^{2n}}f_k^*f_k+
\\[5mm]
\displaystyle
+4\varkappa\hbar\sum_{{\bf k}}\frac{k^{2n}}{a^{2n}}
\sum_\sigma \left(2\langle \Psi_g|\hat c^+_{{\bf k}\sigma}\hat c_{{\bf k}\sigma}
|\Psi_g\rangle f_k^*f_k+\langle \Psi_g|\hat c^+_{{\bf k}\sigma}\hat c^+_{-{\bf k}-
\sigma}|\Psi_g\rangle f_k^{*2}+\langle \Psi_g|\hat c_{-{\bf k}-
\sigma}\hat c_{{\bf k}\sigma}|\Psi_g\rangle f_k^{2}\right)\ .
\end{array}
\label{3.63}
\end{equation}
In the right--hand--side of (\ref{3.63}), the first term is the
functional which is independent of the structure of Heisenberg
state vector. It reads
\begin{equation}
\displaystyle W^{(0)}_{n(grav)}=8\varkappa\hbar\sum_{{\bf
k}}\frac{k^{2n}}{a^{2n}}f_k^*f_k=\frac{4\varkappa\hbar}{\pi^2a^{2n}}\int\limits_0^\infty
k^{2n+2}f_k^*f_kdk\ .
 \label{3.64}
\end{equation}

The integral (\ref{3.64}) describes the contribution of zero
oscillations whose spectrum is deformed by macroscopic
gravitational field. Asymptotic (\ref{3.58}) shows that this
integral is diverges. In such a situation, the usual way is to use
regularization and renormalization procedures. As a result of
these operations, quantum corrections to Einstein equations
appear. These corrections are the conformal anomalies and terms
that came from Lagrangian $\sim R^2\ln (R/\lambda_g^2)$ where
$\lambda_g$ is a scale parameter that comes from renormalization
(see Appendix \ref{ren}). The theory that we present here does not
use such operations. There is a contribution of ghost zero
oscillations in the moments of spectral function. Its sign is
opposite to (\ref{3.64}). It reads
\begin{equation}
\begin{array}{c}
\displaystyle W_{n(ghost)}=-2\sum_{{\bf
k}}\frac{k^{2n}}{a^{2n}}\langle \Psi_{gh}|\bar\theta_{\bf
k}\theta_{\bf k}|\Psi_{gh}\rangle= -8\varkappa\hbar\sum_{{\bf
k}}\frac{k^{2n}}{a^{2n}}f_k^*f_k-
\\[3mm]
\displaystyle -8\varkappa\hbar\sum_{{\bf
k}}\frac{k^{2n}}{a^{2n}}\left(\langle \Psi_{gh}|\hat a^+_{\bf
k}\hat a_{\bf k}+\hat b^+_{\bf k}\hat b_{\bf k}|\Psi_{gh}\rangle
f_k^*f_k+\langle \Psi_{gh}|\hat a^+_{\bf k}\hat b^+_{-{\bf
k}}|\Psi_{gh}\rangle f_k^{*2}+\langle \Psi_{gh}|\hat b_{-{\bf
k}}\hat a_{\bf k}|\Psi_{gh}\rangle f_k^{2}\right)\ .
\end{array}
\label{3.65}
\end{equation}
The observables (\ref{3.61}) are expressed via sums
$W_{n(grav)}+W_{n(ghost)}$. In those sums, the exact
graviton--ghost compensation takes place in the contribution from
zero oscillations.

The final expressions for the moments of spectral function are
obtained by using the explicit form of state vectors \ref{3.48})
and (\ref{3.51}). They read
\begin{equation}
\displaystyle W_n=8\varkappa\hbar\sum_{{\bf
k}}\frac{k^{2n}}{a^{2n}}\left(N_{\bf k}|f_k|^2+U^*_{\bf
k}f_k^{*2}+U_{\bf k}f_k^2\right)\ , \label{3.66}
\end{equation}
where
\begin{equation}
\displaystyle N_{\bf k}=\sum_\sigma\sum_{n_{{\bf
k}\sigma}=1}^\infty |\mathcal{C}_{n_{{\bf k}\sigma}}|^2n_{{\bf
k}\sigma}-\sum_{n_{{\bf k}}=1}^\infty |\mathcal{A}_{n_{{\bf
k}}}|^2n_{{\bf k}}-\sum_{\bar n_{{\bf k}}=1}^\infty
|\mathcal{B}_{\bar n_{{\bf k}}}|^2\bar n_{{\bf k}} \label{3.67}
 \end{equation}
and
\begin{equation}
 \begin{array}{c}
 \displaystyle  U^*_{\bf k}=
\frac12\sum_\sigma\left(\sum_{n_{{\bf
k}\sigma}=0}^\infty\mathcal{C}^*_{n_{{\bf
k}\sigma}+1}\mathcal{C}_{n_{{\bf k}\sigma}}\sqrt{n_{{\bf
k}\sigma}+1}\right)\left(\sum_{n_{{\bf
k'}\sigma'}=0}^\infty\mathcal{C}^*_{n_{{\bf
k'}\sigma'}+1}\mathcal{C}_{n_{{\bf k'}\sigma'}}\sqrt{n_{{\bf
k'}\sigma'}+1}\right)-
 \\[5mm]
 \displaystyle
-\left(\sum_{n_{\bf k}=0}^\infty\mathcal{A}^*_{n_{\bf
k}+1}\mathcal{A}_{n_{\bf k}}\sqrt{n_{\bf k}+1}\right)
\left(\sum_{\bar n_{{\bf k'}}=0}^\infty\mathcal{B}^*_{\bar n_{{\bf
k'}}+1}\mathcal{B}_{\bar n_{{\bf k'}}}\sqrt{\bar n_{{\bf
k'}}+1}\right)
  \end{array}
 \label{3.68}
 \end{equation}
are spectral parameters. They are defined by initial conditions
for the chosen normalized basis of linear independent solutions of
equations (\ref{3.42}). (For sake of brevity, in (\ref{3.68}) and
below we use the following notation ${\bf k'}=-{\bf k},\
\sigma'=-\sigma$.) Note that the relation (\ref{3.66}) does not
contain divergences. Divergences in the relation (\ref{3.66}) may
appear only because of non--physical initial conditions. The
spectrum of real gravitons that slowly decreased for $k\to \infty$
is an example of such a non--physical initial conditions.

The spectral function (\ref{3.59}) depends of three arbitrary
constants as it is averaged over the state vector of general form.
It reads
\begin{equation}
\displaystyle W_{{\bf k}}=8\varkappa\hbar\left(N_{\bf
k}|f_k|^2+U^*_{\bf k}f_k^{*2}+U_{\bf k}f_k^2\right)\ .
\label{3.69}
\end{equation}
In (\ref{3.69}), the basis of normalized linear independent
solutions contains information on the dynamics of operators of
graviton-ghost field; integration constants  $N_{\bf k},\ U^*_{\bf
k}, \ U_{\bf k}$ contain information on the initial ensemble of
this field. Due to the background's homogeneity and isotropy the
moduli of the amplitudes and average occupation numbers do not
depend on the directions of wave vectors and polarizations:
\begin{equation}
\displaystyle \langle n_{k(g)}\rangle=\sum_{n_{{\bf
k}\sigma}=1}^\infty |\mathcal{C}_{n_{{\bf k}\sigma}}|^2n_{{\bf
k}\sigma}\ , \qquad \langle n_{k(gh)}\rangle = \sum_{n_{{\bf
k}}=1}^\infty |\mathcal{A}_{n_{{\bf k}}}|^2n_{{\bf k}}\ ,\qquad
\langle \bar n_{k(gh)}\rangle= \sum_{\bar n_{{\bf k}}=1}^\infty
|\mathcal{B}_{\bar n_{{\bf k}}}|^2\bar n_{{\bf k}}\ .
 \label{3.70}
\end{equation}

Phase of amplitudes, in principle, may depend on the directions
and polarizations. One must bear in mind that in the pure quantum
ensembles, for which the averaging over the state vector is
defined, phases of amplitudes are determined. If the phases are
random, then the additional averaging should be conducted over
them, which corresponds to the density matrix formalism for mixed
ensembles. The question of phases of amplitudes is clearly linked
to the question of the origin of quantum ensembles. In particular,
it is natural to assume that the ensemble of long--wavelength
gravitons arises in the process of restructuring graviton vacuum.
This process is due to conformal non--invariance of the graviton
field and can be described as particle creation. In this case,
there is a correlation between the phases of states with the same
occupation numbers, but mutually opposite momenta and
polarizations: the sum of these phases is zero.

If the typical occupation numbers in the ensemble are large, then
squares of moduli of probability amplitudes are likely to be
described by Poisson distributions. For this ensemble we get
\begin{equation}
\begin{array}{c}
\displaystyle \mathcal{C}_{n_{{\bf
k}\sigma}}=\sqrt{P[n_{k(g)}]}\exp(i\varphi_{n_{{\bf k}\sigma}})\
,\qquad \mathcal{A}_{n_{{\bf
k}}}=\sqrt{P[n_{k(gh)}]}\exp(i\chi_{n_{{\bf k}}})\ ,\qquad
\mathcal{B}_{n_{{\bf k'}}}=\sqrt{P[\bar
n_{k(gh)}]}\exp(i\chi_{\bar n_{{\bf k'}}})\ ,
\\[5mm]
\displaystyle P[n_{k}]=\frac{\langle n_k\rangle^{n_k}}{n_k!}\exp(-\langle n_k\rangle)
\ .
\end{array}
 \label{3.71}
\end{equation}
The substitution of  (\ref{3.70}), (\ref{3.71}) to (\ref{3.67}), (\ref{3.68}) leads to
\begin{equation}
\begin{array}{c}
\displaystyle
N_{\bf k}\equiv N_k=2\langle n_{k(g)}\rangle- \langle n_{k(gh)}\rangle -
\langle \bar n_{k(gh)}\rangle\ .
\end{array}
 \label{3.72}
\end{equation}
\begin{equation}
\begin{array}{c}
\displaystyle U^*_{\bf k}\equiv U^*_k=\langle
n_{k(g)}\rangle\zeta^{(g)}_ke^{i\varphi_k}-\sqrt{\langle
n_{k(gh)}\rangle\langle \bar
n_{k(gh)}\rangle}\zeta_k^{(gh)}e^{i\chi_k}\ ,
\\[5mm]
\displaystyle \zeta_k^{(g)}\leqslant 1,\qquad \zeta_k^{(gh)}\leqslant 1\ ,
\end{array}
 \label{3.73}
\end{equation}
where
\begin{equation}
\begin{array}{c}
\displaystyle
\zeta^{(g)}_ke^{i\varphi_k}=\frac12\sum_\sigma\left(\sum_{n_{{\bf
k}\sigma}}P[n_{k(g)}]\exp(i\varphi_{n_{{\bf
k}\sigma}}-i\varphi_{n_{{\bf
k}+1,\sigma}})\right)\left(\sum_{n_{{\bf
k'}\sigma'}}P[n_{k(g)}]\exp(i\varphi_{n_{{\bf
k'}\sigma'}}-i\varphi_{n_{{\bf k'}+1,\sigma'}})\right)\ ,
\\[5mm]
\displaystyle \zeta^{(gh)}_ke^{i\chi_k}=\left(\sum_{n_{{\bf
k}}}P[n_{k(gh)}]\exp\left(i\chi_{n_{{\bf k}}}-i\chi_{n_{{\bf
k}+1}}\right)\right)\left(\sum_{\bar n_{{\bf k'}}}P[\bar
n_{k(gh)}]\exp\left(i\chi_{\bar n_{{\bf k'}}}-i\chi_{n_{{\bf
k'}+1}}\right)\right)\
\end{array}
 \label{3.74}
\end{equation}
Limit equalities $\zeta_k^{(g)}=\zeta_k^{(gh)}=1$ are satisfied if
the phase difference between states of neighboring occupation
numbers does not depend on values of occupation numbers. It is
also easy to see that (\ref{3.72}) and (\ref{3.73}) apply, with
somewhat different definitions, to any ensemble with
$\zeta^{(g)}_ke^{i\varphi_k}$ and $\zeta^{(gh)}_ke^{i\chi_k}$
parameters.

We already mentioned above that different basis functions that
correspond to different definitions of the virtual graviton can be
used for the same initial physical state. Limitations due to the
prescriptions on the asymptotic expression (\ref{3.58})  allow to
fix only asymptotic expansions of basis functions for
$k\to\infty$. These expansions can be used, however, only for
description of shortwave modes (Section \ref{swg}). Meanwhile, all
non--trivial quantum gravity phenomena take place in spectral
region where characteristic wavelengths are of the order of the
horizon scale. The choice of basis functions to describe these
waves is not unique, and the set of amplitudes of probability
$\mathcal{C}_{n_{{\bf k}\sigma}}$ depends significantly on this
set. At the level of equations (\ref{3.72}), (\ref{3.73}), the
ambiguity in the definition of the virtual graviton reveals itself
in the ambiguity of values of parameters  $\langle
n_{k(g)}\rangle$ and $\zeta^{(g)}_ke^{i\varphi_k}$ . Similar
ambiguity exists in the ghost sector. Two conclusions follow from
that. First, it is necessary to work with the state vector of
general form, at least during the first stage of the study of the
system that contains excitations of long wavelengths.
Concretization of the amplitudes $\mathcal{C}_{n_{{\bf k}\sigma}}$
is possible only after using of additional physical considerations
that are different for each concrete case.  Second, a theory would
be extremely desirable which is invariant with respect to the
choice of linear independent solutions of equation (\ref{3.42}),
and, correspondingly, is invariant with respect to the choice of
amplitudes of probability $\mathcal{C}_{n_{{\bf k}\sigma}}$
defining the structure of Heisenberg's state vector, respectively.
In Section \ref{Bog} we will show that such a formulation of the
theory exists in the form of equation for the spectral function of
gravitons renormalized by ghosts. The mathematically equivalent
formulation of theory exists in the form of infinite BBGKY chain
or hierarchy where joint description of gravitons and ghost is
carried out in terms of moments of spectral function $W_n,\, n=0,\
1,\ 2,\ ...,\ N\to\infty$.

\subsection{Class of Legitimate Gauges}\label{fininv}

The theory presented above is actually the result of
transformations of equations which are set up by the original
gauged path integral (\ref{2.1}). Mostly, these transformations
are mathematically identical. There are only three issues of the
theory that are absent in the original integral. First, the
hypothesis of the existing of the classical spacetime of definite
but self--consistent geometry is introduced. Second, in the
self--consistent system of classical and quantum equations, the
transition to the one--loop approximation is made. Third, a  gauge
is chosen, which automatically provides the one--loop finiteness
of self--consistent theory of gravitons in the isotropic Universe.

The first and second issues were discussed in the process of
incorporating these into the theory formalism. We will return to
an additional discussion of these in Section \ref{con}). The gauge
can be chosen arbitrarily but must be mathematically consistent.
Moreover, the gauge is necessary to strictly define the path
integral as a mathematical object. Nevertheless, the use of a
gauge unavoidably raises a question about gauge dependency of the
results obtained.

Some of the results of the theory presented above are gauge
invariant or can be obtained by use of any gauges. All procedures
used to build the theory in Section \ref{qg} can be done by means
of an arbitrary gauge. The general structure of theory and its key
property (consistency of the system of classic and quantum
equations in each order of the theory of perturbations over the
amplitude of gravitational field) are gauge independent. In
Section \ref{scgt}, the elimination of scalar and vector
fluctuations, quantization of 3--tensor gravitons and the
construction of graviton state vector of the general form are also
gauge invariant operations. Only the ghost sector consisting of
equations of motion and ghost EMT are gauge dependent. Actually,
however, there is no arbitrariness in the choice of ghost sector.

There is almost no doubt that the future theory unites the gravity
with other physical interactions must be finite to all orders of
theory of perturbations. From this point of view, the one--loop
finiteness of self--consistent theory of gravitons is a
manifestation of fundamental properties of quantum gravity and its
generalizations. It is fair to say that the one--loop finiteness
of quantum gravity can be considered as a prototype of the future
comprehensive theory. From the structure of quantum gravity itself
without fields of matter, it follows that in the self--consistent
theory of gravitons (the simplest version of this theory) the
one--loop finiteness can be only achieved by graviton--ghost
compensation of diverged contributions to macroscopic observables.
In this version of theory, there are no other mathematical
algorithms that are able to provide the one--loop finiteness. The
important fact also is the gauge invariance of graviton
contribution to divergences. The mathematical structure of the
ghost sector that is able to compensate these divergences follows
directly from this fact. In the self--consistent theory of
gravitons in the isotropic Universe, which is one--loop finite,
the ghost equations of motion and ghost EMT must agree with that
presented in this work.

We propose the following statement as one of main constructive
principles of the theory. The one--loop quantum gravity as a
theory of gravitons in the macroscopic spacetime with the
self--consistent geometry must be finite by definition and, hence,
by construction. In the framework of this postulate, we present
physical principles that choose {\it the class of legitimate
gauges} that automatically provide one--loop finiteness of
macroscopic quantities.

It directly follows from the definition of the original path
integral (\ref{2.1}) that the gauge is imposed over the full
metric.  This means that {\it the background and fluctuations are
considered in one and the same reference frame} if the transition
to the self--consistent theory is made, which operates with both
classical and quantum variables. The question concerning the class
of legitimate reference frames, i.e. the class of legitimate
gauges, inevitably arises in calculations. It is because the
self--consistent background geometry is actually a non--arbitrary.
It was already mentioned in Section \ref{fac} that the
factorization of measure of path integral (outside of short wave
approximation) is possible only if the background and fluctuations
belong to different representations of group of symmetry of the
background geometry.  Reference frames where the high symmetry of
background spacetime is displayed clearly stand out both
physically and mathematically. It is exactly in these reference
frames that the separation of non--physical degrees of freedom
from physical ones can be done on the basis of simple
classification of quantum fluctuations over irreducible
representations of the symmetry group of the background geometry.
The class of such reference frames is defined by the gauges which
are covariant (form--invariant) with respect to transformation of
the group of background symmetry.

In the self--consistent theory of gravitons in the homogenous and
isotropic Universe we deal with the symmetry of background
spacetime with metric (\ref{3.1}). The form of the (\ref{3.1})
interval is conserved under 3D rotations  with parameters with
arbitrary dependence on time and under arbitrary transformation of
the time itself
\begin{equation}
 \displaystyle
 dx^{\alpha'}=\Omega^\alpha_{\; \beta}(t)dx^\beta\ ,\qquad t'=f(t)\ ,
 \label{3.75}
 \end{equation}
where  $\Omega^\alpha_{\; \beta}(t)$ is the Euler matrix, with
different angles of rotations at different moments of time, and
$f(t)$ is an arbitrary function. Below is the explicit form of two
legitimate gauges that are form--invariant in respect to the
transformations (\ref{3.75})
\begin{equation}
 \displaystyle
 \sqrt{-\hat g}\hat g^{00}=B(t)\ , \qquad \sqrt{-\hat g}\hat
 g^{0\alpha}=0\ ,
 \label{3.76}
 \end{equation}
 \begin{equation}
 \displaystyle
 \frac{\partial\sqrt{-\hat g}\hat g^{i\beta}}{\partial x^\beta}=0\ .
 \label{3.77}
 \end{equation}
There are of course also some non--legitimate gauges. For example,
in the case of $\sqrt{-\hat g}\hat g^{00}=B(t)$,
$\partial\sqrt{-\hat g}\hat g^{\alpha\beta}/\partial x^\beta=0$,
there is no inverse operator $(\hat M^i_{\; k})^{-1}$.

To build the theory in Sections \ref{qg} and \ref{scgt}, we used
the (\ref{3.76}) gauge with $B=1$ that leads to a simple and
elegant ghost sector providing one--loop finiteness automatically.
(Recall that, in final equations of self--consistent theory, there
is a possibility to make a transition from geometrodynamical time
which is defined by condition (\ref{2.29})  to any other time
coordinate (see (\ref{2.69}) in Section \ref{1loop})). The ghost
sector corresponding to the (\ref{3.77}) gauge is of a much more
complex structure. If however we go over to the one--loop
approximation and take into account the explicit form of
background metric (\ref{3.1}), then we get the following simple
expression
 \begin{equation}
 \displaystyle
 \mbox{\rm det}\, ||\hat M^i_{\; k}||=
 \mbox{\rm det}\, ||\partial_k\sqrt{-g}g^{ik}\partial_i||\times
 \mbox{\rm det}\, ||Na\Delta||\times
 \mbox{\rm det}\, ||Na\Delta||\times
 \mbox{\rm det}\, ||Na\Delta||  \
 ,
 \label{3.78}
 \end{equation}
where $\Delta$ is the Laplace operator in the 3D Euclidian space.
As it follows from (\ref{3.78}), one of the ghost fields is
described by the Klein--Gordon--Fock equation, and three other
ghost fields satisfy the Laplace equation $\Delta\chi=0$. The
singular sources of the fields  $\chi$ are absent, therefore the
unique constrained solution of the Laplace equation is actually a
trivial case $\chi=0$. Thus gauge (\ref{3.77}), as well as
(\ref{3.76}), creates a unique non--trivial ghost field. In the
framework of one--loop approximation, the EMT and equations of
motion of this field coincide with (\ref{2.62}) and (\ref{2.66}),
and the conditions of the quantization for the isotropic space
coincide with (\ref{3.39}).

In quantum gravity, the harmonic gauge is often used in
calculations. It reads
 \begin{equation}
 \displaystyle
 \label{3.79} \frac{\partial\sqrt{-\hat g}\hat g^{ik}}{\partial x^k}=0\
 .
 \end{equation}
With regard to the self--consistent theory of gravitons in the
isotropic Universe, note that the group of gauge invariance
(\ref{3.78}) contains within it a subgroup of 3D rotations
parameters of transformation which is independent on time. This
subgroup is sufficient to calibrate the background
\[
\displaystyle \sqrt{-g}g^{00}=1\ ,\qquad \sqrt{-g}g^{0\alpha}=0\
,\qquad \sqrt{-g}g^{\alpha\beta}=-a^4\gamma^{\alpha\beta}\ ,
\]
as well as to separate gauge invariant 3-tensor gravitons from
scalar and vector fluctuations. In the framework of harmonic
gauge, the ghost sector in the one--loop approximation reads
\begin{equation}
\begin{array}{c}
 \displaystyle
 \mbox{\rm det}\, ||\hat M^i_{\; k}||=
 \mbox{\rm det}\, ||\partial_k\sqrt{-g}g^{ik}\partial_i||\times
  \mbox{\rm det}\, ||\partial_k\sqrt{-g}g^{ik}\partial_i||\times
  \mbox{\rm det}\, ||\partial_k\sqrt{-g}g^{ik}\partial_i||\times
  \mbox{\rm det}\, ||\partial_k\sqrt{-g}g^{ik}\partial_i|| \ ,
  \\[5mm]
 \displaystyle L_{ghost}=-\sum_{a=0}^3\sqrt{-g}g^{ik}\bar\theta^a_{,i}\theta^a_{,k}\ .
 \end{array}
 \label{3.80}
 \end{equation}
Each of the four ghost fields, appearing in (\ref{3.80}), has the
dynamic properties of complex scalar field with Grassmann algebra.
One of these fields compensates graviton vacuum divergences, and
the other three fields generate their own divergences of the same
type. The absence of one--loop finiteness for macroscopic
observables in this version of the theory can be linked to the
fact the gauge (\ref{3.79}) does not have a property of
form--invariace with respect to $O_3(t)$  transformation.

Thus, we are ready to formulate the following proposition. {\it
For the theoretical and experimental study of the non--stationary
isotropic Universe, it is preferable to use only such gauges
(reference frames) that satisfy to the following conditions.
First, they must be invariant with respect to transformation of
the symmetry group (\ref{3.75}) of the background spacetime
(\ref{3.1}). Second, they must automatically produce the ghost
sector (\ref{2.62}) and (\ref{2.66}) which provides the one--loop
finiteness of the graviton theory in such space.}

This statement is of the clear interpretation. The first part of
the statement proposes to eliminate {\it fictitious}
inhomogenities and anisotropy that are due to the motion of the
reference frame itself. The second part of the statement proposes
to organize calculations in such a way that allows not to take
into account the {\it fictitious} renorm--group evolution of
observables. Results obtained in the framework of self--consistent
theory of gravitons for the isotropic Universe are independent on
choice of gauge chosen from this class.

Using gauges that do not provide the one--loop finiteness leads to
a fatal internal inconsistency of self--consistent theory of
gravitons. This is the aforementioned fictitious renorm--group
evolution of observables. Such evolution is a direct consequence
of renormalization of divergences. To remove divergences, in the
process of regularization, one needs to introduce counter terms
with the tensor structure in the Lagrangian of theory which are
absent in the original Einstein theory. After modification of
Einstein's theory an additional term of the type of $\hat R^2/f^2$
appears in the Lagrangian of the new theory. This term contains a
new fundamental constant $1/f^2$. The essence of contradiction is
in the fact that after this modification, one can not save the
original properties of the graviton field which actually created a
new post--renormalization Lagrangian (see Appendix \ref{nonren}).
To be free of internal inconsistency, non--renormalizable quantum
gravity (supergravity) must be finite. The one--loop finiteness
off the mass shell (supported by graviton--ghost compensation of
didvirgences) must be considered as a prototype of properties for
future theory.

Thus, the postulate of one--loop finiteness off the mass shell
must be accepted as a condition of internal non--inconsistency of
the self--consistent theory of gravitons. We can conclude also
that the one--loop finiteness (and, accordingly, the use of gauges
of the type (\ref{3.76}), (\ref{3.77})) automatically allows
direct participation of ghosts in the formation of macroscopic
observables.

\subsection{Instability of Trivial Vacuum and Evolution of
Graviton--Ghost Ensembles}\label{vac}

Let us show that the state vector set up at an instant of time
taken at random on the cosmological scale cannot be that of the
trivial vacuum. The term "trivial vacuum"\ refers to the state of
gravitons and ghosts with zero occupation numbers regardless of
their wavelengths.

The operator equation for gravitons (\ref{3.30}) describes the
quantum oscillator of variable frequency. After the conformal
transformation, we get
\begin{equation}
\begin{array}{c}
\displaystyle \hat\psi_{{\bf
k}\sigma}=\frac{1}{a}\sqrt{4\varkappa\hbar}\hat\phi_{{\bf
k}\sigma}\ ,
\\[5mm]
\displaystyle \hat\phi''_{{\bf k}\sigma}+\omega_k^2\hat\phi_{{\bf
k}\sigma}=0\ , \qquad \omega_k^2=k^2-\frac{a''}{a}\ .
\end{array}
\label{3.81}
\end{equation}
Using the method of Ref. \cite{16}, we introduce two new operator
functions $\tilde c_{{\bf k}\sigma}(\eta)$ and $\tilde  c^+_{-{\bf
k}-\sigma}(\eta)$ instead of the original operator function
$\hat\phi_{{\bf k}\sigma}(\eta)$, with the following additional
condition imposed on these functions
\begin{equation}
\begin{array}{c}
\displaystyle \hat\phi_{{\bf
k}\sigma}=\sqrt{\frac{1}{2\omega_k}}\left(\tilde  c_{{\bf
k}\sigma}e^{-i\int\omega_kd\eta}+\tilde c^+_{-{\bf
k}-\sigma}e^{i\int\omega_kd\eta}\right)\ ,
\\[5mm] \displaystyle
\hat\phi'_{{\bf k}\sigma}=-i\sqrt{\frac{\omega_k}{2}}\left(\tilde
c_{{\bf k}\sigma}e^{-i\int\omega_kd\eta}-\tilde c^+_{-{\bf
k}-\sigma}e^{i\int\omega_kd\eta}\right)\ .
\end{array}
\label{3.82}
\end{equation}
Note that (\ref{3.82}) can be regarded as one of definitions of
graviton that is possible if $\omega_k^2>0$. Substitution of
(\ref{3.82}) into (\ref{3.37}) shows that new operator functions
must satisfy to the following time--conserved commutation
relations
\begin{equation}
\displaystyle \left[\tilde  c_{{\bf k}\sigma}(\eta)\ ,\ \tilde
c^+_{{\bf k'}\sigma'}(\eta)\right]_-= \delta_{{\bf k}{\bf
k'}}\delta_{\sigma\sigma'}\ , \qquad \left[\tilde  c_{{\bf
k}\sigma}(\eta)\ ,\ \tilde  c_{{\bf k'}\sigma'}(\eta)\right]_-=0\
, \qquad \left[\tilde  c^+_{{\bf k}\sigma}(\eta)\ ,\ \tilde
c^+_{{\bf k'}\sigma'}(\eta)\right]_-=0 \ .
 \label{3.83}
 \end{equation}
Relations (\ref{3.83}) are consistent with general properties of
solutions of operator equations. The first equation follows from
(\ref{3.82}) and connects derivatives of new operator functions.
Substitution of (\ref{3.82}) to (\ref{3.81}) gives the second
equation. So, the system of equations now reads
\begin{equation}
\begin{array}{c}
 \displaystyle \tilde  c'_{{\bf k}\sigma}=
  \gamma_k \tilde  c^+_{-{\bf k}-\sigma}e^{2i\int\omega_kd\eta}\ ,
 \qquad \tilde  c^{+'}_{-{\bf k}-\sigma}=
  \gamma_k\tilde  c_{{\bf k}\sigma}e^{-2i\int\omega_kd\eta}\ ,
  \\[5mm]
 \displaystyle   \gamma_k=\frac{\omega'_k}{2\omega_k}\ .
\end{array}
 \label{3.84}
 \end{equation}
The commutators (\ref{3.83}) are the integrals of motion for the
system (\ref{3.84}). This fact demonstrates the consistency of the
canonic quantization procedure with the operator dynamics.

In the ghost sector, conformal transformations of ghost fields are
done together with the extraction of Grassman units
\begin{equation}
\begin{array}{c}
\displaystyle \theta_{{\bf
k}}=\frac{1}{a}\sqrt{4\varkappa\hbar}u\tilde\vartheta_{{\bf k}}\
,\qquad \bar\theta_{{\bf k}}=\frac{1}{a}\sqrt{4\varkappa\hbar}\bar
u\tilde\vartheta^+_{{\bf k}}\ ,
\\[5mm]
\displaystyle \tilde\vartheta''_{{\bf
k}}+\omega_k^2\tilde\vartheta_{{\bf k}}=0\ , \qquad
\tilde\vartheta^{+''}_{{\bf k}}+\omega_k^2\tilde\vartheta^+_{{\bf
k}}=0\ .
\end{array}
\label{3.85}
\end{equation}
Analogous to (\ref{3.82}), the operator structure and additional
condition are presented by the following relations
\begin{equation}
\begin{array}{c}
\displaystyle \tilde\vartheta_{{\bf
k}}=\sqrt{\frac{1}{2\omega_k}}\left(\tilde a_{{\bf
k}}e^{-i\int\omega_kd\eta}+\tilde b^+_{-{\bf
k}}e^{i\int\omega_kd\eta}\right)\ ,
\\[5mm] \displaystyle
\tilde\vartheta'_{{\bf k}}=-i\sqrt{\frac{\omega_k}{2}}\left(\tilde
a_{{\bf k}}e^{-i\int\omega_kd\eta}-\tilde b^+_{-{\bf
k}}e^{i\int\omega_kd\eta}\right)\ .
\end{array}
\label{3.86}
\end{equation}
From (\ref{3.86}), we get the following system
\begin{equation}
\begin{array}{c}
 \displaystyle \tilde  a'_{{\bf k}}=
  \gamma_k \tilde  b^+_{-{\bf k}}e^{2i\int\omega_kd\eta}\ ,
 \qquad \tilde  b^{+'}_{-{\bf k}}=
  \gamma_k\tilde  a_{{\bf k}\sigma}e^{-2i\int\omega_kd\eta}\ ,
  \\[5mm]
 \displaystyle   \tilde  a^{+'}_{{\bf k}}=
  \gamma_k \tilde  b_{-{\bf k}}e^{-2i\int\omega_kd\eta}\ ,
 \qquad \tilde  b'_{-{\bf k}}=
  \gamma_k\tilde  a^+_{{\bf k}\sigma}e^{2i\int\omega_kd\eta}\ .
\end{array}
 \label{3.87}
 \end{equation}
Commutators are obtained from (\ref{3.50}) by replacement of
operator constants $\hat a_{\bf k},\ \hat b^+_{\bf k}$ by operator
functions $\tilde a_{\bf k}(\eta),\ \tilde b^+_{\bf k}(\eta)$.
They are the integrals of motion of the system (\ref{3.87}).

The following operators are taken at the initial instant of time
\begin{equation}
\displaystyle \hat n_{{\bf k}\sigma}(\eta_0)=\tilde c^+_{{\bf
k}\sigma}(\eta_0)\tilde c_{{\bf k}\sigma}(\eta_0)\ ,\qquad \hat
n_{{\bf k}}(\eta_0)=\tilde a^+_{{\bf k}}(\eta_0)\tilde a_{{\bf
k}}(\eta_0)\ ,\qquad \hat n_{{\bf k}}(\eta_0)=\tilde b^+_{{\bf
k}}(\eta_0)\tilde b_{{\bf k}}(\eta_0)\ , \label{3.88}
 \end{equation}
They give rise to the Fock basis. The Heisenberg state vector of
the general form is constructed over this basis. Macroscopic
quantities are expressed via normal distribution functions
(\ref{3.89}) and anomalous distribution functions (\ref{3.90})
that read
 \begin{equation}
\begin{array}{c}
\displaystyle F^{(g)}_{{\bf
k}}(\eta)=\frac12\sum_\sigma\langle\Psi_g|\tilde c^+_{{\bf
k}\sigma}(\eta)\tilde c_{{\bf k}\sigma}(\eta)+\tilde c^+_{-{\bf
k}-\sigma}(\eta)\tilde c_{-{\bf k}-\sigma}(\eta)|\Psi_g\rangle\ ,
 \\[5mm]
\displaystyle F^{(gh)}_{{\bf k}}(\eta)=\langle\Psi_{gh}|\tilde
a^+_{{\bf k}}(\eta)\tilde a_{{\bf k}}(\eta)+\tilde b^+_{-{\bf
k}}(\eta)\tilde b_{-{\bf k}}(\eta)|\Psi_{gh}\rangle\ ,
 \end{array}
\label{3.89}
 \end{equation}
\begin{equation}
\begin{array}{c}
\displaystyle G^{*(g)}_{{\bf
k}}(\eta)=\frac12\sum_\sigma\langle\Psi_g|\tilde c^+_{{\bf
k}\sigma}(\eta)\tilde c^+_{-{\bf k}-\sigma}(\eta)|\Psi_g\rangle\
,\qquad G^{(g)}_{{\bf
k}}(\eta)=\frac12\sum_\sigma\langle\Psi_g|\tilde c_{-{\bf
k}-\sigma}(\eta)\tilde c_{{\bf k}\sigma}(\eta)|\Psi_g\rangle\ ,
 \\[5mm]
\displaystyle G^{*(gh)}_{{\bf k}}(\eta)=\langle\Psi_{gh}|\tilde
a^+_{{\bf k}}(\eta)\tilde b^+_{-{\bf k}}(\eta)|\Psi_{gh}\rangle\ ,
\qquad G^{(gh)}_{{\bf k}}(\eta)=\langle\Psi_{gh}|\tilde b_{-{\bf
k}}(\eta)\tilde a_{{\bf k}}(\eta)|\Psi_{gh}\rangle\ ,
 \end{array}
\label{3.90}
 \end{equation}

{\it Inhomogeneous} ordinary differential equation of the third
order for (\ref{3.89}) functions (taken from equations
(\ref{3.84}) and (\ref{3.87})) can be derived, taking into account
commutation relations
\begin{equation}
\displaystyle \frac{d^2}{d\eta^2}\frac{1}{\gamma_k}\frac{dF_{{\bf
k}}}{d\eta}-
2\gamma_k\frac{d}{d\eta}\frac{1}{\gamma_k}\frac{dF_{{\bf
k}}}{d\eta}-4\frac{d}{d\eta}\left(\gamma_kF_{\bf
k}\right)+8\gamma^2_kF_{\bf
k}+4\frac{\omega_k^2}{\gamma_k}\frac{dF_{{\bf
k}}}{d\eta}=4\gamma'_k-8\gamma^2_k\ .
 \label{3.91}
 \end{equation}
Anomalous and normal distribution functions are connected by the
following equations
\begin{equation}
\begin{array}{c}
 \displaystyle \frac{dF_{{\bf k}}}{d\eta}=2\gamma_k\left(G^{*}_{{\bf
k}}e^{2i\int\omega_kd\eta}+G_{{\bf
k}}e^{-2i\int\omega_kd\eta}\right)\ ,
\\[5mm]
 \displaystyle
 \frac{dG^*_{\bf k}}{d\eta}=\gamma_k\left(F_{\bf
k}+1\right)e^{-2i\int\omega_kd\eta}\ ,\qquad
 \frac{dG_{\bf k}}{d\eta}=\gamma_k\left(F_{\bf k}+1\right)e^{2i\int\omega_kd\eta}\ .
\end{array}
\label{3.92}
 \end{equation}

If gravitons and ghosts are defined by conditions (\ref{3.82}) and
(\ref{3.86}) at an arbitrary instant of time, the equation
(\ref{3.91}) (or the system of equations (\ref{3.92})) describes
the evolution of quantum ensemble. The normal distribution
functions $F^{g}_{\bf k}(\eta),\ F^{gh}_{\bf k}(\eta)$  are the
averages of graviton and ghost occupation numbers as functions of
time. If the initial state is defined as trivial vacuum, then
$F^{g}_{\bf k}(\eta_0)=0,\ F^{gh}_{\bf k}(\eta_0)=0$. The
inhomogeneity of equations (\ref{3.91}) and (\ref{3.92}) leads to
instability of trivial vacuum: for $\eta_1>\eta_0$ it clearly
leads to $F^{g}_{\bf k}(\eta_1)\ne 0,\ F^{gh}_{\bf k}(\eta_1)\ne
0$. In addition, the equations (\ref{3.91}) and (\ref{3.92}) show
that the graviton--ghost ensemble of the general form inevitably
appears in the process of evolution. We get the product of
normalized superpositions at the instant of time  $\eta=\eta_0$
for an arbitrary state vector that is built by operators
(\ref{3.88}) at the instant of time $\eta_1>\eta_0$. As a matter
of fact, such a state vector can be obtained automatically. To
construct it, one needs to use the following operators that are
solutions of operator equations (\ref{3.82}) and (\ref{3.85})
 \begin{equation}
\displaystyle \hat n_{{\bf k}\sigma}(\eta_1)=\tilde c^+_{{\bf
k}\sigma}(\eta_1)\tilde c_{{\bf k}\sigma}(\eta_1)\ ,\qquad \hat
n_{{\bf k}}(\eta_1)=\tilde a^+_{{\bf k}}(\eta_1)\tilde a_{{\bf
k}}(\eta_1)\ ,\qquad \hat n_{{\bf k}}(\eta_1)=\tilde b^+_{{\bf
k}}(\eta_1)\tilde b_{{\bf k}}(\eta_1)\ ,
\label{3.93}
 \end{equation}
Any different definition of gravitons and ghost leads to the same
conclusion.

Thus, states with graviton and ghost zero occupation numbers are
degenerative and unstable (for a given determination). During the
evolution of the Universe, such states can appear only
incidentally at  a single specific instant of time. The status of
ghosts as a quantum field with non--zero occupation numbers
results only from the internal properties of the theory. First,
there are no ghostless gauges in the quantum gravity. Second, the
one--loop finiteness determines dynamic properties of ghost field,
specifically its conformal non--invariance and zero rest mass.
Third, in cosmology (in distinction to the scattering problem),
the trivial ghost vacuum is unstable in principle.

The properties of quantum gravity listed above differentiate the
gravity from the Yang--Mills gauge fields. To emphasize these
differences, let us mention that the Yang--Mills theory has the
ghostless gauges. This fact deprives ghosts of status of physical
objects. Furthermore, in the  $S$--matrix theory, all observables
are calculated for asymptotical states in which all objects of the
theory are situated at their mass shells. The vacuum and particles
in these states are unambiguously defined, and the vacuum is
stable. It allows defining ghost states as vacuum states if ghost
gauges are in use. Note that the elimination of ghosts from the
physical sector of the theory is the procedure of restoration of
gauge invariance of  $S$--matrix.  This statement is relevant to
both the Yang--Mills theory and theory of graviton scattering. The
theory of gravitons in the cosmological space of self--consistent
geometry is different. Its specific feature is that there are no
asymptotical states in the non--stationary Universe; vacuum is
unstable and conformal non--invariant gravitons and ghosts of zero
rest mass are situated off the mass shell. In the ultra--short
wave limit only, ghosts can gain the status in cosmology similar
to the status they have in $S$--matrix theory. This is because all
non--stationary effects are negligible in this limit in the
Universe (Section  \ref{swg}). There is no physical or/and
mathematical basis to exclude ghosts from the set of physical
objects of the theory if we deal with long wavelength modes.
Direct participation of long wave ghosts in the formation of
macroscopic observables leads to creation of physical states that
have no analogies in the classic theory of gravity (Sections
\ref{lgw} and \ref{exact}). We name these states as "vacuum
graviton--ghost condensate"\ .

Note, that the ghosts and ghost condensates that might play a
possible role in the formation of properties of the early Universe
appear in some generalizations of the gravity theory and were
discussed in many papers (see e.g. \cite{17, 18, 19, R1}). In this
paper, we emphasize the fact that the long wave ghosts gain the
status of physical objects even in the framework of one--loop
quantum gravity.

\section{Approximate Solutions}\label{approx}

\subsection{Gas of Short Wave Gravitons}\label{swg}

Let us consider the gas of gravitons of wavelength that is much
shorter than the distance to the cosmological horizon. We exclude
the long waves from the model. Also, the calculation of
observables is done approximately, so that non--adiabatic
evolution of quantum ensemble is not taken into account. In the
framework of these approximations, it is possible to save the pure
vacuum status of ghosts because their role is just  to provide the
one--loop finiteness of macroscopic quantities. Long wave
excitations we will consider in Section \ref{lgw}.

The calculation of observables for the gas of short wave gravitons
can be done by general formulas (\ref{3.61}), (\ref{3.66}) ---
(\ref{3.68}) after the definition of basis functions and the state
vector. For the short wave approximation, the full asymptotic
expansion of basis functions exists that satisfies the
normalization condition (\ref{3.45}) and asymptotes (\ref{3.58}).
Of course, to use the method of asymptotic expansions, basis
functions must be taken in the following form
\begin{equation}
\displaystyle f_k=\frac{1}{a\sqrt{2\epsilon_k}}e^{-i\phi_k}\ ,
\qquad f^*_k=\frac{1}{a\sqrt{2\epsilon_k}}e^{i\phi_k}\ , \qquad
\phi_k=\int\limits_{\eta_0}^\eta\epsilon_kd\eta\ , \label{4.1}
\end{equation}
where
\[
\displaystyle \epsilon_k=\epsilon_k(\rho,\ \rho',\ \rho'',\ ...)\ ,\qquad \rho=-\frac{a''}{a}
\]
is a real functional of scale factor and its derivatives. In the
short wave approximation, this functional is expanded into the
local asymptotic series, which satisfies to the following boundary
condition\footnote{ Note that the $\rho^{(n)}(\eta)\to 0$
asymptotic exists for cosmological solutions of usual interest.
For instance, $\rho^{(n)}(\eta_0)=0$ as $\eta_0=-\infty$ for the
inflation solution. For $\eta_0=+\infty$ it takes place for the
FRW solution for the Universe filled with ordinary matter.}
\begin{equation}
\displaystyle \epsilon_k=\epsilon_k(\rho,\ \rho',\ \rho'',\ ...)\to k\ ,\qquad
\rho,\ \rho',\ \rho'',\ ... \ \to 0\ .
\label{4.2}
\end{equation}
There are no arbitrary constants in this expansion if the (\ref{4.2}) condition is satisfied.

The following linear ordinary differential equation of the third
order with respect to $1/\epsilon_k$ functional follows from the
equation (\ref{3.57}) for $y_k=af_k,\ af^*_k$ functions
\begin{equation}
 \begin{array}{c}
  \displaystyle
  \frac12\left(\frac{1}{\epsilon_k}\right)'''
  +2\omega_k^2\left(\frac{1}{\epsilon_k}\right)'+
  \left(\omega_k^2\right)'\frac{1}{\epsilon_k}=0\ ,
  \\[5mm]
 \displaystyle \omega_k^2=k^2+\rho\ .
 \end{array}
   \label{4.3}
\end{equation}
The solution of equation (\ref{4.3}) satisfying to the asymptotic condition (\ref{4.2}) reads
 \begin{equation}
\displaystyle
  \frac{1}{\epsilon_k}=\frac{1}{\omega_k}\sum_{s=0}^{\infty}(-1)^s\hat
 J_k^s\cdot 1\ .
 \label{4.4}
\end{equation}
Powers of $\hat J_k$ operator from (\ref{4.4}) are defined as follows
 \begin{equation}
 \begin{array}{c}
  \displaystyle \hat J_k\cdot \varphi=
\frac14\int\limits_{\eta_0}^\eta\frac{d\eta}{\omega_k}\left(\frac{\varphi}{\omega_k^3}\right)'''\
  ,
    \\[5mm]
    \displaystyle
\hat J_k^0\cdot 1\equiv 1\ ,\qquad J_k\cdot
1=\frac18\left(-\frac{\rho''}{\omega_k^4}+\frac54\frac{\rho^{'2}}{\omega_k^6}\right)\
, \qquad J_k^2\cdot 1=J_k\cdot (J_k\cdot 1)\ , \qquad  J_k^s\cdot
1=J_k^{s-1}\cdot (J_k\cdot 1)\ .
  \end{array}
   \label{4.5}
\end{equation}
The integral is calculated explicitly for arbitrary $s$, so that
$J_k^s\cdot 1$ is a local functional of $\rho$  and its
derivatives. It follows from (\ref{4.5}) that a small parameter of
asymptotic expansion is of the order of $\sim 1/k^2$. The
(\ref{4.4})  solution is approximate because non--local effects
are not included to the local asymptotical series. Calculation of
these effects is beyond of limits of this method.

The asymptotic expansion (\ref{4.4}), (\ref{4.5}) defines the
$1/\epsilon_k$  functional, and hence, it defines basis functions
(\ref{4.1}). The substitution of (\ref{4.1}) to (\ref{3.66})
produces asymptotic expansions of moments of spectral function
that read
\begin{equation}
 \begin{array}{c}
  \displaystyle
  W_n=
\frac{4\varkappa\hbar}{a^{2+2n}}\sum_{{\bf
k}}\frac{k^{2n}}{\epsilon_k} \left\{\sum_\sigma\langle
\Psi_g|c^+_{{\bf k}\sigma}c_{{\bf k}\sigma}|\Psi_g\rangle -\langle
\Psi_{gh}|a^+_{{\bf k}}a_{{\bf k}}+b^+_{{\bf k}}b_{{\bf
k}}|\Psi_{gh}\rangle +\right.
  \\[5mm]
   \displaystyle  \left.
+\left[\frac12\sum_\sigma\langle\Psi_g|c^+_{{\bf
k}\sigma}c^+_{-{\bf k}-\sigma}|\Psi_g\rangle -
\langle\Psi_{gh}|a^+_{{\bf k}}b^+_{-{\bf
k}}|\Psi_{gh}\rangle\right] e^{2i\phi_k}+\right.
\\[5mm]  \displaystyle
\left.+\left[\frac12\sum_\sigma\langle\Psi_g|c_{-{\bf
k}-\sigma}c_{{\bf k}\sigma}|\Psi_g\rangle -
  \langle\Psi_{gh}|b_{-{\bf k}}a_{{\bf k}}|\Psi_{gh}\rangle\right] e^{-2i\phi_k}\right\}\ .
  \end{array}
\label{4.6}
\end{equation}

State vectors from (\ref{4.6}) can be concretized from the general
considerations. It was mentioned in Section \ref{stvec} that such
terms as vacuum, zero oscillations and quantum wave excitations
are well defined for the $\rho^{(n)}(\eta)\to 0$ condition. Under
the same condition, state vectors that are built on basis vectors
of the Fock space are easily interpreted. First of all, this
statement is relevant to gravitons. Eigenvalues $n_{{\bf
k}\sigma}$ and eigenvectors $|n_{{\bf k}\sigma}\rangle$ of $\hat
n_{{\bf k}\sigma}=\hat c^+_{{\bf k}\sigma}\hat c_{{\bf k}\sigma}$
operator describe real gravitons in asymptotic states. In the
short wave approximation, the concept of real gravitons is valid
for all other stages of the Universe evolution. Thus, in this
particular case, the state vector of the general form can be
reduced to the product of vectors corresponding to states with
definite graviton numbers $n_{{\bf k}\sigma}=0,\ 1,\ 2, ...$
possessing definite momentum and polarization. It reads
\begin{equation}
\displaystyle |\Psi_g\rangle=\prod\limits_{{\bf k}\sigma}|n_{{\bf
k}\sigma}\rangle\ .
\label{4.7}
\end{equation}
In asymptotic states, short wave ghosts are only used to
compensate non--physical vacuum divergences. In accordance with
such an interpretation of the ghost status, we suppose that ghosts
and anti--ghosts sit in vacuum states that read
 \begin{equation}
\displaystyle |\Psi_{gh}\rangle=\prod\limits_{{\bf k}}\prod\limits_{{\bf k'}}|0_{{\bf
k}}\rangle|\bar 0_{{\bf k'}}\rangle\ .
 \label{4.8}
 \end{equation}
Averaging over the quantum state that is defined by (\ref{4.7})
and (\ref{4.8}) vectors, we get
\[
 \begin{array}{c}
  \displaystyle
\langle \Psi_g|c^+_{{\bf k}\sigma}c_{{\bf
k}\sigma}|\Psi_g\rangle=n_{{\bf k}\sigma}, \qquad \langle
\Psi_{gh}|a^+_{{\bf k}}a_{{\bf k}}|\Psi_{gh}\rangle=\langle
\Psi_{gh}|b^+_{{\bf k}}b_{{\bf k}}|\Psi_{gh}\rangle=0,
 \\[5mm] \displaystyle
 \Psi_g|c^+_{{\bf k}\sigma}c^+_{-{\bf k}-\sigma}|\Psi_g\rangle=
 \Psi_g|c_{-{\bf k}-\sigma}c_{{\bf k}\sigma}|\Psi_g\rangle=
 \langle\Psi_{gh}|a^+_{{\bf k}}b^+_{-{\bf k}}|\Psi_{gh}\rangle=
 \Psi_{gh}|b_{-{\bf k}}a_{{\bf k}}|\Psi_{gh}\rangle=0,
 \\[5mm]\displaystyle
  W_n=
 \frac{4\varkappa\hbar}{a^{2+2n}}\sum_{{\bf k}\sigma}\frac{k^{2n}}{\epsilon_k}
n_{{\bf k}\sigma}.
  \end{array}
\]

To calculate macroscopic observables in this approximation, it is
sufficient to keep only the first terms of expansion of moments of
spectral function that contain no higher than second derivative of
scale factor.  In this approximation, moments of spectral function
read
\begin{equation}
 \begin{array}{c}
 \displaystyle W_0=\frac{4\varkappa\hbar}{a^2}\sum_{{\bf
k}\sigma}\frac{n_{{\bf k}\sigma}}{k}\ ,\qquad D=-
\frac{8\varkappa\hbar}{a^2}\left(\dot H+H^2\right)\sum_{{\bf
k}\sigma}\frac{n_{{\bf k}\sigma}}{k}\ ,
\\[5mm]\displaystyle W_1=
\frac{4\varkappa\hbar}{a^4}\sum_{{\bf k}\sigma}kn_{{\bf
k}\sigma}+\frac{2\varkappa\hbar}{a^2}\left(\dot
H+2H^2\right)\sum_{{\bf k}\sigma}\frac{n_{{\bf k}\sigma}}{k}\ ,
 \end{array}
 \label{4.9}
\end{equation}
Taking into account (\ref{4.9}), we get energy density and
pressure of high--frequency graviton gas from (\ref{3.61}) that
read
 \begin{equation}
 \begin{array}{c}
\displaystyle
\varkappa\varepsilon_g=\frac{\varkappa\hbar}{a^4}\sum_{{\bf
k}\sigma}kn_{{\bf
k}\sigma}+\frac{\varkappa\hbar}{2a^2}H^2\sum_{{\bf
k}\sigma}\frac{n_{{\bf k}\sigma}}{k}\ ,
\\[5mm] \displaystyle
\varkappa p_g=\frac{\varkappa\hbar}{3a^4}\sum_{{\bf
k}\sigma}kn_{{\bf k}\sigma}-\frac{\varkappa\hbar}{6a^2}\left(2\dot
H+H^2\right)\sum_{{\bf k}\sigma}\frac{n_{{\bf k}\sigma}}{k}\ .
\end{array}
 \label{4.10}
\end{equation}
Relations (\ref{4.9}) and (\ref{4.10}) are valid if $a^2/\bar
k^2\sim \bar \lambda^2\ll H^{-2},\, |\dot H|^{-1}$, i.e. the
square of ratio of graviton wavelength to horizon distance is much
less than unity. In case of large occupation numbers, these
results are of the quasi--classical character and can be obtained
by the classical theory of gravitational waves \cite{20}.

As can be seen from (\ref{4.10}), the high--frequency graviton gas
differs from the ideal gas with the equation of state
$p=\varepsilon/3$ by only so--called post--hydrodynamic
corrections. In accordance with the approximation used, these
corrections are of the order of $\bar \lambda^2H^2\ll 1$ in
comparison with main terms. Thus, the following simple formula can
be used
\begin{equation}
\displaystyle \varkappa\varepsilon_g\simeq 3\varkappa p_g\simeq
\frac{C_{g1}}{a^4}\ , \qquad C_{g1}=\varkappa\hbar\sum_{{\bf
k}\sigma}kn_{{\bf k}\sigma}\ .
 \label{4.11}
\end{equation}

\subsection{Quantized Gravitons and Ghosts of Super-Long Wavelengths}\label{lgw}

In the framework of this theory, it is possible to describe the
ensemble of super--long gravitational waves ($k^2\ll |a''/a|$) by
an approximate analytical method. Such an ensemble corresponds to
the Universe whose observable part is in the chaotic bunch of
gravitational waves of wavelengths greater than the horizon
distance. The chaotic nature of the bunch is provided by non--zero
wave vectors of these waves, so that observable properties of the
Universe are formed by superposition of waves of different
polarizations and orientations in the space. Such a wave system
can produce an the isotropic spectrum and isotropic polarization
ensemble consistent with the homogeneity and isotropy of the
macroscopic space.

Such an ensemble of super--long waves can be formed only if the
size of causally--bounded region is much greater than the horizon
distance, which is possible in the framework of the hypothesis of
early inflation (or other scenarios (see, e.g. \cite{21}).
However, the problem of kinematical stability of an ensemble
exists even in the framework of the hypothesis of early inflation.
The case is due to the fact that the ensemble of long waves is
destroyed during the post--inflation epoch if the Universe is
expanded with a deceleration. When long waves come out of horizon,
they are transformed to the short waves. Below we show that the
kinematical self--stabilization of an ensemble is possible in the
framework of self--consistent theory of long waves.

Long waves under discussion correspond to virtual gravitons. To
describe them approximately, one needs to use asymptotic
expansions of basis functions over the small parameter $\sim k^2$.
As well as in the case of short waves, the basis can be chosen in
the representation of self--conjugated functions that are
parameterized by the universal real functional $\epsilon_k(a)$.
This preserves the definition (\ref{4.1}) and equation
(\ref{4.3}). However, due to of our interest in the asymptotical
expansion $1/\epsilon_k(a)$ over $\sim k^2$ parameter, it is
necessary to rewrite equation (\ref{4.3}) in the following form
 \begin{equation}
 \begin{array}{c}
 \displaystyle
\left\{a^2\left[a^2\left(\frac{1}{a^2\epsilon_k}\right)'\right]'\right\}'=
-4k^2a^2\left(\frac{1}{\epsilon_k}\right)'\
.
\end{array}
 \label{4.12}
\end{equation}
Let us introduce the geometric--dynamic time $d\tau=d\eta/a^2$ and
the following functional
\[
\displaystyle \frac{1}{2\epsilon_ka^2}\equiv\xi_k=\sum_{n=0}^\infty\xi_k^{(n)}
\]
Note that the $\tau$ time coordinate corresponds to the original
gauge $\sqrt{-g}g^{00}=1$. Equation (\ref{4.12}) and the spectral
function (\ref{3.69}) now read
\begin{equation}
 \begin{array}{c}
  \displaystyle
\frac{d^3\xi_k}{d\tau^3}=-4k^2a^2\frac{d}{d\tau}(a^2\xi_k)\ ,
\end{array}
 \label{4.13}
\end{equation}
\begin{equation}
\displaystyle W_{{\bf k}}=8\varkappa\hbar\xi_k\left(N_{\bf
k}+U^*_{\bf k}e^{i\Phi_k}+U_{\bf k}e^{-i\Phi_k}\right)\ , \qquad
\Phi_k=\int\limits_{\tau_k}^\tau\frac{d\tau}{\xi_k} \ ,
\label{4.14}
\end{equation}
where $\tau_k$ is a numerical parameter. Its value is unimportant
because the constant's contribution to phases of basis functions
is absorbed by phases of contributors that form vectors of the
general form (\ref{3.48}) and (\ref{3.51}). Observables
(\ref{3.61}) are expressed via moments of spectral function. The
latter read
\begin{equation}
\begin{array}{c}
\displaystyle  D=\frac{1}{a^6}\sum_{\bf k}\frac{d^2W_{\bf k}}{d\tau^2}=
\\[5mm]
\displaystyle =\frac{8\varkappa\hbar}{a^6}\sum_{\bf
k}\left[\frac{d^2\xi_k}{d\tau^2}N_{\bf
k}+\left(\frac{d^2\xi_k}{d\tau^2}-\frac{1}{\xi_k}\right)\left(U^*_{\bf
k}e^{i\Phi_k}+U_{\bf k}e^{-i\Phi_k}\right)
+\frac{i}{\xi_k}\frac{d\xi_k}{d\tau}\left(U^*_{\bf
k}e^{i\Phi_k}-U_{\bf k}e^{-i\Phi_k}\right)\right]\ ,
\\[5mm]
\displaystyle W_1=\frac{1}{a^2}\sum_{\bf k}k^2W_{\bf
k}=\frac{8\varkappa\hbar}{a^2}\sum_{\bf k}k^2\xi_k\left(N_{\bf
k}+U^*_{\bf k}e^{i\Phi_k}+U_{\bf k}e^{-i\Phi_k}\right)\ .
\end{array}
\label{4.15}
\end{equation}

The iteration procedure over $\sim k^2$ parameter for equation
(\ref{4.13}) is constructed accordingly to the following rules
\begin{equation}
 \begin{array}{c}
 \displaystyle
\frac{d^3\xi_k^{(0)}}{d\tau^3}=0\ , \qquad
\xi_k^{(0)}\equiv\frac{1}{a^2\epsilon_k^{(0)}}=P_{k}+R_{k}\tau+Q_{k}\tau^2\ ,
\\[5mm]
 \displaystyle
\frac{d^3\xi_k^{(n)}}{d\tau^3}=-4k^2a^2\frac{d}{d\tau}(a^2\xi^{(n-1)}_k)\
, \qquad n\geqslant 1\  .
\end{array}
 \label{4.16}
\end{equation}
In particular, we get
\begin{equation}
 \begin{array}{c}
\displaystyle \frac{d^2\xi_k^{(1)}}{d\tau^2}=-2k^2P_ka^4+...\ .
\end{array}
 \label{4.17}
 \end{equation}
The virtual graviton is defined by integration constants $P_{k},\
Q_{k},\ R_{k}$  of the main term of asymptotic expansion. Because
the  $\epsilon_k$ functional of (\ref{4.1}) is real (and therefore
the $\xi_k^{(0)}$ functional of (\ref{4.16}) is also real), we
obtain following inequality
\begin{equation}
\displaystyle 4P_kQ_k-R_k^2>0, \qquad P_k>0,\qquad Q_k>0\ .
\label{4.18}
\end{equation}
The dependence of constants $P_{k},\  Q_{k},\ R_{k}$ and phase
$\Phi_k$ on $k$ for  $k\to 0$ is defined by the finiteness
condition for $k^2W_{\bf k}$ and $d^2W_{\bf k}/d\tau^2$, and
taking into account the inequality (\ref{4.18}) we obtain
\[
\displaystyle P_k=\mathcal{O}(k^{-2}),\qquad
R_k=\mathcal{O}(k^{-1}),\qquad Q_k=\mathcal{O}(k^{0}),\qquad
\Phi_k=\mathcal{O}(k^{2})\ .
\]
The main terms of asymptotical expansions of moments (\ref{4.15}),
energy density and pressure of long wave gravitons can be obtained
from (\ref{4.16}) for $\xi_k^{(0)}$ and (\ref{4.17}) for
$\xi_k^{(1)}$. They read
\begin{equation}
\begin{array}{c}
 \displaystyle D=-\frac{16C_{g2}}{a^2}+\frac{16C_{g3}}{a^6}\ ,\qquad
 W_1=\frac{8C_{g2}}{a^2}\ ,
 \\[5mm]
\displaystyle \varkappa\varepsilon_g =
\frac{C_{g2}}{a^2}+\frac{C_{g3}}{a^6}\ , \qquad \varkappa p_g =
-\frac{C_{g2}}{3a^2}+\frac{C_{g3}}{a^6}\ .
\end{array}
\label{4.19}
\end{equation}
where
\begin{equation}
 \begin{array}{c}
 \displaystyle
C_{g2}=\varkappa\hbar\sum_{\bf k}k^2P_k\left(N_{\bf k}+U^*_{\bf k}+U_{\bf k}\right)\ ,
\\[5mm]
 \displaystyle
C_{g3}=\varkappa\hbar\sum_{\bf k}Q_k\left(N_{\bf k}+U^*_{\bf k}+U_{\bf k}\right)\ .
\end{array}
 \label{4.20}
\end{equation}
For the first time, approximate solutions for the energy density
and pressure in the (\ref{4.19}) form were obtained for classical
long gravitational waves in \cite{42,43}. In the theory of
classical gravitational waves \cite{42,43}, the constants of
integration $C_{g2}$  and $C_{g3}$ must be positive. The crucial
formal difference between classical and quantum long gravitational
waves is in the fact that the last ones allow an arbitrary sign of
$C_{g2}$  and $C_{g3}$ (negative as well as positive). The physics
of this crucial difference will be discussed below (Section
\ref{sce}).

In a particular case of  $\delta$--type graviton spectrum, which
is localized at the region of very small conformal wave numbers,
(\ref{4.19}) can be considered as exact solutions. One needs to to
go over from summation to integration
\[
\sum_{\bf k}... \to \frac{1}{(2\pi)^3}\int d^3k...=\frac{1}{2\pi^2}\int\limits_0^\infty k^2dk...\ .
\]
After that, these solutions can be obtained by the following limits
\begin{equation}
 \begin{array}{c}
 \displaystyle  k^2P_k\to \frac{k_1}{a_1}=const(k),\qquad Q_k\to Q=const(k)\ ,
 \\[5mm] \displaystyle
 N_{\bf k}+U^*_{\bf k}+U_{\bf k}\to \frac{2\pi^2}{k^2}\mathcal{N}_0\delta(k-\kappa_0)\ ,
 \qquad \mathcal{N}_0=const(k)\ ,\qquad \kappa_0\to 0\ .
 \end{array}
 \label{4.21}
\end{equation}
In (\ref{4.21}) $k_1$ and $a_1$ are the constants of dimension of
conformal wave number and scale factor, respectively. They provide
the correct dimension to parameter $\displaystyle\lim_{k\to
0}k^2P_k$.

\subsection{Scenarios of Macroscopic Evolution}\label{sce}

In accordance with (\ref{4.19}), the system of long wave gravitons
behaves as a medium consisting of two subsystems whose equations
of state are $p_1=-\varepsilon_1/3$ and $p_2=\varepsilon_2$. But,
the internal structure of this substratum cannot be determined by
measurements that are conducted under the horizon of events. The
substratum effect (\ref{4.19}) on evolution of the Universe, is
seen by an observer as an energy density and pressure of the
"empty"\ (non--structured) spacetime, i.e. vacuum. The question
is: {\it does the graviton vacuum have a quasi--classic nature, or
has its quantum gravity origin been revealed in some cases?}

Let us review the situation. First, the superposition of quantum
states in state vectors of the general form (\ref{3.48}) and
(\ref{3.51}) could be essentially non--classical. Second, the
clearly non--classical ghost sector is inevitably presented in the
theory. Its properties are determined by the condition of
one--loop finiteness of macroscopic quantities (Section
\ref{fininv}). The ghost sector is directly relevant to the
(\ref{4.19}) solution. Let us consider (\ref{3.72}) and
(\ref{3.73}), assuming for the sake of simplicity  $\langle
n_{k(gh)}\rangle=\langle \bar n_{k(gh)}\rangle$. Parameters of
solution (\ref{4.19}) are expressed via parameters of
graviton--ghost ensemble as follows
\begin{equation}
\begin{array}{c}
\displaystyle C_{g2}=2\varkappa\hbar\sum_{\bf
k}k^2P_k\left[\langle
n_{k(g)}\rangle(1+\zeta^{(g)}_k\cos\varphi_k)- \langle
n_{k(gh)}\rangle(1+\zeta^{(gh)}_k\cos\chi_k)\right]\ ,
\\[5mm]
\displaystyle C_{g3}=2\varkappa\hbar\sum_{\bf k}Q_k\left[\langle
n_{k(g)}\rangle(1+\zeta^{(g)}_k\cos\varphi_k)- \langle
n_{k(gh)}\rangle(1+\zeta^{(gh)}_k\cos\chi_k)\right]\ .
\end{array}
 \label{4.22}
\end{equation}
It follows from (\ref{4.22}) that  $C_{g2}>0, \ C_{g3}>0$ if the
graviton contribution dominates over ghosts in the quantum
condensate. We will name such a condensate "quasi-classical"\ .
Its energy density is positive, and it can be formed by usual
super--long gravitational waves. If the ghost contribution
dominates over gravitons in the quantum condensate, then
$C_{g2}<0, \ C_{g3}<0$. Such a condensate of negative energy
density has no classical analogy.

Summarizing the results of Sections \ref{swg} and \ref{lgw}, we
see that in cosmological applications of one--loop quantum gravity
we deal with the multi--component system consisting of short wave
graviton gas  $g1$ and two subsystems of graviton--ghost
condensate $g2,\ g3$. Taking into account (\ref{4.11}) and
(\ref{4.19}), we get the following equation for the scale factor
\begin{equation}
 \begin{array}{c}
 \displaystyle 3\frac{a'^2}{a^4}=\frac{C_{g1}}{a^4}+\frac{C_{g2}}{a^2}+\frac{C_{g3}}{a^6}\ .
 \end{array}
 \label{4.23}
\end{equation}
In the first scenario, the long wavelength condensate is of
negative energy, which means that the contribution of ghost
dominates over gravitons. The evolution of such a Universe is of
oscillating type. The solution reads
\begin{equation}
 \begin{array}{c}
\displaystyle
a^2=
\frac{C_{g1}}{2|C_{g2}|}+\frac{\sqrt{C_{g1}^2-
4C_{g2}C_{g3}}}{2|C_{g2}|}\sin\sqrt{\frac{4|C_{g2}|}{3}}\eta\
,
 \\[5mm]
\displaystyle
a^2_{1,2}=\frac{C_{g1}}{2|C_{g2}|}\mp\frac{\sqrt{C_{g1}^2-4C_{g2}C_{g3}}}{2|C_{g2}|}
 \end{array}
 \label{4.24}
\end{equation}
There is no classic analogy to the solution (\ref{4.24}). It can
be used for scenarios of evolution of the early quantum Universe.
In the region of minimal values of the scale factor $a_{min}=a_1$,
the $g3$ condensate bounces the Universe back from a singularity.
The transition from the expansion to the contraction epoch at the
region of maximal scale factor $a_{max}=a_2$ is provided by $g2$
condensate. Because of correlation of signs of  $C_{g2}<0$ and
$C_{g3}<0$, the non--singular Universe oscillates. Recent
scenarios of oscillating Universes based on condensates of
hypothetical ghost fields are under discussion in the current
literature as an alternative to the idea of inflation (see, e.g.
\cite{21})). Actually, we have shown that the same--type scenario
is constructed with the standard building blocks of quantum
gravity the well-known De Witt--Faddeev--Popov's ghosts located
far from the mass shell. Thus, a very attractive idea is that one
and the same mechanism of graviton--ghost condensate formations in
the framework of one--loop quantum gravity based on the
"standard"\ Einstein equations (without hypothetical fields and
generalizations of Einstein's general relativity) could be
responsible for both Dark Energy effect (see Section \ref{de}) and
cyclic evolution of the early Universe (instead of inflation).

The second type of scenario applies if gravitons dominate over
ghosts in the condensate of positive energy. The solution reads
\begin{equation}
 \begin{array}{c}
\displaystyle
2\sqrt{C_{g2}(C_{g2}a^4+C_{g1}a^2+C_{g3})}+2C_{g2}a^2+C_{g1}=
\left(2\sqrt{C_{g2}C_{g3}}+C_{g1}\right)
 \exp\left(\sqrt{\frac{4|C_{g2}|}{3}}\eta\right)\ .
  \end{array}
 \label{4.25}
\end{equation}
The $g3$ condensate forms the regime of evolution in the vicinity
of singularity; meanwhile the asymptote of cosmological solution
for $\eta\to \infty$ is formed by $g2$ condensate. Short wave
gravitons $g1$ dominate during the intermediate epoch. The ratio
of graviton wavelength to horizon distance is constant during the
following asymptotical regime
\[
 \displaystyle
a\sim \exp\left(\sqrt{\frac{|C_{g2}|}{3}}\eta\right)\sim t,
\]
This means that the long wave condensate $g2$ forms the
self--consistent regime of evolution that provides its kinematic
stability.

\section{BBGKY Hierarchy (Chain)
and Exact Solutions of One--Loop Quantum Gravity
Equations}\label{Bog}

\subsection{Constructing the  Chain}\label{chain}

Approximate methods used in Sections \ref{swg} and \ref{lgw}
provide an opportunity to describe only limit cases which are
ultra shortwave gravitons and ghosts against the background of
almost stable Fock vacuum and super--long wave modes, forming
nearly stable graviton--ghost condensate. Now we are examining
self--consistent theory of gravitons and ghosts with the
wavelengths of the order of distance to the horizon:
 \begin{equation}
  \begin{array}{c}
  \displaystyle \frac{k^2}{a^2}\sim H^2,\ |\dot H|\ .
  \end{array}
  \label{5.1}
\end{equation}
When describing modes (\ref{5.1}), one should keep in mind two
factors. First, in the area of the spectrum (\ref{5.1}), there are
no reasonable approximations, which could be used to solve
equations (\ref{3.30}) and (\ref{3.31}), if the law of
cosmological expansion $a(t)$, $H(t)$ is not known in advance.
Second, the (\ref{5.1}) modes are quasi--resonant. Quantum gravity
processes of vacuum polarization, spontaneous graviton creation by
self--consistent field and graviton--ghost condensation are the
most intensive in this region of spectrum. From (\ref{5.1}) it is
also obvious that the threshold for quantum gravitational
processes involving zero rest mass gravitons and ghosts is absent.
These processes at the scale of horizon occur at any stage of
evolution of the Universe, including, in the modern Universe.

The theory that allows quantitatively describe quasi--resonant
quantum gravitational effects is constructed in the following way.
For the spectral function of gravitons and ghosts $W_{{\bf k}}$,
as defined in (\ref{3.59}), a differential equation is derived.
For this, the first equation (\ref{3.30}) is multiplied by the
$\hat\psi^+_{{\bf k}\sigma}$ (and then by the
$\dot{\hat\psi}^+_{{\bf k}\sigma}$), conjugated equation
(\ref{3.3}) is multiplied by the $\hat\psi_{{\bf k}\sigma}$ (and
then by the $\dot{\hat\psi}_{{\bf k}\sigma}$); and the equations
obtained are averaged and added. Similar action is carried out
with equations for ghosts, after which the equations for ghosts
are subtracted from the equations for gravitons. These operations
yield:
\begin{equation}
\begin{array}{c}
   \displaystyle
\ddot W_{{\bf k}}-2F_{{\bf k}}+3H\dot W_{{\bf
k}}+\frac{2k^2}{a^2}W_{{\bf k}}=0\ ,
   \end{array}
  \label{5.2}
 \end{equation}
 \begin{equation}
\begin{array}{c}
   \displaystyle
   \dot F_{{\bf k}}=-6HF_{{\bf k}}-\frac{k^2}{a^2}\dot W_{{\bf
   k}}\ ,
   \end{array}
  \label{5.3}
 \end{equation}
where
\[
\begin{array}{c}
   \displaystyle
W_{{\bf k}}=\sum_\sigma\langle
\Psi_g|\hat\psi^+_{{\bf k}\sigma}\hat\psi_{{\bf
 k}\sigma}|\Psi_g\rangle-2\langle\Psi_{gh}|{\bar\theta}_{{\bf k}}\theta_{{\bf
 k}}|\Psi_{gh}\rangle\ ,
\\[5mm]
\displaystyle
F_{{\bf k}}=\sum_\sigma\langle
\Psi_g|\dot{\hat\psi}^+_{{\bf k}\sigma}\dot{\hat\psi}_{{\bf
 k}\sigma}|\Psi_g\rangle-2\langle\Psi_{gh}|\dot{\bar\theta}_{{\bf k}}\dot\theta_{{\bf
 k}}|\Psi_{gh}\rangle\ .
 \end{array}
 \]
Further, equation (\ref{5.2}) is differentiated.  Expressions  for
$F_{{\bf k}},\ \dot F_{{\bf k}}$ via $W_{{\bf k}}$  are
substituted into the results of differentiation. For the spectral
function the third--order equation is produced
\begin{equation}
 \begin{array}{c}
    \displaystyle \stackrel{...}{W}_{{\bf k}} +9H\ddot W_{{\bf k}}
 +3\left(\dot H+6H^2\right)\dot W_{{\bf k}}
  +\frac{4k^2}{a^2}\left(
 \dot W_{{\bf k}}+2H W_{{\bf k}}\right)=0.
 \end{array}
 \label{5.4}
 \end{equation}

It is now necessary to draw attention to the fact that $W_{{\bf
k}}(t)$  is Fourier image of the two--point function, taken at
$t=t'$:
\begin{equation}
 \begin{array}{c}
    \displaystyle
    W(t,t'; {\bf{x}}-{\bf{x'}})=\langle
\Psi|\hat\psi_i^k(t,{\bf{x}})\hat\psi^i_k(t',{\bf{x'}})-2
\bar\theta(t,{\bf{x}})\theta(t',{\bf{x'}})|\Psi\rangle\ ,
\\[5mm]
\displaystyle W_{{\bf k}}(t)=\frac{1}{V}\int d^3y W(t,t;
{\bf{y}})e^{-i{\bf ky}}\ .
 \end{array}
 \label{5.5}
 \end{equation}
An infinite set of Fourier images is mathematically equivalent to
the infinite set of moments of the spectral function
\begin{equation}
\begin{array}{c}
\displaystyle W_n=\sum_{{\bf
k}}\frac{k^{2n}}{a^{2n}}\left(\sum_\sigma\langle
\Psi_g|\hat\psi^+_{{\bf k}\sigma}\hat\psi_{{\bf
k}\sigma}|\Psi_g\rangle-2\langle \Psi_{gh}|\bar\theta_{{\bf
k}}\theta_{{\bf k}}|\Psi_{gh} \rangle\right),\qquad
 n=0,\,1,\,2,\,...,\infty.
\end{array}
 \label{5.6}
 \end{equation}
Therefore, from the equation for Fourier images (\ref{5.4}), we
can move to an infinite system of equations for the moments. For
this, equation (\ref{5.4}) is multiplied by $(k/a)^{2n}$ followed
by summation over wave numbers. The result is a
Bogoliubov--Born--Green--Kirkwood--Yvon (BBGKY) chain. Each
equation of this chain connects the neighboring moments:
\begin{equation}
\displaystyle  \dot D+6HD+4\dot W_1+16HW_1=0\ ,
\label{5.7}
\end{equation}
 \begin{equation}
\begin{array}{c}
 \displaystyle
B(1,2)\equiv \stackrel{...}{W}_1 +15H\ddot{W}_1  +3\left(22H^2
+3\dot{H} \right)\dot{W}_1+ 2\left( 40H^3
+18H\dot{H}+\ddot{H} \right)W_1
+4\dot{W}_2+24HW_{2}=0,
\end{array}
\label{5.8}
\end{equation}
 \begin{equation}
\begin{array}{c}
 \displaystyle
B(n,n+1)\equiv\stackrel{...}{W}_n +3(2n+3)H\ddot{W}_n  +3\left[ \left(
4n^2+12n+6\right)H^2
+(2n+1)\dot{H} \right]\dot{W}_n+ \vspace{5mm} \\
\displaystyle +2n\left[ 2\left(2n^2+9n+9\right)H^3
+6(n+2)H\dot{H}+\ddot{H} \right]W_n
+4\dot{W}_{n+1}+8(n+2)HW_{n+1}=0, \qquad \displaystyle
n=2,\,...,\,\infty\, .
\end{array}
\label{5.9}
\end{equation}
Equations (\ref{5.7}) --- (\ref{5.9}) have to be solved jointly
with the following macroscopic Einstein equations
\begin{equation}
 \begin{array}{c}
\displaystyle
\dot H=-\frac{1}{16}D-\frac16W_1\ ,
 \\[5mm]
\displaystyle
3H^2=\frac{1}{16}D+\frac{1}{4}W_1+\varkappa\Lambda\ .
 \end{array}
 \label{5.10}
\end{equation}

Note that an infinite chain of equations (\ref{5.7}) ---
(\ref{5.9}) contains information not only on the space--time
dynamics of field operators, but also about the quantum ensemble,
over which the averaging is done. The multitude of solutions of
the equations of the chain includes all possible self--consistent
solutions of the operator equations, averaged over all possible
quantum ensembles. Theory of gravitons presented by BBGKY chain,
conceptually and mathematically corresponds to the axiomatic
quantum field theory in the Wightman formulation (see Chapter 8 in
monograph \cite{6}). Here, as in Wightman, full information on the
quantum field is contained in an infinite sequence of averaged
correlation functions, definitions of which simply relate to the
symmetry properties of manifold, on which this field determines.

In BBGKY chain (\ref{5.7}), (\ref{5.8}) and (\ref{5.9}), unified
graviton--ghost objects appear which are moments of the spectral
function, renormalized by ghosts. The ghosts are not explicitly
labeled so that the chain is can be built formally in the model
not containing ghost fields. Mathematical incorrectness of such a
model is obvious only with a microscopic point of view because in
the quantum theory all the moments of spectral function diverge
the stronger, the more the moment number is. The system of
equations (\ref{5.7}) --- (\ref{5.9}) does not "know"\ , however,
that without the involvement of ghosts (or something other
renormalization procedure) it applies to the mathematically
non--existent quantities. The three following mathematical facts
are of  principal importance.

(i) {\it In one--loop quantum gravity, the BBGKY chain can be
formally introduced at an axiomatic level;}

(ii) {\it The internal properties of equations (\ref{5.7}) ---
(\ref{5.10}) provide the existence of finite solutions to this
system;}

(iii) {\it In finite solutions, there are solutions which do not
meet the "classic"\ condition of positiveness of moments  (see
Sections \ref{ggh_cond} and \ref{Sitt}).}

It follows from these facts that there should be an opportunity
and the need to implement a renormalization procedure to the
theory. This procedure should be able to redefine the moments of
the spectral function to finite values, but that leaves them
sign--undefined. As it can be seen from the theory which is
presented in Sections \ref{qg} and \ref{scgt}, in the one--loop
quantum gravity such a procedure is contained within the theory
under condition that the ghost sector automatically provides the
one--loop finiteness.

We found three exact self--consistent solutions of the system of
equations consisting of   the BBGKY chain (\ref{5.7}) ---
(\ref{5.9}) and macroscopic given below in Sections \ref{ggh_cond}
and \ref{Sitt}. The existence of exact solutions can be obtained
through direct substitution into the original system of equations.
The microscopic nature of these solutions, i.e. dynamics of
operators and structure of state vector is described in Sections
\ref{exact}, \ref{inst}.

\subsection{Graviton--Ghost Condensates of Constant Conformal Wavelength}\label{ggh_cond}

In Section \ref{lgw} the exact solution was found for the
graviton--ghost condensate, consisting of spatially uniform modes
(see (\ref{4.19}) --- (\ref{4.21})). This solution satisfies to
the first two BBGKY equations (\ref{5.7}), (\ref{5.8}) for an
arbitrary law of evolution $H(t)$ and under condition that $W_n=0$
for $n\geqslant 2$. (Recall that in this solution $D$ and  $W_1$
must be understood as the result of limit transition $k^2\to 0$;
and equality to zero of higher moments follows from the spatial
uniformity of modes.) Now we describe the exact self--consistent
solutions for the system, in which in addition to spatially
uniform modes, quasi--resonant modes with a wavelength equal to
the distance to the horizon of events are taken into account. In
terms of moments of the spectral function, the structure of
solutions under discussion is
\begin{equation}
 \begin{array}{c}
\displaystyle
D=D(g2)+D(g3)+D(g4)\ ,\qquad W_1=W_1(g2)+W_1(g4)\ , \qquad W_{n}=W_{n}(g4),\; n\geqslant 2\ ,
 \\[5mm]
\displaystyle D(g3)=\frac{16C_{g3}}{a^6}\ ,\qquad
D(g2)=-\frac{16C_{g2}}{a^2}\ , \qquad W_1(g2)=\frac{8C_{g2}}{a^2}\
,
\\[5mm]
\displaystyle
D(g4)=-\frac{48C_{g4(1)}}{a^2}\ln\frac{a_0}{e^{1/4}a}\ , \qquad
W_n(g4)=\frac{24C_{g4(n)}}{a^{2n}}\ln\frac{a_0}{a},\quad
n=1,\,...,\,\infty\ .
 \end{array}
 \label{5.11}
\end{equation}
Here $C_{g3},\ C_{g2},\ C_{g4(n)},\ a_0$  are numerical
parameters. Restrictions on their values follow from the condition
of the existence of the exact self--consistent solution.

The solution is found by using of the consistency of functions
(\ref{5.11}) with the relations arising from the macroscopic
Einstein's equations (we are discussing model with $\Lambda=0$):
\begin{equation}
 \begin{array}{c}
\displaystyle
H^2=\frac{C_{g3}}{3a^6}+\frac{C_{g2}}{3a^2}+\frac{C_{g4}}{a^2}\ln\frac{e^{1/4}a_0}{a}\
,
\\[5mm]
\displaystyle
\dot H= -\frac{C_{g3}}{a^6}-\frac{C_{g2}}{3a^2}-\frac{C_{g4}}{a^2}\ln\frac{e^{3/4}a_0}{a}\ ,
\\[5mm]
\displaystyle \ddot
H=2H\left(\frac{3C_{g3}}{a^6}+\frac{C_{g2}}{3a^2}+
\frac{C_{g4}}{a^2}\ln\frac{e^{5/4}a_0}{a}\right)\
.
\end{array}
 \label{5.12}
\end{equation}
In (\ref{5.12}) as well as further, we use notation
$C_{g4(1)}\equiv C_{g4}$. Functions $D$ and $W_1$ from
(\ref{5.11}) transform the equation (\ref{5.7}) to an identity.
The substitution of  $W_1$ and $W_2$ into (\ref{5.8}), taking into
account (\ref{5.12}), leads to the following expression
\begin{equation}
 \begin{array}{c}
 \displaystyle
B(1,2)= H\frac{48}{a^4}\left[
4(C_{g4(2)}-C_{g4}^2)\ln\frac{a_0}{a}-\frac43C_{g2}C_{g4}+C_{g4}^2-2C_{g4(2)}\right]=0\
.
\end{array}
\label{5.13}
\end{equation}
The infinite chain (\ref{5.9}), in contrast to the equation
(\ref{5.8}), contains moments of spectral functions of
quasi--resonant modes. Nevertheless, it does result, only
including (\ref{5.13}) as a particular case
\begin{equation}
 \begin{array}{c}
 \displaystyle
B(n,n+1)= H\frac{48}{a^{2n+2}}\left[
4(C_{g4(n+1)}-C_{g4}C_{g4(n)})\ln\frac{a_0}{a}-\frac43C_{g2}C_{g4(n)}+C_{g4}C_{g4(n)}-2C_{g4(n+1)}\right]=0\
,
\\[5mm]
n=2,\,...,\,\infty.
\end{array}
\label{5.14}
\end{equation}
The following relations between parameters follow from
(\ref{5.13}) and (\ref{5.14})
\begin{equation}
 \begin{array}{c}
 \displaystyle
 C_{g4(n)}=C_{g4}^n\ , \qquad C_{g2}=-\frac34C_{g4}\ .
\end{array}
\label{5.15}
\end{equation}
Thus, moments of the spectral function of quasi--resonant modes
satisfy to the following recurrent relation
\begin{equation}
 \begin{array}{c}
 \displaystyle
 W_{n+1}(g4)=\frac{C_{g4}}{a^2}W_n(g4)=\left(\frac{C_{g4}}{a^2}\right)^{n}W_1(g4)\ .
\end{array}
\label{5.16}
\end{equation}
Comparison of (\ref{5.16}) with (\ref{5.6}) shows that in the
exact solution under discussion all quasi--resonant modes have the
same wavelength $\lambda=a/\sqrt{|C_{g4}|}\equiv a/k_0$. In other
words, in the space of conformal wave numbers the spectrum of
quasi--resonant wave modes is localized in the vicinity of the
fixed value $|{\bf k}|=k_0$.

Depending on the sign of $C_{g4}$, we get two exact solutions to
the macroscopic observables of graviton--ghost media in the form
of functionals of scale factor.

(i) {\it Oscillating Universe.}

Suppose that $C_{g4}>0$. In accordance with (\ref{5.15}), in this
case all  $C_{g4(n)}>0$. The positive sign of all moments
$W_n(g4)>0$ suggests that gravitons dominate over ghosts in the
ensemble of quasi--resonant modes.  We also see that the parameter
of spatially uniform mode $g2$ is negative, i.e. $C_{g2}<0$.  As
was shown in Section \ref{sce}, signs of parameters of $g2$ and
$g3$  modes are the same, so $C_{g3}<0$. From this it follows that
ghosts are dominant in case of spatially uniform modes. The energy
density and pressure of graviton--ghost substratum read
\begin{equation}
\displaystyle
\varkappa\varepsilon_g=-\frac{|C_{g3}|}{a^6}+\frac{3C_{g4}}{a^2}\ln\frac{a_0}{a}\
,\qquad \varkappa
  p_g=-\frac{|C_{g3}|}{a^6}-\frac{C_{g4}}{a^2}\ln\frac{a_0}{ea}\ .
\label{5.17}
\end{equation}
The parameter $C_{g2}$ is not explicitly showed up in (\ref{5.17})
because it is expressed via $C_{g4}$ in accordance with
(\ref{5.15}). There is an oscillating solution to the Einstein
equation $3H^2=\varkappa\varepsilon_g$ if solutions for the
turning points $a_m=a_{min},\, a_{max}$ exist, i.e.
\begin{equation}
\displaystyle b=\frac{3C_{g4}a_0^4}{4|C_{g3}|}>e,\qquad
\left(\frac{a_0}{a_m}\right)^4=b\ln\left(\frac{a_0}{a_m}\right)^4\
. \label{5.18}
\end{equation}
In the vicinity of turning points energy density is formed by
contributions of ghosts and gravitons, which are comparable in
their absolute values, but have opposite signs. Far from turning
points, graviton quasi--resonant modes dominate. Simplifying the
situation, we can say that in the oscillating Universe spatially
uniform modes have essentially quantum nature, and quasi--resonant
modes allow semi--classical interpretation.

In the absence of a spatially homogeneous subsystem  $g3$, the
infinite sequence of oscillations degenerates into one
semi--oscillation. Indeed, with  $C_{g3}=0$ the scale factor, as a
function of cosmological time, reads
\begin{equation}
\displaystyle a(\eta)=a_0\exp\left(-\frac{C_{g4}\eta^2}{4}\right)\ ,\qquad C_{g4}>0\ .
\label{5.19}
\end{equation}
In accordance with (\ref{5.19}), the Universe originates from a
singularity, reaches the state of maximal scale factor
$a_{max}=a_0$ and then collapses again to singularity.

(ii)    {\it Birth in Singularity and Accelerating Expansion.}

Accordingly to (\ref{5.16}), moments of the spectral function of
quasi--resonant modes form an alternating sequence if $C_{g4}<0$.
It reads
\begin{equation}
\displaystyle W_n(g4)=-(-1)^{n}\frac{24|C_{g4}|^n}{a^{2n}}\ln\frac{a}{a_0}\ ,
\qquad n=1,\,...,\,\infty \ .
\label{5.20}
\end{equation}
It is clear that the result (\ref{5.20}) can not be obtained for
the quasi--classical ensemble of gravitational waves. The
microscopic nature of this solution is discussed in Section
\ref{exact}. It is appropriate here to emphasize one more time
that the theory, which is formulated in the most common way in the
BBGKY form, captures the existence of such a solution.

It is not difficult to notice that the solution which we are now
discussing is in a sense, an alternative to the previous solution.
With $C_{g4}<0$, parameters of spatially homogeneous modes are
positive $C_{g2}>0,\ C_{g3}>0$. Thus, spatially uniform modes
admit semi--classical interpretation, but quasi--resonant modes
have essentially quantum nature. The energy density and pressure
of graviton--ghost substratum are
 \begin{equation}
\displaystyle
\varkappa\varepsilon_g=\frac{C_{g3}}{a^6}+\frac{3|C_{g4}|}{a^2}\ln\frac{a}{a_0}\
,\qquad \varkappa
  p_g=\frac{C_{g3}}{a^6}-\frac{|C_{g4}|}{a^2}\ln\frac{ea}{a_0}\ .
\label{5.21}
\end{equation}
Specific properties of solutions to Einstein's equations
$3H^2=\varkappa\varepsilon_g$ depend on initial conditions and
relations between the parameters of graviton--ghost substratum.
First of all, let us mention a scenario that corresponds to a
singular origin with the strong excitation of spatially uniform
modes
\begin{equation}
\displaystyle C_{g3}\ne 0,\qquad H>0, \qquad \frac{3|C_{g4}|a_0^4}{4C_{g3}}<e\ .
\label{5.22}
\end{equation}
In the case  (\ref{5.22}), the Universe is born in the singularity
and fairly quickly reaches the area of large scale factor values,
where it expands with the acceleration:
\begin{equation}
\displaystyle a\simeq |C_{g4}|^{1/2}t\ln^{1/2}\frac{t}{t_0}\ ,
\qquad \frac{\ddot{a}}{a}\simeq \frac{|C_{g4}|}{2a^2},\qquad a\gg a_0,\;
\left(\frac{C_{3g}}{3|C_{4g}|}\right)^{1/4}\ .
\label{5.23}
\end{equation}
Branch of the same solution, with $H<0$  describes the collapsing
Universe with a singular end--state.

Two other scenarios correspond to the weak excitation of graviton spatially uniform modes
\begin{equation}
\displaystyle C_{g3}\ne 0, \qquad \frac{3|C_{g4}|a_0^4}{4C_{g3}}>e\ .
\label{5.24}
\end{equation}
In the case of  (\ref{5.24}), the region of legitimate values of
the scale factor is divided into two sub--regions $0\leqslant
a\leqslant  a_1$ and $a_2\leqslant a<\infty$  separated by a
barrier of finite width $a_2>a_1$. In the sub--region of small
values of the scale factor, the Universe is born in a singularity,
reaches the state with a maximum value of $a=a_1$, and then
returns to the singularity. In the  limit  $C_{3g}\to 0$ the
possibility of such an evolution disappears because of $a_1\to 0$.
In sub--region of the large scale factor, the evolution of the
Universe starts at the infinite past from the state of zero
curvature. At the stage of compression, the Universe reaches the
state with a minimum value of $a=a_2$, and then turns into an
accelerated mode of expansion. With $C_{g3}=0$, this branch of
cosmological solution is described by the following function of
cosmological time
\begin{equation}
\displaystyle a(\eta)=a_0\exp\left(\frac{|C_{g4}|\eta^2}{4}\right)\ ,\qquad C_{g4}<0\ .
\label{5.25}
\end{equation}
Note that degenerate solutions (\ref{5.19}) and (\ref{5.25})
differ only in the sign under of exponent.

\subsection{Self--Polarized Graviton--Ghost Condensate in De Sitter Space}\label{Sitt}

It is easy to find that the system of equations (\ref{5.7}) ---
(\ref{5.10}) has a simple stationary solution  $H=const$,
$D=const$,  $W_n=const$. This solution describes the highly
symmetrical graviton--ghost substratum that fills the De Sitter
space. It reads
 \begin{equation}
 \begin{array}{c}
 \displaystyle  H^2=\frac{1}{36}W_1+\frac13\varkappa\Lambda\ ,\qquad
 a=a_0e^{Ht}\ ,
 \\[5mm]
\displaystyle \varepsilon_{g}=-p_{g}=\frac{1}{12}W_{1}\ .
 \end{array}
\label{5.26}
\end{equation}
This solution exists both for the $\Lambda=0$ case and for
$\Lambda\ne 0$. The first moment of the spectral function
satisfies the inequality $W_1>-12\varkappa\Lambda$ is the only
independent parameter of the solution. The remaining moments are
expressed through  by recurrence relations:
 \begin{equation}
 \begin{array}{c}
  \displaystyle D=-\frac{8}{3}W_1\ ,\qquad
 W_{n+1}=-\frac{n(2n+3)(n+3)}{2(n+2)}H^2W_n\ ,
 \qquad n\geqslant 1\ .
\end{array}
\label{5.27}
 \end{equation}

From  (\ref{5.26}) and (\ref{5.27}) it clearly follows that the
solution has essentially {\it vacuum and quantum} nature. The
first can be seen from the equation of state $p_g=-\varepsilon_g$.
The second can be seen from the fact that the signs of the moments
$W_{n+1}/W_n<0$ alternate. Another sign of the quantum nature of
the effect is contained in the properties of graviton spectrum.
The first of recurrence relations allows estimating of wavelengths
of gravitons and ghosts that play a dominant part in the formation
of observables
\begin{equation}
\displaystyle \lambda\sim \frac{a}{\overline{k}}\sim \sqrt{
 \frac{W_1}{|W_2|}}=\frac{1}{H}\sqrt{\frac{3}{10}}=const\ .
 \label{5.28}
\end{equation}
As can be seen from (\ref{5.28}), during the exponential expansion
of the Universe typical values of $\overline{k}$ rapidly shift to
the region of exponentially large conformal wave numbers. The
physical wavelength and macroscopic observables are unchanged in
time. Such a situation occurs if the following two conditions
apply.

(i) In the  ${\bf k}-$ space of conformal wave numbers spectra of
graviton vacuum fluctuations are flat;

(ii) In the integration over the flat spectrum, divergent
components of integrals excluded for reason to be discussed in
Section \ref{S}. Observables are formed by finite residuals of
these integrals.

In Section \ref{S} we will show that these conditions are actually
satisfied on the exact solution of operator equations of motion,
with special choice of Heisenberg's state vector of
graviton--ghost vacuum. Microscopic calculation also allows
expressing the first moment of spectral function through the
curvature of De Sitter space
\begin{equation}
 \displaystyle W_1=\frac{9\varkappa\hbar N_g}{2\pi^2}H^4\ ,
 \label{5.29}
 \end{equation}
where  $N_g$ is a functional of parameters of state vector, which
is of the order of the number of virtual gravitons and ghosts that
are situated under the horizon of events. Their wavelengths are of
the order of the distance to the horizon. It must be stressed that
the number of gravitons and ghosts $N_g$ is a macroscopic value.

The order of magnitude of $N_g$ is determined by graviton and
ghost numbers in the condensate. Let us emphasize that {\it
numbers of gravitons and ghosts and hence, $N_g$ parameters are
macroscopic qualities.} Further down in this section it is assumed
that the gravitons dominate in the condensate and that the
parameter $N_g>0$.

Note that the result (\ref{5.29})  can be easily predicted from
the general considerations, including considerations of dimension.
Indeed, the general formula (\ref{3.66}) shows that the moment
$W_1$ is of dimension  $[W_1]=[l]^{-2}$ ($[l]$ is of dimension of
length). It also contains the square of the Planck length as a
coefficient. Because $W_1$ is a functional of the metric, desired
dimension can be obtained only using metric's derivatives. It
follows from this that $W_1=C\cdot \varkappa\hbar H^4$  where $C$
dimensionless constant that contains parameters of vacuum
condensate. Given (\ref{5.29}), the solution in its final form is
as follows:
 \begin{equation}
 \begin{array}{c}
 \displaystyle
 D=-\frac{12\varkappa\hbar N_g}{\pi^2}H^4\ ,
 \qquad
 W_n=\frac{(-1)^{n+1}}{2^{2n}}
 (2n-1)!(2n+1)(n+2)\times
  \frac{2\varkappa\hbar N_g}{\pi^2}H^{2n+2} ,\qquad n\geqslant 1\
  .
   \end{array}
 \label{5.30}
 \end{equation}
\begin{equation}
\displaystyle \varepsilon_g=-p_g=\frac{3\hbar
 N_g}{8\pi^2}H^4\ ,
 \label{5.31}
 \end{equation}
The macroscopic Einstein's equation is transformed into the
equation for the inflation exponent
\begin{equation}
\displaystyle 3H^2=\frac{3\varkappa\hbar
 N_g}{8\pi^2}H^4+\varkappa\Lambda\ .
 \label{5.32}
 \end{equation}

Because $N_g$ is a macroscopic parameter, the solution under
discussion can be directly relevant to the asymptotic future of
the Universe. In this case, the number of gravitons and ghosts
under the horizon of events and $\Lambda$--term in the equation
(\ref{5.32}) should be considered as parameters, whose values were
formed during the earlier stages of cosmological evolution.
According to Zel'dovich \cite{22},  $\Lambda$--term is the total
energy density of equilibrium vacuum subsystems of
non--gravitational origin. The problem of the $\Lambda$--term
formation is so complex that little has changed since the
excellent review of Weinberg \cite{23}. We are limited only to
showing the order of magnitude of $\Lambda\sim 3 \cdot
10^{-47}\hbar^{-3}$ GeV$^4$ allowed by observational data. (See
also Appendix \ref{Lambda}.)

Some possibilities of co--existence of graviton condensate and
$\Lambda$--term will be discussed for $\Lambda\geqslant 0,\
N_g>0$. (For other possibilities see Section \ref{S}.) The
curvature of the De Sitter space for the asymptotical state of the
Universe is calculated by means of the solution to the equation
(\ref{5.32}). It reads
 \begin{equation}
\displaystyle H_\infty^2=\frac{4\pi^2}{\varkappa\hbar}
  \left(\frac{1}{N_g}\pm\sqrt{\frac{1}{N_g^2}-\frac{\varkappa^2\hbar\Lambda}{6\pi^2N_g}}\right)\ ,\qquad R=-12H_\infty^2\ .
 \label{5.33}
 \end{equation}
The energy density of vacuum in this state contains contributions
of subsystems formed by all physical interactions including the
gravitational one
\begin{equation}
\displaystyle \varepsilon_{vac}^{(\infty)}=\frac{3\hbar
 N_g}{8\pi^2}H_{\infty}^4+\Lambda\ .
 \label{5.34}
 \end{equation}
The relative input of graviton--ghost condensate into asymptotic
energy density of the vacuum depends on parameters of the
Universe. If the following inequality
 \begin{equation}
\displaystyle \frac{\varkappa^2\hbar\Lambda N_g}{6\pi^2}\ll 1\ ,
\label{5.35}
 \end{equation}
applies because of a small number of gravitons and ghosts, then
the quantum--gravitational term is small and one must use the
following solution
 \begin{equation}
\displaystyle H_{\infty}^2\simeq
\frac13\varkappa\Lambda\left(1+\frac{\varkappa^2\hbar\Lambda}
{24\pi^2}N_g\right)\ .
 \label{5.36}
 \end{equation}
If the inequality (\ref{5.35}) is satisfied because of a small
$\Lambda$--term then the asymptotic state is mostly formed by the
graviton--ghost condensate
\begin{equation}
\displaystyle H_{\infty}^2\simeq \frac{8\pi^2}{\varkappa\hbar
N_g}-\frac{\varkappa\Lambda}{3}\ .
 \label{5.37}
 \end{equation}
It can be seen from (\ref{5.33}) for $\Lambda>0$,  the number of
gravitons and ghosts that can appear in the Universe is limited by
maximum value
 \begin{equation}
\displaystyle N_{g(max)}=\frac{6\pi^2}{\varkappa^2\hbar\Lambda}\sim 10^{122}\
.
  \label{5.38}
 \end{equation}
In this limiting case (\ref{5.38}), the equipartition of the
vacuum energy takes place between graviton--ghost and
non--gravitational vacuum subsystems
\begin{equation}
\displaystyle H_\infty^2=\frac{4\pi^2}{\varkappa\hbar
 N_{g(max)}}=\frac23\varkappa\Lambda\ ,\qquad\qquad \varepsilon_g^{(\infty)}=\Lambda=\frac12
 \varepsilon_{vac}^{(\infty)}\ .
 \label{5.39}
 \end{equation}

\subsection{The Problem of Quantum--Gravity Phase Transitions}\label{pt}

Three exact solutions of the equations of quantum gravity (with no
matter fields and in the absence of $\Lambda$--term) are, in our
view, impressive illustrations of physical content of the theory.
(Of course, we can not exclude the existence of other exact
solutions). Before the integral
\begin{equation}
\displaystyle
t=\int\limits_{t_0}^tda\sqrt{\frac{3}{\varkappa\varepsilon_g(a)}}\
\label{5.40}
 \end{equation}
is calculated, three solutions are given by three different
functionals $\varkappa\varepsilon_g(a)$ that are displayed in the
right--hand--side of the macroscopic Einstein equation
\begin{equation}
 \begin{array}{c}
\displaystyle
H^2=-\frac{|C^{(I)}_{g3}|}{3a^6}+\frac{C^{(I)}_{g4}}{a^2}\ln\frac{a^{(I)}_0}{a}\
, \qquad (I)
\\[5mm]
\displaystyle
H^2=\frac{C^{(II)}_{g3}}{3a^6}+\frac{|C^{(II)}_{g4}|}{a^2}\ln\frac{a}{a^{(II)}_0}\
, \qquad (II)
\\[5mm]
\displaystyle
H^2=\frac{8\pi^2}{\varkappa\hbar N_g}\ . \qquad\qquad \qquad\qquad\qquad (III)
 \end{array}
 \label{5.41}
\end{equation}
If each of solutions (\ref{5.41}) is considered as independent of
the others, then one can note that (\ref{5.41}.I) and
(\ref{5.41}.II) are one--parameter solutions, meanwhile
(\ref{5.41}.III) does not contain any free parameter. After
multiplicative transformations of scale factor $a\to aa_0$ and
time $t\to ta_0/|C_{g4}|^{1/2}$ in equations (\ref{5.41}.I) and
(\ref{5.41}.II), and time transformation $t\to t(\varkappa\hbar
N_g/8\pi^2)^{1/2}$ in the equation (\ref{5.41}.III), we get
\begin{equation}
 \begin{array}{c}
\displaystyle
H^2=-\frac{|C|}{3a^6}-\frac{\ln a}{a^2}\
, \qquad \; (I)
\\[5mm]
\displaystyle
H^2=\frac{|C|}{3a^6}+\frac{\ln a}{a^2}\
, \qquad\; (II)
\\[5mm]
\displaystyle
H^2=1\ .  \qquad\qquad\qquad\; (III)
 \end{array}
 \label{5.42}
\end{equation}

Formulas (\ref{5.42}.I), (\ref{5.42}.II) and (\ref{5.42}.III) are
special solutions of nonlinear system of equations allocated by
special initial conditions. The first step is to determine what
relationship they have to a general solution of equations
(\ref{5.7}) --- (\ref{5.10}), corresponding to fairly arbitrary
initial conditions. (Recall that the initial conditions are set by
definitions of virtual gravitons and ghosts and structure of the
state vector). Immediately note that we do not have an answer to
this question in the form of strictly proven mathematical
theorems. The mathematical problem is that we are dealing with a
non--linear system, the number of degrees of freedom of which is
infinite. This fact is reflected both in the operator formalism
(an infinite number of modes, interacting through self--consistent
field) and the BBGKY formalism (an infinite number of equations
for moments of the spectral function).

In examining the problem, using numerical experiments, the
infinite system of equations (\ref{5.7}) --- (\ref{5.10}) is
transformed into a finite system by breaking the chain
(\ref{5.9}). In the framework of this method, three questions are
raise. (i) The choice of approximation of higher moment $W_{N+1}$
through the lower ones $W_1,\ W_2,\ ...,\ W_N$; (ii) dependence of
solution asymptotes of initial conditions; (iii)  dependence of
the solution of the number of equations $N$ in the chain. The
third of these issues is trivial enough, in the sense that the
response to it is produced by a mere repetition of numerical
experiments with the sequential increase in $N$. In all the
experiments that we conducted, there was convergence of
observables for $N>10\; -\; 12$  to some final functions
\begin{equation}
\displaystyle a(t),\qquad H(t),\qquad D(t),\qquad W_1(t)\ .
 \label{5.43}
\end{equation}
(The experiments are mostly completed for $N=20$, but some of them
held up to $N=50$.)

We assumed that the three exact solutions are independent
attractors of nonlinear system of equations. Under this
assumption, the mathematical classification of attractors
corresponds to the physical classification of possible asymptotic
regimes of the Universe evolution. Breaking the chain (\ref{5.9})
is governed by a choice of asymptotics, and this is our proposed
response to the first of the above issues. To truncate the chain,
recurrence relations from exact solutions are used
\begin{equation}
 \begin{array}{c}
\displaystyle
W_{N+1}=\frac{|C_{g4}|}{a^2}W_N\
, \qquad\qquad\qquad\qquad \; (I)
\\[5mm]
\displaystyle
W_{N+1}=-\frac{|C_{g4}|}{a^2}W_N\
, \qquad\qquad\qquad\qquad\; (II)
\\[5mm]
\displaystyle
W_{N+1}=-\frac{N(2N+3)(N+3)}{2(N+2)}H^2W_N\ .  \qquad\; (III)
 \end{array}
 \label{5.44}
\end{equation}

We found by means of numerical experiments that all the exact
solutions are stable with respect to small perturbations. Indeed
experiments themselves have been limited to small variations of
initial conditions at the vicinity of values fixing the exact
solutions. In all cases, small perturbations quickly died out. It
is necessary, of course, to remember that the statements about the
stability are made on the basis of numerical experiments using
approximations (\ref{5.44}).

Next, we conducted experiments with the initial conditions that
have nothing to do with the conditions relevant to any of the
three exact solutions. Nevertheless, the exact solution was used
in selecting the appropriate method for truncating the chain. In
all cases we saw a clear line between the ways truncating the
chain (\ref{5.44}.I), (\ref{5.44}.II), (\ref{5.44}.III) and
asymptotics of numerical solutions (\ref{5.41}.I),
(\ref{5.41}.II), (\ref{5.41}.III). In fact, fixing the asymptotics
by way of truncating the chain does not depend on initial
conditions. However, all stages of evolution depend on the initial
conditions, including the initial stage (which is natural), the
intermediate stage of evolution and the nature of transition
processes before the system reaches its asymptotic state.

The intermediate stage of evolution in all cases was related to
one of asymptotic (\ref{5.44}.I) or (\ref{5.44}.II). In Section
\ref{I} it will be shown that such character at intermediate
stages of evolution stems from the general properties of the
equations of the theory. The type of solution
$\varkappa\varepsilon_g\sim a^{-2}f(a)$, where $f(a)$ is a slow
function of scale factor, is formed as a result of excitation of
graviton--ghost modes primarily on quasi--resonant frequencies.
The nontrivial fact is that the regime of transition to the
asymptotics determined by approximations (\ref{5.44}.I) or
(\ref{5.44}.II) depends on initial conditions. In numerical
experiments, we have witnessed either a smooth transition or a
transition, accompanied by non--linear oscillations of physical
quantities (\ref{5.43}). In the case of approximation
(\ref{5.44}.III), the transition to the asymptotic De Sitter space
always proceeds in the regime of non--linear oscillations.

According to the results of numerical experiments, we came to the
following conclusions.

(i) {\it Three exact solutions describe three different stable (at
least, meta--stable) phases of graviton--ghost vacuum. The
physical assumption that the Universe must be in one of these
phases in the process of evolution, is formalized by the choice of
the way of truncating the BBGKY chain.}

(ii) {\it Arbitrary enough initial conditions correspond to the
non--equilibrium state of vacuum in phases or $I$ or $II$. These
conditions are divided into two classes: a consistent and not
consistent with the equilibrium phase, given by the
approximation.}

(iii) {\it If the physical nature of initial non--equilibrium
phase matches the equilibrium phase chosen as asymptotic, the
solution, starting with the intermediate stage, describes smooth
relaxation of the graviton-ghost vacuum to an equilibrium state.
If initial and asymptotic states do not match, a phase transition
is initiated in the system. The signs of such a transition are
nonlinear oscillations of physical quantities.}

(iv) {\it The self--polarized graviton--ghost condensate in the De
Sitter space can emerge only as a result of quantum gravity phase
transition.}

The rationale for introducing of the notion of phases of
graviton--ghost vacuum is the fact that three exact solutions
match spaces with different symmetry. The solution
(\ref{5.44}.III) describes 4--space of constant curvature, with
the highest possible symmetry. Solution (\ref{5.44}.II) (in the
version of appropriate unlimited expansion) describes 4--space,
the geometry of which tends asymptotically to the geometry of the
Milln space. Finally, the solution (\ref{5.44}.I) (in the version
corresponding to oscillations) describes 3--geometry, which is
translation--invariant along the axis of time. Taking into account
considerations of symmetry (see above), the term "phase of
graviton--ghost vacuum"\ that we introduced seems mathematically
and physically justified.

Representations of phase transitions are, of course, only
heuristic nature. In the one--loop quantum gravity,
multi--particle correlations in the system of gravitons and ghosts
are not taken into account. For this reason, in this theory it is
impossible to define the order parameter that plays the role of
the master parameter when choosing a phase state. Phase
transitions that were discussed above, were actually initiated by
disparity between the choice of the asymptotic state and set of
the initial conditions. Of course, such operations are meaningful
only within the suggestion that the effect of {\it
non--equilibrium} phase transition will be contained in future
theory.

Staying on the heuristic level, we can use the exact solutions
(\ref{5.41}.I),  (\ref{5.41}.II),  (\ref{5.41}.III) to demonstrate
in principle the possibility of the existence of equilibrium phase
transitions. Let us consider the exact solutions as the various
branches of a general solution. A rough phase transition model is
the passage from one branch to another while maintaining
continuity of scale factor and its first derivative. As can be
seen from (\ref{5.41}.I),  (\ref{5.41}.II),  (\ref{5.41}.III),
these conditions provide the equality of energies of
graviton--ghost systems on both sides of the transition point. The
second derivative of the scale factor and vacuum pressure are
discontinued (have a jump) at the point of transition. The
microscopic theory also makes it possible to see  that at the
point of transition the internal structure of graviton--ghost
substratum is changed (see Sections \ref{exact}, \ref{inst}).

Consider consistently simplified models of all of the phase
transitions. The condition of the sewing together solutions
(\ref{5.41}.I) and (\ref{5.41}.II) has the form:
\begin{equation}
 \begin{array}{c}
\displaystyle C_{g4}\ln\frac{a_c}{a_0}+\frac{C_{g3}}{3a_c^4}=0\
\qquad \to \qquad \frac{3C_{g4}a_0^6}{4C_{g3}}> e\ ,
 \end{array}
 \label{5.45}
\end{equation}
where
\[
 \begin{array}{c}
\displaystyle C_{g4}=|C^{(I)}_{g4}|+|C^{(II)}_{g4}|\ ,\qquad
C_{g3}=|C^{(I)}_{g3}|+|C^{(II)}_{g3}|\ ,
\\[5mm]
\displaystyle a_0=\left[a_0^{(I)}\right]^{|C^{(I)}_{g4}|/C_{g4}}\cdot
\left[a_0^{(II)}\right]^{|C^{(II)}_{g4}|/C_{g4}}\ ,
 \end{array}
 \]
$a_c$ is the value of the scale factor value at the fitting point,
common to the two branches. The condition of the existence of
transition between phases $I$ and $II$  is reduced to the
inequality shown in (\ref{5.45}). As we know, in phase $I$
gravitons dominate in quasi--resonant modes, and ghosts dominate
in spatially uniform modes. Following the transition, in phase
$II$ quasi--resonant modes are dominated by ghosts, but spatially
uniform modes are dominated by gravitons.

Further, the condition of sewing together of solutions
(\ref{5.41}.I) and (\ref{5.41}.III) reads
\begin{equation}
 \begin{array}{c}
\displaystyle \varkappa\varepsilon_g^{(I)}\equiv
\frac{3|C^{(I)}_{g4}|}{a_c^2}\ln\frac{a^{(I)}_0}{a_c}-
\frac{|C^{(I)}_{g3}|}{a_c^6}=\frac{24\pi^2}{\varkappa\hbar
N_g}\ .
 \end{array}
 \label{5.46}
\end{equation}
It must be borne in mind that the energy density in phase $I$
$\varkappa\varepsilon_g^{(I)}$ is limited above and below.
Therefore, in phase $III$ the number of gravitons under the
horizon of events must lie in a certain interval, whose borders
are defined by parameters of phase $I$. The phase transition looks
like a "freezing" of the distance to the horizon and of the value
of the physical wavelength of quasi-resonant modes.

Finally, the third possible transition is illustrated by sewing
together of solutions (\ref{5.41}.II) and (\ref{5.41}.III):
\begin{equation}
 \begin{array}{c}
\displaystyle \varkappa\varepsilon_g^{(II)}\equiv
\frac{3|C^{(II)}_{g4}|}{a_c^2}\ln\frac{a_c}{a^{(II)}_0}+\frac{|C^{(II)}_{g3}|}{a_c^6}=
\frac{24\pi^2}{\varkappa\hbar
N_g}\ .
\end{array}
 \label{5.47}
\end{equation}
In the most general case, the solution (\ref{5.41}.II) describes
the birth of the Universe from the singularity and its further
expansion with the acceleration. In this scenario for any preset
value of energy density there is a corresponding point on the
evolutionary path. Therefore, the transition from phase $II$ to
phase $III$ can occur at any point by choosing the appropriate
value $N_g$.

\section{Exact Solutions: Dynamics of Operators and Structure of State Vectors}\label{exact}

In this section, we get the exact solutions for field operators
and expressions for the state vectors that correspond to exact
analytical solutions of BBGKY chain (\ref{5.41}.I) and
(\ref{5.41}.III). Microscopic studies of exact solutions allow
greater detail to identify their physical content. Solutions
(\ref{5.41}.I) and (\ref{5.41}.III) are formed as a result of
certain spectrally dependent correlations between graviton and
ghost contributions to the observables. These are full
graviton--ghost compensation of contributions of zero oscillations
(one--loop finiteness); full compensation of contributions in all
parts of the spectrum, except the region of quasi--resonant (QR)
and spatially homogeneous (SH) modes; incomplete compensation of
contributions of QR and SH modes with non--zero occupation
numbers; correlations between excitation levels and
graviton--ghost contents of  QR and SH modes, and, finally, some
correlations of phases in quantum superpositions of graviton and
ghost state vectors.

The physical nature of solution (\ref{5.41}.II) turned out to be
unexpected and nontrivial. In Section \ref{inst} it will be shown
that mathematically this solution describes instanton condensate,
which physically corresponds to the system of correlated
fluctuations arising during tunneling of graviton--ghost medium
between states with fixed difference of graviton and ghost
numbers. We explain also that self--polarized graviton--ghost
condensate in the De Sitter space also allows instanton
interpretation.

\subsection{Condensate of Constant Conformal Wavelength}\label{K}

Let us consider the solution (\ref{5.41}.I) for $C_{3g}=0, \,
C_{4g}=k_0^2$:
 \begin{equation}
 \begin{array}{c}
\displaystyle
H^2=\frac{k_0^2}{a^2}\ln\frac{a_0}{a}\ , \qquad a=a_0\exp\left(-\frac{k_0^2\eta^2}{4}\right)\ .
 \end{array}
 \label{6.1}
\end{equation}
The graviton wave equation with the (\ref{6.1}) background reads
 \begin{equation}
 \begin{array}{c}
\displaystyle \hat\psi''_{{\bf k}\sigma}-k^2_0\eta\hat\psi'_{{\bf k}\sigma}+
k^2\hat\psi_{{\bf k}\sigma}=0\ .
 \end{array}
 \label{6.2}
\end{equation}
The equation for the ghosts looks similar. Fundamental solutions
of equation (\ref{6.2}) are degenerate hypergeometric functions.
It is unnecessary to consider those solutions for all possible
values of the parameter $k^2$. First of all, it is obvious that
the macroscopic observables can be formed only by simplest
hypergeometric functions. Values   $k^2$  that are $k^2=0$
(spatially uniform modes) and  $k^2=k_0^2$ (quasi--resonant modes)
stand out. For all other modes there is a precise graviton--ghost
compensation. The reason why it is a mathematically possible
follows from the general formulas  (\ref{3.69}), (\ref{3.72}),
(\ref{3.73}) \footnote{Formally, all modes except with  $k^2=0$
and $k^2=k_0^2$, look like "frozen"\ degrees of freedom, which are
excluded from consideration by the model postulate. By virtue of
the principle of uncertainty, postulates of this type are outside
the formalism of quantum field theory. We want to emphasize that
in the finite one--loop quantum gravity there is no need to
"freeze"\ degrees of freedom not participating in the formation of
particular exact solutions. Instead of mathematically incorrect
operation of "freezing"\ , the formalism of the theory offers
mathematically consistent operations of graviton--ghost
compensations.}.

Let us start with quasi--resonant modes. Exact solutions of the
equation (\ref{6.1}) and similar equation for ghosts for
$k^2=k_0^2$ read
\begin{equation}
\begin{array}{c}
\displaystyle
\hat\psi_{{\bf k}\sigma}=
\frac{\sqrt{4\varkappa\hbar k_0}}{a_0}\left[-\eta\left(\hat Q_{{\bf k}\sigma}+
k_0\hat P_{{\bf k}\sigma}\int\limits_0^\eta e^{k_0^2\eta^2/2}d\eta\right)+
\frac{\hat P_{{\bf k}\sigma}}{k_0}e^{k_0^2\eta^2/2}\right]=
\\[5mm]
\displaystyle =-\sqrt{\frac{16\varkappa\hbar}{k_0a_0^2}}\left[\hat Q_{{\bf k}\sigma}+
\hat P_{{\bf k}\sigma}F(a)\right]\ln^{1/2}\frac{a_0}{a} \ ,
\end{array}
 \label{6.3}
\end{equation}
\begin{equation}
\begin{array}{c}
\displaystyle
\hat\vartheta_{{\bf k}}=\frac{\sqrt{4\varkappa\hbar k_0}}{a_0}\left[-\eta\left(\hat q_{{\bf k}}+
k_0\hat p_{{\bf k}}\int\limits_0^\eta e^{k_0^2\eta^2/2}d\eta\right)+
\frac{\hat p_{{\bf k}}}{k_0}e^{k_0^2\eta^2/2}\right]=
\\[5mm]
\displaystyle =-\sqrt{\frac{16\varkappa\hbar}{k_0a_0^2}}\left[\hat q_{{\bf k}}+
\hat p_{{\bf k}}F(a)\right]\ln^{1/2}\frac{a_0}{a} \ ,
\end{array}
 \label{6.4}
\end{equation}
where $\hat Q_{{\bf k}\sigma},\, \hat P_{{\bf k}\sigma}$ and $\hat
q_{{\bf k}},\, \hat p_{{\bf k}}$ are operators whose properties
are defined in (\ref{3.53}), (\ref{3.54}), (\ref{3.50});
\[
 \displaystyle F(a)=a_0^2\int\limits_{a_0}^a\frac{da}{a^3\displaystyle\ln^{1/2}\frac{a_0}{a}}-
 \frac{a_0^2}{2a^2\displaystyle\ln^{1/2}\frac{a_0}{a}}\ .
 \]

Note that one of fundamental solutions to equation (\ref{6.2}) is
the Hermite polynomial $H_1(\eta)$, which corresponds to positive
eigenvalue $k^2/k_0^2=1$. In the reproduction of solutions
(\ref{5.41}.I) at the microscopic level, this fact is crucial. We
will show that the choice of a state vector, satisfying the
condition of coherence leads to the fact that only this solution
takes part in the formation of the observables. The second
solution, containing a function $F(a)$, is a mathematical
structure that does not correspond to the exact solution to the
BBGKY chain.

Averaging of bilinear forms of operators (\ref{6.3}) and
(\ref{6.4}) over the state vector of the general form leads to the
following spectral function
\begin{equation}
\begin{array}{c}
\displaystyle
W_{{\bf k}}=\sum_\sigma\langle\Psi_g|\hat\psi_{{\bf k}\sigma}^+\hat\psi_{{\bf k}\sigma}
|\Psi_g\rangle-2\langle\Psi_{gh}|\hat\vartheta_{{\bf k}}^+\hat\vartheta_{{\bf k}}
|\Psi_{gh}\rangle=
\frac{16\varkappa\hbar}{ k_0a_0^2}\left[A_{\bf k}+B_{\bf k}F^2(a)+
C_{\bf k}F(a)\right]\ln\frac{a_0}{a}
\ .
\end{array}
 \label{6.5}
\end{equation}
The constants appearing in (\ref{6.5}) are expressed through
averaged quadratic forms of operators of generalized coordinates
and momentums:
\begin{equation}
\begin{array}{c}
\displaystyle A_{\bf k}=
\sum_\sigma\langle\Psi_g|\hat Q^+_{{\bf k}\sigma}Q_{{\bf k}\sigma}|\Psi_g\rangle-
2\langle\Psi_{gh}|\hat q_{{\bf k}}^+\hat q_{{\bf k}}|\Psi_{gh}\rangle\ ,
\\[5mm]
\displaystyle B_{\bf k}=
\sum_\sigma\langle\Psi_g|\hat P^+_{{\bf k}\sigma}P_{{\bf k}\sigma}|\Psi_g\rangle-
2\langle\Psi_{gh}|\hat p_{{\bf k}}^+\hat p_{{\bf k}}|\Psi_{gh}\rangle\ ,
\\[5mm]
\displaystyle  C_{\bf k}=
\sum_\sigma\langle\Psi_g|\left(\hat Q^+_{{\bf k}\sigma}P_{{\bf k}\sigma}+
\hat P^+_{{\bf k}\sigma}Q_{{\bf k}\sigma}\right)|\Psi_g\rangle-
2\langle\Psi_{gh}|\left(\hat q_{{\bf k}}^+\hat p_{{\bf k}}+
p_{{\bf k}}^+\hat q_{{\bf k}}\right)|\Psi_{gh}\rangle\ .
\end{array}
 \label{6.6}
\end{equation}
Following the transition to the ladder operators in formulas
(\ref{3.54}) and calculations, carried out similar to (\ref{3.66})
--- (\ref{3.74}), we get
\begin{equation}
\begin{array}{c}
\displaystyle
A_{\bf k}=2\langle n_{k(g)}\rangle(1+\zeta^{(g)}_k\cos\varphi_k)-
2\langle n_{k(gh)}\rangle(1+\zeta_k^{(gh)}\cos\chi_k)\ ,
\\[5mm]
\displaystyle B_{\bf k}=2\langle n_{k(g)}\rangle(1-\zeta^{(g)}_k\cos\varphi_k)-
2\langle n_{k(gh)}\rangle(1-\zeta_k^{(gh)}\cos\chi_k)\ ,
\\[5mm]
\displaystyle C_{\bf k}=0\ .
\end{array}
 \label{6.7}
\end{equation}
For sake of simplicity, in (\ref{6.7}) average numbers of ghosts
and anti--ghosts are assumed to be the same:  $\langle
n_{k(gh)}\rangle=  \langle \bar n_{k(gh)}\rangle$.

Let us go back to the expression (\ref{6.5}). Obviously, the
spectral function (\ref{6.5}) creates moments (\ref{5.16}) only if
$B_{\bf k}=C_{\bf k}=0$. The condition $C_{\bf k}=0$ is satisfied
automatically as a consequence of isotropy of macroscopic state,
i.e. because of independence of average occupation numbers of the
direction of vector ${\bf k}$. $B_{\bf k}=0$ imposes the
conditions on amplitudes and phases of quantum superpositions of
state vectors with different occupation numbers. It is necessary
to draw attention to the fundamental fact: {\it the solution under
discussion does not exist, if phases of superpositions are
random.} Indeed, averaging the expression (\ref{6.7}) over phases,
we see that condition $B_{\bf k}=0$ is satisfied only if $\langle
n_{k(g)}\rangle=\langle n_{k(gh)}\rangle$. The last equality
automatically leads to $A_{\bf k}=0$, i.e. which eliminates the
nontrivial solution.

Thus, the condition of the existence of the solution under
discussion is the coherence of the quantum state. It is easy to
notice (see (\ref{3.74})), that equality  $B_{\bf k}=0$, as a
condition of coherence, is satisfied for zero phase difference of
states with the neighboring occupation numbers of gravitons and
ghosts:
\begin{equation}
\displaystyle \zeta^{(g)}_k\cos\varphi_k=\zeta_k^{(gh)}\cos\chi_k=1 \quad \to \quad
\zeta^{(g)}_k=\zeta_k^{(gh)}=1,\qquad \cos\varphi_k=\cos\chi_k=1\ .
\label{6.8}
\end{equation}
Taking into account (\ref{6.8}), we get the following final
expression (\ref{6.9}) for the spectral function of
quasi--resonant gravitons and ghosts
\begin{equation}
\begin{array}{c}
\displaystyle
W_{{\bf k}}\equiv W_k=
\frac{64\varkappa\hbar}{ k_0a_0^2}\left(\langle n_{k(g)}\rangle-
\langle n_{k(gh)}\rangle\right)\ln\frac{a_0}{a}
\ .
\end{array}
 \label{6.9}
\end{equation}
In calculating moments, summation over wave numbers is replaced by
integration. Account is taken of that the spectrum as the
delta--form with respect to the modulus of $k=|{\bf k}|$.  Also a
new parameter  $N_g$ is introduced where $N_g$ is the difference
of numbers of gravitons and ghosts in the unit volume of $V=\int
d^3x=1$  in the 3--space, which is conformally similar to the
3--space of expanding Universe. Index "$g$"\ in designation of
$N_g$  parameter indicates the dominance of gravitons in
quasi--resonant modes. In accordance with this definition, the
following replacement is performed
\begin{equation}
\displaystyle \langle n_{k(g)}\rangle-\langle n_{k(gh)}\rangle \to
\frac{2\pi^2}{k^2}N_g\delta(k-k_0)\ ,
 \label{6.10}
\end{equation}
Results of calculating of moments are equated to the relevant
expressions of (\ref{5.11}) and (\ref{5.16}), which were obtained
by exact solution of the BBGKY chain:
\begin{equation}
\begin{array}{c}
\displaystyle W_n(g4)=\frac{1}{2\pi^2a^{2n}}\int\limits_0^\infty W_kk^{2n+2}dk=
\frac{64\varkappa\hbar N_g k_0^{2n-1}}{a^2_0a^{2n}}\ln\frac{a_0}{a}=
\frac{24k_0^{2n}}{a^{2n}}\ln\frac{a_0}{a}\ ,
\\[5mm]
\displaystyle D(g4)=\frac{1}{a^2}\left(W_0''+2\frac{a'}{a}W_0'\right)=
-\frac{128\varkappa\hbar N_g k_0}{a^2_0a^{2}}\ln\frac{a_0}{e^{1/4}a}=
-\frac{48k_0^{2}}{a^{2}}\ln\frac{a_0}{e^{1/4}a}\ .
\end{array}
\label{6.11}
\end{equation}
In accordance with (\ref{6.11}), there is a relation between
parameters $k_0,\ a_0$   and $N_g$ that  appear in the microscopic
solution
\begin{equation}
 \displaystyle N_g=\frac{3k_0a_0^2}{8\varkappa\hbar}\ .
 \label{6.12}
\end{equation}
Recall that in the solution under discussion, the Universe was
born in singularity, expands to a state with a maximum scale
factor  $a_{max}=a_0$, and then is again compressed to the
singularity. In this scenario, value $a_0$ can be defined as the
size of the Universe, accessible for observation in the end stage
of expansion. As can be seen from (\ref{6.12}), if  $a_0$ is a
macroscopic value, the difference in numbers gravitons and ghosts
$N_g\gg 1$ is also a macroscopic value.

Contributions of SH modes to the expressions for the moments are
shown in (\ref{5.11}), and the relation between the parameters
$C_{2g}$ and $C_{4g}$ is shown in (\ref{5.15}). As a part of the
microscopic approach, the construction of exact solutions for
these modes is performed by the method of transaction to the
limit, described at the end of Section \ref{lgw}. The parameter of
spatially homogeneous condensate is introduced similarly to
(\ref{6.10}):
\begin{equation}
\displaystyle
\langle n_{0(gh)}\rangle(1+\zeta^{(gh)}_0\cos\varphi_0)-
\langle n_{0(g)}\rangle(1+\zeta^{(g)}_0\cos\chi_0)
\to \frac{2\pi^2}{k^2}N_{gh}\delta(k-\kappa_0)\ ,\qquad \kappa_0\to 0\ .
 \label{6.13}
\end{equation}
The index "$gh$"\ in $N_{gh}>0$ indicates the dominance of ghosts
over the gravitons in the spatially homogeneous  condensate. The
moments are:
\begin{equation}
\displaystyle W_1(g2)=-\frac{16\varkappa\hbar k_1 N_{gh}}{a_1^2a^2}\ ,
\qquad D(g2)=\frac{32\varkappa\hbar k_1N_{gh}}{a_1^2a^2}\ .
 \label{6.14}
\end{equation}
Definitions of parameters $k_1$ and  $a_1$ are given in
(\ref{4.21}). The energy density and pressure of the system of QR
and SH modes are given by (\ref{6.11}) and (\ref{6.14}):
\begin{equation}
 \begin{array}{c}
 \displaystyle
\varkappa\varepsilon_g=\frac{8\varkappa\hbar
k_0N_g}{a_0^2a^2}\ln\frac{a_0}{a}+
\frac{2\varkappa\hbar}{a^2}\left(\frac{k_0N_g}{a^2_0}-
\frac{k_1N_{gh}}{a^2_1}\right)=\frac{8\varkappa\hbar
k_0N_g}{a_0^2a^2}\ln\frac{a_0}{a}\ ,
\\[5mm]
\displaystyle \varkappa p_g=-\frac{8\varkappa\hbar
k_0N_g}{3a_0^2a^2}\ln\frac{a_0}{ea}-
\frac{2\varkappa\hbar}{3a^2}\left(\frac{k_0N_g}{a^2_0}-
\frac{k_1N_{gh}}{a^2_1}\right)=-\frac{8\varkappa\hbar
k_0N_g}{3a_0^2a^2}\ln\frac{a_0}{ea}\ .
\end{array}
 \label{6.15}
\end{equation}
In formulas (\ref{6.15}), the terms in brackets are eliminated by
the condition (\ref{5.15}), which is rewritten in terms of
macroscopic parameters
\begin{equation}
\displaystyle \frac{k_0N_g}{a^2_0}=\frac{k_1N_{gh}}{a^2_1}\ .
\label{6.16}
\end{equation}
The solution (\ref{6.15}), (\ref{6.16}) describes a quantum
coherent condensate of quasi--resonant modes with graviton
dominance, parameters of which are consistent with that of
spatially homogeneous  condensate with the ghost dominance.

\subsection{Condensate of Constant Physical Wavelength}\label{S}

The De Sitter solution for plane isotropic Universe reads
\begin{equation}
\displaystyle a=a_0e^{Ht}=-\frac{1}{H\eta}\ , \qquad H=const\ .
\label{6.17}
\end{equation}
For the background (\ref{6.17}), the gravitons and ghost equations
and their solutions read
\begin{equation}
 \begin{array}{c}
 \displaystyle
\hat \psi''_{{\bf k}\sigma}-\frac{1}{\eta}\hat \psi'_{{\bf k}\sigma}+
k^2\hat \psi_{{\bf k}\sigma}
=0\ ,
\qquad
\hat \psi_{{\bf k}\sigma}=\frac{1}{a}\sqrt{\frac{2\varkappa\hbar}{k}}
 \left[c_{{\bf k}\sigma}f(x)+c^+_{{\bf -k}-\sigma}f^*(x)\right]\ ,
\end{array}
 \label{6.18}
\end{equation}
\begin{equation}
 \begin{array}{c}
 \displaystyle
\hat \vartheta''_{{\bf k}}-\frac{1}{\eta}\hat \vartheta'_{{\bf k}}+k^2\hat \vartheta_{{\bf k}}=0\ ,
\qquad
\hat \vartheta_{{\bf k}}=\frac{1}{a}\sqrt{\frac{2\varkappa\hbar}{k}}
 \left[a_{{\bf k}}f(x)+b^+_{{\bf -k}}f^*(x)\right]\ ,
\end{array}
 \label{6.19}
\end{equation}
where
\[
\displaystyle f(x)=\left(1-\frac{i}{x}\right)e^{-ix}\ , \qquad x=k\eta\ .
\]
Ladder operators in (\ref{6.18}), (\ref{6.19}), have the standard
property of (\ref{3.47}), (\ref{3.50}), which allow their use of
in constructing build basic vectors for the Fock space from which
the general state vectors are constructed.

The self--consistent dynamics of gravitons and ghosts in the De
Sitter space are not trivial in the sense that the averaged
bilinear forms of operators (\ref{6.18}), (\ref{6.19}) which are
explicitly and essentially depending on time, must lead to
time--independent macroscopic observables. It must be emphasized,
that the existence of such, at first glance unlikely solution, is
guaranteed by the existence of the solution for the BBGKY chain.
The key to the solution lies in the structure of the state vectors
of gravitons and ghosts.

Substitution of operator functions (\ref{6.18}), (\ref{6.19}) into
the general expression for the moments (\ref{5.6}) yields:
\begin{equation}
  \begin{array}{c}
 \displaystyle
 W_n=\frac{2\varkappa\hbar}{\pi^2}H^{2n+2}\int\limits_0^{\infty}dxx^{2n+1}
 \biggl\{U_{{\bf k}(wave)}|f(x)|^2 +U_{{\bf k}(cr)}[f^*(x)]^2+U_{{\bf k}(ann)}[f(x)]^2\biggr\}
 \ ,
\end{array}
 \label{6.20}
 \end{equation}
where
 \begin{equation}
 \begin{array}{c}
 \displaystyle
  N_{{\bf k}}\equiv U_{{\bf k}(wave)}=\sum_\sigma\langle \Psi_g|c^+_{{\bf k}\sigma}c_{{\bf
 k}\sigma}|\Psi_g\rangle-
  \langle \Psi_{gh}|a^+_{{\bf k}}a_{{\bf
 k}}|\Psi_{gh}\rangle-\langle \Psi_{gh}|b^+_{{\bf k}}b_{{\bf
 k}}|\Psi_{gh}\rangle\ ;
\end{array}
\label{6.21}
 \end{equation}
\begin{equation}
 \begin{array}{c}
\displaystyle U^*_{{\bf k}}\equiv U_{{\bf k}(cr)}=\frac12\sum_\sigma\langle \Psi_g| c^+_{{\bf
k}\sigma}c^+_{{\bf
 -k}-\sigma}|\Psi_g\rangle-\langle\Psi_{gh}| a^+_{{\bf
k}}b^+_{{\bf
 -k}}|\Psi_{gh} \rangle\ ;
\\[5mm]
\displaystyle U_{{\bf k}}\equiv U_{{\bf k}(ann)}=\frac12\sum_\sigma\langle \Psi_g| c_{{\bf
-k}-\sigma}c_{{\bf
 k}\sigma}|\Psi_g\rangle-\langle \Psi_{gh}| b_{{\bf
-k}}a_{{\bf
 k}}|\Psi_{gh}\rangle\equiv U^*_{{\bf k}(cr)}\ .
\end{array}
\label{6.22}
 \end{equation}
Here $U_{{\bf k}(wave)}$  is the spectral parameter of quantum
waves, which become real gravitons if $k\eta\gg 1$; $U_{{\bf
k}(cr)}$, $U_{{\bf k}(ann)}$ are the spectral parameters of
quantum fluctuations that emerge in the processes of graviton (and
ghost) creation from the vacuum and graviton (and ghost)
annihilation to the vacuum.

Obviously, at the first stage of calculations we assume that the
averaging in (\ref{6.21}), (\ref{6.22}) is conducted over the
state vectors of the general form (\ref{3.48}), (\ref{3.51}). This
allows us to go to formulas (\ref{3.67}), (\ref{3.68}) or
(\ref{3.72}) --- (\ref{3.74}). Then it is necessary to take into
account that the moments  $W_n$ must not depend on time, and that
they also should be free of divergences. When analyzing the
conditions for these demands, the specific form of the expression
(\ref{6.20}) plays an important part. The measure of integration
and the dependence of field operators on the wave number and time
can be represented in the terms of the variable $x=k\eta$. A
separate (additional) dependence on the wave number can be
connected with the structure of spectral parameters. After
substitution of the variable $k=x/\eta$   in the equation
(\ref{6.21}), it is seen that the first term in (\ref{6.20}) is
time-independent only if $U_{{\bf k}(wave)}$ is independent on the
wave number. This means that the graviton and ghost spectra must
be flat. However, with the flat spectrum there is danger of
divergences: if $U_{{\bf k}(wave)}=const\ ({\bf k})\ne 0$, then
the first integral in (\ref{6.20}) does not exist, because
$|f(x)|^2\to 1$ with $x\to \infty$.

The divergences can be avoided only with exact compensation of
contributions from gravitons and ghosts to the spectral parameter
$U_{{\bf k}(wave)}$. Let us point out, that in that case we are
not talking about zero oscillations but about the contributions
from the states with non--zero occupation numbers. The
compensation condition leading to $U_{{\bf k}(wave)}=0$ is:
\begin{equation}
\displaystyle  |\mathcal{C}_{n_{{\bf k}\sigma}}|=|\mathcal{A}_{n_{{\bf
k}}}|=|\mathcal{B}_{n_{-{\bf k}}}|\ .
\label{6.23}
\end{equation}
The result (\ref{6.23}) has a simple physical interpretation. The
quantum waves of gravitons and ghosts with the equation of state
which differs from $p=-\varepsilon$ can not be carriers of energy
in the De Sitter space with the self--consistent geometry. The
total energy of quantized waves is equal to zero due to exactly
the same number of gravitons and ghosts in all regions of the
spectrum:
\begin{equation}
\displaystyle  \langle n_{{\bf k}\sigma_1}\rangle +\langle n_{{\bf k}\sigma_2}\rangle=
\langle n_{{\bf k}}\rangle+\langle\bar n_{{\bf k}}\rangle\ .
\label{6.24}
\end{equation}
With equal polarizations of gravitons and the equality of numbers
of ghosts and anti--ghosts, it follows from (\ref{6.24})  that
$\langle n_{{\bf k}(g)}\rangle=\langle n_{{\bf k}(gh)}\rangle$.
Exact equality of the average number of gravitons and ghost is a
characteristic feature of the De Sitter space with the
self--consistent geometry. Let us mention that for the solution
discussed in the previous section \ref{K}, that equality is absent
in principle. It means that different solutions have different
microscopic structures of the graviton--ghost condensate.

Based on the reasoning analogous to the one described above,
spectrum parameters $U_{{\bf k}(cr)}$, $U_{{\bf k}(ann)}$  also
must not depend on the wave vector ${\bf k}$. However, the
corresponding integrals in the second and third terms of
(\ref{6.20}) are not divergent. The absence of divergences is due
to the fact that with $x\to \infty$  the integration is taken over
the fast oscillating functions $\sim e^{\pm 2ix}$. To calculate
these integrals, they should be additionally defined as follows:
\begin{equation}
 \begin{array}{c}
 \displaystyle
 \lim_{\zeta\to \ 0}\int\limits_0^\infty dxx^{2n\pm 1}e^{-(\zeta-2i)x}=
 \mp (-1)^n\frac{(2n\pm 1)!}{2^{2n+1\pm 1}}\ ,
 \\[3mm]
  \displaystyle
 2i\lim_{\zeta\to \ 0}\int\limits_0^\infty dxx^{2n}e^{-(\zeta-2i)x}=
  (-1)^{n+1}\frac{(2n)!}{2^{2n}}\ .
\end{array}
\label{6.25}
 \end{equation}
At every instant of time, the procedure of re-definitions of
integrals (\ref{6.25}) selects the contributions from virtual
gravitons and ghosts with a characteristic wavelength (\ref{5.28})
and eliminate the contributions of all other graviton--ghost
modes. This redefining procedure provides the existence of
recursive relations (\ref{5.27}) in the exact solution of the
BBGKY chain.

Thus, in (\ref{6.20}) we have a flat spectrum of gravitons and
ghosts, $U_{{\bf k}(wave)}\equiv 0$,  $U_{{\bf k}(cr)}=U^*_{{\bf
k}(ann)}=U=const(k)$. The expression for the spectral parameter
takes the form:
\begin{equation}
 \begin{array}{c}
 \displaystyle  U=\left(\sum_n\mathcal{C}^*_{n+1}\mathcal{C}_n\sqrt{n+1}\right)^2-
 \left(\sum_n\mathcal{A}^*_{n+1}\mathcal{A}_n\sqrt{n+1}\right)
 \left(\sum_n\mathcal{B}^*_{n+1}\mathcal{B}_n\sqrt{n+1}\right)\ ,
 \\[5mm]
 \displaystyle  |\mathcal{C}_n|=|\mathcal{A}_n|=|\mathcal{B}_n|\equiv \sqrt{\mathcal{P}_n} \ ,
 \end{array}
 \label{6.26}
 \end{equation}
where  $\mathcal{P}_n$ is a normalized statistical distribution.
The average value of the number of gravitons and ghosts, having
the wavelength in the vicinity of characteristic values
(\ref{5.28}), are calculated by the formula
\begin{equation}
\displaystyle\langle n_{g}\rangle=\langle n_{gh}\rangle=\langle n\rangle=
\sum_{n=0}^\infty n\mathcal{P}(n)\ .
\label{6.27}
 \end{equation}
Using the Poisson distribution in (\ref{6.26}), (\ref{6.27}), the
values of integrals (\ref{6.25}) and the formulas (\ref{3.73}),
(\ref{3.74}), we get the moments
\begin{equation}
 \begin{array}{c}
 \displaystyle
 D=-\frac{12\varkappa\hbar N_g}{\pi^2}H^4\ ,
 \qquad
 W_n=\frac{(-1)^{n+1}}{2^{2n}}
 (2n-1)!(2n+1)(n+2)\times
  \frac{2\varkappa\hbar N_g}{\pi^2}H^{2n+2} ,\qquad n\geqslant 1\ ,
\end{array}
 \label{6.28}
 \end{equation}
where
\begin{equation}
\displaystyle N_g= \langle n\rangle(\zeta_{g}\cos\varphi- \zeta_{gh}\cos\chi)\  .
\label{6.29}
\end{equation}
Zero moment $W_0$, which has an infrared logarithmic singularity,
is not contained in the expressions for the macroscopic
observables, and for that reason, is not calculated. In the
equation for  $W_0$, the functions are differentiated in the
integrand and the derivatives are combined in accordance with the
definition $D=\ddot W_0+3H\dot W_0$. At the last step the
integrals that are calculated, already posses no singularities.

Averaging of the parameter (\ref{6.29}) over the phases yields
$N_g=0$.  Therefore {\it the solution under discussion does not
exist if the superposition of the phases are random.} The
coherence  of the quantum ensemble, i.e. the correlation of phases
in the quantum superposition of the basic vectors, corresponding
to the different occupation numbers, points to the fact that the
medium is in the graviton--ghost condensate state. The gravitons
are dominant in the condensate if $N_g>0$, and the ghosts are
dominant if $N_g<0$.

The duality of the condensate and the indeterminate sign of the
$\Lambda$--term create different evolutional scenarios. Of course,
all these scenarios are present in the expression (\ref{5.33}),
which is obtained as a solution of the macroscopic Einstein
equation (\ref{5.32}). In addition to the scenarios described in
the Section \ref{Sitt}, we will show the possibility of strong
renormalization of energy of non--gravitational vacuum subsystems
by the energy of the graviton--ghost
condensate\footnote{Mechanisms that are able to drive the
cosmological constant to zero have been discussed for decades (see
\cite{23,79} for a review). Any particular scenarios were
considered in \cite{Dolgov1983, Ford1987, Ford2002, R2, ST}.
Renormalizations of cosmological constant by gravitons in the
framework of one--loop quantum gravity were also considered in
\cite{MV, 48,49,50} where the effect of reconstruction of zero
oscillations of gravitational field in the self--consistent De
Sitter space, i.e. effect of conformal anomalies was discussed.
Conformal anomalies that arise due to regularization and
renormalization procedures do not apply to this work (see also
Appendix \ref{an-d})}.

We have in mind a situation, in which the modulus of
$\Lambda$--term exceeds the density of vacuum energy in the
asymptotic state of the Universe by many orders of magnitude:
\begin{equation}
\displaystyle \frac{|\Lambda|}{\varepsilon_{vac}^{(\infty)}}\equiv
\frac{\varkappa |\Lambda|}{3H^2_\infty}={\mathcal{N}}\gg 1\ ,
\label{6.30}
 \end{equation}
where ${\mathcal{N}}$ is a huge macroscopic number. From
(\ref{5.33}) it follows that the effect of  strong renormalization
takes place if
\begin{equation}
\displaystyle \frac{\Lambda}{N_g}<0\ , \qquad
|N_g|\gg \frac{6\pi^2}{\varkappa^2\hbar |\Lambda|}\ ,
\qquad \varepsilon_{vac}^{(\infty)}\simeq 2\pi\sqrt{\frac{6|\Lambda|}{\varkappa^2\hbar |N_g|}}
\label{6.31}
\end{equation}
Let us mention that the strong renormalization of the positive
$\Lambda$--term is provided by a condensate in which the ghosts
are dominant, and for the negative  $\Lambda$--term --- by a
condensate for which the gravitons are dominant.

For clarity and for the evaluations let us introduce the Plank
scale $M_{Pl}=(8\pi\hbar/\varkappa)^{1/2}=1.22\cdot 10^{19}$ GeV,
the scale of  $\Lambda$--term
$M_\Lambda=(\hbar^3|\Lambda|)^{1/4}$, and the scale of the density
of Dark Energy in the asymptotical state of the Universe,
$M_{{\scriptscriptstyle D}{\scriptscriptstyle
E}}=(\hbar^3\varepsilon_{vac}^{(\infty)})^{1/4}$. We discuss the
case when $M_{{\scriptscriptstyle D}{\scriptscriptstyle E}}\ll
M_\Lambda$.

If non--gravitational contributions to $\Lambda$--term are
self--compensating, then a realistic estimate of the
$M_\Lambda$--scale can be based on the Zeldovich remark \cite{22}.
According to \cite{22}, non--gravitational $\Lambda$--term is
formed by gravitational exchange interaction of quantum
fluctuations on the energy scale of hadrons. In terms of
contemporary understanding of hadron's vacuum, the focus should be
on non--perturbative fluctuations of quark and gluon fields,
forming a quark--gluon condensate (see Appendix \ref{Lambda}). In
this case, $\Lambda$--term is expressed only through the minimum
and maximum scales of particle physics which are the QCD scale
$M_{{\scriptscriptstyle Q}{\scriptscriptstyle
C}{\scriptscriptstyle D}}\simeq 215\ \text{MeV}$ and Planck scale
$M_{Pl}=1.22\cdot 10^{19}\ \text{GeV}$:
\begin{equation}
\displaystyle \hbar^3|\Lambda|=M_\Lambda^4=
\frac{M_{{\scriptscriptstyle Q}{\scriptscriptstyle C}{\scriptscriptstyle D}}^6}{M^2_{Pl}}\simeq
10^{-42}\ \text{GeV}^4\ .
\label{6.32}
\end{equation}
In terms of these scales, it is turns out that a large number of
${\mathcal{N}}=M_\Lambda^4/M_{{\scriptscriptstyle
D}{\scriptscriptstyle E}}^4\sim 10^5$, which is defined in
(\ref{6.30}), can be obtained by the huge number of $|N_g|^{1/2}$,
for the same number of orders of magnitude greater than the ratio
$(M_{Pl}/M_\Lambda)^2$. Indeed, choosing ${\mathcal{N}}$, we find
the value of $|N_g|$, which determines the ratio of vacuum energy
density to the true cosmological constant in the asymptotic state:
\begin{equation}
\displaystyle {\mathcal{N}}=\frac{M_\Lambda^2}{M^2_{Pl}}\sqrt{\frac{2|N_g|}{3}}\ .
\label{6.33}
\end{equation}
The vacuum energy density of asymptotical state is calculated as
follows
\begin{equation}
\begin{array}{c}
\displaystyle \varepsilon_{vac}^{(\infty)}\simeq
\hbar^{-3}M_{Pl}^2M_\Lambda^2\sqrt{\frac{3}{2|N_g|}}\ .
   \end{array}
 \label{6.34}
 \end{equation}

Thus, {\it the macroscopic effect of quantum gravity --- the
condensation of gravitons and ghosts into the state with a certain
wavelength of the order of the horizon scale --- plays a
significant role in the formation of the asymptotic values of
energy density of cosmological vacuum.} The current theory
explains {\it how} the strong renormalization of the vacuum energy
occurs, but, unfortunately, it does not explain {\it why} this
happens and {\it why} the quantitative characteristics of the
phenomenon are those that are observed in the modern Universe. Of
the general considerations one can suggest that the coherent
graviton--ghost condensate occurs in the quantum--gravitational
phase transition (see Section \ref{pt}), and the answers to
questions should be sought in the light of the circumstances.

\section{Gravitons and Ghosts as Instantons}\label{inst}

\subsection{Self--Consistent Theory of Gravitons in Imaginary Time}\label{git}

\subsubsection{Invariance of Equations of the Theory
 with Respect to Wick Rotation of Time Axis}\label{vick}

As has been repeatedly pointed out, the complete system of
equations of the theory consists of the BBGKY chain (\ref{5.7})
--- (\ref{5.9}) and macroscopic Einstein's equations (\ref{5.10}).
On the basis of common mathematical considerations, it can be
expected that  solutions to these equations covers every possible
self--consistent states of quantum subsystem of gravitons and
ghosts and the classical subsystem of macroscopic geometry as
well. In examining the model that operates with the pure gravity
(no matter fields and $\Lambda$--term), one can identify the
following unique property of the theory. {\it Equations of the
theory (\ref{5.7}) --- (\ref{5.10}) are invariant with respect to
the Wick time axis rotation, conducted jointly with the
multiplicative transformation of moments of the spectral
function}:
\begin{equation}
\begin{array}{c}
\displaystyle t\to i\tau,\qquad H\to -i\mathcal{H},\qquad D\to
-\mathcal{D},\qquad W_n\to (-1)^n\mathcal{W}_n\ .
   \end{array}
 \label{7.1}
 \end{equation}
Rules of transformation of time derivatives are obtained from
(\ref{7.1})
\begin{equation}
\begin{array}{c}
\displaystyle \dot H \to -{\mathcal{H}}^{^\centerdot}\ ,
\qquad \ddot H \to i{\mathcal{H}}^{{^\centerdot}{^\centerdot}}\ ,
\qquad \dot D\to i{\mathcal{D}}^{^\centerdot}\ ,
\\[5mm]
\displaystyle \dot W_n\to -i(-1)^n{\mathcal{W}}_n^{^\centerdot}\ , \qquad
\ddot W_n \to -(-1)^n{\mathcal{W}}_n^{{^\centerdot}{^\centerdot}}\ , \qquad
\stackrel{...}{W}_n\to i(-1)^n{\mathcal{W}}_n^{{^\centerdot}{^\centerdot}{^\centerdot}}\ .
\end{array}
\label{7.2}
 \end{equation}
In (\ref{7.2}) and further on we use the notation
${\mathcal{F}}^{^\centerdot}=d\mathcal{F}/d\tau$. The statement
about the invariance of the theory can proved by direct
calculations. As a matter of fact, transformations of quantities
that appear in (\ref{5.7}) --- (\ref{5.10}) by the use of the
rules (\ref{7.1}) and (\ref{7.2}) lead to the BBGKY chain with
imaginary time
 \begin{equation}
\displaystyle  {\mathcal{D}}^{^\centerdot}+6\mathcal{H}\mathcal{D}+
4{\mathcal{W}}_1^{^\centerdot}+16\mathcal{H}\mathcal{W}_1=0\ ,
\label{7.3}
\end{equation}
\begin{equation}
\begin{array}{c}
 \displaystyle
{\mathcal{W}}_n^{{^\centerdot}{^\centerdot}{^\centerdot}} +
3(2n+3)\mathcal{H}{\mathcal{W}}_n^{{^\centerdot}{^\centerdot}}  +3\left[ \left(
4n^2+12n+6\right)\mathcal{H}^2
+(2n+1){\mathcal{H}}^{^\centerdot} \right]{\mathcal{W}}_n^{^\centerdot}+ \vspace{5mm} \\
\displaystyle +2n\left[ 2\left(2n^2+9n+9\right)\mathcal{H}^3
+6(n+2)\mathcal{H}{\mathcal{H}}^{^\centerdot}+
{\mathcal{H}}^{{^\centerdot}{^\centerdot}} \right]{\mathcal{W}}_n
+4{\mathcal{W}}^{^\centerdot}_{n+1}+8(n+2)\mathcal{H}{\mathcal{W}}_{n+1}=0,
\qquad \displaystyle
n=1,\,...,\,\infty\, ,
\end{array}
\label{7.4}
\end{equation}
and to macroscopic Einstein's equations with imaginary time
\begin{equation}
 \begin{array}{c}
\displaystyle
{\mathcal{H}}^{^\centerdot}=-\frac{1}{16}\mathcal{D}-\frac16\mathcal{W}_1\ ,
 \\[5mm]
\displaystyle
3\mathcal{H}^2=\frac{1}{16}\mathcal{D}+\frac{1}{4}\mathcal{W}_1\ .
 \end{array}
 \label{7.5}
\end{equation}
It is easy to see that for  $\Lambda=0$ equations (\ref{5.7}) ---
(\ref{5.10}) identically coincide with (\ref{7.3}) --- (\ref{7.5})
after some trivial renaming.

The invariance of the theory with respect to the Wick rotation of
the time axis leads to the nontrivial consequence.  {\it Having
only self--consistent solution of the BBGKY chain and macroscopic
Einstein's equations, we can not say whether this solution is in
real or imaginary time.} Nevertheless, having a concrete solution
of BBGKY chain, we can view the status of time during further
study. To do so, it is necessary to explore the opportunity to
obtain the same solution at the level of operator functions and
state vectors. If this opportunity exists, the  appropriate
self--consistent solution of BBGKY chain and macroscopic
Einstein's equations is recognized as existing in real time. In
the previous Section \ref{exact}, we showed that two exact
solutions (\ref{5.41}.I) and (\ref{5.41}.III) really exist at the
level of operators and vectors, and thus have a physical
interpretation of standard notions of quantum theory.

The problem is: {\it What a physical reality reflects the
existence of solutions to the equations (\ref{5.7}) ---
(\ref{5.10}) (or that the same thing, (\ref{7.3}) ---
(\ref{7.5})), not reproducible in real time at the level of
operators and vectors?} The existence of the problem is explicitly
demonstrated by the example of exact solutions (\ref{5.41}.II).
Assume that this solution for  $C_{3g}=0, \, C_{4g}=-k_0^2<0, \,
C_{g2}=3k_0^2/4$ exists in real time:
 \begin{equation}
 \begin{array}{c}
\displaystyle
H^2=\frac{k_0^2}{a^2}\ln\frac{a}{a_0}\ , \qquad a=a_0\exp\left(\frac{k_0^2\eta^2}{4}\right)\ .
 \end{array}
 \label{7.6}
\end{equation}
The wave equation for gravitons with the (\ref{7.6}) background
reads
 \begin{equation}
 \begin{array}{c}
\displaystyle \hat\psi''_{{\bf k}\sigma}+k^2_0\eta\hat\psi'_{{\bf k}\sigma}+
k^2\hat\psi_{{\bf k}\sigma}=0\ .
 \end{array}
 \label{7.7}
\end{equation}
The equation for the ghosts looks similar. Equation (\ref{7.7})
differs from (\ref{6.2}) just in the sign of coefficient before
the first derivative. However, this difference is crucial: if
$k^2/k_0^2>0$ it is impossible to allocate the finite Hermit
$H_1(\eta)$  polynomial from degenerate hypergeometric functions
that correspond to solutions of the equation (\ref{7.7}). We have
been left with the infinite series only. These series and
integrals over spectrum of products of these series can not be
made consistent with the simple mathematical structure of the
exact solutions (\ref{5.41}.II). For this reason the solution
(\ref{5.41}.II), as the functional of scale factor is not relevant
to solving operator equations in real time.

\subsubsection{Imaginary Time Formalism}\label{formit}

As is known, the imaginary time formalism is used in
non--relativistic Quantum Mechanics (QM) (examples see, e.g., in
book \cite{24}), in the instanton theory of Quantum Chromodynamics
(QCD) \cite{25, 26, 27, 28, Sh} and in the axiomatic quantum field
theory (AQFT) (See Chapter 9 in the monograph \cite{6}). The
instanton physics in Quantum Cosmology was discussed in \cite{R3,
R4}.

In QM and QCD the imaginary time formalism is a tool for the study
of tunnelling, uniting classic independent states that are
degenerate in energy, in a single quantum state. In AQFT, the
Schwinger functions are defined in the four--dimensional Euclidian
space --- Euclid analogues of Wightman functions defined over the
Minkowski space. It is believed that using properties of
Euclid--Schwinger functions after their analytical continuation to
the Minkowski space, one can reconstruct the properties of
Wightman functions, and thereby restore the physical meaning of
the appropriate model of quantum field theory.

All prerequisites for the use of the formalism of imaginary time
in the QM and QCD on the one hand, and in AQFT, on the other hand,
are united in the self--consistent theory of gravitons.
Immediately, however, the specifics of the graviton theory under
discussion should be noted. Macroscopic space--time in
self--consistent theory of gravitons, unlike the space--time in
the QM, QCD  and AQFT, is a classical dynamic subsystem, which
actually evolved in real time. If in QCD and AQFT Wick's turn is
used to examine the significant properties of quantum system
expressed in {\it the probabilities}  of quantum processes, then
in relation to {\it the deterministic } evolution of classical
macroscopic subsystem this turn makes no sense. Therefore, after
solving equations of the theory in imaginary time, we are obliged
to apply (to the solution obtained) the operation of analytic
continuation of the space for the positive signature to the space
of negative signature. It is clear from the outset that the
operation is not reduced to the opposite Wick turn, but is an
independent postulate of the theory.

Before discussing the physical content of the theory, let us
define its formal mathematical scheme. The theory is formulated in
the space with metric
\begin{equation}
\begin{array}{c}
\displaystyle ds^2=-d\tau^2-a^2(\tau)(dx^2+dy^2+dz^2)\ .
\end{array}
 \label{7.8}
\end{equation}
Note that in our theory, that is suppose to do with cosmological
applications (as opposed to QCD and AQFT), one of the coordinates
is singled out simply because the scale factor depends on it. This
means that in the classical sector of the theory time $\tau$,
despite the fact that it is imaginary, is singled out in
comparison with the 3--spatial coordinates. In the quantum sector
the $\tau$ coordinate also has a special status. {\it Operators}
of graviton and ghost fields with nontrivial commutation
properties are defined over the space (\ref{7.8}). Symmetry
properties of space (\ref{7.8}) allow us to define the Fourier
images of the operators by coordinates $x,\ y,\ z$, and to
formulate the canonical commutation relations in terms of
derivatives of operators with respect to the imaginary time
$\tau$:
 \begin{equation}
\begin{array}{c}
 \displaystyle  \frac{a^3}{4\varkappa}\left[\frac{d\hat\psi^+_{{\bf k}\sigma}}{d\tau}\ ,\
\hat\psi_{{\bf k'}\sigma'}\right]_{-}=-i\hbar \delta_{{\bf k}{\bf
k'}}\delta_{\sigma\sigma'}\ .
\end{array}
\label{7.9}
 \end{equation}
 \begin{equation}
\begin{array}{c}
 \displaystyle
\frac{a^3}{4\varkappa}\left[\frac{d{\hat\vartheta}^+_{{\bf k}}}{d\tau}\ ,\
\hat\vartheta_{{\bf k'}}\right]_{-}=-i\hbar \delta_{{\bf k}{\bf k'}}\ ,
\qquad
\frac{a^3}{4\varkappa}\left[\frac{d{\hat\vartheta}_{{\bf k}}}{d\tau}\ ,\
\hat\vartheta^+_{{\bf k'}}\right]_{-}=-i\hbar \delta_{{\bf k}{\bf k'}}\ .
\end{array}
 \label{7.10}
 \end{equation}
Note that(\ref{7.9}), (\ref{7.10}) are introduced by the newly
independent postulate of the theory, and not derived from standard
commutation relations (\ref{3.37}), (\ref{3.41}) by conversion of
$t\to i\tau$. (Such a conversion would lead to the disappearance
of the imaginary unit from the right hand sides of the commutation
relations.) Thus, the imaginary time formalism can not be regarded
simply as another way to describe the graviton and ghost fields,
i.e. as a mathematically equivalent way for real time description.
In this formalism the new specific class of quantum phenomena is
studied.

The system of self--consistent equations is produced by variations
of action, as defined in 4--space with a positive signature:
\begin{equation}
\begin{array}{c}
 \displaystyle S=\frac{1}{\varkappa}\int d\tau\left\{3\left[\frac{a^2}{N}\frac{d^2a}{d\tau^2}-
 \frac{a^2}{N^2}\frac{dN}{d\tau}\frac{da}{d\tau}
 +\frac{a}{N}\left(\frac{da}{d\tau}\right)^2\right]+\right.
 \\[5mm]
 \displaystyle \left.
 +\frac{1}{8}\sum_{{{\bf k}\sigma}}
 \left(\frac{a^3}{N}\frac{d{\hat \psi}_{{\bf k}\sigma}^+}{d\tau}
 \frac{d{\hat \psi}_{{\bf k}\sigma}}{d\tau}+Nak^2\hat \psi_{{\bf k}\sigma}^+
 \hat \psi_{{\bf k}\sigma}\right)-
 \frac{1}{4}\sum_{\bf k}
 \left(\frac{a^3}{N}\frac{d{\hat\vartheta}_{\bf k}^+}{d\tau}
 \frac{d{\hat\vartheta}_{\bf k}}{d\tau}+
 Nak^2\hat\vartheta_{\bf k}\hat\vartheta_{\bf k}\right)\right\}\ .
 \end{array}
\label{7.11}
 \end{equation}
Note that the full derivative with respect to the imaginary time
is not excluded from Lagrangian. In (\ref{7.11}) the integrand
contains the density of  invariant $\sqrt{\hat g}\hat R$. The
Lagrange multiplier $N$ after the completion of the variation
procedure is assumed to be equal to unity. The system of equations
corresponding to the action (\ref{7.11}) can also be obtained from
the system of equations in real time by conversion of $t\to
i\tau$. Quantum equations of motion for field operators in the
imaginary time read
 \begin{equation}
 \displaystyle
 \frac{d^2\hat\psi_{{\bf
k}\sigma}}{d\tau^2}+3\mathcal{H}\frac{d\hat\psi_{{\bf
k}\sigma}}{d\tau}-\frac{k^2}{a^2}\hat\psi_{{\bf k}\sigma}=0\ ,
 \label{7.12}
 \end{equation}
\begin{equation}
 \displaystyle
 \frac{d^2\hat\vartheta_{{\bf k}}}{d\tau^2}+3\mathcal{H}\frac{d\hat\vartheta_{{\bf
k}}}{d\tau}-\frac{k^2}{a^2}\hat\vartheta_{{\bf k}}=0\ ,
 \label{7.13}
\end{equation}
where ${\mathcal{H}}=a^{^\centerdot}/a$.

Equations (\ref{7.12}), (\ref{7.13}) differ from (\ref{3.30}),
(\ref{3.31}) by only replacement of $k^2\to -k^2$. At the level of
analytic properties of solutions of the equations this difference,
of course, is crucial. However, formal transformations, not
dependent on the properties of analytic solutions to equations
(\ref{3.30}), (\ref{3.31}) and (\ref{7.12}), (\ref{7.13}), look
quite similar. Therefore, all operations to construct the equation
for the spectral function in imaginary time (analogue to equation
(\ref{5.4})) and the subsequent construction of BBGKY chain
coincide with that described in Section \ref{chain} with the
replacement of $k^2\to -k^2$. Replacing  $k^2\to -k^2$ changes the
definition of moments only parametrically: instead of (\ref{5.6})
we get
 \begin{equation}
\begin{array}{c}
\displaystyle \mathcal{W}_n=\sum_{{\bf
k}}\left(\frac{-k^2}{a^2}\right)^n\left(\sum_\sigma\langle
\Psi_g|\hat\psi^+_{{\bf k}\sigma}\hat\psi_{{\bf
k}\sigma}|\Psi_g\rangle-2\langle \Psi_{gh}|\hat\vartheta^+_{{\bf
k}}\hat\vartheta_{{\bf k}}|\Psi_{gh} \rangle\right),\qquad
 n=0,\,1,\,2,\,...,\infty\ ,
 \\[3mm]
 \displaystyle \mathcal{D}=
 \frac{d^2\mathcal{W}_0}{d\tau^2}+3\mathcal{H}\frac{d\mathcal{W}_0}{d\tau}\ .
\end{array}
 \label{7.14}
 \end{equation}
Further actions lead obviously to the BBGKY chain (\ref{7.3}),
(\ref{7.4}) and to the macroscopic Einstein equations (\ref{7.5}).

To solve equations (\ref{7.12}) and (\ref{7.13}), we will be using
only the real linear--independent basis
\begin{equation}
\begin{array}{c}
 \displaystyle \hat\psi_{{\bf k}\sigma}=
 \sqrt{4\varkappa\hbar}\left(\hat Q_{{\bf k}\sigma}g_k+\hat P_{{\bf k}\sigma}h_k\right)\ ,
 \qquad \hat\vartheta_{\bf k}=\sqrt{4\varkappa\hbar}\left(\hat q_{\bf k}g_k+
 \hat p_{\bf k}h_k\right)\ ,
 \\[3mm]
 \displaystyle  g_k h_k^{^\centerdot}-h_k g_k^{^\centerdot}=\frac{1}{a^3}\ .
 \end{array}
 \label{7.15}
 \end{equation}
As will be seen below, one of the basic solutions satisfies the
known definition of instanton: an instanton is a solution to the
classical equation, which is localized in the imaginary time and
corresponds to the finite action in the 4--space with a positive
signature. We will call the operator functions (\ref{7.15}) {\it
the quantum instanton fields of gravitons and ghosts.} Operator
constants of integration  $\hat Q_{{\bf k}\sigma},\ \hat P_{{\bf
k}\sigma}$  and  $q_{\bf k},\ p_{\bf k}$ satisfy commutation
relations (\ref{3.55}). Ladder operators are imposed by the the
equations (\ref{3.54}) and then used in the procedure for
constructing the state vectors over the basis of occupation
numbers. State vectors of the general form in graviton and ghost
sectors are already familiar structure (\ref{3.48}) and
(\ref{3.51}). Only the interpretation of occupation numbers is
changed: now it is number of instantons $n_{{\bf k}\sigma}$,
$n_{\bf k}$, $\bar n_{\bf k}$  of graviton, ghost and anti--ghost
types, respectively.

Direct calculation of the moments of the spectral function leads
to the expression:
\begin{equation}
\begin{array}{c}
\displaystyle
\mathcal{W}_n=4\varkappa\hbar (-1)^n\sum_{\bf k}\left(\frac{k^2}{a^2}\right)^n\left(A_kg_k^2+
B_kh_k^2\right)\ ,
\end{array}
 \label{7.16}
\end{equation}
where
\begin{equation}
\begin{array}{c}
\displaystyle A_k=
\sum_\sigma\langle\Psi_g|\hat Q^+_{{\bf k}\sigma}Q_{{\bf k}\sigma}|\Psi_g\rangle-
2\langle\Psi_{gh}|\hat q_{{\bf k}}^+\hat q_{{\bf k}}|\Psi_{gh}\rangle=
\\[5mm]
\displaystyle =2\langle n_{k(g)}\rangle(1+\zeta^{(g)}_k\cos\varphi_k)-
2\langle n_{k(gh)}\rangle(1+\zeta_k^{(gh)}\cos\chi_k)\ ,
\end{array}
 \label{7.17}
\end{equation}
\begin{equation}
\begin{array}{c}
\displaystyle
B_k=\sum_\sigma\langle\Psi_g|\hat P^+_{{\bf k}\sigma}P_{{\bf k}\sigma}|\Psi_g\rangle-
2\langle\Psi_{gh}|\hat p_{{\bf k}}^+\hat p_{{\bf k}}|\Psi_{gh}\rangle=
\\[5mm]
\displaystyle  =2\langle n_{k(g)}\rangle(1-\zeta^{(g)}_k\cos\varphi_k)-
2\langle n_{k(gh)}\rangle(1-\zeta_k^{(gh)}\cos\chi_k)\ .
\end{array}
 \label{7.18}
\end{equation}
The term containing products of basis functions $g_kh_k$ is
eliminated from (\ref{7.16}) by the condition of homogeneity of
3--space. In (\ref{7.17}) and (\ref{7.18}) average values of
numbers of instantons of ghost and anti--ghost types are assumed
to be equal: $\langle n_{k(gh)}\rangle=  \langle \bar
n_{k(gh)}\rangle$. One needs to pay attention to the multiplier
$(-1)^n$ in (\ref{7.16}): {\it the alternating sign  of moments is
a common symptom of instanton nature of the spectral function.}

Instanton equations of motion (\ref{7.12}), (\ref{7.13}) are of
the hyperbolic type. This fact determines the form of asymptotics
of basis function for $k|\xi|\gg 1$ where $\xi=\int d\tau/a$ is
conformal imaginary time. One of basis functions is localized in
the imaginary time and the other is increasing without limit with
the increasing of modulus of the imaginary time
\begin{equation}
\displaystyle g_k\sim \frac{e^{-k\xi}}{a\sqrt{2k}}\ ,
\qquad h_k\sim \frac{e^{k\xi}}{a\sqrt{2k}}\ , \qquad k|\xi|\gg 1\ .
 \label{7.19}
\end{equation}
In this situation, it is necessary to differentiate between stable
and unstable instanton configurations. We call a configuration
stable, if moments of the spectral function are formed by
localized basis functions only. Without limiting generality, we
assigned $h_k$ to the class of increasing functions. It is easy to
see that the condition of stability $B_k=0$ that eliminates
contributions of $h_k$ from (\ref{7.16}) {\it is reduced to the
condition of quantum coherence of instanton condensate}:
\begin{equation}
\displaystyle \zeta^{(g)}_k\cos\varphi_k=\zeta_k^{(gh)}\cos\chi_k=1 \quad \to \quad
\zeta^{(g)}_k=\zeta_k^{(gh)}=1,\qquad \cos\varphi_k=\cos\chi_k=1\ .
\label{7.20}
\end{equation}
Expressions for the moments are simplified and read
\begin{equation}
\begin{array}{c}
\displaystyle
\mathcal{W}_n=4\varkappa\hbar (-1)^n\sum_{\bf k}A_{\bf k}\left(\frac{k^2}{a^2}\right)^ng_k^2\ .
\end{array}
 \label{7.21}
\end{equation}
Exact solutions, with the stable instanton configurations, are
described in the following Sections \ref{ins1} and \ref{ins2}. In
principle, for a limited imaginary time interval, there might be
unstable configurations, but in the present work such
configurations are not discussed. (The example of the unstable
instanton configuration see in \cite{R5}.)

Note that moments (\ref{7.21}) can be obtained within the
classical theory, limited, as generally accepted, to the solutions
localized in imaginary time. In doing so, $A_{k}$ acts as a
constant of integration of classical equation.

The above approach is the quantum theory of instantons in
imaginary time. Here are present all the elements of quantum
theory: operator nature of instanton field; quantization on the
canonical commutation relations; basic vectors in the
representation of instanton occupation numbers; state vectors of
physical states in the form of superposition of basic vectors.
With the quantum approach, a significant feature of instantons is
displayed, which clearly is not visible in the classical theory.
It is the nature of instanton stable configurations as coherent
quantum condensates.

Construction of the formalism of the theory is completed by
developing a procedure to transfer the results of the study of
instantons to real time. It is clear that this procedure is
required to match the theory with the experimental data, i.e. to
explain the past and predict the future of the Universe. As
already noted, the procedure of transition to real time is not an
inverse Wick rotation. This is particularly evident in the quantum
theory: in (\ref{7.9}), (\ref{7.10}) the reverse Wick turn leads
to the commutation relations for non--Hermitian operators, which
can not be used to describe the graviton field.

The procedure for the transition to real time has the status of an
independent theory postulates. We will formulate this postulate as
follows.

(i) Results of solutions of quantum equations of motion
(\ref{7.12}), (\ref{7.13}), together with the macroscopic
Einstein's equations (\ref{7.5}) {\it after calculating of the
moments} (that is, after averaging over the instanton state
vector) should be represented in the functional form
\begin{equation}
\begin{array}{c}
\displaystyle \mathcal{D}=\mathcal{D}(a, \mathcal{H}, \mathcal{H}^{^\centerdot},...)\ ,
\qquad \mathcal{W}_n=\mathcal{W}_n(a, \mathcal{H}, \mathcal{H}^{^\centerdot},...)\ .
\end{array}
 \label{7.22}
\end{equation}

(ii) It is postulated that {\it functional} dependence of the
moments of the spectral function on functions describing the
macroscopic geometry must be identical in the real and imaginary
time. Thus, {\it at the level of the moments of the spectral
function}, the transition to the real time is reduced to a change
of notation
\begin{equation}
\begin{array}{c}
\displaystyle  \mathcal{D}(a, \mathcal{H}, \mathcal{H}^{^\centerdot},...)
\qquad \to \qquad D(a,H,\dot H,...)\ ,
\\[5mm]
\displaystyle \mathcal{W}_n(a, \mathcal{H}, \mathcal{H}^{^\centerdot},...)
\qquad \to \qquad W_n(a,H,\dot H,...)\ .
\end{array}
 \label{7.23}
\end{equation}

(iii) Moments $D(a,H,\dot H,...)$ and $W_1(a,H,\dot H,...)$
obtained by operations (\ref{7.23}), are substituted to right hand
side of macroscopic Einstein equations that are considered now as
equations in real time. Formally this means that the transition to
the real time in the left hand side of equations (\ref{7.5}) is
reduced to changing of the following notations
\begin{equation}
\begin{array}{c}
\displaystyle \mathcal{H}^{^\centerdot} \qquad \to \qquad \dot H,\qquad\qquad
{\mathcal{H}}^2 \qquad \to \qquad  H^2\ .
\end{array}
 \label{7.24}
\end{equation}

Thus, the acceptance of postulates (\ref{7.22}) --- (\ref{7.24})
is equivalent to the suggestion that in real time the
self--consistent evolution of classic geometry and quantum
instanton system is described by the following equations
\begin{equation}
 \begin{array}{c}
\displaystyle
\dot H=-\frac{1}{16}D(a,H,\dot H,...)-\frac16W_1(a,H,\dot H,...)\ ,
 \\[5mm]
\displaystyle
3H^2=\frac{1}{16}D(a,H,\dot H,...)+\frac{1}{4}W_1(a,H,\dot H,...)\ ,
 \end{array}
 \label{7.25}
\end{equation}
under the condition that the form of functionals in right hand
sides of  (\ref{7.25}) is established by microscopic calculations
in imaginary time. It is obvious also that in the framework of
these postulates {\it any solution of equations consisting of
BBGKY chain and macroscopic Einstein equations (obtained without
use of microscopic theory) can be considered as the solution in
real time.}

\subsubsection{Physics of Imaginary Time}\label{physit}

Mathematical and physical motivation to look for the formalism of
imaginary time comes from the fact that there are degenerate
states separated by the classical impenetrable barrier. In
non--relativistic quantum mechanics the barriers are considered,
that have been formed by classical force fields and for that
reason they have the obvious interpretation. It is well known,
that the calculation of quantum tunnelling across the classical
impenetrable barrier can be carried out in the following order:
(1) the solution of classical equation of motion inside the
barrier area is obtained with imaginary time; (2) from the
solution obtained for the tunnelling particle, one calculates the
action $S$ for the imaginary time; (3) the tunnelling probability,
coinciding with the result of the solution for Schrodinger
equation in the quasi--classical approximation, is equal
$w=e^{-S}$. Obviously, the  sequence described bears a formal
character and cannot be interpreted operationally. Nevertheless, a
strong argument toward the use of the formalism of imaginary time
in the quantum mechanics is the agreement between the calculations
and experimental data for the tunnelling micro--particles.

A new class of phenomena arises in the cases when tunnelling
processes form a macroscopic quantum state. The Josephson effect
is a characteristic example: fluctuations of the electromagnetic
field arise  when a superconductive condensate is tunnelling
across the classically impenetrable non--conducting barrier. Here,
{\it the tunnelling can be formally described as a process
developing in imaginary time, but the fluctuations arise and exist
in the real space--time.} Experimental data show that regardless
of the description, the tunnelling process forms a physical
subsystem in the real space--time, with perfectly real
energy--momentum.

In Quantum Chromodynamics (QCD) physically similar phenomena are
studied by similar methods. The vacuum degeneration is an internal
property of QCD: different classical vacuums of gluon field are
not topologically equivalent. In the framework of the classic
dynamics any transitions between different vacuums are impossible.
In that sense the topological non--equivalence plays role of the
classical impenetrable barrier. There is an heuristic hypothesis
in quantum theory --- that the probability of tunnelling
transition between different vacuums can be calculated as
$w=e^{-S}$, where $S$ is the action of the classical instanton.
The instanton is defined as a solution of gluon--dynamic equations
localized in the Euclidian space--time connecting configurations
with different topologies. As in the case of Josephson Effect, it
is assumed that the tunnelling processes between topologically
non--equivalent vacuums are accompanied by  generation of
non--perturbative fluctuations of gluon and quark fields in real
space--time. Let us notice that {\it in QCD the instanton
solutions, analytically continued into real space--time, are used
to evaluate the amplitude of fluctuations.} The fluctuations in
real space--time are considered as a quark--gluon condensate
(QGC). The existence of QGC with different topological structure
in "off-adrons"\ and "in-adrons"\ vacuums, is confirmed by
comparison of theoretical predictions with experimental data. One
of remarkable facts is that the carrier of approximately the half
of nucleon mass is in fact the energy of the reconstructed QGC.

Now let us go back to the self--consistence theory of gravitons.
In that theory, due to its one--loop finiteness, all observables
are formed by the difference between graviton and ghost
contributions. That fact is obvious both from the general
expressions for the observables (see (\ref{3.72}), (\ref{3.73})),
and from the exact and approximate solutions (described in the
previous sections) as well. The same final differences of
contributions may correspond to the totally different graviton and
ghost contributions themselves. All quantum states are degenerated
with respect to mutually consistent transformations of gravitons
and ghosts occupation numbers, but providing unchanged values of
observable quantities. Thus {\it the multitude of state vectors of
the general form, averaging over which leads to the same values of
spectral function, is a direct consequence of the internal
mathematical structure of the self--consistent theory of
gravitons, satisfying the one--loop finiteness condition.}

In that situation, it is very natural to introduce a hypothesis
about the tunnelling of the graviton--ghost system between quantum
states corresponding to the same values of macroscopic
observables. By the analogy with the effects described above, one
may suggest that 1) the tunnelling processes unite degenerate
quantum states into a single quantum state; 2) tunnelling is
accompanied by creation of specific quantum fluctuations of
graviton and ghost fields in real space--time. With regard to the
mathematical method used to describe these phenomena, today we may
use only those methods that have been tested in adjacent brunches
of quantum theory. It is easy to see that this program has been
realized in Sections \ref{vick}, \ref{formit}. We solve the
equations of the theory for imaginary time, but the amplitude of
the arising fluctuations we evaluate by the analytical
continuation (\ref{7.22}) --- (\ref{7.24}), analoguos to the ones
used in QCD. The specific of our theory lie in the fact that at
the final step of calculations we use the classical Einstein
equations (\ref{7.25}) describing the evolution of the macroscopic
space in real time. The possibility of using these equations is
determined by the action (\ref{7.11}), which, when calculated by
means of the instanton solutions and averaged over the state
vector of instantons, is identically equal zero. As a matter of
fact, after using instanton equations (\ref{7.12}), (\ref{7.13})
and averaging, the action (\ref{7.11}) is reduced to the form:
\begin{equation}
\begin{array}{c}
\displaystyle \langle\Psi|S|\Psi\rangle=
\frac{1}{\varkappa}\int d\tau a^3\left[3\left({\mathcal{H}}^{^\centerdot}+{\mathcal{H}}^2\right)
+\frac{1}{16}\mathcal{D}\right]\ .
 \end{array}
\label{7.26}
\end{equation}
The integrand in (\ref{7.26}) is equal zero in the Einstein
equations with imaginary time (\ref{7.5}). The fact that
$w=\exp(-\langle\Psi|S|\Psi\rangle)=1$  means that the macroscopic
evolution of the Universe is determined. That feature allows the
use of equations (\ref{7.25}), after the moments are analytically
continued into the real time.

\subsection{ Instanton Condensate in the De Sitter Space}\label{ins1}

Among exact solutions of the one--loop quantum gravity, a special
status is given to De Sitter space if the space curvature of this
space is self--consistent with the quantum state of gravitons and
ghosts. In Section \ref{S} it was shown that in the
self--consistent solution, gravitons and ghosts can be interpreted
as quantum wave fields in real space--time. Nevertheless, it
should be mentioned, that the alternating sign of the moments
(\ref{6.28}) points to a possibility of instanton interpretation
of that solution. Methods described in Sections \ref{vick},
\ref{formit}, when applied to De Sitter space, show that such
interpretation is really possible.

We will work with the imaginary conformal time $\xi=\int d\tau/a$.
The cosmological solution is:
\begin{equation}
\displaystyle a=a_0e^{\mathcal{H}\tau}=-\frac{1}{\mathcal{H}\xi}\ ,
\qquad -\infty<\xi\leqslant 0\ .
\label{7.27}
\end{equation}
At the level of the BBGKY chain, due to the fact that the theory
is invariant with respect to the Wick rotation, the calculations
performed to get the solutions coincide  with the those described
in Section \ref{Sitt}. At the microscopic level we use the exact
solutions (\ref{7.12}), (\ref{7.13}) with the background
(\ref{7.27}):
\begin{equation}
 \begin{array}{c}
 \displaystyle
\hat \psi_{{\bf k}\sigma}=\frac{1}{a}\sqrt{\frac{2\varkappa\hbar}{k}}
 \left[Q_{{\bf k}\sigma}g(x)+P_{{\bf k}\sigma}h(x)\right]\ ,
\qquad
\hat \vartheta_{{\bf k}}=\frac{1}{a}\sqrt{\frac{2\varkappa\hbar}{k}}
 \left[q_{{\bf k}}g(x)+p_{{\bf k}}h(x)\right]\ ,
\end{array}
 \label{7.28}
\end{equation}
where
\[
\displaystyle g(x)=\left(1-\frac{1}{x}\right)e^{x}\ ,\qquad h(x)=
\left(1+\frac{1}{x}\right)e^{-x} \qquad x=k\xi<0\ .
\]
The expressions for the moments of the spectral function are reduced to the form:
\begin{equation}
\begin{array}{c}
\displaystyle
\mathcal{W}_n=(-1)^{n+1}\frac{\varkappa\hbar}{\pi^2}{\mathcal{H}}^{2n+2}\int\limits^0_{-\infty}
dxx^{2n+2}\left(A_kg^2+B_kh^2\right)\ .
\end{array}
 \label{7.29}
\end{equation}
Equations for  $A_k,\ B_k$ are given in (\ref{7.17}),
(\ref{7.18}). From (\ref{7.29})  it is obvious that the
self--consistent values $\mathcal{W}_n=const$  can be obtained
only for a flat spectrum of instantons. However, with the flat
specter and $B_k\ne 0$, the second term in (\ref{7.29}) creates a
meaningless infinity. Therefore $B_k=0$, and that, in turn, leads
to the condition (\ref{7.20}), i.e. to quantum coherence of the
instanton condensate.  The quantitative characteristics of the
condensate are formed by instantons only, localized in imaginary
time.

It is easy to calculate of the converging integrals in
(\ref{7.29}):
\begin{equation}
\begin{array}{c}
\displaystyle \int\limits_{-\infty}^0dxx^{2n+2}\left(1-\frac{1}{x}\right)^2e^{2x}=
\frac{1}{2^{2n+1}}(2n-1)!!(2n+1)(n+2)\ .
\end{array}
 \label{7.30}
\end{equation}
After  analytical continuation into the real space--time,
following the rules (\ref{7.22}) --- (\ref{7.24}), we obtain the
final result:
\begin{equation}
 \begin{array}{c}
 \displaystyle
 D=-\frac{12\varkappa\hbar N_{inst}}{\pi^2}H^4\ ,
 \qquad
 W_n=\frac{(-1)^{n+1}}{2^{2n}}
 (2n-1)!(2n+1)(n+2)\times
  \frac{2\varkappa\hbar N_{inst}}{\pi^2}H^{2n+2} ,\qquad n\geqslant 1\ ,
\end{array}
 \label{7.31}
 \end{equation}
where
\begin{equation}
\displaystyle N_{inst}= \langle n_{g}\rangle-\langle n_{gh}\rangle\  .
\label{7.32}
\end{equation}

The comparison of the two models of graviton--ghost condensate in
the De Sitter space reveals some interesting features. In both
cases we deal with the effect of quantum coherence. Expressions
(\ref{7.31}) differ from (\ref{6.28}) only in the formal
substitution $N_g\to N_{inst}$. However the conditions leading to
the quantum coherence are different in these models. According to
(\ref{6.29}), in the condensate of virtual gravitons and ghosts,
the average value of graviton and ghost occupational numbers are
the same, and the non--zero effect appears due to the fact that
the phase correlation in the quantum superposition in the
graviton's and ghost's sectors are formed differently. As it
follows from (\ref{7.20}), (\ref{7.32}), in the instanton
condensate the phases in the graviton and ghost sectors correlate
similarly, but the non--zero effect appears due to the difference
of average occupation numbers for graviton's and ghost's
instantons. The absence of the macroscopic structure of the
condensates does not allow the detection of the differences by
macroscopic measurements. In both cases the graviton--ghost vacuum
possess equal energy--momentum characteristics.

The question about the actual nature of the De Sitter space is
lies in the formal mathematical domain. In these circumstances one
should pay attention to the following facts. While describing the
condensate of virtual gravitons and ghosts, we were forced to
introduce an additional definition of the mathematically
non--existent integrals (\ref{6.25}), i.e. to introduce into the
theory some operations that were not present from the beginning.
It is the additional operations that have provided a very specific
property of the solution --- the alternating signs in the sequence
of the moments of the spectral function. By contrast, the theory
of the instanton condensate has a completely different formal
mathematics. The theory is motivated by the concrete property of
the graviton--ghost system which is degeneration of quantum
states, and the construction of the theory is constructed by the
introduction of mathematically non--contradictory postulates. The
moments of the spectral function's with alternating signs is an
internal property of the graviton--ghost instanton theory.  When
we considered the instanton condensate in the De Sitter space, no
additional mathematical redefinitions were necessary (compare the
formulas (\ref{6.25}) and (\ref{7.30})). We have the impression
that the instanton version of the De Sitter space is more
mathematically comprehensive. Therefore, one may suggest that {\it
the key role in the formation of the De Sitter space (the
asymptotic state of the Universe) belongs to the instanton
condensate, appearing in the tunnelling processes between
degenerated states of the graviton--ghost vacuum.}

\subsection{Instanton Condensate of Constant Conformal Wavelength}\label{ins2}

The exact solution (\ref{5.41}.II) has a pure instanton nature.
Now we will obtain that solution with the value $C_{3g}=0$. One
can rewrite the formulas (\ref{7.6}), (\ref{7.7}) for the
imaginary time:
 \begin{equation}
 \begin{array}{c}
\displaystyle
{\mathcal{H}}^2=\frac{k_0^2}{a^2}\ln\frac{a}{a_0}\ ,
\qquad a=a_0\exp\left(\frac{k_0^2\xi^2}{4}\right)\ .
 \end{array}
 \label{7.33}
\end{equation}
 \begin{equation}
 \begin{array}{c}
\displaystyle \frac{d^2\hat\psi_{{\bf k}\sigma}}{d\xi^2}+k^2_0\xi
\frac{d\hat\psi_{{\bf k}\sigma}}{d\xi}-k^2\hat\psi_{{\bf k}\sigma}=0,\qquad
\frac{d^2\hat\vartheta_{{\bf k}}}{d\xi^2}+
k^2_0\xi\frac{d\hat\vartheta_{{\bf k}}}{d\xi}-k^2\hat\vartheta_{{\bf k}}=0\ .
 \end{array}
 \label{7.34}
\end{equation}
As we already know, the spatially homogeneous modes participate in
the formation of the solution for the equation (\ref{5.41}.II). As
follows from (\ref{7.34}), when $k^2\to 0$, the description of the
spatially homogeneous modes in imaginary time does not differ from
their description in real time. The contribution from modes $g2$
is present in (\ref{7.33}), with the relations $C_{g4}=-k_0^2<0,
\, C_{g2}=3k_0^2/4$ taken into account. These relations are
necessary to provide the existence of the self--consistent
solution. In what follows we are considering the quasi--resonant
modes only.

For   $k^2=k^2_0$, the signs of the last terms in the equations
(\ref{7.34}) provide the existence of instanton solutions we are
looking for:
\begin{equation}
\begin{array}{c}
\displaystyle
\hat\psi_{{\bf k}\sigma}=
\frac{\sqrt{4\varkappa\hbar k_0}}{a_0}\left[\xi\left(\hat Q_{{\bf k}\sigma}+
k_0\hat P_{{\bf k}\sigma}\int\limits_0^\xi e^{-k_0^2\xi^2/2}d\xi\right)+
\frac{\hat P_{{\bf k}\sigma}}{k_0}e^{-k_0^2\xi^2/2}\right]=
\\[5mm]
\displaystyle =\sqrt{\frac{16\varkappa\hbar}{k_0a_0^2}}\left[\hat Q_{{\bf k}\sigma}+
\hat P_{{\bf k}\sigma}F(a)\right]\ln^{1/2}\frac{a}{a_0} \ ,
\end{array}
 \label{7.35}
\end{equation}
\begin{equation}
\begin{array}{c}
\displaystyle
\hat\vartheta_{{\bf k}}=\frac{\sqrt{4\varkappa\hbar k_0}}{a_0}\left[\xi\left(\hat q_{{\bf k}}+
k_0\hat p_{{\bf k}}\int\limits_0^\xi e^{-k_0^2\xi^2/2}d\xi\right)+
\frac{\hat p_{{\bf k}}}{k_0}e^{-k_0^2\eta^2/2}\right]=
\\[5mm]
\displaystyle =\sqrt{\frac{16\varkappa\hbar}{k_0a_0^2}}\left[\hat q_{{\bf k}}+
\hat p_{{\bf k}}F(a)\right]\ln^{1/2}\frac{a}{a_0} \ ,
\end{array}
 \label{7.36}
\end{equation}
where
\[
 \displaystyle F(a)=a_0^2\int\limits_{a_0}^a\frac{da}{a^3\displaystyle\ln^{1/2}\frac{a}{a_0}}+
 \frac{a_0^2}{2a^2\displaystyle\ln^{1/2}\frac{a}{a_0}}\ .
 \]
Calculations which follow contain the same mathematical operations
we have already described several times in the previous sections.
After we remove contributors to the spectral function which
contains $F(a)$, we obtain the condition for the coherence of the
condensate. Some details of the calculations is related to the
alternating signs  of the moments, i.e. with the multiplier
$(-1)^n$, characteristic for the instanton theory. Particularly,
in the expression for $\mathcal{W}_1(g4)$, there is a general sign
"minus"\ . But, according to the Einstein equations in imaginary
time  $\mathcal{W}_1(g4)>0$. The positive sign of the first moment
is provided by the dominant contribution of ghost instantons over
the contribution of graviton instantons. With that taken into
account, we obtain the final equations for the moments of
quasi--resonant modes, obtained after the analytic continuation
into the real space--time:
\begin{equation}
\begin{array}{c}
\displaystyle W_n(g4)= (-1)^{n+1}\frac{64\varkappa\hbar
N_{inst}^{(gh)} k_0^{2n-1}}{a^2_0a^{2n}}\ln\frac{a}{a_0}=
(-1)^{n+1}\frac{24k_0^{2n}}{a^{2n}}\ln\frac{a}{a_0}\ ,
\\[5mm]
\displaystyle D(g4)=
-\frac{128\varkappa\hbar N_{inst}^{(gh)} k_0}{a^2_0a^{2}}\ln\frac{e^{1/4}a}{a_0}=
-\frac{48k_0^{2}}{a^{2}}\ln\frac{e^{1/4}a}{a_0}\ .
\end{array}
\label{7.37}
\end{equation}
Here the following definition has been used:
\[
\displaystyle \langle n_{k(gh)}\rangle-\langle n_{k(g)}\rangle
\to \frac{2\pi^2}{k^2}N_{inst}^{(gh)} \delta(k-k_0)\ , \qquad N_{inst}^{(gh)}=
\frac{3k_0a_0^2}{8\varkappa\hbar}\ .
\]
The graviton instantons are dominant for the spatially homogeneous modes:
\begin{equation}
\displaystyle W_1(g2)=\frac{16\varkappa\hbar k_1
N_{inst}^{(g)}}{a_1^2a^2}\ , \qquad D(g2)=-\frac{32\varkappa\hbar
k_1N_{inst}^{(g)}}{a_1^2a^2}\ .
 \label{7.38}
\end{equation}
The parameter of the spatially homogeneous condensate is defined as follows:
\[
\displaystyle \langle
n_{0(g)}\rangle(1+\zeta^{(g)}_0\cos\varphi_0)- \langle
n_{0(gh)}\rangle(1+\zeta^{(gh)}_0\cos\chi_0) \to
\frac{2\pi^2}{k^2}N_{inst}^{(g)}\delta(k-q_0)\ ,\qquad q_0\to 0\ .
 \]
From expressions (\ref{7.37}), (\ref{7.38}), one gets energy
density and pressure for the system of quasi--resonant and
spatially homogeneous instantons:
\begin{equation}
 \begin{array}{c}
 \displaystyle
\varkappa\varepsilon_g=\frac{8\varkappa\hbar
k_0N_{inst}^{(gh)}}{a_0^2a^2}\ln\frac{a}{a_0}+
\frac{2\varkappa\hbar}{a^2}\left(\frac{k_0N_{inst}^{(gh)}}{a^2_0}-
\frac{k_1N_{inst}^{(g)}}{a^2_1}\right)=\frac{8\varkappa\hbar
k_0N_{inst}^{(gh)}}{a_0^2a^2}\ln\frac{a}{a_0}\ ,
\\[5mm]
\displaystyle \varkappa p_g=-\frac{8\varkappa\hbar
k_0N_{inst}^{(gh)}}{3a_0^2a^2}\ln\frac{ea}{a_0}-
\frac{2\varkappa\hbar}{3a^2}\left(\frac{k_0N_{inst}^{(gh)}}{a^2_0}-
\frac{k_1N_{inst}^{(g)}}{a^2_1}\right)=-\frac{8\varkappa\hbar
k_0N_{inst}^{(gh)}}{3a_0^2a^2}\ln\frac{ea}{a_0}\ .
\end{array}
 \label{7.39}
\end{equation}
In formulas (\ref{7.39}), the terms in brackets are eliminated by
the condition (\ref{5.15}), which is rewritten in terms of
macroscopic parameters
\begin{equation}
\displaystyle
\frac{k_0N_{inst}^{(gh)}}{a^2_0}=\frac{k_1N_{inst}^{(g)}}{a^2_1}\
. \label{7.40}
\end{equation}
Solutions (\ref{7.39}), (\ref{7.40}) describe a quantum coherent
condensate of quasi--resonant instantons with the ghost dominance.
The parameters of the condensate are in accordance with parameters
of a spatially homogeneous condensate with graviton dominance.

\section{Gravitons in the Presence of Matter}\label{gm}

\subsection{Nonlinear Representation of the BBGKY Chain and Integral Identities}\label{I}

The full system of equations of self--consistent theory of
gravitons in the isotropic Universe consists of the BBGKY chain
(\ref{5.7}) --- (\ref{5.9}) and macroscopic Einstein equations. In
equations (\ref{5.7}) --- (\ref{5.9}), the Hubble function $H$ and
its derivatives $\dot H,\ \ddot H$ are coefficients multiplied by
the moments of the spectral function. In such a form the chain
conserves its form even if besides of gravitons, other physical
fields are also sources of the macroscopic gravitational field. We
are interesting in the evolution of the flat isotropic Universe at
a stage when the contributions of gravitons and non--relativistic
particles, baryons and neutralinos, are quantitatively
significant. (The latter are presumably carriers of the mass of
Dark Matter.) We assume also that non--gravitational physical
interactions created the equilibrium vacuum subsystems with full
energy (an effective $\Lambda$--term) of the order of $\Lambda\sim
3 \cdot 10^{-47}\hbar^{-3}$ GeV$^4$.   The macroscopic Einstein
equations containing all sources mentioned above read
 \begin{equation}
 \begin{array}{c}
\displaystyle
R_0^0-\frac12R=\varkappa\varepsilon_{tot}\quad\to\quad H^2=\frac{1}{48}D+
\frac{1}{12}W_1+\frac{\varkappa}{3}\left(\Lambda+\frac{M}{a^3}\right)\
,
 \end{array}
 \label{8.1}
\end{equation}
\begin{equation}
 \begin{array}{c}
\displaystyle
R_0^0-\frac14R=\frac{3\varkappa}{4}\left(\varepsilon_{tot}+p_{tot}\right)\quad\to\quad
 \dot
H=-\frac{1}{16}D-\frac16W_1-\frac{\varkappa M}{2a^3}\ .
 \end{array}
 \label{8.2}
\end{equation}
Equation (\ref{8.2}) should be differentiated with respect to
time, and then $\dot D$ from (\ref{5.7}) should be substituted
into the result of differentiation. These operations produce one
more equation
\begin{equation}
 \begin{array}{c}
\displaystyle
 \ddot
H=H\left(\frac{3}{8}D+W_1+\frac{3\varkappa M}{2a^3}\right)+\frac{1}{12}\dot W_1\ .
 \end{array}
 \label{8.3}
\end{equation}

The BBGKY chain (\ref{5.7}) --- (\ref{5.9}) takes into account the
interaction of gravitons with the self--consistent classical
gravitational field which is represented by the Hubble function
and its derivatives. According to Einstein equations (\ref{8.1})
--- (\ref{8.3}), a self--consistent gravitational field is created
by gravitons and other components of cosmological medium, i.e. by
the matter and non--gravitational vacuum subsystems. Therefore,
the self--consistent gravitational field is a way of describing of
significantly non--linear properties of the system that are the
result of gravitational interaction of elements of the system.
After excluding higher derivatives of the metric from the BBGKY
chain (\ref{5.8}) and (\ref{5.9}), the true non--linear character
of the theory emerges. Substitution of (\ref{8.1}) --- (\ref{8.3})
into (\ref{5.8}) and (\ref{5.9}) gives the non--linear
representation of BBGKY chain:
\begin{equation}
\begin{array}{c}
\displaystyle  \dot D+6HD+4\dot W_1+16HW_1=0\ , \\[3mm]
 \displaystyle
\stackrel{...}{W}_n +3(2n+3)H\ddot{W}_n+
\\[5mm]
\displaystyle +\left[\frac{1}{16}(4n^2+6n+3)D+(n+1)^2W_1+
(8n^2+18n+9)\frac{\varkappa
M}{2a^3}+2(2n^2+6n+3)\varkappa\Lambda\right]\dot{W}_n
+ \vspace{5mm} \\
\displaystyle +\frac{n}{3}\left\{\frac12\dot
W_1+H\left[\frac{n^2}{2}D+(2n^2+3n+3)W_1+(8n^2+18n+9)\frac{\varkappa
M}{a^3}+4(2n^2+9n+9)\varkappa\Lambda\right]\right\}W_n+
\\[5mm]
\displaystyle +4\dot{W}_{n+1}+8(n+2)HW_{n+1}=0, \qquad
\displaystyle n=1,\,...,\,\infty\, .
\end{array}
\label{8.4}
\end{equation}

In the general case, the system of equations (\ref{8.2}) and
(\ref{8.4}) (to which the definition $\dot a/a=H$ is added) should
be solved numerically with initial conditions determined by the
scale factor, moments of the spectral function and their
derivatives
 \begin{equation}
\displaystyle a(0);\quad D(0); \quad W_n(0),\quad \dot
W_n(0),\quad \ddot W_n(0),\qquad n=1,\,...,\,\infty\ .
 \label{8.5}
\end{equation}
The initial condition for the Hubble function should be calculated
via the equation of the constraint (\ref{8.1})
\begin{equation}
\displaystyle H(0)=+\sqrt{\frac{1}{48}D(0)+\frac{1}{12}W_1(0)+
\frac13\varkappa\Lambda+\frac{\varkappa M}{3a^3(0)}}\ .
 \label{8.6}
\end{equation}
Any solution of equations (\ref{8.2}) and (\ref{8.4}), which
corresponds to initial conditions (\ref{8.5}), (\ref{8.6}),
satisfies the identity which is local in time
\begin{equation}
\displaystyle H^2(t)=\frac{1}{48}D(t)+\frac{1}{12}W_1(t)+
\frac13\varkappa\Lambda+\frac{\varkappa M}{3a^3(t)}\ ,
 \label{8.7}
\end{equation}
From the original equations of the chain (\ref{5.7}) ---
(\ref{5.9}) one can obtain the chain of integral identities
 \begin{equation}
 \begin{array}{c}
 \displaystyle
 D(t)=-2W_1(t)+\frac{1}{\tilde a^6}\left\{D(0)+2W_1(0)-
 2\int\limits_0^tdt_1\tilde a^4\frac{d}{dt_1}(\tilde a^2W_1)\right\}\ ,
 \\[5mm]
  \displaystyle
  W_n(t)=\frac{1}{\tilde
 a^{2n}}\left\{C_{n(1)}+C_{n(2)}\int\limits_0^t\frac{dt_1}{\tilde
 a^3}+C_{n(3)}\int\limits_0^t\frac{dt_1}{\tilde a^3}\int\limits_0^{t_1}\frac{dt_2}{\tilde
 a^3}-\right.
 \\[5mm]
 \displaystyle \left. -
 4\int\limits_0^t\frac{dt_1}{\tilde
a^3}\int\limits_0^{t_1}\frac{dt_2}{\tilde
a^3}\int\limits_0^{t_2}\frac{dt_3}{\tilde
a^2}\frac{d}{dt_3}\displaystyle\left(\tilde
a^{2n+4}W_{n+1}\right)\right\}\ , \qquad n=1,\,...,\,\infty\ ,
   \end{array}
 \label{8.8}
 \end{equation}
where
 \begin{equation}
 \begin{array}{c}
\displaystyle
 \tilde a =\frac{a(t)}{a(0)}\ ,\qquad C_{n(1)}=W_n(0),\qquad C_{n(2)}=\dot W_n(0)+2nH(0)W_n(0),
 \\[5mm]
 \displaystyle C_{n(3)}=\ddot W_n(0)+(4n+3)H(0)\dot W_n(0)+2n[\dot
 H(0)+(2n+3)H^2(0)]W_n(0)\ .
  \end{array}
 \label{8.9}
 \end{equation}
According to (\ref{8.8}), in the region attached to the point
where initial conditions are defined, the solution can be
represented in following form
\begin{equation}
\displaystyle -\frac12D\approx W_1\sim \frac{1}{a^2}f_1(a)\
,\qquad W_n\sim \frac{1}{a^{2n}}f_n(a)\ ,
 \label{8.10}
 \end{equation}
where $f_n(a)$ are slow--changing functions of the scale factor.
Comparison of (\ref{8.8}) with (\ref{5.11}) shows that the system
tends to form a graviton--ghost condensate with constant conformal
wavelength.

\subsection{Equation of State of Cosmological Medium Consisting
of Dark Energy and Non-Relativistic matter. $\Lambda$GCDM
Model}\label{LGCDM}

In the framework of the theoretical model under discussion, the
cosmological medium consists of three subsystems, each of these is
described by its own energy density and pressure. They are the
non--relativistic matter $\varepsilon_{mat}=M/a^3\ ,\ p_{mat}=0$;
graviton--ghost coherent quantum condensate $\varepsilon_g\{a\},\
p_g\{a\}$; and, possibly, the non--gravitational contribution of
the $\Lambda$--term to the equilibrium vacuum energy. The original
equations of the theory (\ref{8.1}), (\ref{8.2}) can be written in
the form (after simple and obvious transformations)
\begin{equation}
\begin{array}{c}
\displaystyle 3H^2=\varkappa\varepsilon_{tot}\{a\}\equiv
\varkappa\left(\varepsilon_g\{a\}+\Lambda+\frac{M}{a^3}\right)\ ,
\\[5mm] \displaystyle
 -2\dot H-3H^2=\varkappa p_{tot}\{a\}\equiv
 \varkappa \left(p_g\{a\}-\Lambda\right)\ ,
 \end{array}
  \label{8.11}
\end{equation}
where
\begin{equation}
\begin{array}{c}
\displaystyle
\varkappa\varepsilon_g\{a\}=\frac{1}{16}D+ \frac{1}{4}W_1\ ,
\\[5mm] \displaystyle \varkappa p_g\{a\}=\frac{1}{16}D+\frac{1}{12}W_1
\end{array}
 \label{8.12}
 \end{equation}
are {\it functionals} of scale factor. Explicit forms of these
functionals are determined by the solution of the BBGKY chain in
the nonlinear representation (\ref{8.4}). The energy density and
pressure of cosmological medium as a whole
$\varepsilon_{tot}\{a\},\ p_{tot}\{a\}$ which are defined in
(\ref{8.11}) are also functionals of the scale factor. It is
generally accepted that the combination of two formulae
$\varepsilon_{tot}\{a\},\ p_{tot}\{a\} $ is called the equation of
state of cosmological medium in parametrical form.

The exact particular solutions of the equations of the theory are
described in detail in Sections \ref{ggh_cond}, \ref{Sitt},
\ref{K}, \ref{S}, \ref{ins1},\ref{ins2}. We will bear in mind
these results during the discussion of the dependence of
functionals (\ref{8.12}) on the scale factor. The most important
is the result (\ref{8.10}) which follows from the integral
identities (\ref{8.8}), which means that {\it independently of
concrete dynamics of the scale factor, the self--organization of
the graviton--ghost medium leads to the formation of a condensate
whose equation of state is of the form}
\begin{equation}
\begin{array}{c}
\displaystyle \varepsilon_g(a)\approx \frac{C_g}{a^2}f_1(a)\ ,
\\[5mm] \displaystyle  p_g(a)\approx-\frac{C_g}{3a^2}f_1(a)\ ,
\end{array}
 \label{8.13}
 \end{equation}
where $C_g$ is a constant and $f_1(a)$ is a slow varying function
of the scale factor.  Approximate expressions (\ref{8.13}) can be
obtained by a simple substitution of approximate relations
(\ref{8.10}) to the exact expressions (\ref{8.12}).  According to
(\ref{8.8}), the value and sign of the constant $C_g$ and the form
of the function $f_1(a)$ depend of initial conditions (\ref{8.9}).
Let us mention also that the formation of the graviton--ghost
medium of the approximate equation of state (\ref{8.13}) is
confirmed by numerical experiments that we conducted.

The comparison of the characteristic dependence
$\varepsilon_g(a)\sim 1/a^2$ from  (\ref{8.13}) with the energy
density of non--relativistic matter $\varepsilon_{mat}(a)=M/a^3$
leads to two conclusions. First, the non--relativistic matter
dominates over the condensate and the non--gravitational
$\Lambda$--term during the epoch of sufficiently small scale
factor,
\begin{equation}
\begin{array}{c}
\displaystyle
\frac{M}{a^3}\gg\frac{C_g}{a^2}f_1(a)\ ,
\qquad
\frac{M}{a^3}\gg \Lambda  \ .
\end{array}
\label{8.14}
\end{equation}
Second, the epoch of the dominance of non--relativistic matter
must be replaced by the epoch of condensate domination during
which the inequality $C_gf_1(a)/a^2\gg M/a^3$ is satisfied. This
change of epochs is inevitable because in comparison to
$\varepsilon_{mat}(a)$, the energy density of graviton--ghost
condensate $\varepsilon_g(a)$ is a more slowly varying  function
of the scale factor, which increases with time. Further evolution
of the cosmological medium depends on the relation between the
energy density of the condensate and the value of the
non--gravitational $\Lambda$--term.

Let us assume that during the epoch of the dominance of the
condensate over matter, the condensate (in some time interval)
also dominates additionally over the $\Lambda$--term. This means
that  the following two inequalities are satisfied simultaneously
\begin{equation}
\begin{array}{c}
\displaystyle
\frac{C_g}{a^2}f_1(a)\gg \frac{M}{a^3}\ ,
\qquad \frac{C_g}{a^3}f_1(a)\gg \Lambda
\end{array}
 \label{8.15}
 \end{equation}
In turns, this means that the cosmological model, which represents
a medium consisting only of the condensate of the equation of
state such as $p_g\sim \varepsilon_g/3$, has good accuracy. This
sort of equation of state follows from (\ref{8.8}), (\ref{8.10})
and can be clearly seen from (\ref{8.13}). Such a model
(graviton--ghost condensate of constant conformal wavelength) was
studied in Sections \ref{ggh_cond}, \ref{K},\ref{ins2}. The exact
solutions of the system of equations for the scale factor and
graviton--ghost field obtained in these Sections are attractors.
This mathematical status of these solutions is confirmed by
numerical experiments. From this it follows that during the epoch
of the Universe evolution when the inequalities (\ref{8.14}) are
satisfied, the graviton--ghost condensate relaxes to the state in
which its energy density and pressure are described by expressions
that correspond to the attractor\footnote{In model (\ref{8.16}),
the additive contribution of the spatial homogeneous condensate of
the equation of state $p_g=\varepsilon_g\sim 1/a^6$ is not taken
into account. Because of the rapidly decreasing energy density
with increasing scale factor, such a condensate can play some role
only in the vicinity of cosmological singularity, while the epochs
close to the contemporary Universe are of interest in this work.
This is the reason why we also do not take into account the
contribution of radiation whose equation of state is
$p_r=\varepsilon_r/3\sim 1/a^4$.} (see (\ref{5.17}), (\ref{5.21}),
(\ref{6.15}), (\ref{6.16}), (\ref{7.39}), (\ref{7.40})):
\begin{equation}
\begin{array}{c}
\displaystyle
\varepsilon_{g}(a)=\frac{C_g}{a^2}\ln\frac{a_0}{a}\  ,
\qquad
p_{g}(a)=-\frac{C_g}{3a^2}\ln\frac{a_0}{ea}\ .
\end{array}
\label{8.16}
\end{equation}
The sign of the $C_g$ parameter in (\ref{8.16}) depends on the
microstructure of the graviton--ghost condensate. In the case of
the condensate of virtual particles with graviton dominance
$C_g>0$, while in the case of instanton condensate with the ghost
dominance $C_g<0$. The type of  condensate corresponding to the
observable Dark Energy effect is determined by  comparison of the
theory with  observational data. It will be shown in the next
Section that $C_g>0$.

To discuss the future of the cosmological medium, it is necessary
to bear in mind the fact of the existence of the De Sitter space.
The geometry of this space is consistent with the vacuum states of
all physical fields; it is stable and its symmetry is the highest
among all possible symmetries. These are the reasons why the De
Sitter space of the self--consistent geometry can be considered as
an asymptotical state of the Universe. The results of comparison
of the theory with observational data (see Section \ref{de}) as
well as internal properties of the theory itself are in favor of
this assumption. The De Sitter solution of the BBGKY chain
obtained in Section  \ref{Sitt} allows the interpretation in terms
of virtual particles (Section \ref{S}) as well as in terms of
instantons (Section \ref{ins1}). The vacuum energy density in the
asymptotic state $\varepsilon_{vac}^{(\infty)}\equiv
\Lambda_\infty$ acquires the status of a fundamental cosmological
parameter. The formulae to calculate
$\varepsilon_{vac}^{(\infty)}$ are given in (\ref{5.31}),
(\ref{5.33}), (\ref{5.34}).

The self--polarized condensate in the De Sitter space is formed in
the process of quantum--gravity phase transition (Section
\ref{pt}). The energy density of graviton--ghost vacuum in the
asymptotical state (with no $\Lambda$--term of non--gravitational
nature) is
\begin{equation}
\begin{array}{c}
\displaystyle
\Lambda_{\infty}=\frac{24\pi^2}{\varkappa^2\hbar N_g}\ .
 \end{array}
  \label{8.17}
\end{equation}
If $\Lambda\ne 0$, it is necessary to make some assumptions on the
absolute value and sign of $\Lambda$--term. For reasonable
(relatively small) absolute values of the $\Lambda$--term, the
theory proposes several more or less natural scenarios to form the
asymptotical value
$\varepsilon_{vac}^{(\infty)}\equiv\Lambda_\infty$ consistent with
existing observational data. In Section \ref{Sitt}, the
possibility of equipartition of the vacuum energy between
graviton--ghost and non--gravitational vacuum subsystems was
shown. In this case, in accordance with (\ref{5.38}),
(\ref{5.39}), the total vacuum energy density in the asymptotical
state is
\begin{equation}
\begin{array}{c}
\displaystyle
\Lambda_{\infty}=\frac{12\pi^2}{\varkappa^2\hbar N_{g(max)}}=2\Lambda\ .
\end{array}
\label{8.18}
\end{equation}
If the absolute value of the non--gravitational $\Lambda$--term is
significantly greater than the value which is acceptable from
phenomenological considerations, then the scenario described in
Section \ref{S} makes some sense. The theory predicts that the
strong renormalization of the non--gravitational $\Lambda$--term
of any sign by a graviton--ghost condensate (decreasing its
original absolute value by several orders of magnitude) is
possible. In this case, accordingly to (\ref{6.31}), the
asymptotical value of the total vacuum energy density is positive
and can be estimated by the formula
\begin{equation}
\displaystyle
\Lambda_{\infty}\simeq 2\pi\sqrt{\frac{6|\Lambda|}{\varkappa^2\hbar |N_g|}}\ .
\label{8.19}
\end{equation}
The results (\ref{8.17}), (\ref{8.18}), (\ref{8.19}) illustrate
mathematically the existence of physical states of the condensate
of constant energy density. From this fact it follows that under
the following condition (when the inequality is satisfied)
\begin{equation}
\displaystyle \varepsilon_g\{a\}+\Lambda\sim \Lambda_{\infty}\gg \frac{M}{a^3}
\label{8.20}
\end{equation}
the asymptotic behavior of the solution of the original equations
(\ref{8.11}) describes the De Sitter space of constant 4-curvature
$R_\infty=-4\varkappa\Lambda_{\infty}$.

Thus, the internal mathematical properties of equations
(\ref{8.11}) allow the following classification of epochs of
cosmological evolution:

1. The epoch (\ref{8.14}), during which the non--relativistic
matter dominates over the graviton--ghost condensate and
non--gravitational $\Lambda$--term. During this epoch the equation
of state of the cosmological medium is $\varepsilon_{tot}(a)\simeq
M/a^3\ , \ p_{tot}(a)\simeq 0$. Note that the existence of this
epoch follows directly from WMAP data \cite{31}.)

2. The epoch (\ref{8.15}), during which the graviton--ghost
condensate of constant conformal wavelength dominates over the
non--relativistic matter and non--gravitational $\Lambda$--term.
During this epoch the equation of state of the cosmological medium
is $\varepsilon_{tot}(a)\simeq \varepsilon_g(a)\ , \
p_{tot}(a)\simeq p_g(a)$, where functions $\varepsilon_g(a),\
p_g(a)$ are defined by expressions (\ref{8.16}).

3. The epoch (\ref{8.20}), during which the equilibrium vacuum
consisting of non--gravitational $\Lambda$--term and
graviton--ghost condensate of constant physical wavelength
dominates over the non--relativistic matter. During this epoch the
equation of state of the cosmological medium is
$\varepsilon_{tot}(a)\simeq \Lambda_\infty \ , \ p_{tot}(a)\simeq
-\Lambda_\infty$.

It follows from inequalities (\ref{8.14}), (\ref{8.15}),
(\ref{8.20}), that each epoch is clearly identified with the
relevant region of the cosmological time scale. In these regions
equations of state of the cosmological medium are very similar to
equations of state obtained from exact solutions taking into
account only the dominant component of the medium. An orderly
sequence of exact solutions that take into account only the
dominant terms, represents the most important properties of the
exact solution of complete equations (\ref{8.11}), augmented by
BBGKY chain (\ref{8.4}). Significant differences between the
ordered sequence of exact solutions and solution of complete
equations can be expected only in those parts of the cosmological
scale where change of epochs takes place.

The above presentation, based on the internal properties of the
theory, clearly leads to the interpolation formulae for the energy
density and pressure of cosmological medium:
\begin{equation}
\begin{array}{c}
\displaystyle
\varepsilon_{tot}(a)=\Lambda_{\infty}+\frac{C_g}{a^2}\ln\frac{a_0}{a}+\frac{M}{a^3}\  ,
\qquad  p_{tot}(a)=-\Lambda_{\infty}-\frac{C_g}{3a^2}\ln\frac{a_0}{ea}\ .
\end{array}
\label{8.21}
\end{equation}
The Universe filled by the medium with the equation of state
(\ref{8.21}), described by Einstein's equations
\begin{equation}
\displaystyle
3\frac{{\dot a}^2}{a^2}=\varkappa\varepsilon_{tot}(a)\equiv
\varkappa\left(\Lambda_{\infty}+\frac{C_g}{a^2}\ln\frac{a_0}{a}+\frac{M}{a^3}\right)\ ,
\label{8.22}
\end{equation}
\begin{equation}
\displaystyle
6\frac{{\ddot a}}{a}=
-\varkappa\left(\varepsilon_{tot}(a)+3p_{tot}(a)\right)\equiv
\varkappa \left(2\Lambda_{\infty}-\frac{C_g}{a^2}-\frac{M}{a^3}\right)\ ,
\label{8.23}
\end{equation}
In the terms of the quantities contained in the interpolation
formulae (\ref{8.21}), the classification of epochs is performed
according to the following inequalities and definition:
\begin{equation}
\displaystyle
\frac{M}{a^3}\gg  \Lambda_{\infty}+\frac{C_g}{a^2}\ln\frac{a_0}{a}
\label{8.24}
\end{equation}
is the definition of the epoch of the matter dominance;
\begin{equation}
\displaystyle
\frac{C_g}{a^2}\ln\frac{a_0}{a}\gg \Lambda_{\infty}+ \frac{M}{a^3}\ .
\label{8.25}
\end{equation}
is the definition of the epoch of the dominance of the condensate
of constant conformal wavelength;
\begin{equation}
\displaystyle  \Lambda_{\infty}\gg \frac{C_g}{a^2}\ln\frac{a_0}{a}+\frac{M}{a^3}
\label{8.26}
\end{equation}
is the definition of the epoch of the dominance of the equilibrium
vacuum.

It is not difficult to notice the following fact. If one takes
into account only the dominant component of the cosmological
medium for each epoch, the streamlined set of approximate
solutions obtained from (\ref{8.22}), (\ref{8.23}), practically
coincides with the ordered collection of approximate solutions of
the equations of one--loop quantum gravity, i.e. equations
(\ref{8.11}), augmented by BBGKY chain (\ref{8.4}).

The exact solution of equations (\ref{8.22}), (\ref{8.23})
interpolates the exact solution of equations (\ref{8.11}),
(\ref{8.4}). Differences between the exact solutions may be
visible only in narrow transitional areas at the intersections of
epochs. The most noticeable differences occur at the transition
from the era of dominance of the condensate of a constant
conformal wavelength to the era of dominance of the equilibrium
vacuum. As was mentioned in Section \ref{pt}, numerical
experiments show that the transition from pre--asymptotic stage to
the asymptotic stage is followed by nonlinear fluctuations of
physical observables.  The latter forces us to assume that the
asymptotic state of the graviton--ghost condensate is formed as a
result of quantum--gravity phase transition on the scale of the
whole Universe. It is worth mentioning that comparison of the
model (\ref{8.22}), (\ref{8.23}) with the observed data shows that
at the stage (\ref{8.25}) (or what is the same, at the stage
(\ref{8.15})) the Universe expanded with deceleration. Thus, the
currently observed acceleration of expansion may be a consequence
of a phase transition  into an asymptotic state, i.e. a state of
4--space of constant curvature, with the highest symmetry among
all possible symmetries.

The issue of experimental verification of the phase transition is
very complicated, and we will not discuss it in this paper. It is
worth mentioning only that, in order to clarify the issue, it is
necessary to conduct detailed computer experiments. However,
comparison of the results of computer experiments with real
observational data can be done only on the condition that the
observational data to meet very rigid standards of completeness
and accuracy.

At the level of accuracy of the modern cosmological observations,
transition processes can be neglected. Thus, it gives
interpolation formulae (\ref{8.21}) the status of a model which
follows only from the first principles of the one--loop quantum
gravity. These principles are three: {\ it (i) the inevitable
appearance of the sector of nontrivial ghosts interacting with the
macroscopic gravity; (ii) conformal non--invariance and zero rest
mass of gravitons and ghosts; (iii)  one--loop finiteness off the
mass shell.}

The difference between the model (\ref{8.21}) and the known
$\Lambda$CDM model is in the additional term which is relevant to
the variable component of the Dark Energy. We propose the name
"$\Lambda$GCDM model"\  for the Graviton--Cold--Dark--Matter model
(\ref{8.21}). We assume that the equations  (\ref{8.21}) are able
to represent the energy density and pressure of the cosmological
medium after the separation of matter and radiation, including the
epoch of the formation of galaxies and the contemporary Universe,
including its future.

Currently, the priority task is the experimental verification of
the variable component of Dark Energy and accordance with the
properties of this component predicted by the $\Lambda$GCDM model.

\section{Graviton--Ghost Condensates in the Role of Dark Energy.
Comparison of Theory with Observational Data}\label{de}

\subsection{Approaches to the Dark Energy Problem}\label{prob}

Acceleration of the Universe and, consequently, the existence of
Dark Energy, causing the phenomenon, was discovered experimentally
in the works \cite{03,04}. Today, we can identify three aspects of
the Dark Energy problem. The first is obtaining reliable
observational data on the density of Dark Energy
$\varepsilon_{{\scriptscriptstyle D}{\scriptscriptstyle E}}(a)$ as
a function of the scale factor, or which is the same thing, as a
function of the redshift $z=1/a-1$.  It is well known that the
simplest Dark Energy model is the cosmological constant.
Observational aspect of the problem of Dark Energy lies in the
answer to the question {\it is it true that the density
$\varepsilon_{{\scriptscriptstyle D}{\scriptscriptstyle
E}}=\Lambda=const$ or does Dark Energy contain a variable
component?} From the point of view of available data, some
preferred models are those with variable component, but, of
course, further detailed study is needed (see, e.g., the SNAP
project \cite{29}). The second aspect is that the physical nature
of Dark Energy is a problem of fundamental physics. The statement
of the problem will be determined to a substantial extent by
existing observational data (that we expect is to be provided with
the necessary completeness and accuracy). The third aspect is the
astrophysical aspect. It will become particularly relevant if the
variable component of Dark Energy will be reliably identified
during the era of large--scale structure formation in the
Universe.

It is common to consider that the problem of the physical nature
of Dark Energy is a window into new physics. These expectations
suggest that the theoretical model of Dark Energy is not limited
to the cosmological constant. A cosmological constant could hardly
be called new physics. Indeed, the most common principles of
geometrized theory of gravitation say only that in a weak
gravitational field the Lagrangian of the theory should be
presented in a series of powers of invariants of the curvature.
From this perspective,  $\Lambda$--term is a zero term of
expansion, and the scalar curvature, generating the left hand side
of Einstein's equations, is the first term of the expansion of the
Lagrangian of the theory of gravity. In other words, the presence
of $\Lambda$--term in Einstein's equations is not a conceptual
problem. On the contrary, the theory of gravity, taking into
account the  $\Lambda$--term looks more natural than the theory
without it.

The problem is the numerical value of  $\Lambda$--term (it is
anticipated that the value of  $\Lambda$--term can be calculated
in principle). In this regard we wish to point out that the
problem of calculating of $\Lambda$--term conceptually does not
differ from the problem of calculating other fundamental constants
of physics, e.g., electron charge, the Fermi constant of weak
interaction, scale of Quantum Chromodynamics and others.
Variations of these constants, as is known, also lead to a change
in the properties of the Universe. The  $\Lambda$--term problem is
singled out rather emotionally. With the variation of the
cosmological constant catastrophic changes in the Universe will
follow, as can be seen by a simple analysis of a simple equation.
We will continue the discussion of  $\Lambda$--term in the
Appendix \ref{Lambda}.

Possible deviation of the Dark Energy density of the cosmological
constant is being actively studied by theorists (for a catalogue
of models see review \cite{30}) and is the subject of planned
experiments (see SNAP project in \cite{29}). A quintessence
(evolving scalar field) was widely discussed in
\cite{60,61,62,63,64}. A free scalar field of a small mass as a
model of Dark Energy was discussed in \cite{65}. Models with
violation of the weak energy condition ("phantom energy"\ ) were
proposed in \cite{66,67}. Models with  Chaplygin gas and its
modifications were considered in \cite{68} (see review \cite{30}
and references therein). The massless scalar field and Casimir
effect were proposed in \cite{70,71}. Some brane world models were
proposed in \cite{72,73,74,75}. The discussion of these and other
ideas and models, including the "new physics"\ proposals can be
found in \cite{30,70,76,78} and other reviews. All of these ideas
and approaches have no bearing on this paper. Current prospects
for the dynamical theory of the graviton vacuum with applications
to the Dark Energy problem can be found in the review \cite{02}.
Technical approaches to the vacuum energy and the problem of
cosmological constant in terms of quantum field theory in curved
spacetime can be found in review \cite{79}.  The state of the art
of observational data and theoretical prospects for the Dark
Energy problem can be found in \cite{35,A}.

The problem of choosing a model for dealing with experimental data
becomes very relevant. In essence, the choice of a model is the
choice of the physical interpretation of the nature of Dark
Energy. The problem is complicated by a lack of completeness and
accuracy of data available on SNIa. The intervals of redshift
$0<z<1.8$, available from observations of SNIa, are too narrow to
choose between alternative models. In that situation, priority is
given to models based on common and experimentally motivated
principles of fundamental physics. Of course,  $\Lambda$--term
belongs to the first--principle models. In our view, a model based
on the effect of condensation of quasi--resonant graviton and
ghost modes in the self--consistent field of the isotropic
Universe has the same status. The emergence of graviton--ghost
condensates in the process of cosmological evolution is a
consequence of only the most general properties of one--loop
quantum gravity.  These are conformal non--invariance and zero
rest mass of gravitons; inevitability of appearance of the ghost
sector; and the one--loop finiteness off the mass shell.
Interpolation formulae of $\Lambda$GCDM model (\ref{8.21}),
describing this effect, are based on exact solutions and are
justified by integral identities (\ref{8.8}).

With $C_g\to 0$ $\Lambda$GCDM (\ref{8.21}) turns into $\Lambda$CDM
model. It is therefore obvious that the model of Dark Energy,
based on one--loop quantum gravity, is consistent with existing
observational data as well as the $\Lambda$--term model. In this
situation, one must first determine whether the $\Lambda$GCDM
model has statistically significant advantages over the
$\Lambda$CDM model.

\subsection{Observational Data}\label{data}

The set of formulas used for observational data processing follows
below. The normalization of the scale factor in equations
(\ref{8.22}), (\ref{8.23}) can always be chosen in such a way that
$a(t_{\scriptscriptstyle U})=1$ in the contemporary Universe. To
normalize other physical quantities, we use the Hubble constant in
the contemporary epoch $H_0=100h$\ km/(s$\cdot$Mpc), where
$h=0.73^{+0.04}_{-0.03}$. In the dimensionless form, equations of
$\Lambda$GCDM model read
\begin{equation}
 \begin{array}{c}
 \displaystyle
 \tilde H^2=\tilde\varepsilon_{g}+\Omega_{\scriptscriptstyle M}(1+z)^3\ ,
 \\[5mm]
\displaystyle \tilde Q=-\frac12\left[\tilde\varepsilon_{g}+3\tilde p_{g}+
\Omega_{\scriptscriptstyle
 M}(1+z)^3\right]=\Omega_\Lambda-\frac12\Omega_g(1+z)^2-\frac12\Omega_{\scriptscriptstyle
 M}(1+z)^3\ ,
    \end{array}
 \label{9.1}
 \end{equation}
where
 \[
  \begin{array}{c}
  \displaystyle \tilde H=\frac{H}{H_0}\equiv \frac{\dot a}{H_0a}, \qquad
  \tilde Q=\frac{Q}{H^2_0}\equiv\frac{\ddot a}{H_0^2a}\ ,
  \\[5mm]
\displaystyle  \Omega_{\scriptscriptstyle M}=
 \frac{M}{3H_0^2}\ ,\qquad \Omega_\Lambda=\frac{\Lambda_{\infty}}{3H_0^2}\ ,
 \qquad \Omega_g=\frac{C_g}{3H_0^2}\ .
  \end{array}
\]
The values of parameters
\[
 \displaystyle \Omega_{\scriptscriptstyle
M}=0.24^{+0.03}_{-0.04}, \qquad    h=0.73^{+0.04}_{-0.03}
\]
are known from WMAP data \cite{31}. Below are expressions for the
density and pressure of Dark Energy:
\begin{equation}
 \begin{array}{c}
 \displaystyle
\tilde \varepsilon_{{\scriptscriptstyle D}{\scriptscriptstyle E}}(z)= \tilde\varepsilon_{g}(z)=
\Omega_\Lambda+\Omega_g(1+z)^2\ln[a_0(1+z)]\ ,
 \\[5mm]
 \displaystyle
\tilde p_{{\scriptscriptstyle
D}{\scriptscriptstyle E}}(z)=\tilde p_{g}(z)=-\Omega_\Lambda-
\frac13\Omega_g(1+z)^2\left(\ln[a_0(1+z)]-1\right)\ .
\end{array}
 \label{9.2}
 \end{equation}
In the flat Universe, four parameters of  $\Lambda$GCDM model are
connected by the following condition
 \begin{equation}
\displaystyle
 \Omega_\Lambda+\Omega_g\ln a_0+\Omega_{\scriptscriptstyle M}=1\ .
\label{9.3}
 \end{equation}

The information on the normalized density of Dark Energy is
contained in the dependence of conformal distance to supernovae
SNIa of their redshifts. It reads
\begin{equation}
\begin{array}{c}
\displaystyle  R(z)=\int_0^z\limits\frac{dz}{\tilde H(z)}\ ,
\\[5mm]
\displaystyle \tilde H(z)= \sqrt{\Omega_{\scriptscriptstyle
M}(1+z)^3+ \tilde \varepsilon_{{\scriptscriptstyle
D}{\scriptscriptstyle E}}(z)}\ ,
\end{array}
\label{9.4}
\end{equation}
Observational data on SNIa are usually presented in terms of
distant moduli,  which reads
\begin{equation}
\begin{array}{c}
\displaystyle
 \mu_{\scriptscriptstyle B}(z)=\mathcal{M}+5\lg[(1+z)R(z)],
 \\[3mm]
\displaystyle \mathcal{M}=-5\lg h+42.3841+(M_{\scriptscriptstyle
B}+19.31)\ ,
 \end{array}
\label{9.5}
\end{equation}
where $M_{\scriptscriptstyle B}$ is the absolute bolometric
brightness of a standard candle. The full set, containing data on
all known supernovae (292 points), is given in \cite{32}. In
\cite{33}, a  "Gold Data Set"\ was built from the data of
\cite{32,34}\footnote{New versions of SNIa data see in \cite{NSN1,
NSN2}}. In works \cite{32,33,34}, it is anticipated that
$M_{\scriptscriptstyle B}=-19.31$. In the equation (\ref{9.5}),
$\mathcal{M}=43.09_{-0.12}^{+0.09}$ for $h,\ M_{\scriptscriptstyle
B}$ values given above. Actually, however, questions about the
precise values of normalized Hubble constant and the brightness of
a standard candle remain open. For this reason, values
$\mu_{\scriptscriptstyle B}(z)$  quoted in \cite{32,33,34} are
treated as data in arbitrary units. The value of $\mathcal{M}$ is
determined in the process of processing these data.

A set of observational data on supernovae SNIa is presented in
Figure 1 (280 points selected from 292 points \cite{32}).
Solid lines in Figure 1 show a fit corresponding to the $\Lambda$CDM
model (curve 1) and  $\Lambda$GCDM model (curve 2).
\begin{figure}
\begin{minipage}[b]{.60\linewidth}
\centering\epsfig{figure=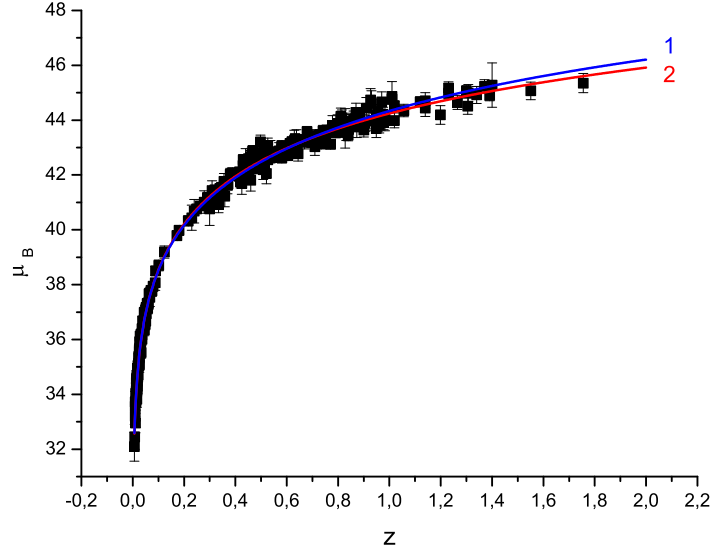,width=\linewidth}
\caption{Distant modulii of supernovae SNIa. Observational data
--- 280 counts points from \cite{32}. Fitting curves:  1 ---
$\Lambda$CDM model (\ref{9.9}), (\ref{9.12}); 2 --- $\Lambda$GCDM
model (\ref{9.14}), (\ref{9.15}).}
\end{minipage}
 \label{Fig1}
\end{figure}

In the verification procedure of theoretical models an important
role plays  the redshift $z_0$, at which the deceleration of the
expansion of the Universe is changed to an accelerated expansion.
The value below is recovered from the Hubble diagram
\begin{equation}
 \displaystyle
  1+z_0=1.46\pm 0.13\ .
\label{9.6}
\end{equation}
From the condition $\tilde Q(z_0)=0$ for a given value of $1+z_0$,
one more condition is obtained for the  $\Lambda$GCDM model which
reads
\begin{equation}
\displaystyle
 2\Omega_\Lambda-\Omega_g(1+z_0)^2-\Omega_{\scriptscriptstyle M}(1+z_0)^3=0\ .
\label{9.7}
 \end{equation}
Note the following circumstance. If one fixes
$\Omega_{\scriptscriptstyle M}$ and  $1+z_0$ parameters by their
average values and substitutes these values into (\ref{9.3}) and
(\ref{9.7}), then only one independent parameter remains in the
$\Lambda$GCDM model. The value of this parameter will be found by
fitting of the Hubble diagram as a whole. It is convenient to
select $\Omega_g$ as such an independent fit parameter
characterizing the variable component of Dark Energy. We hope that
in the future, with increasing accuracy of measurements of
$\Omega_{\scriptscriptstyle M}$ and   $1+z_0$ the implementation
of this program becomes possible.

Currently, we can use intervals of acceptable values of parameters
$\Omega_{\scriptscriptstyle M}$  and $1+z_0$ to assess the status
of the model of  $\Lambda$--term. Assuming that $\Omega_g=0$ in
(\ref{9.3}) and (\ref{9.7}), we get
\begin{equation}
 \displaystyle
 0.73\ <\ \Omega_\Lambda\ <\ 0.8\ , \qquad
 (1+z_0)_{\Lambda \text{CDM}}=
 \left(\frac{2\Omega_\Lambda}{\Omega_{\scriptscriptstyle M}}\right)^{1/3}=1.85^{+0.15}_{-0.10}\ .
\label{9.8}
\end{equation}
As we can see, $1\sigma$ interval predictions of $\Lambda$CDM
model (\ref{9.8}) does not overlap the  $1\sigma$ interval of
measured values (\ref{9.6}). On its own, this fact does not reject
the  $\Lambda$--term as a carrier of Dark Energy. For this, it
would be necessary to detect the contradiction at the level of
more than three standard deviations. In addition, the
contradiction at the level of $1\sigma$  is smoothed out if one
takes into account systematic errors. Discussion of this range of
issues can be found in the Weinberg book \cite{35}. Nevertheless,
when comparing (\ref{9.6}) and (\ref{9.8}) doubts about the
adequacy of $\Lambda$CDM model emerge. In this situation,
consideration of $\Lambda$GCDM model strongly indicated.

\subsection{Results of the Data Fit and its Discussion}\label{fit}

Below, the best--fit parameters of multi--parameter models were
calculated using the standard LSM statistical approach based on
the calculation of the Fisher matrix \cite{A,B}. To get the sense
of the error margins for the parameters as they come from the
errors in the current observational data, we show the square root
of diagonal elements of the covariance matrix (1-sigma). Of
course, for any further use of the calculated parameters the full
covariance matrix must be used to take into account the
correlation between parameters.

We evaluate the  $\Lambda$GCDM model in the comparison with the
$\Lambda$CDM model. Therefore, first, let us introduce the results
of the fit of the Hubble diagram using formulas of the
$\Lambda$CDM model. The value of (\ref{9.6}) is included in the
database of the fitting data, and the interval of allowed values
of the parameter $\Omega_{\scriptscriptstyle M}$ is defined by
WMAP data. The results of the fit are shown on Figure 1, curve 1.
The values of fitting parameters are:
\begin{equation}
\begin{array}{c}
 \displaystyle
 \Omega_{\scriptscriptstyle M}=0.27\pm 0.03\ ,\qquad \Omega_\Lambda=0.73\pm 0.03\ , \qquad
 (1+z_0)=1.75\pm 0.07\ ,
 \\[5mm]
 \displaystyle \mathcal{M}=43.37\pm 0.02\ , \qquad \chi^2/dof=0.97\ .
 \end{array}
\label{9.9}
\end{equation}
We will use also the value of the statistical sum and the formula
to calculate the number of degrees of freedom with their
statistical fluctuations taken into account:
\begin{equation}
\begin{array}{c}
 \displaystyle
 S=\frac12\sum_{i=1}^N\left[\frac{\mu_{fit}(z_i)-\mu_{exp}(z_i)}{\Delta\mu_{exp}(z_i)}\right]^2+
  \frac12\left[\frac{z_{0fit}-z_{0exp}}{\Delta z_{0exp}}\right]^2\ ,
  \qquad \chi^2/dof=\frac{2S}{N+1-n}\ ,
  \\[5mm]
 \displaystyle \frac12\left(N_{dof}\pm \Delta N_{dof}\right)=
 \frac{N+1-n}{2}\pm \sqrt{\frac{N+1-n}{2}}\ .
 \end{array}
\label{9.10}
\end{equation}
In (\ref{9.10}), $N+1-n$ is the number of degrees of freedom; $N$
is the number of experimental points of distant moduli
$\mu_{exp}(z_i)$;  term "1"\  corresponds to the experimental
value $1+z_0$; $n$  is the number of independent fitting
parameters (which are $\Omega_\Lambda$ and $\mathcal{M}$ for
$\Lambda$CDM model). The interval of values of the statistical
sum, characterizing the fit $\Lambda$CDM model (in the framework
of appropriate  $\Lambda$CDM's formulae) is calculated as follows:
\begin{equation}
 \displaystyle S\pm \Delta S= \chi^2/dof\left(\frac{N+1-n}{2}\pm \sqrt{\frac{N+1-n}{2}}\right)\ .
 \label{9.11}
\end{equation}
\begin{equation}
 \displaystyle (S\pm \Delta S)_{\Lambda CDM}= 136\pm 11\ .
 \label{9.12}
\end{equation}

The question whether the existing set of experimental data could
pinpoint a single preferable model is reduced to the comparison of
the intervals of the statistical sums. Let us assume that there is a
set of models $X_a$, $a=1\ ,2\ ,...,\ M$, and the results of the
fits by the formulas of each model are characterized by the
respective intervals of the statistical sums $S_a\pm \Delta S_a$.
The models $X_a$ and  $X_b$ are statistically equivalent if the
intervals of the statistical sums $S_a\pm \Delta S_a$  and $S_b\pm
\Delta S_b$ overlap. If the intervals do not overlap, the
statistically preferable model is a model that corresponds to the
minimal $\chi^2/dof$. Certainly, in addition to the statistical
criteria, it is necessary to take into account how the models
correspond to general principles, how hypothetical the concepts
are, etc. Assuming that the model of the $\Lambda$--term is the
simplest one among all of the first--principle models, the interval
of values of the statistical sum (\ref{9.12}) could be treated as
the reference interval.

The fit of data in Figure 1 using equations of the $\Lambda$GCDM
model shows that the results practically do not depend on the
parameter  $\Omega_{\scriptscriptstyle M}$ chosen within the
interval  $0.2<\Omega_{\scriptscriptstyle M}<0.27$ which is in
accordance with WMAP data. As is shown below, that observation is
connected with a specific prediction of the $\Lambda$GCDM model:
in the area of the Hubble diagram for supernovae SNIa,
$0.007<z<1.755$ , the density of the Dark Energy is on the par
with the density of non--relativistic matter. In that situation,
$\Omega_{\scriptscriptstyle M}$ cannot have the status of fitting
parameter. We have conducted three fitting procedures with three
fixed values of  $\Omega_{\scriptscriptstyle M}=0.20; \; 0.24;\;
0.27$.  In all three cases
\begin{equation}
\displaystyle
 \chi^2/dof=0.91\ , \qquad \mathcal{M}=43.33\pm 0.03\ , \qquad
1+z_0=1.34\pm 0.04\ . \label{9.13}
 \end{equation}
The values of other parameters are as follows:
 \begin{equation}
 \begin{array}{c}
\displaystyle \Omega_{\scriptscriptstyle M}=0.20: \qquad
   \Omega_\Lambda=2.25\pm 0.53 , \qquad  \Omega_g=2.23\pm 0.68\ ,\qquad a_0=0.52\pm 0.02\ ;
\\[5mm]
 \displaystyle
 \Omega_{\scriptscriptstyle M}=0.24:\qquad
 \Omega_\Lambda=2.23\pm 0.54\ , \qquad  \Omega_g=2.15\pm 0.68\ ,\qquad a_0=0.51\pm 0.02\ ;
\\[5mm]
 \displaystyle
 \Omega_{\scriptscriptstyle M}=0.27:\qquad
 \Omega_\Lambda=2.21\pm 0.55\ , \qquad  \Omega_g=2.09\pm 0.68\
,\qquad a_0=0.49\pm 0.02\ .
\end{array}
 \label{9.14}
 \end{equation}
The interval of values of the statistical sum is calculated by
equations (\ref{9.10}) and (\ref{9.11}) with $n=3$  (the fitting
parameters are  $\Omega_g,\ a_0,\ \mathcal{M}$):
\begin{equation}
 \displaystyle (S\pm \Delta S)_{\Lambda GCDM}= 127\pm 11\ .
 \label{9.15}
\end{equation}

As follows from (\ref{9.12}) and (\ref{9.15}), the intervals for
the statistical sums for $\Lambda$CDM and $\Lambda$GCDM models are
largely overlapping. Therefore, from the point of view of
available observational data, these models should be considered
statistically equivalent, with small advantage of $\Lambda$GCDM
model over $\Lambda$CDM in accordance with the
$\chi^2$--criterion.

$\Lambda$CDM and $\Lambda$GCDM models start diverging from each
other with increasing completeness and accuracy of
observational data. We can even make them statistically
distinguishable from each other on the basis of available
observational data by data smoothing (using a simple operation of
noise reduction). The first smoothing operation is conducted by
averaging of distant moduli with the redshift conserved. In
the second step, at the interval up to $z=1.4$, the averaging is
done by a moving window, the size of three neighboring points.
Thus, the errors in the averaged data are reduced by a factor of
$\sqrt{3}$. To maintain the statistical weight of the
transition point, its error is also reduced by a factor of
$\sqrt{3}$: $1+z_0=1.460\pm 0.075$. 211 points obtained by that
operation, $\mu_{exp}(z_i)$ are shown in Figure 2. The fit of
smoothed points by the formulas of $\Lambda$CDM model provides the
following results:
\begin{equation}
\begin{array}{c}
 \displaystyle
 \Omega_{\scriptscriptstyle M}=0.27\pm 0.02\ ,\qquad \Omega_\Lambda=0.73\pm 0.02\ , \qquad
 (1+z_0)=1.75\pm 0.05\ ,
 \\[5mm]
 \displaystyle \mathcal{M}=43.34\pm 0.01\ , \qquad \chi^2/dof=1.15\,
 \end{array}
\label{9.16}
\end{equation}
\begin{equation}
\begin{array}{c}
 \displaystyle (S\pm \Delta S)_{\Lambda CDM}^{AA}= 121\pm 12\ .
 \end{array}
\label{9.17}
\end{equation}
With the fit of the same points by using the $\Lambda$GCDM model,
again we observe the insensitivity of the  $\chi^2$--criterion to
the variations of the parameter  $\Omega_{\scriptscriptstyle M}$
within the experimentally allowed interval:
\begin{equation}
\displaystyle
 \chi^2/dof=0.93\ , \qquad \mathcal{M}=43.32\pm 0.02\ , \qquad
1+z_0=1.34\pm 0.03\ .
 \label{9.18}
 \end{equation}
The values of other parameters change slightly with variation of
the parameter  $\Omega_{\scriptscriptstyle M}$:
\begin{equation}
 \begin{array}{c}
\displaystyle \Omega_{\scriptscriptstyle M}=0.20: \qquad
   \Omega_\Lambda=2.19\pm 0.35 , \qquad  \Omega_g=2.16\pm 0.45\ ,\qquad a_0=0.53\pm 0.02\ ;
\\[5mm]
 \displaystyle
 \Omega_{\scriptscriptstyle M}=0.24:\qquad
 \Omega_\Lambda=2.16\pm 0.36\ , \qquad  \Omega_g=2.08\pm 0.45\ ,\qquad a_0=0.51\pm 0.01\ ;
\\[5mm]
 \displaystyle
 \Omega_{\scriptscriptstyle M}=0.27:\qquad
 \Omega_\Lambda=2.15\pm 0.36\ , \qquad  \Omega_g=2.02\pm 0.45\
,\qquad a_0=0.50\pm 0.01\ .
\end{array}
 \label{9.19}
 \end{equation}
The statistical sum of the smoothed data has the following value
and deviation:
\begin{equation}
 \displaystyle (S\pm \Delta S)_{\Lambda GCDM}^{AA}= 97.6\pm 9.5\ .
 \label{9.20}
\end{equation}

\begin{figure}
\begin{minipage}[b]{.60\linewidth}
\centering\epsfig{figure=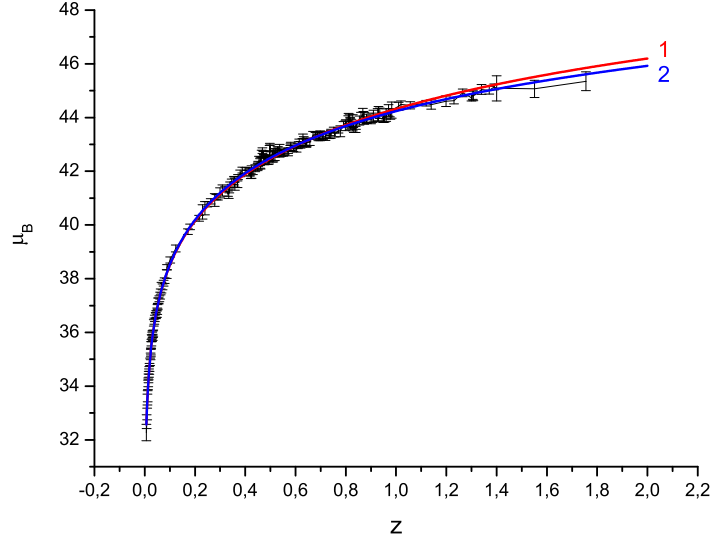,width=\linewidth}
\caption{Distant modulii of supernovae SNIa. Smoothed
observational data
--- 211 points. Fitting curves: 1 --- $\Lambda$CDM model
(\ref{9.16}), (\ref{9.17}); 2 --- $\Lambda$GCDM model
(\ref{9.19}), (\ref{9.20}).} \label{Fig2}
\end{minipage}
\end{figure}
\begin{figure}
    \begin{minipage}[b]{.60\linewidth}
        \centering\epsfig{figure=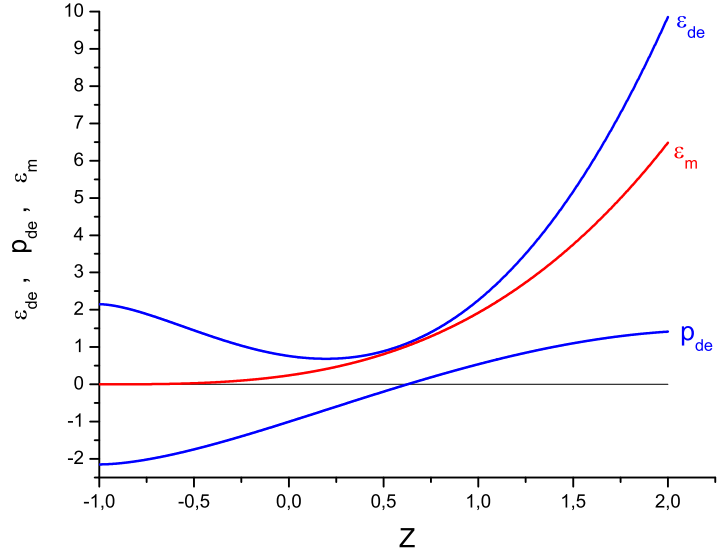,width=\linewidth}
        \caption{Density and pressure of Dark Energy and density of non--relativistic
        matter in the region of observations of SNIa and in the future Universe.}
        \label{Fig3}
        \end{minipage}
\end{figure}
\begin{figure}
    \begin{minipage}[b]{.60\linewidth}
        \centering\epsfig{figure=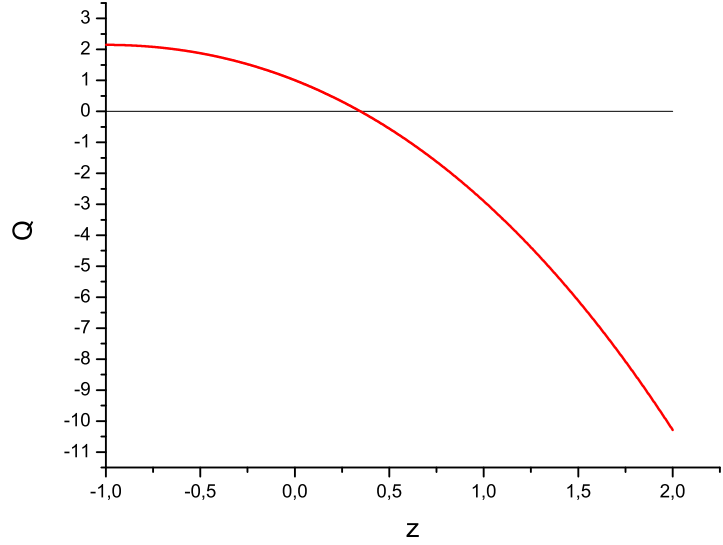,width=\linewidth}
        \caption{Acceleration of cosmological expansion in the region
        of observations of SNIa and in the future Universe.}
        \label{Fig4}
        \end{minipage}
\end{figure}

Comparing the results (\ref{9.9}) and (\ref{9.16}), belonging to
$\Lambda$CDM model with the results (\ref{9.12}), (\ref{9.13}) and
(\ref{9.18}), (\ref{9.19}), belonging to the $\Lambda$GCDM model, it
is easy to see that {\it in the framework of the given model} the
values of parameters obtained from the fits of the original and
smoothed data, are statistically equivalent. This fact is not
something amazing because the smoothing procedure and the fitting
are mathematically related. The use of smoothed data is motivated
by the necessity to compare and make a preliminary choice between
{\it two different models} using the criterion of statistical
likelihood.

When the two models are compared, two facts appear. First, the
application of the $\chi^2$--criterion to the smoothed data of the
$\Lambda$GCDM model shows that the model gains a more significant
advantage. Second, comparison of (\ref{9.17}) and (\ref{9.20})
shows that on the smoothed data the intervals of statistical sums
are not overlapping, with the lesser value of the statistical sum
belonging to the $\Lambda$GCDM model. Therefore, {\it by statistical
criteria obtained from the smoothed Hubble diagram for supernovae
SNIa, the $\Lambda$GCDM model has an advantage.}

In addition to the Hubble diagram for supernovae SNIa, the
information about Dark Energy is contained in the Hubble diagram
for radio--galaxies \cite{RG} and gamma--ray bursts \cite{GB}. It
is also contained in the cosmological parameters extracted from
the CMB data and correlation functions characterizing the large
scale structure of the Universe. Results of processing of the full
data set will be reported in a separate paper. Because of big
statistical errors of appropriate data, we believe that the use of
Hubble diagrams for radio--galaxies and gamma--ray bursts is to no
purpose in this work where we discuss statistical criteria to
choose between  $\Lambda$CDM and $\Lambda$GCDM models. As to the
third type of data (following from CMB and correlation functions),
we would like to mention the following. Information about the
density of Dark Energy on the cosmological scale from the instant
of the last scattering $Z_{ls}$ up to the present time is
contained in the shift parameter \cite{31}
\begin{equation}
 \displaystyle  \mathcal{R}=\sqrt{\Omega_{{\scriptscriptstyle M}}}
 \int\limits_0^{Z_{ls}}\frac{dz}{\tilde H(z)}=1.716\pm 0.062\ .
 \label{9.21}
\end{equation}
The (\ref{9.21}) parameter can be used under condition that the
Dark Energy model, used for the interpretation of SNIa data in the
$0<z<2$ interval, can be extrapolated up to $z\sim 1000$. In our
view, such an extrapolation (three orders of magnitude for
redshifts and nine orders of magnitude for curvature and energy
density) looks improbable. In any case, the interpolation
$\Lambda$GCDM model is not allowed to be extrapolated in such a
way as can be seen from the algorithm of its construction (see
Section \ref{LGCDM}). This is the reason why we do not discuss
constraints following from (\ref{9.21}).  The parameter
$\mathcal{A}$ extracted from the acoustic peak data is of a
different status. Acoustic oscillations in the photon--baryon
plasma prior to recombination give rise to a pick in the
correlation function of galaxies. This effect was recently been
measured in a sample of luminous red galaxies and leads to the
value \cite{CMB2}
\begin{equation}
 \displaystyle  \mathcal{A}=\frac{\sqrt{\Omega_{{\scriptscriptstyle
 M}}}}{Z_{\scriptscriptstyle
 L\scriptscriptstyle R\scriptscriptstyle G}}\left[\frac{Z_{\scriptscriptstyle
 L\scriptscriptstyle R\scriptscriptstyle G}}{\tilde H(Z_{\scriptscriptstyle
 L\scriptscriptstyle R\scriptscriptstyle G})}
 \left(\int\limits_0^{Z_{\scriptscriptstyle
 L\scriptscriptstyle R\scriptscriptstyle G}}\frac{dz}{\tilde
 H(z)}\right)^2\right]^{1/3}
 =0.469\left(\frac{n}{0.98}\right)^{-0.35}\pm 0.017\ ,
 \label{9.22}
\end{equation}
where $Z_{\scriptscriptstyle L\scriptscriptstyle
R\scriptscriptstyle G}=0.35$ is the redshift at which the acoustic
scale has been measured and $n=0.961$ is the spectral index of the
primordial power spectrum.

The $\Lambda$GCDM model with the (\ref{9.19}) parameters, which
were determined from the SNIa data, {\it predicts} the following
interval for the allowed values of the $\mathcal{A}$ parameter
$0.413<\mathcal{A}<0.479$ if the parameter
$\Omega_{\scriptscriptstyle M}$ is taken from the allowed interval
$0.20<\Omega_{\scriptscriptstyle M}<0.27$. The $\Lambda$GCDM
model, which is precisely consistent with the value
$\mathcal{A}=0.472$, is as follows
\begin{equation}
\begin{array}{c}
\displaystyle
 \chi^2/dof=0.93\ , \qquad \mathcal{M}=43.32\pm 0.02\ , \qquad
1+z_0=1.34\pm 0.03\ ,
\\[5mm]
  \displaystyle
 \Omega_{\scriptscriptstyle M}=0.26\pm 0.01:\qquad
 \Omega_\Lambda=2.15\pm 0.27\ , \qquad  \Omega_g=2.04\pm 0.40\ ,\qquad a_0=0.50\pm 0.02\
 .
\end{array}
 \label{9.23}
 \end{equation}

The procedure of noise reduction in the Hubble diagram and the
obtained results bear, of course, an illustrative character. In
fact, we have demonstrated only the near term possibilities of the
experiment: i.e. possibility to get the variable component of the
Dark Energy and compare the results of the measurements with the
prediction of the $\Lambda$GCDM model derived from the exact results
of one--loop quantum gravity.

The density and pressure of the Dark Energy and the energy density
of non--relativistic matter are plotted in Figure 3. The graphs
are calculated by the formulas of $\Lambda$GCDM model (\ref{9.2})
with parameters corresponding to  $\Omega_{\scriptscriptstyle
M}=0.240$ in (\ref{9.19}). The acceleration of the Universe,
calculated by the formula (\ref{9.1}) with the same parameters, is
shown in Figure 4. Two features on the graphs in Figure 3 are of
special interest: (i) in the entire area of observation of SNIa
the density of non--relativistic  matter and Dark Energy are
compatible in their values; (ii) the density of Dark Energy first
decreases with decreasing $z$, and then increases, starting
approximately from $z\sim 0.2$. Let us mention the astrophysical
aspect of these predictions of the  $\Lambda$GCDM model: (i)
at the epoch of creation of large scale structures in the
Universe,  Dark Energy has played a quantitatively important role
in the global cosmological dynamics; (ii) the area of relatively
small values of redshifts represents significant interest for the
observations, because in that area the reconstruction of the
graviton--ghost condensate is happening: pre--asymptotic state of
the condensate with constant conformal wavelength is
transforming into the asymptotic state of the condensate with the
constant physical wavelength. With the increasing accuracy of
experiments, it is possible that in this area, the nonlinear
fluctuations of Dark Energy could be discovered (see
Section\ref{pt}).

\section{Conclusion}\label{con}

From the formal mathematical point of view, the above theory is
identical to transformations of equations, determined by the
original gauged path integral (\ref{2.1}), leading to exact
solutions for the model of self--consistent theory of gravitons in
the isotropic Universe. To assess the validity of the theory, it
is useful to discuss again but briefly the three issues of the
theory that are missing in the original path integral.

(i) {\it The hypothesis of the existence of classic spacetime with
deterministic, but self--consistent geometry is introduced into
the theory.} It is not necessary to discuss in detail this
hypothesis because it simply reflects the obvious experimental
fact (region of Planck curvature and energy density is not a
subject of study in the theory under discussion). Note, however,
that the introduction of this hypothesis into the formalism of the
theory leads to a rigorous mathematical consequence: the strict
definition of the operation of separation of classical and quantum
variables uniquely captures the exponential parameterization of
the metric.

(ii) {\it The transfer to the one--loop approximation is conducted
in the self--consistent classical and quantum system of
equations.} Formally, this approximation is of a technical nature
because the equations of the theory are simplified only in order
to obtain specific approximate solutions. After classical and
quantum variables are identified, the procedure of transition to
the one--loop approximation is of a standard and known character
\cite{7}. In reality, of course, the situation in the theory is
much more complex and paradoxical. On the one hand, the quantum
theory of gravity is a non--renormalized theory (see, e.g.
\cite{36}). Specific quantitative studies of effects off one--loop
approximation are simply impossible. On the other hand, the
quantum theory of gravity without fields of matter is finite in
the one--loop approximation \cite{3}. The latter means that the
results obtained in the framework of one--loop quantum gravity
pose limits to its applicability that is mathematically clear and
physically significant. The existence of a range of validity for
the one--loop quantum gravity without fields of matter is a
consequence of two facts. First, there are supergravity theories
with fields of matter which are finite beyond the limits of
one--loop approximation. Second, the quantum graviton field is the
only physical field with a unique combination of such properties
as conformal non--invariance and zero rest mass. For this field
only there is no threshold for the vacuum polarization and
particle creation in the isotropic Universe. Therefore, in the
stages of evolution of the Universe, where $H^2,\ |\dot H|\ll m^2$
($m$ is mass of any of the elementary particles), quantum
gravitational effects can occur only in the subsystem of
gravitons. It is also clear that in any future theory that unifies
gravity with other physical interactions, equations of theory of
gravitons in one--loop approximation will not be different from
those we discuss in this work. Therefore the self--consistent
theory of gravitons has the right to lay claim be a reliable
description of the most significant quantum gravity phenomena in
the isotropic Universe.

(iii) {\it Dynamic properties of ghost fields are captured by the
condition of one--loop finiteness of the theory off mass shell of
gravitons and ghosts.} The class of legitimate gauges picked out
by this condition includes gauges that are form--invariant with
respect to transformation of the symmetry group of the background
geometry. This point is the most nontrivial part of the theory
because it is essentially an additional mathematical condition on
the theory ensuring its internal consistency. The condition of
one--loop finiteness off the mass shell largely determines the
mathematical and physical content of the theory. Given that the
main results of this work are exact solutions and exact
transformations, the evaluation of he proposed approach is reduced
to a discussion of this point of the theory. Let us enumerate once
more logical and mathematical reasons, forcing us to include the
condition of one--loop finiteness off the mass shell into the
structure of the theory.

a)  Future theory that will unify quantum gravity with the theory
of other physical interactions may not belong to renormalizability
theories. If such a theory exists, it may only be a finite theory.
One--loop finiteness of quantum gravity with no fields of matter
that is fixed on the mass shell \cite{3} can be seen as the
prototype of properties of the future theory.

b) Because of their conformal non--invariance and zero rest mass,
gravitons and ghosts fundamentally can not be located exactly on
the mass shell in the real Universe. Therefore, the problem of
one--loop finiteness off the mass shell is contained in the
internal structure of the theory.

c) In formal schemes, which do not meet the one--loop finiteness,
divergences arise in terms of macroscopic physical quantities. To
eliminate these divergences, one needs to modify the Lagrangian of
the gravity theory, entering quadratic invariants. This, in turn,
leads to abandonment of the original definition of the graviton
field that generates these divergences. The logical inconsistency
of such a formal scheme is obvious. (The mathematical proof of
this claim is contained at Appendix \ref{nonren}.)

d) In the self--consistent theory of gravitons, one--loop
finiteness off the mass shell can be achieved only through mutual
compensation of divergent graviton and ghost contributions in
macroscopic quantities. The existence of gauges, automatically
providing such a compensation, is an intrinsic property of the
theory.

From our perspective, the properties of the theory identified in
points a), b), c) and d), clearly dictate the need to use only the
formulation of self--consistent theory of gravitons, in which the
condition of one--loop finiteness off the mass shell (the
condition of internal consistency of the theory) is performed
automatically. We also want to emphasize that, as it seems to us,
the scheme of the theory given below has no alternative both
logically and mathematically.

{\it Gauged path integral $\Longrightarrow$  factorization of
classic and quantum variables, which ensures the existence of a
self--consistent system of equations $\Longrightarrow$  transition
to the one--loop approximation, taking into account the
fundamental impossibility of removing the contributions of ghost
fields to observables $\Longrightarrow$ choice of the ghost
sector, satisfying the condition of one--loop finiteness off the
mass shell} --- appears to us logically and mathematically as the
only choice.

As part of the theories preserving macroscopic spacetime being
clearly one of its components, we see two topics for further
discussions. The first of these is the replication of the results
of this work by mathematically equivalent formalisms of one--loop
quantum gravity. Here we can note that, for example, in the
formalism of the extended phase space with BRST symmetry, our
results are reproduced, even though the mathematical formalism is
more cumbersome. The second topic is the reproduction of our
results in more general theories than the one--loop quantum
gravity without fields of matter. Here is meant a step beyond the
limits of one--loop approximation as well as a description of
quantum processes involving gravitons, while taking into account
the existence of other quantum fields of spin $J\leqslant 3/2$. In
the framework of discussion on this topic, we can make only one
assertion: in the one--loop $N=1$  supergravity containing
graviton field and one gravitino field, the results of our work
are fully retained. This is achieved by two internal properties of
$N=1$ supergravity: (i) The sector of gravitons and graviton
ghosts in this theory is exactly the same as in the one--loop
quantum gravity without fields of matter; (ii) The physical
degrees of freedom of gravitino with chiral $h=\pm 3/2$ in the
isotropic Universe are dynamically separated from the
non--physical degrees of freedom and are conformally invariant;
(iii) The gauge of gravitino field can be chosen in such a way
that the gravitino ghosts automatically provide one--loop
finiteness of $N=1$ supergravity. As for multi--loop calculations
in the $N=1$ supergravity and more advanced theoretical models, we
have not explored the issue.

Of course, a rather serious problem of the physical nature of
ghosts remains. The present work makes use in practice only of
formal properties of quantum gravity of Faddeev--Popov--De Witt,
which point to the impossibility in principle of removing
contributions of ghosts to observable quantities off the mass
shell. A deeper analysis undoubtedly will address the foundations
of quantum theory. In particular, one should point out the fact
that the formalism of the path integral of Faddeev--Popov--De Witt
is mathematically equivalent to the assumption that observable
quantities can be expressed through derivatives of
operator--valued functions defined on the classical spacetime of a
given topology. On the other hand, finiteness of physical
quantities is ensured in the axiomatic quantum field theory by
invoking limited field operators smoothed over certain small areas
of spacetime. Extrapolation of this idea to quantum theory of
gravity immediately brings up the question on the role of
spacetime foam \cite{37} (fluctuations of topology on the
microscopic level) in the formation of smoothed operators, and
consequently, observable quantities. To make this problem more
concrete, a question can be posed on collective processes in a
system of topological fluctuations that form the foam. It is not
excluded that the non--removable Faddeev--Popov ghosts  in
ensuring the one--loop finiteness of quantum gravity are at the
same time a phenomenological description of processes of this
kind.

Study of equations of self--consistent theory of gravitons,
automatically satisfying the condition of one--loop finiteness,
leads to the discovery  of a new class of physical phenomena which
are {\it macroscopic effects of quantum gravity.} Like the other
two macroscopic quantum phenomena of superconductivity and
superfluidity, macroscopic effects of quantum gravity occur  on
the macroscopic scale of the system as a whole, in this case, on
the horizon scale of the Universe. Interpretation of these effects
is made in terms of {\it gravitons--ghost condensates arising from
the interference of quantum coherent states.} Each of coherent
states is a state of gravitons (or ghosts) with a certain
wavelength of the order of the distance to the horizon and a
certain occupation number. The vector of the physical state is a
coherent superposition of vectors with different occupation
numbers.

A key part in the formalism of self--consistent theory of
gravitons is played by the BBGKY chain for the spectral function
of gravitons, renormalized by ghosts. It is important that
equations of the chain may be introduced at an axiomatic level
without specifying explicitly field operators and state vectors.
It is only necessary to assume the preservation of the structure
of the chain equations in the process of elimination of
divergences of the moments of the spectral function. Three exact
solutions of one--loop quantum gravity are found in the framework
of BBGKY formalism. The invariance of the theory with respect to
the Wick rotation is also shown. This means that the solutions of
the chain equations, in principle, cover two types of condensates:
condensates of virtual gravitons and ghosts and condensates of
instanton fluctuations.

All exact solutions, originally found in the BBGKY formalism, are
reproduced at the level of exact solutions for field operators and
state vectors. It was found that exact solutions correspond to
various condensates with different graviton--ghost microstructure.
Each exact solution we found is compared to a phase state of
graviton--ghost medium; quantum--gravity phase transitions are
introduced.

We suspect that the manifold of exact solutions of one--loop
quantum gravity is not exhausted by three solutions described in
this paper. Search for new exact solutions and development of
algorithms for that search, respectively, is a promising research
topic within the proposed theory. Of great  interest will also be
approximate solutions, particularly those that describe
non--equilibrium and unstable graviton--ghost and instanton
configurations.

Self--consistent theory of gravitons allows an easy generalization
that takes into account participation of non--relativistic matter
in the formation of common self--consistent gravitational field.
From the equations of this theory it follows that the era of
dominance of non--relativistic matter should be replaced by an era
of dominance of graviton--ghost condensate. This result is of
direct relevance to the physics of Dark Energy. The
pre--asymptotic condition of Dark Energy is interpreted as a
condensate of virtual gravitons and ghosts of a constant conformal
wavelength. As an asymptotic condition, the theory predicts
self--polarized graviton--ghost condensate of constant physical
wavelength in the De Sitter space.

The view of the nature of Dark Energy is formulated in the form
$\Lambda$GCDM model which interpolates exact solutions of the
one--loop quantum gravity. The proposed theory is consistent with
existing observational data on Dark Energy extracted from the
Hubble diagram for supernovae SNIa. According to the criteria for
statistical reliability, the $\Lambda$GCDM model has certain
advantages over the  $\Lambda$CDM model. A graviton--ghost
condensate lays claim to being a variable component of Dark
Energy. Result of observational data processing suggests that
during the era of large--scale structure formation in the
universe, a graviton--ghost condensate played a measurably
significant part in shaping global cosmological dynamics.

Further applications of our theory to the physics of Dark Energy
will be to conduct numerical experiments with the BBGKY chain,
taking into account the non--relativistic matter. The results of
these experiments may be able to explain the finer details of the
Hubble diagram for supernovae SNIa, which will be identified with
increasing completeness and accuracy of observation data. Future
observations, in our view, must focus on the variable component of
Dark Energy. For reliable identification of this component it is
necessary, first move as far as possible into the area of large
redshift, and second to explore in detail the area of small red
shift, in which, quite possibly, there is a transition from the
pre--asymptotic state of graviton--ghost condensate to its
asymptotic state.
\[
\]
{\bf ACKNOWLEDGMENT.}  We would like to express our deep
appreciation to our friend and colleague Dr. Walter Sadowski for
invaluable advise and help in the preparation of the manuscript.

\newpage

\section{APPENDIX I. Cosmological Constant as Energy Density of Equilibrium Vacuum}\label{Lambda}

The cosmological constant in Einstein equations does not
contradict the general principles of geometrized gravity theory.
Moreover, the gravity theory which includes the $\Lambda$--term
looks more natural than a theory without $\Lambda$--term.
According to Zeldovitch \cite{20}, $\Lambda$--term is interpreted
as the energy density of the vacuum. Today, that definition should
be "updated" a little bit: the subject is the energy density of
the {\it equilibrium} vacuum subsystems of non--gravitational
origin in 4--D space--time: $\varepsilon_{vac}^{(0)}=\Lambda$. The
calculation of $\varepsilon_{vac}^{(0)}$ is the task of the future
"Theory of Everything"\ , based, possibly, on the superstring
theory. The existing experimentally verified theory (the Standard
Model of quark and leptons interactions) allows making some
general conclusions. The most important of these is the fact that
{\it for every fundamental physical interaction there is an
associated vacuum subsystem with non--zero energy density}. They
are well--known Higgs condensate in the theory of electro--week
interaction and quark--gluon condensate in Quantum Chromodynamics.

The Higgs condensate is forming as a result of spontaneous
breaking of electro--week symmetry $U(1)\times SU_L(2)$ down to
the electromagnetic symmetry $U_{em}(1)$. The energy density of
Higgs condensate can be estimated as (in that section $\hbar=1$):
\begin{equation}
\begin{array}{c}
 \displaystyle \varepsilon_{e{\scriptscriptstyle W}}=
 -\frac{M^2_{\scriptscriptstyle  H}M^2_{\scriptscriptstyle W}}{2g^2}
 -\frac{1}{128\pi^2}\left(M^4_{\scriptscriptstyle H}+3M^4_{\scriptscriptstyle Z}+
 6M^4_{\scriptscriptstyle W}-12M^4_t\right)\  ,
\end{array}
 \label{11.1}
\end{equation}
where $M_{\scriptscriptstyle H},\ M_{\scriptscriptstyle Z},\
M_{\scriptscriptstyle W},\ M_t$ are masses of Higgs boson,
intermediate vector bosons and $t$--quark; $g^2 =4\pi/29$ is the
gauge constant of $SU_L(2)$ group on $W$-boson mass--shell. The
first term in (\ref{11.1}) is the energy density of spatially
homogeneous (vacuum) component of the scalar Higgs field
$\langle0|H|0\rangle=v/\sqrt{2}$; the second term is the change of
the energy of zero--fluctuations of quantum fields, which have
obtained nonzero rest mass from the interaction with the vacuum
field $v=246\ \text{GeV}$. Masses of all particles $M\sim v$,
therefore
\begin{equation}
\begin{array}{c}
 \displaystyle \varepsilon_{e{\scriptscriptstyle W}}(v)=-
 \frac14\lambda v^4\simeq -(120\ \text{GeV})^4\ ,
\end{array}
 \label{11.2}
\end{equation}
where $\lambda$ is a constant or (if high radiation corrections
are taken into account) is a very slow function of vacuum field.
The numerical value used in (\ref{11.2}) corresponds to the mass
of Higgs boson $M_{\scriptscriptstyle H}=2M_{\scriptscriptstyle
W}\simeq 160\ \text{GeV}$. There is no doubt in the existence of
the electro--week vacuum subsystem at the scale
$\lambda_{e{\scriptscriptstyle W}}\sim 120\ \text{GeV}$, however a
fact that this subsystem is specifically formed by the Higgs
mechanism, is not yet experimentally confirmed. A decisive step in
that direction would be experimental discovery of the Higgs boson
by the LHC.

The quark--gluon condensate is a system of mutually correlated
non--perturbative fluctuations, arising during
quantum--topological tunnel transitions between degenerated states
of gluon vacuum. Energy density of the condensate is
\begin{equation}
\begin{array}{c}
 \displaystyle \varepsilon_{{\scriptscriptstyle Q}{\scriptscriptstyle C}{\scriptscriptstyle D}} =
 -\frac{b}{32} \langle 0|\frac{\alpha_s}{\pi}G_{ik}^aG^{ik}_a|0\rangle\ ,
 \\[5mm]
\displaystyle b\simeq 9+8T_g(m_u+m_d+0.8m_s)\simeq 9.6\ ,
\end{array}
 \label{11.3}
\end{equation}
where
\begin{equation}
\displaystyle \langle 0|\frac{\alpha_s}{\pi}G_{ik}^aG^{ik}_a|0\rangle=
u^4\simeq (360\ \text{MeV})^4
\label{11.4}
\end{equation}
is the main energy--momentum parameter of the quark--gluon
condensate; $T_g\simeq (1.5\ \text{GeV})^{-1}$ is a characteristic
spacetime scale of fluctuations; $m_u,\ m_d,\ m_s$ are the masses
of light quarks. According to the modern paradigm, the
quark--gluon condensate has several phase states, in each of which
the fluctuations have there own specific microstructure. The value
of the parameter, used in (\ref{11.4}), refers to the confinement
phase of the out--of--hadron vacuum. In that case
\begin{equation}
\displaystyle \varepsilon_{{\scriptscriptstyle
Q}{\scriptscriptstyle C}{\scriptscriptstyle D}}(u) =
 -\frac{b}{32}u^4\simeq -(265\ \text{MeV})^4
\label{11.5}
\end{equation}

As follows from (\ref{11.2}) and (\ref{11.5}), the scale of the
electro--week vacuum is of the order of $M_{e{\scriptscriptstyle
W}}\sim 120\ \text{GeV}$, and the QCD scale is
$M_{{\scriptscriptstyle Q}{\scriptscriptstyle
C}{\scriptscriptstyle D}}\simeq 265\ \text{MeV}$. The scale
$M_{{\scriptscriptstyle D}{\scriptscriptstyle E}} \sim 10^{-12}\
\text{GeV}$, corresponding to the current density of the Dark
Energy $\varepsilon_{{\scriptscriptstyle D}{\scriptscriptstyle
E}}\sim M^4_{{\scriptscriptstyle D}{\scriptscriptstyle E}}$, is
not on a pair  with $M_{e{\scriptscriptstyle W}},\
M_{{\scriptscriptstyle Q}{\scriptscriptstyle C}{\scriptscriptstyle
D}}$. It means that for the high--energy expansions of the
Standard Model, the vacuum is of a more complicated structure.
First, there should be vacuum subsystems with positive energy
density; Second, the theory should contain mechanisms of mutual
compensation of the contributions from different vacuum subsystems
to the full energy density of the equilibrium vacuum.

The Standard Model, when applied to the description of the
cosmological vacuum and cosmological plasma of elementary
particles, forecasts that the vacuum subsystems (\ref{11.1}),
(\ref{11.3}) are of evolutionary origin. The energy density of
these subsystems is not constant by the definition. Higgs' and
quark--gluon condensates appear in the early Universe in the
processes of relativistic phase transitions. The energy density of
the vacuum is a functional of non--equilibrium parameters
$\mathcal{V}\ne v$, $\mathcal{U}\ne u$:
\begin{equation}
\displaystyle \varepsilon_{vac}(\mathcal{V},  \mathcal{U})=
\Lambda_0+\varepsilon_{e{\scriptscriptstyle W}}(\mathcal{V})+
\varepsilon_{{\scriptscriptstyle Q}{\scriptscriptstyle C}{\scriptscriptstyle D}}(\mathcal{U})\ ,
\label{11.6}
\end{equation}
where $\Lambda_0=const$ is the energy density of the vacuum at the
evolutionary stage before the electro--week transition. In the
areas of phase transitions, functions
$\varepsilon_{e{\scriptscriptstyle W}}(\mathcal{V})$ and
$\varepsilon_{{\scriptscriptstyle Q}{\scriptscriptstyle
C}{\scriptscriptstyle D}}(\mathcal{U})$ are changing from zero
values to equilibrium values given in (\ref{11.2}) and
(\ref{11.5}). The characteristic transition times and the times of
relaxations of vacuum subsystems to the equilibrium states are
incommensurably shorter than the characteristic times of the
Universe evolution. By that reason, almost immediately after the
phase transition, one might talk about the contributions of the
respective condensates of the Standard Model to the cosmological
constant. After the quark--hadron transition, we have the current
value of the non--gravitational contributions to $\Lambda$--term:
\begin{equation}
\displaystyle \Lambda=
\Lambda_0+\varepsilon_{e{\scriptscriptstyle W}}(v)+
\varepsilon_{{\scriptscriptstyle Q}{\scriptscriptstyle C}{\scriptscriptstyle D}}(u)\ .
\label{11.7}
\end{equation}

Thus, the theory of fundamental interactions, particularly, the
Standard Model, leads to the conclusion that {\it the treatment of
the cosmological constant as an energy density of the equilibrium
vacuum, is related to the late stages of the Universe evolution,
situated on the cosmological scale after all the relativistic
phase transitions in the vacuum and in the plasma of elementary
particles are completed.}

Inevitability of the "fine tuning"\ of different vacuum
subsystems, with the result that the asymptotical value of the
cosmological constant (\ref{11.7}) turns out to be very small or
even equal to zero, is actually the experimental fact, resulting
from the very existence of the modern Universe. Actually, without
the "fine tuning"\ any of the phase transitions from the Standard
Model will lead either to the collapse of the Universe, or to the
exponential inflation, preventing the formation of large--scale
structures. At the same time, strong inequality
$|\varepsilon_{e{\scriptscriptstyle W}}(v)|\gg
|\varepsilon_{{\scriptscriptstyle Q}{\scriptscriptstyle
C}{\scriptscriptstyle D}}(u)|$ forces us to consider separately
the problem of the "fine tuning"\ at the electro--week and QCD
scales.

The compensation of the energy of the Higgs condensate
(\ref{11.1}) at the fundamental level can, in principle, be
provided by the supersymmetry of particles and interactions.
Unfortunately, in the superstring scenarios of the "Theory of
everything"\ , the effective algorithms to develop that kind of
theory are absent at the present time. The supersymmetry, if it
actually exists in Nature, is strongly broken, but the mechanisms
of destruction, it seems, are not spontaneous and 4--D spacetime
is not an arena of an action of these mechanisms. It is not
excluded, that the reduction of the "Theory of Everything"\ to the
low--energy Standard Model is provided by physical phenomena
working in the extra spatial dimensions. Experimental test of the
supersymmetry, as a realistic concept of the elementary particle
physics, will be conducted at LHC. The discussion about the
compensation mechanisms for the energy density of Higgs condensate
(\ref{11.1}) should continue after the results of the experiments
start arriving.

Supersymmetry has no relation to the problem of compensation of
the quark--gluon condensate (\ref{11.3}). We think the problem
should be solved in the framework of low--energy physics of strong
interactions. One of the possible approaches is to assume that at
the QCD scale there are additional contributions to (\ref{11.3}),
and the sum turns to zero due to the tuning of QCD parameters. The
discussion of these scenarios is out of the scope of the current
work.

The hypothesis about the mutual compensations of
non--gravitational contributions to the $\Lambda$--term do not
remove the questions regarding the final value of $\Lambda\ne 0$,
which is in agreement with observed data. At the order--of--value
estimations, $\Lambda$--term, with its value equal to the modern
Dark Energy density, satisfies Zeldovitch's relationship \cite{22}
with the Kardashev's modification \cite{38}
 \begin{equation}
 \displaystyle \varepsilon_{{\scriptscriptstyle D}{\scriptscriptstyle
 E}}\sim \Lambda=
 \frac{m_\pi^6}{(2\pi)^4M_{Pl}^2}=3\times
10^{-47}\ \text{GeV}^4 \
 ,
 \label{11.8}
\end{equation}
where $m_\pi=138\ \text{MeV}$ --- pion mass; $M_{Pl}=1.22\cdot
10^{19}\ \text{GeV}$ --- Planck mass. As it is known, the pion
mass is an object of non--perturbative QCD. Therefore, in formula
(\ref{11.8}) the observed density of Dark Energy is expressed only
via the combination of minimal $\Lambda_{{\scriptscriptstyle
Q}{\scriptscriptstyle C}{\scriptscriptstyle D}}\simeq 2m_\pi$ and
maximal scales $M_{Pl}$ of the elementary particle physics.
Adopting also the Sakharov's idea of Einstein equations
modification by the effect of gravitational exchange interaction
of identity particles \cite{39}, we may suggest that the energy
density of vacuum (\ref{11.8}) is formed by the gravitational
exchange interaction of quantum--topological fluctuations in the
hadron vacuum. This idea leads to a formula
\begin{equation}
 \displaystyle
 \varepsilon^{(grav)}_{{\scriptscriptstyle Q}{\scriptscriptstyle C}{\scriptscriptstyle D}}
 \sim \Lambda=
 \frac{\pi T_g^2}{2M_{Pl}^2}
 |\varepsilon_{{\scriptscriptstyle Q}{\scriptscriptstyle C}{\scriptscriptstyle D}}|^2
 \ln^2
 \frac{T_g^{-1}}{\Lambda_{{\scriptscriptstyle Q}{\scriptscriptstyle C}{\scriptscriptstyle D}}}
 =4\times
10^{-43}\ \text{GeV}^4 \ .
 \label{11.9}
\end{equation}
The difference of four orders between values (\ref{11.8}) and
(\ref{11.9}) is no any longer a catastrophe for the theory (see
formula (\ref{11.10}) below).

The nature of the cosmological constant discussed above, (more
precisely, the contributions to the energy density of the vacuum
controlled by the modern theory of elementary particles), suggests
a set of interesting analogies. First, because the vacuum
subsystems of electromagnetic, week and strong interactions are
known from experiments, it is appropriate to ask a question: {\it
what is the vacuum subsystem connected with the fundamental
gravitational interaction?}  We hope that an answer is provided in
our work: the existence of the graviton--ghost condensate, i.e.
the Dark Energy, is a direct consequence of applying the first
principles of quantum theory of gravitation. The second analogy is
in the fact that all vacuum subsystems are the subjects of
evolution as Universe progresses, and the asymptotical states
appear as results of relativistic phase transitions. From the
evolutional criterion, the difference between graviton--ghost
vacuum subsystem and non--gravitational vacuum subsystems is of a
quantitative character only. Non--gravitational vacuum subsystems
relax on the respective characteristic scales of elementary
particle physics, while due to the weakness of gravitational
interaction, the graviton--ghost condensate relaxation takes place
at the time intervals which are of the order the age of the
Universe. Finally, the third general property of all vacuum
subsystems is that the vacuum energy density in the asymptotical
(equilibrium) state acquires a constant value. It is worth
mentioning that the graviton--ghost condensate in its equilibrium
state, as well the vacuum of QCD, is of the instanton
microstructure.

Of course, the problem of mutual compensation of different
contributions to the energy density of non--gravitational vacuum
subsystems remains unsolved. However, we believe that the
consideration of the graviton--ghost component of the physical
vacuum somewhat reduces the magnitude of the problem. Actually, as
shown at the end of the section \ref{S}, the mathematical
structure of the self--consistent theory of gravitons itself
possesses an effective renormalization procedure applied to the
energy density of non--gravitational vacuum by virtue of the
graviton--ghost contributions. The possibility of such
renormalization is here regardless of the sign of the
$\Lambda$--term. Particularly, the renormalization of
$\Lambda$--term (\ref{11.9}), formed by the gravitational exchange
interactions of non-perturbative quark--gluon fluctuations, is
conducted by formula (\ref{6.34}) and leads to the result:
\begin{equation}
 \displaystyle \varepsilon^{(\infty)}_{vac}=
 \sqrt{\frac{3\pi}{4|N_g|}} T_gM_{Pl}|
 \varepsilon_{{\scriptscriptstyle Q}{\scriptscriptstyle C}{\scriptscriptstyle D}}|
 \ln\frac{T_g^{-1}}{\Lambda_{{\scriptscriptstyle
Q}{\scriptscriptstyle C}{\scriptscriptstyle
D}}}=\frac{10^{17}}{\sqrt{N_g}}\ \text{GeV}^4 \ .
 \label{11.10}
\end{equation}
As follows from (\ref{11.10}), $\varepsilon^{(\infty)}_{vac}\sim
10^{-46}\ \text{GeV}^4$ with $|N_g|\sim 10^{126}$. Unfortunately,
the question regarding the concordance of the graviton--ghost
condensate's parameter $|N_g|$ with the numerical values of
parameters of a different physical nature (example (\ref{11.10}),
with QCD parameters), remains open. But in that case one always
may appeal to the anthropic principle.

\section{APPENDIX II. Renormalizations and Anomalies}\label{an}

The problem of calculating the anomalies in the energy-momentum
tensor of gravitons (quantum field with spin $J=2$) was discussed
in \cite{37, 46, 47}. A large number of works are devoted to the
study of graviton quantum field in De Sitter space. Here we
provide only some links. In some works \cite{48, 49} the
non--ghost models were considered; in other works \cite{50, 51}
the ghost fields with harmonic gauge were taken into account.
Common feature of all known versions of quantum theory of
gravitons in the isotropic Universe is the lack of one--loop
finiteness off the mass shell.

In Sections \ref{ren}, \ref{nonren} we discuss a self--consistent
theory of gravitons in isotropic Universe with the ghost sector
not taken into account. As has been repeatedly stated, we believe
that such a model is not mathematically sound. Gauges, completely
removing the degeneracy, are absent in the theory of gravity.
Thus, in the self--consistent theory of gravitons the ghost sector
is inevitable present. Now, however, let us assume for the moment
that the self--consistent theory of gravitons without ghosts is
worth at least as a model of mathematical physics. The purpose of
this Section is to get the properties of this model and to show
that it is mathematically and physically internally inconsistent.
Note also that the self--consistent theory of gravitons in the
harmonic gauge (\ref{3.79}) has qualitatively the same status as
the non--ghost model: the one--loop non--renormalizability is
hindering the attempts to obtain definitive results within this
theory.

\subsection{Gravitons with no Ghosts. Vacuum Einstein Equations
with Quantum Logarithmic Corrections}\label{ren}

It clear from the outset that in the non--ghost model the
calculation of observables will be accompanied by the emergence of
divergences. It is therefore necessary to formulate the theory in
such a way that the regularization and renormalization operations
are to be contained in its mathematical structure from the very
beginning. We talk here about changes in the mathematical
formulation of the theory. The relevant operations should be
introduced into the theory with care: first, in the amended
theory, coexistence of classical and quantum equations should be
ensured automatically; second, the enhanced theory should not
contain objects initially missing from the theory of gravity.

The dimensional regularization satisfies both above--mentioned
conditions. Important, however, is the following fact: the use of
dimensional regularization suggests that the self--consistent
theory of gravitons in the isotropic Universe is originally
formulated in a spacetime of dimension $D=1+d$, where $1$ is the
dimension of time; $d=3-2\varepsilon$ is the dimension of space.
The special status of the time is due to the two factors: (i) all
the events in the Universe, regardless of its actual dimension,
are ordered along the one--dimensional temporal axis; (ii) the
canonical quantization of the graviton field in terms of the
commutation relations for generalized coordinates and generalized
momenta also presuppose the existence of the one--dimensional
time. As for the space dimension, the limit transition to the true
dimension $d=3$ is implemented after the regularization and
renormalization.

Thus, we are working in a space with a metric
 \begin{equation}
\begin{array}{c}
 \displaystyle
 ds^2=a^2(\eta)(d\eta^2-\gamma_{\alpha\beta}dx^\alpha dx^\beta)\ ,
 \qquad \gamma^{\alpha\beta}\gamma_{\alpha\beta}=d\ ,
 \\[5mm]
\displaystyle \sqrt{|g_{(d)}|}=a^{d+1}\ ,\qquad R_{(d)}=
-\frac{d}{a^2}\left(
2\frac{a''}{a}+(d-3)\frac{a^{'2}}{a^2}\right)\ .
\end{array}
 \label{12.1}
\end{equation}
To avoid mathematical contradictions that could arise at the limit
$d\to 3$, Einstein equations in $D$-dimensional spacetime should
be written down in exactly the form in which they were obtained
from the variational principle:
 \begin{equation}
 \begin{array}{c}
\displaystyle
 \frac{1}{\varkappa_d}\sqrt{|g_{(d)}|}\left(R_{0(d)}^0-\frac12R_{(d)}\right)\equiv
 \\[5mm]
\displaystyle \frac1{2\varkappa_d}d(d-1)a^{d-3}{a'}^2=
 \frac{1}{8\varkappa_d}a^{d-1}\sum_{{{\bf k}\sigma}}
 \langle\Psi_g|\hat \psi_{{\bf k}\sigma}^{+'}\hat \psi'_{{\bf k}\sigma}+
 k^2\hat \psi_{{\bf k}\sigma}^+\hat \psi_{{\bf k}\sigma}|\Psi_g\rangle\ ,
 \\[5mm]
 \displaystyle
 -\frac{d-1}{2\varkappa_d}\sqrt{|g_{(d)}|}R_{(d)}\equiv
 \\[5mm]
\displaystyle
\frac1{2\varkappa_d}d(d-1)\left[2a^{d-2}{a''}+(d-3)a^{d-3}{a'}^2\right]=
 -\frac{d-1}{8\varkappa_d}a^{d-1}\sum_{{{\bf k}\sigma}}\langle\Psi_g|
 \hat \psi_{{\bf k}\sigma}^{+'}\hat \psi'_{{\bf k}\sigma}-
 k^2\hat \psi_{{\bf k}\sigma}^+\hat \psi_{{\bf k}\sigma}|\Psi_g\rangle\ ,
 \end{array}
 \label{12.2}
\end{equation}
 \begin{equation}
  \displaystyle \hat\psi''_{{\bf k}\sigma}+(d-1)\frac{a'}{a}\hat\psi'_{{\bf k}\sigma}+
  k^2\hat\psi_{{\bf k}\sigma}=0\ .
   \label{12.3}
\end{equation}
Here $\varkappa_d$ is the Einstein gravitational constant in
$D$--dimensional spacetime. (Dimension
$[\varkappa_d\hbar]=[l]^{D-2}$.) The left hand sides of equations
(\ref{12.2}) satisfy the Bianchi identity:
\begin{equation}
 \begin{array}{c}
\displaystyle
 \frac1{2\varkappa_d}d(d-1)\left[a^{d-3}{a'}^2\right]'-
\frac1{2\varkappa_d}d(d-1)\frac{a'}{a}\left[2a^{d-2}{a''}+(d-3)a^{d-3}{a'}^2\right]\equiv 0\ .
 \end{array}
 \label{12.4}
\end{equation}
In the right hand side of equations (\ref{12.2}), the identity
(\ref{12.4}) generates condition of the graviton EMT  conservation
that satisfies if the equations of motion (\ref{12.3}) are taken
into account. Regarding the origin of the system of equations
(\ref{12.2}) and (\ref{12.3}), we should  make the following
comment. In this case it is inappropriate to invoke the reference
to the path integral and factorization of its measures because the
path integral inevitably leads to the theory of ghosts interacting
with the macroscopic gravity. We can only mention a heuristic
recipe: one should refer to the density of Einstein equations with
mixed indices, define the exponential parameterization of the
metric, and expand the equations into a series of metric
fluctuations with an accuracy of the second--order terms.
Deviations from this recipe (for example, linear parameterization
$\hat g_{ik}=g_{ik}+\hat h_{ik}$) lead to a system of inconsistent
classical and quantum equations. To remove this sort of
inconsistency, one is forced to use artificial transactions
outside the formalism of the theory (see, for example,
\cite{42}).

While working with the system of equations (\ref{12.2}),
(\ref{12.3}), we face with two mathematical problems. The first
problem is that in the framework of that system of equations,
except in very special cases, it is impossible to formulate the
dynamics of operators on a given background that is to get the
solution of the equation (\ref{12.3}) as an accurate operator
function of time. This is due to the fact that formulae of
(\ref{12.2}) in reality are not yet specific equations. They are
only a layout of Einstein equations with radiation corrections.
These equations can only be obtained after regularization and
renormalizations of the ultraviolet divergences. In addition, the
functional form of equations depends on which quantum
gravitational effects are to be taken into account outside the
sector of vacuum (i.e. zero) fluctuations of the graviton field.
The only possible way to study the system of equations
(\ref{12.2}) and (\ref{12.3}) is (i) to obtain the solution of
operator equation (\ref{12.3}) in a form of a functional of the
scale factor without specifying the dependence on $a(\eta)$ {\it
with a clear emphasis on zero fluctuations in this functional},
(ii) to substitute the obtained functional in (\ref{12.2}) under
certain assumptions about the state vector; (iii) to regularize
and renormalize and finally (iv) to solve the macroscopic Einstein
equations, obtained after these operations. Implementation of the
program, an essential element of which is the allocation of zero
fluctuations generating ultraviolet divergences, is possible only
when using the method of asymptotic expansions of solutions of
operator equation in the square of wavelength of the graviton
modes. Thus, the problem of the lack of macroscopic Einstein
equations in the original formulation of this theory with
divergences limits the methods of this theory to the short--wave
approach. Note that this fact was clearly indicated by DeWitt
\cite{7}.

The second problem is related to the infrared instability of the
theory, with the object of the theory being a conformal
non--invariant massless quantum field. The problem is due to the
fact that not every representation of the asymptotic series can be
substituted into energy--momentum tensor to perform the summation
over the wave numbers. For example, if in the explicit form, a
term in the asymptotic series contains a large parameter $k^{2n}$
in the denominator, then starting from $n=2$ in the integration
over the wave numbers the infrared divergences will appear. Such
an asymptotic series can not be used even for the renormalization
of ultraviolet divergences, because when it is used in the space
of the physical dimension $d=3$, the logarithmic divergences arise
simultaneously at the ultraviolet and the infrared limits. In the
method of dimensional regularization the problem is reduced to the
fact that it is impossible to choose an interim dimension $d$ in a
way such that the integral exists at both limits.

Formally, the technical problem described above is partly solved
by reformatting the asymptotic series. In particular, the
following method will be used, in which parameter of the
asymptotic expansion is the effective frequency
 \begin{equation}
 \displaystyle \omega_k^2=k^2+\rho\ ,
 \qquad \rho=\frac{d-1}{4d}a^2R_{(d)}=-\frac{d-1}{4}\left[
2\frac{a''}{a}+(d-3)\frac{a^{'2}}{a^2}\right]\ .
 \label{12.5}
\end{equation}
In this method, the integrals over the wave numbers can be defined
in terms of the principal value. Contributions of the poles at
$k=\sqrt{-\rho}$ can not be mathematically verified if only
because there are such contributions from each term of the
infinite asymptotic series. The inability to describe infrared
effects is the principal disadvantage of a theory with
divergences, which uses only asymptotic expansions with respect to
the wavelength. Meanwhile, as general considerations and the
results of this work show, in the physics of conformal
non--invariant massless field the most interesting and innovative
effects occur in the infrared spectrum. The method of describing
these effects, based on the exact BBGKY chain, can not be used in
the theory with divergences, because a method regularizing the
infinite chain of moments of the spectral function does not exist.

The above problems automatically reduces the interest toward the
theory with divergences. However, given that all previous works in
this area have been implemented in the framework of regularization
and renormalization, let us conduct our analysis to the end. In
calculations, it is enough to consider the equation for the
convolution. After identity transformations, using the equation of
motion (\ref{12.3}), we get
\begin{equation}
 \begin{array}{c}
\displaystyle
\frac1{2\varkappa_d}d(d-1)\left[2a^{d-2}{a''}+(d-3)a^{d-3}{a'}^2\right]=
 -\frac{d-1}{16\varkappa_d}\sum_{{{\bf k}\sigma}}\left(W'_{{\bf k}\sigma}a^{d-1}\right)'\ ,
 \end{array}
 \label{12.6}
\end{equation}
where
\[
\displaystyle W_{{\bf k}\sigma}=
\langle\Psi_g|\hat \psi_{{\bf k}\sigma}^+\hat \psi_{{\bf k}\sigma}|\Psi_g\rangle
\]
is the spectral function of gravitons. The calculation of the
spectral function by the method of asymptotic expansion with
respect to the square of wavelength was described in Section
\ref{swg}. Now we need to repeat this calculation excluding the
ghosts, but with input from zero fluctuations in the spacetime of
dimension $D=d+1$. The relevant calculations do not require
additional comments. A spectral function is represented as:
\begin{equation}
\displaystyle W_{{\bf k}\sigma}=W^{(vac)}_{{\bf k}\sigma}+W^{(exc)}_{{\bf k}\sigma}\ ,
\label{12.7}
\end{equation}
where $W^{(vac)}_{{\bf k}\sigma}$ is the vacuum component of the
spectral function and $W^{(exc)}_{{\bf k}\sigma}$ is the spectral
function of excitations. After passage to the limit $d\to 3$, the
contribution of $W^{(exc)}_{{\bf k}\sigma}$ to the EMT of short
gravitons is exactly the same as (\ref{4.9}), (\ref{4.10}). In the
future, we discuss only the contribution from vacuum components of
the spectral function. In the calculations, we must keep in mind
that in the $d$--dimensional space the number of internal degrees
of freedom of transverse gravitons is $w_g=(d+1)(d-2)/2$. The
solution for the vacuum spectral function is expressed in terms of
the functional (\ref{4.4}):
\begin{equation}
\displaystyle \sum_\sigma W^{(vac)}_{{\bf k}\sigma}=
\frac{4\varkappa_d\hbar}{a^{d-1}}\cdot\frac{(d+1)(d-2)}{4\epsilon_k}=
\frac{4\varkappa_d\hbar}{a^{d-1}}\cdot
\frac{(d+1)(d-2)}{4\omega_k}\sum_{s=0}^{\infty}(-1)^s\hat J_k^s\cdot 1\ ,
\label{12.8}
\end{equation}
The powers of operator $\hat J_k^s\cdot 1$ are defined by formulas
(\ref{4.5}), in which $\omega_k^2$ has the form (\ref{12.5}).
After substitution of (\ref{12.8}) into (\ref{12.6}), the
zero--term in the asymptotic expansion creates an integral,
calculated by the rules of dimensional regularization:
\begin{equation}
  \displaystyle \sum_{{\bf
  k}}\frac{1}{\omega_k}=
  \frac{1}{(2\pi)^d}\frac{2\pi^{d/2}}{\Gamma(d/2)}
  \int\limits_0^{\infty}\frac{k^{d-1}dk}{(k^2+\rho)^{1/2}}=
  \frac{\displaystyle \Gamma\left[(3-d)/2\right]}
  {\displaystyle 2^{d-1}\pi^{(d+1)/2}(1-d)}\rho^{(d-1)/2}\ .
  \label{12.9}
\end{equation}
The $\Gamma-$function in (\ref{12.9}) diverges for $d\to 3$.
Therefore, calculation of the integral (\ref{12.9}) and
transformation of expressions with $\Gamma-$functions are carried
out with those values of $d$ which provide the existence of the
integral and $\Gamma-$functions. At the final stage, the result of
these calculations is analytically continued to the vicinity
$d=3$. All other terms of the asymptotic expansion (\ref{12.8})
generate finite integrals and do not require a dimensional
regularization. For reasons of heuristic rather than mathematical
nature, it is considered that these terms are negligible compared
to the contribution of the principal term of the asymptotic
expansion (see below the effective Lagrangian (\ref{12.19})).
Convolution of $D$--dimensional Einstein's equations (\ref{12.6}),
containing the main term of the vacuum EMT of gravitons, has the
form:
\begin{equation}
 \begin{array}{c}
\displaystyle
\frac1{2\varkappa_d}d(d-1)\left[2a^{d-2}{a''}+(d-3)a^{d-3}{a'}^2\right]=
\frac{\hbar
(d+1)(d-2)}{2^{d+3}\pi^{(d+1)/2}}\Gamma\left(\frac{3-d}{2}\right)
\left[\left(\frac{\rho^{(d-1)/2}}{a^{d-1}}\right)'a^{d-1}\right]'\
.
 \end{array}
 \label{12.10}
\end{equation}
Other Einstein equations can be obtained using the Bianchi
identities. A complete system of Einstein vacuum equations is
written in $D$--covariant form:
\begin{equation}
 \begin{array}{c}
\displaystyle R_{i(d)}^k-\frac12\delta_i^kR_{(d)}+
\\[5mm]
\displaystyle +
\frac{\varkappa_d\hbar(d+1)(d-2)(d-1)^{\frac{d-1}{2}}}{2^{2d+2}(d\pi)^{\frac{d+1}{2}}}
\Gamma\left(\frac{3-d}{2}\right)
\left[\left(R_{(d)}^{\frac{d-1}{2}}\right)^{;k}_{;i}-
\delta_i^k\left(R_{(d)}^{\frac{d-1}{2}}\right)^{;l}_{;l}-
\left(R_{i(d)}^k-\frac1{d+1}\delta_i^kR_{(d)}\right)R_{(d)}^{\frac{d-1}{2}}\right]=0\ .
 \end{array}
 \label{12.11}
\end{equation}
Equation (\ref{12.11}) are obtained by the variation of action
\begin{equation}
 \begin{array}{c}
\displaystyle S_{vac}=\int\sqrt{|g_{(d)}|}d^{\scriptscriptstyle D}x
\left[-\frac{1}{2\varkappa_d}R_{(d)}+
\frac{\hbar(d-2)(d-1)^{\frac{d-1}{2}}}{2^{2d+2}(d\pi)^{\frac{d+1}{2}}}
\Gamma\left(\frac{3-d}{2}\right)
R_{(d)}^{\frac{d+1}{2}}\right]\ .
 \end{array}
 \label{12.12}
\end{equation}
It is obvious from (\ref{12.11}), (\ref{12.12}) that the method of
dimensional regularization retains overall covariance of the
theory. Of course, quantum corrections, appearing in
(\ref{12.11}), satisfy the condition of conservation.

Renormalization and removal of regularization (limit $d\to 3$) are
held at the level of action. A parameter with the dimension of
length, which will eventually acquire the status of
renormalization scale, is contained within the theory. This
parameter, referred to as $L_g$, is appears in the
$D$--dimensional constant of gravity:
\begin{equation}
\displaystyle \varkappa_d=\varkappa\cdot L_g^{d-3}\ .
\label{12.13}
\end{equation}
The technique of removal the regularization assumes conservation
of dimensionality for those objects in which the limit operation
is performed. There are two such objects: the measure of
integration $d\mu$ and the density of the Lagrangian
$\mathcal{L}$. As can be seen from (\ref{12.12}), (\ref{12.13}),
the first (Einstein) term of the action is written down as
\begin{equation}
\begin{array}{c}
\displaystyle S^{(1)}_{vac}=\int\mathcal{L}^{(1)}d\mu\ ,
\\[5mm]
\displaystyle \mathcal{L}^{(1)}=-\frac{1}{2\varkappa}R_{(d)}\ ,\qquad d\mu=\sqrt{|g_{(d)}|}L_g^{4-D}d^Dx\ ,
 \end{array}
 \label{12.14}
\end{equation}
where $D$--dimensional objects $\mathcal{L}$ and $d\mu$ have the
same dimensions as the corresponding 4--dimensional objects. In
this sector of the theory the limit transition is trivial:
$R_{(d)}\to R$, $d\mu\to \sqrt{-g}d^4x$. In the sector of quantum
corrections to the Einstein theory, we introduce the same measure
and obtain the density of the Lagrangian:
\begin{equation}
\displaystyle
\mathcal{L}^{(2)}=
\frac{\hbar L_g^{d-3}(d-2)(d-1)^{\frac{d-1}{2}}}{2^{2d+2}(d\pi)^{\frac{d+1}{2}}}
\Gamma\left(\frac{3-d}{2}\right)
R_{(d)}^{\frac{d+1}{2}}
\label{12.15}
\end{equation}
It is necessary to emphasize that  {\it the operations of
renormalizations and removal of regularization have to be
mathematically well--defined and generally--covariant.} The
condition of mathematical certainty assumes that {\it the
renormalization is conducted before the lifting of
regularization.} At the same time, the general--covariance of the
procedure is automatically fulfilled if the counter--terms imposed
in the Lagrangian are the $D$--dimensional invariants. Note also
that if the mathematical value is finite at $d=3$, then the above
formulated conditions do not prevent the expansion of this
quantity in a Taylor series over the parameter $(3-d)/2$. In
particular, we can write:
\begin{equation}
 \begin{array}{c}
\displaystyle L_g^{d-3}R_{(d)}^{\frac{d+1}{2}}\equiv R_{(d)}^2\left(L_g^2R_{(d)}\right)^{\frac{d-3}{2}}=
R_{(d)}^2\left(1+\frac{3-d}{2}\ln\frac{\mu_g^2}{R_{(d)}}+...\right)\ ,
 \end{array}
 \label{12.16}
\end{equation}
where $\mu_g=1/L_g$; ellipsis designate the terms which do not
contribute to the final result. The substitution (\ref{12.16}) in
(\ref{12.15}) provides:
\begin{equation}
\displaystyle
\mathcal{L}^{(2)}=\frac{\hbar (d-2)(d-1)^{\frac{d-1}{2}}}{2^{2d+2}(d\pi)^{\frac{d+1}{2}}}
\Gamma\left(\frac{3-d}{2}\right)R_{(d)}^2+
\frac{\hbar (d-2)(d-1)^{\frac{d-1}{2}}}{2^{2d+2}(d\pi)^{\frac{d+1}{2}}}
\Gamma\left(\frac{5-d}{2}\right)R_{(d)}^2\ln\frac{\mu_g^2}{R_{(d)}}+...\ .
\label{12.17}
\end{equation}
According to (\ref{12.17}), the source Lagrangian of the theory
requires a $D$--invariant counter--term, which removes the
contribution proportional to the diverging $\Gamma$--function:
\begin{equation}
\displaystyle
\mathcal{L}_0^{(2)}=-\frac{\hbar (d-2)(d-1)^{\frac{d-1}{2}}}{2^{2d+2}(d\pi)^{\frac{d+1}{2}}}
\Gamma\left(\frac{3-d}{2}\right)R_{(d)}^2+\frac{\hbar}{4f^2}R_{(d)}^2\ .
\label{12.18}
\end{equation}
In (\ref{12.18}), there is a new finite constant of the theory of
gravity  $1/f^2$. The removal of the regularization in the
renormalized Lagrangian is conducted by the regular transition:
\begin{equation}
\begin{array}{c}
\displaystyle
\mathcal{L}_{ren}=\lim_{d\to 3}\left(\mathcal{L}^{(1)}+\mathcal{L}^{(2)}+
\mathcal{L}^{(2)}_0\right)=
\\[5mm]
\displaystyle
=-\frac{1}{2\varkappa}R+\frac{\hbar}{4f^2}R^2+\frac{\hbar}{1152\pi^2}
R^2\ln\frac{\mu_g^2}{R}=-\frac{1}{2\varkappa}R+\frac{\hbar}{1152\pi^2}
R^2\ln\frac{\lambda_g^2}{R}\ ,
\end{array}
\label{12.19}
\end{equation}
where
\[
\displaystyle \lambda_g^2=\mu_g^2\exp\frac{288\pi^2}{f^2}
\]
is the renorm--invariant scale. There is a heuristic argument
allowing to use the obtained expression: quantum corrections in
the Lagrangian (\ ref (12.19)) dominate over all other neglected
terms of the asymptotic series over The logarithmic parameter
$\ln(\lambda_g^2/R)\gg 1$.

The renormalized Einstein vacuum equations with quantum
corrections obtained from the Lagrangian (\ref{12.19}) are as
follows:
\begin{equation}
 \begin{array}{c}
\displaystyle R_i^k-\frac12\delta_i^kR+
\\[5mm]
\displaystyle +
\frac{\varkappa\hbar}{288\pi^2}\left\{
 \left[R\ln\frac{\lambda_g^2}{R}\right]_{;i}^{;k}-
 \delta_i^k\left[R\ln\frac{\lambda_g^2}{R}\right]_{;l}^{;l}-
 \left(RR_i^k-\frac14\delta_i^kR^2\right)\ln\frac{\lambda_g^2}{R}-
 \frac18\delta_i^kR^2\right\}=0\ .
 \end{array}
 \label{12.20}
\end{equation}
Note that exactly the same equations are obtained from
$D$--dimensional equations (\ref{12.11}), provided that the
operations are performed in the same sequence: {\it first a
renormalization with the introduction of $D$--covariant
counter--terms is conducted, and then a limit transition to the
physical dimension is performed.}

\subsection{Intrinsic Contradiction of Theory with no Ghosts:
Impossibility of One-Loop Renormalization }\label{nonren}

We are still discussing a formal model --- self--consistent theory
of gravitons with no ghosts. In the previous section it was shown
that the renormalization of divergences, that inevitably arise in
this model, requires the imposition of an additional term
quadratic in the curvature in the Lagrangian. It is now necessary
to draw attention to two mathematical facts: (i) the need for a
modification of Einstein theory is caused by quantum effects
contained in the Lifshitz operator equation (\ref{12.3}); (ii) the
original Lagrangian and operator equations of the modified theory
have the form:
\begin{equation}
 \begin{array}{c}
\displaystyle \mathcal{L}=\int\left(-\frac{1}{2\varkappa}\hat R+
\frac{\hbar}{4f^2}\hat R^2\right)\sqrt{-\hat g}d^4x\ ,
 \end{array}
 \label{12.21}
\end{equation}
\begin{equation}
 \begin{array}{c}
\displaystyle
\sqrt{-\hat g}\left[\frac{1}{\varkappa}\left(\hat R_i^k-
\frac12\delta_i^k\hat R\right)+\frac{\hbar}{f^2}\left(
 \hat D_i\hat D^k\hat R-
 \delta_i^k\hat D_l\hat D^l\hat R-
 \hat R\hat R_i^k+\frac14\delta_i^k\hat R^2\right)\right]=0\ ,
 \end{array}
 \label{12.22}
\end{equation}
where $\hat D_i$ is a covariant derivative in a space with the
operator metric $\hat g_{ik}$. It is quite obvious that these
facts contradict each other: the quantum effects in the Lifshitz
equation lead to a theoretical model that contradicts the Lifshitz
equation. Let us demonstrate that the contradiction is a direct
consequence of the non--renormalizability of the model
(\ref{12.21}) off the graviton mass shell.

Equations (\ref{12.22}), after their linearization describe
quantized waves of two types --- tensor and scalar. It makes sense
to discuss the problem of the scalar modes only in the event that
at least preliminary criteria for consistency of modified theory
will be obtained. Therefore, first of all, we should reveal
properties of the tensor modes. Here is an expression for the
Lagrangian of a system consisting of self--consistent cosmological
field and tensor gravitons:
\begin{equation}
\begin{array}{c}
 \displaystyle
 S=\int dtNa^3\left\{-\frac{3}{\varkappa N^2}\frac{{\dot a}^2}{a^2}+
 \frac{9\hbar}{f^2N^4}\left(\frac{\ddot a}{a}-
 \frac{\dot N}{N}\frac{\dot a}{a}+\frac{{\dot a}^2}{a^2}\right)^2+\right.
  \\[5mm]
 \displaystyle \left.
 +\frac{1}{8}\left[\frac{1}{\varkappa}+
 \frac{6\hbar}{N^2f^2}\left(\frac{\ddot a}{a}-
 \frac{\dot N}{N}\frac{\dot a}{a}+
 \frac{{\dot a}^2}{a^2}\right)\right]
 \sum_{{{\bf k}\sigma}}\left(\frac{1}{N^2}
 \frac{d{\hat \psi}_{{\bf k}\sigma}^+}{dt}\frac{d{\hat \psi}_{{\bf k}\sigma}}{dt}-
 \frac{k^2}{a^2}\hat \psi_{{\bf k}\sigma}^+\hat \psi_{{\bf k}\sigma}\right)\right\}\ .
 \end{array}
\label{12.23}
 \end{equation}
The equation for gravitons is produced either by the linearization
of the equation (\ref{12.22}), or from (\ref{12.23}) by the
variation procedure:
\begin{equation}
\begin{array}{c}
 \displaystyle \left(1-\frac{\varkappa\hbar}{f^2}R\right)\left(\hat \psi''_{{\bf k}\sigma}
 +2\frac{a'}{a}\hat \psi'_{{\bf k}\sigma} +k^2\hat \psi_{{\bf k}\sigma}\right)-
 \frac{\varkappa\hbar}{f^2}R'\hat \psi'_{{\bf k}\sigma}=0\ .
 \end{array}
\label{12.24}
 \end{equation}
Please note that the last term in (\ref{12.24}) makes it
impossible to retain the Lifshitz equation. After the
transformation
\[
 \displaystyle \hat \psi_{{\bf k}\sigma}=a^{-1}
 \left(1-\varkappa\hbar R/f^2\right)^{-1/2}\hat\varphi_{{\bf k}\sigma}
\]
equation (\ref{12.24}) has a form
\begin{equation}
\begin{array}{c}
 \displaystyle \hat\varphi''_{{\bf k}\sigma} +
 \left[k^2+a^2\left(\frac{R}{6}+P\right)\right]\hat\varphi_{{\bf k}\sigma}=0\ .
 \end{array}
\label{12.25}
 \end{equation}
In (\ref{12.25}), the deviation from the Lifshitz equation is
manifested in the effective frequency of gravitons --- the latter
contains an additional function of curvature's derivatives
 \begin{equation}
\begin{array}{c}
\displaystyle P=-\frac12\left[\ln\left(1-\varkappa\hbar R/f^2\right)\right]^{;l}_{;l}-
\frac14
\left[\ln\left(1-\varkappa\hbar R/f^2\right)\right]^{;l}
\left[\ln\left(1-\varkappa\hbar R/f^2\right)\right]_{;l}\ .
 \end{array}
\label{12.26}
 \end{equation}
When calculating quantum corrections to the macroscopic equations,
the modification of the effective frequency leads to additional
divergences. Averaged vacuum equations (\ref{12.22}), after their
polynomial expansion in powers of curvature, look as follows
(finite logarithmic corrections are omitted):
\begin{equation}
 \begin{array}{c}
\displaystyle R_i^k-\frac12\delta_i^kR+
\varkappa\hbar\left(\frac{\Gamma(\varepsilon)}{288\pi^2}+\frac{1}{f_0^2}\right)\left(
 R_{;i}^{;k}-
 \delta_i^kR_{;l}^{;l}-
 RR_i^k+\frac14\delta_i^kR^2\right)+
 \\[5mm]
\displaystyle +(\varkappa\hbar)^2\frac{\Gamma(\varepsilon)}{48\pi^2f_0^2}\left(
 R_{\; ;l;i}^{;l\; \;\; ;k}-
 \delta_i^kR_{\; ;l;m}^{;l\; \;\; ;m}-
 R_i^kR_{;l}^{;l}+\frac12R_{;i}R^{;k}-\frac14\delta_i^kR_{;l}R^{;l}\right)=0\ .
 \end{array}
 \label{12.27}
\end{equation}
Here $\Gamma(\varepsilon)\sim 1/\varepsilon$ is a divergent
$\Gamma$--function obtained by dimensional regularization;
$1/f_0^2$ is a seed constant of a theory with quadratic invariant.
The complete quantum Lagrangian corresponding to equations
(\ref{12.27}) has the form:
\begin{equation}
 \begin{array}{c}
\displaystyle \mathcal{L}=\int\left[-\frac{1}{2\varkappa}\hat R+
\hbar\left(\frac{\Gamma(\varepsilon)}{1152\pi^2}+\frac{1}{4f_0^2}\right)\hat R^2+
\varkappa\hbar^2\frac{\Gamma(\varepsilon)}{192\pi^2f_0^2}
\hat R^{;l}\hat R_{;l}\right]\sqrt{-\hat g}d^4x\ .
 \end{array}
 \label{12.28}
\end{equation}
Renormalization of the second term in (\ref{12.28}) is performed
by selecting the seed constant:
\[
\displaystyle \frac{1}{f_0^2}=-\frac{\Gamma(\varepsilon)}{288\pi^2}+\frac{1}{f^2}\ .
\]
However, a divergent coefficient forms before the third term. To
overcome this divergence, it is necessary to introduce a new
seeding "fundamental"\ constant of the modified theory of gravity
$1/h_0^2$ with a renormalization rule:
\[
\displaystyle \displaystyle \frac{1}{h_0^2}=\frac{\Gamma(\varepsilon)}{48\pi^2}
\left(\frac{\Gamma(\varepsilon)}{288\pi^2}-\frac{1}{f^2}\right)+\frac{1}{h^2}\ .
\]

Further actions are obvious and pointless: Lifshitz equation is
the subject of the next modification; quantum corrections generate
another new divergence; to renormalize the new divergence a new
theory of gravity is introduced, etc. The only conclusion to be
drawn from this procedure is that {\it based on the criteria of
quantum field theory, the one--loop self--consistent theory of
gravitons in the isotropic Universe, and not possessing the
property of one--loop finiteness outside of  mass shell, does not
exists as a mathematical model. In such a theory it is impossible
to quantitatively analyze any physical effect.} The theory of
gravitons without ghosts is non--renormalizable even in the
one--loop approximation. It is also important to stress that the
correct alternative to a {\it non--renormalizable} theory is only
a {\it finite} theory with the graviton--ghost compensation of
divergences.

In the future, from our perspective, the method of regularization
and renormalization in general will be excluded from the arsenal
of quantum theory of gravity, including one from the theory of
one--loop quantum effects involving matter fields. Correct
alternatives to existing methods of analysis of these effects to
be found in extended supergravities, finite at least in one--loop
approximation.

The situation prevailing in the scientific literature is a
paradoxical one. On the one hand, inadequate nature of the
regularization and renormalization methods in the quantum theory
of gravity should be obvious from the latest development trends in
the theories of supergravity and superstrings. On the other hand,
however, in all works we know on cosmological applications of
one--loop quantum gravity theoretical models are used, which,
according to the criteria of quantum field theory, do not exist.
We cannot comment on the specific results obtained in these models
by the reasons clear from the content of this Section. Once again
we should emphasize that {\it the self--consistent theory of
gravitons, if it exists as a theoretical model, must be finite
outside the mass shell of gravitons.} Effects arising in the
finite theory are described in the main text of this work.

\subsection{Dimensional Transmutation of Finite Theory}\label{an-d}

One--loop finiteness is the central important feature of the
quantum theory of gravity, defining its mathematical structure and
the algorithms of concrete calculations. Nevertheless, one should
bear in mind that in the calculations we are dealing with, {\it
compensation} of divergences comes from the graviton and ghost
sectors. Therefore, the question arises: how are the structure and
the results of the theory changed when an intermediate
regularization of divergences is applied? Comparison of results
obtained by different methods of computation will allow us to
judge the stability or volatility of the results of one--loop
quantum gravity with respect to the intermediate regularization.

We will continue to use the method of dimensional regularization.
The passage to the limit $D=4$ will be applied only in the final
expressions. General considerations show that after the transition
to $D=4$ in the expressions for the observed values, some terms
may appear, whose mathematical structure has no analogues in the
equations of the original theory. In such cases we are forced to
talk about quantum anomalies that have arisen as a result of
dimensional transmutation. If there are no such terms, the theory
is stable with respect to the dimensional transmutation, and has
no anomalies. Here we show that one--loop quantum gravity (without
the matter fields) not only finite, but is void of anomalies.

Let us turn to the convolutions of Einstein equations, assuming
that the latter are written exactly in the form in which they were
obtained from the path integral:
 \begin{equation}
\displaystyle
 -\frac1{2\varkappa_{(d)}}(d-1)\sqrt{|g_{(d)}|}R_{(d)}=\sqrt{|g_{(d)}|}
 \langle\Psi|T_{(d)}|\Psi\rangle\ .
 \label{12.29}
\end{equation}
In the study of dimensional transmutation, one can not divide the
left and right hand sides of equation (\ref{12.29}) by the common
multiplier $\sqrt{|g_{(d)}|}$, because the limit $d\to 3$ on the
left hand side is a regular one, but on the right hand side the
same limit applies under conditions of compensation of
divergences. In carrying out the operations in the right hand
side, one must take into account the total dependence on the
parameter $d$.

The metric of $d$--dimensional isotropic Universe and its scalar
curvature are represented in the form of (\ref{12.1}). A theory of
gravitons formulated in that spacetime has undergone a preliminary
investigation. In doing so, it was established that (i) the ghost
sector still consists of one complex Grassmann field, satisfying
the Klein-Gordon-Fock equation in the space with a metric
(\ref{12.1}). Thus, the number of internal degrees of freedom of
the ghost field is $w_{gh}=2$; (ii) the number of $d-$tensor gauge
invariant degrees of freedom (the number of transverse
polarizations) of the graviton field does not match the number of
internal degrees of freedom of ghosts:
  \begin{equation}
  \displaystyle \sum_\sigma\equiv w_g=\frac12\left(d^2-d-2\right)=
  2+\frac12\left(d+2\right)\left(d-3\right)\ ;
   \label{12.30}
\end{equation}
(iii) the equations for the wave functions of gravitons and ghosts
have the same form:
 \begin{equation}
  \displaystyle \psi''_{{\bf k}\sigma}+(d-1)\frac{a'}{a}\psi'_{{\bf k}\sigma}+
  k^2\psi_{{\bf k}\sigma}=0\ ,
  \qquad \theta''_{{\bf k}}+(d-1)\frac{a'}{a}\theta'_{{\bf k}}+
  k^2\theta_{{\bf k}}=0\ ;
   \label{12.31}
\end{equation}
(iv) the trace of the energy-momentum tensor of gravitons and
ghosts in (\ref{12.29}) is as follows:
 \begin{equation}
  \displaystyle \sqrt{|g_{(d)}|}
 \langle\Psi|T_{(d)}|\Psi\rangle=-\frac{d-1}{16\varkappa_{(d)}}\left\{
 a^{d-1}\sum_{{\bf k}}\left[\sum_\sigma\langle \Psi_g|
   \psi^+_{{\bf k}\sigma}\psi_{{\bf k}\sigma}|\Psi_g\rangle-
  2\langle \Psi_{gh}|\bar\theta_{{\bf k}}\theta_{{\bf k}}|\Psi_{gh}\rangle\right]'\right\}'\
  .
   \label{12.32}
\end{equation}

Each of two terms on the right hand side (\ref{12.32}) contains
the contribution of zero fluctuations of quantum fields, which,
because (\ref{12.31}), give rise to divergences of the same type.
With $d=3$ these divergences exactly cancel out each other out,
because $w_g=w_{gh}=2$ --- this comes from the nature of one--loop
finiteness of the quantum theory of gravitons in the isotropic
Universe. However, at $d=3-2\varepsilon$ there is no exact
compensation: as seen from (\ref{12.30}),
$w_g-w_{gh}=(d+2)(d-3)/2\simeq 5\varepsilon$. The difference in
the numbers of internal degrees of freedom is multiplied by a
divergent coefficient proportional to $1/\varepsilon$. As a
result, in the limit of $d\to 3$, a conformal anomaly arises,
caused by the spontaneous dimensional transmutation in the theory
of gravitons and ghosts.

The goal is to calculate the conformal anomaly as a functional of
the spacetime metric. Clearly, this requires solutions of operator
equations (\ref{12.32}) in the form functionals of the same--type.
As we know, when examining the zero fluctuations by the methods of
regularization and renormalization, it is enough to have the
solutions in the form of asymptotic expansion in powers of
curvature. Virtually all the computations coincide with the
computations already described in Section \ref{ren} down to the
formula (\ref{12.10}) --- with the only difference being that it
is now necessary to take into account the additive  contributions
of gravitons and ghosts. The trace of the energy--momentum tensor
(\ref{12.32}) is divided into two terms:
 \begin{equation}
  \displaystyle \sqrt{|g_{(d)}|}
 \langle\Psi|T_{(d)}|\Psi\rangle=\sqrt{|g_{(d)}|}
 T^{(vac)}_{(d)}+\sqrt{|g_{(d)}|}
 \langle\Psi|T^{(exc)}_{(d)}|\Psi\rangle
  .
   \label{12.33}
\end{equation}
The first of these contributors describes zero vacuum
fluctuations, "deformed" by the self--consistent gravitational
field, with the graviton--ghost compensation (which is incomplete
with $d\ne 3$) taken into account:
 \begin{equation}
  \displaystyle \sqrt{|g_{(d)}|}
 T^{(vac)}_{(d)}=-\frac{\hbar}{16}(d-1)(d+2)(d-3)\left\{
 \sum_{{\bf k}}a^{d-1}
 \left[\frac{1}{\omega_ka^{d-1}}\left(1+\sum_{s=1}^{\infty}(-1)^s\hat J_k^s\cdot 1\right)\right]'\right\}'\
  .
   \label{12.34}
\end{equation}
Note that the relation (\ref{12.30}) was used to obtain
(\ref{12.34}). The second term in (\ref{12.33}) is a trace of
energy--momentum tensor of the excitations. Because of the obvious
limitations on the range of excitations, the term has no divergent
integrals, so the limit is taken in the regular manner. In doing
so, of course, an expression is obtained that exactly coincides
with the trace of the energy--momentum tensor in 4--dimensional
finite theory not containing the contribution of zero
fluctuations:
 \begin{equation}
  \displaystyle \sqrt{-g}
 \langle\Psi|T^{(exc)}|\Psi\rangle= -\frac1{8\varkappa}\sqrt{-g}D\ ,
   \label{12.35}
\end{equation}
where $D$ is a function appearing in the 4--dimensional finite
BBGKY chain (\ref{5.7}) --- (\ref{5.9}).

Note that previous calculations in this section and further
operations with the expression (\ref{12.34}) are formally
accurate. As for the (\ref{4.5}), we should notice that all
integrals of expression
\[
\displaystyle \frac{1}{\omega_k}\sum_{s=1}^{\infty}(-1)^s\hat J_k^s\cdot 1
\]
have regular limit at $d\to 3$, so the multiplication of these
integrals by the factor $d-3$, would provide zero. The effect of
dimensional transmutation is contained entirely in the expression
 \begin{equation}
  \displaystyle \sqrt{-g}
 T^{(vac)}=-\lim_{n\to
 3}\frac{\hbar}{16}(n-1)(n+2)(n-3)\left[a^{d-1}\left( \sum_{{\bf k}}\frac{1}{\omega_ka^{d-1}}
 \right)'\right]'\ .
   \label{12.36}
\end{equation}
The integral in $d-$dimensional space of wave numbers is defined
in (\ref{12.9}). Substituting this expression into (\ref{12.36}),
we get the final result:
\begin{equation}
 \begin{array}{c}
  \displaystyle \sqrt{-g} T^{(vac)}=-\lim_{d\to 3}\frac{\hbar(d+2)\Gamma\left[(5-d)/2\right]}
 {2^{d+2}\pi^{(d+1)/2}}\left(\frac{d-1}{4d}\right)^{(d-1)/2}
  \left[a^{d-1}\left(R_{(d)}^{(d-1)/2}\right)'\right]'=
  -\frac{5\hbar}{192\pi^2}\sqrt{-g}R^{;l}_{;l}\  .
  \end{array}
   \label{12.37}
\end{equation}
From (\ref{12.37}), using the Bianchi identities, one can
reconstruct all components of the anomalous energy--momentum
tensor of gravitons and ghosts:
 \begin{equation}
\displaystyle  T_i^{k(vac)}=\frac{5\hbar}{576\pi^2}\left(
 R_{;i}^{;k}-\delta_i^kR_{;l}^{;l}-RR_i^k+\frac14\delta_i^kR^2\right)
 \ .
 \label{12.38}
\end{equation}
Expression (\ref{12.38}) describes a concrete quantum effects,
which is the deformation of the spectrum of zero fluctuations of
gravitons and ghosts in the self--consistent classical
gravitational field. In obtaining (\ref{12.38}), the
renormalization with the removal of divergences by introducing the
counter--terms was not used. The existence of this effect does not
require modification of the original quantum Lagrangian.

In an isotropic space with the metric (\ref{12.1}), two types of
conformal anomalies of the energy--momentum tensor of quantum
fields were repeatedly discussed:
\begin{equation}
\displaystyle
T^{k(1)}_{iJ}=\frac{C_J^{(1)}\hbar}{2880\pi^2}\left(
 R_{;i}^{;k}-\delta_i^kR_{;l}^{;l}-RR_i^k+\frac14\delta_i^kR^2\right)
 \ ,
 \label{12.39}
\end{equation}
\begin{equation}
\displaystyle
T^{k(2)}_{iJ}=\frac{C_J^{(2)}\hbar}{2880\pi^2}\left[R_{i}^lR_l^{k}-\frac23RR_i^k-
\frac12\delta_i^k\left(R_{l}^mR_m^{l}-\frac12R^2\right)\right]\ .
 \label{12.40}
\end{equation}
(Numerical coefficients $C_J^{(1)}$ and $C_J^{(2)}$ for the fields
for the fields with spin $J=0,\ 1/2,\ 1$ are given, for example,
in the monograph \cite{52}.) The anomaly of second type has exited
an interest, in particular because the quantum corrections
(\ref{12.40}), added to the Einstein equation, are capable of
providing a self--consistent De Sitter solution in the vicinity of
the Plank's values of the curvature \cite{53}. As we can see from
(\ref{12.38}), the dimensional transmutation of the finite theory
of gravitons and ghosts generates only the first type anomaly in
the energy--momentum tensor. The anomaly of the second type simply
does not arise under the dimensional transmutation. It means that
in the one--loop quantum gravity (without matter fields) the
effects of zero fluctuations are not able to sustain an
inflationary expansion with the constant parameter of inflation.
{\it In the finite theory, the De Sitter solution can be formed
only by the graviton--ghost condensate.}

The last question is this: does the emergence of anomalies
(\ref{12.38}) change of the mathematical structure of the system
of equations of the theory consisting of the BBGKY chain
(\ref{5.7}) --- (\ref{5.9}) and the macroscopic Einstein's
equations (\ref{5.10})? The answer is negative. The fact is that
the BBGKY chain (\ref{5.7}) --- (\ref{5.9}) is form--invariant
with respect to the additive transformation of the moments of the
spectral function:
 \begin{equation}
\displaystyle W_n\to W_n + \frac{b\hbar(-1)^n}{2\pi^2a^{2n+2}}\hat
K_\rho^{n+1}\cdot 1\ ,\qquad n=0,\ 1,\ 2,\ ...,\ \infty\ ,
 \label{12.41}
\end{equation}
where $b$ is an arbitrary numerical parameter; $\hat K_\rho$ is an
integral--differential operator, functionally dependent on
$\rho=-a''/a$.  The operator is defined as follows:
 \begin{equation}
 \begin{array}{c}
\displaystyle
\hat K_\rho\cdot f=\frac14\left(\frac{d^2}{d\eta^2}f+2\rho f+
2\int\limits_{-\infty}^\eta d\eta\rho\frac{d}{d\eta}f\right)\ ,
\\[5mm]
\displaystyle \hat K_\rho\cdot 1=\frac12\rho\ , \qquad
\hat K^2_\rho\cdot 1=\frac38\left(\rho^2+\frac13\rho''\right)\ ,
\\[5mm]
\displaystyle \hat K^3_\rho\cdot 1=
\frac{5}{16}\left(\rho^3+\frac12\rho'^2+\rho\rho''+\frac{1}{10}\rho''''\right)\ ,\qquad  ...\ ,
\qquad  \hat K^{n+1}_\rho\cdot 1=\hat K_\rho\cdot\left(\hat K_\rho^n\cdot 1\right)\ .
 \end{array}
 \label{12.42}
\end{equation}
The transformation (\ref{12.41}) can be seen as a trace of
renorm--group symmetry of a theory with the quadratic invariant
--- the divergences have disappeared, but their existence in the
graviton and ghost sectors was separately recorded in the symmetry
properties of the BBGKY chain. This transformation leaves the
chain (\ref{5.7}) --- (\ref{5.9}) unchanged, but the same
transformation $D$ and $W_1$ with a coefficient $b=-5$ eliminates
the conformal anomaly from Einstein equations (\ref{5.10}). Thus,
a simple renaming of the moments of the spectral function returns
the system of equations to its former view. The result means that
{\it the one--loop quantum gravity not only finite, but
anomaly--free as well}.

The anomaly (\ref{12.38}) is contained within the BBGKY chain
along with all other quantum gravitational effects. In the general
solution, the relative role of the anomaly is exclusively governed
by the initial conditions. Of course, the particular solution of
the chain (i.e. a solution containing only the anomaly) also has a
meaning. In that particular solution, the value $b=5$, obtained by
the method of dimensional transmutation, is not a special one. In
examining particular solutions of this type in the vacuum
energy--momentum tensor (\ref{12.38}), the coefficient $5$ should
be replaced by an arbitrary constant $b_0$. Different values of
the constant $b_0$ physically correspond to different versions of
incomplete compensation of energy of zero--fluctuations of
gravitons and ghosts. The solution of such vacuum equations at
$b_0<0$ ("physics of the scalarons"\ ) was discussed in \cite{53}.

\end{document}